%% file: tr_main.tex
\documentclass[a4paper,11pt]{article}
\usepackage[utf8]{inputenc}
\usepackage[T1]{fontenc} 
\usepackage[english]{babel} 
\usepackage{graphicx}
\usepackage{tabularx}
\usepackage{longtable}
\usepackage{xltabular}
\usepackage{amsmath}
\usepackage{booktabs}
\usepackage{amsfonts}
\usepackage{amssymb}
\usepackage{multirow}
\usepackage{physics}
\usepackage{tikz}
\usetikzlibrary{snakes}
\usepackage[hyphens]{url}
\usepackage{caption}
\usepackage{subcaption}
\usepackage{fancyhdr}
\usepackage{pdflscape}
\usepackage[capposition=top]{floatrow}
\usepackage[colorlinks=true, citecolor=blue]{hyperref}
\usepackage{makecell}

\usepackage[backend=biber,natbib=true,style=authoryear]{biblatex}
\title{Who are the arbitrageurs? Empirical evidence from Bitcoin traders in the Mt.\ Gox exchange platform}
\author{
Pietro Saggese%
\thanks{
IMT School for Advanced Studies Lucca, Italy, and AIT, Austria. 
\texttt{pietro.saggese@imtlucca.it}
}
\and
Alessandro Belmonte%
\thanks{
IMT School for Advanced Studies Lucca, Italy, and CAGE, UK. 
\texttt{alessandro.belmonte@imtlucca.it}
}
\and
Nicola Dimitri%
\thanks{
Università di Siena, Italy. 
\texttt{nicola.dimitri@unisi.it}
}
\and 
Angelo Facchini%
\thanks{
IMT School for Advanced Studies Lucca, Italy. 
\texttt{angelo.facchini@imtlucca.it}
}
\and 
Rainer B\"{o}hme%
\thanks{
Department of Computer Science, Universit\"at Innsbruck, Austria.
\texttt{rainer.boehme@uibk.ac.at}
}
}
\date{\today}

\def\sym#1{\ifmmode^{#1}\else\(^{#1}\)\fi}

\fancypagestyle{headings}{  \fancyhf{}
}

\newcommand{\myfrac}[3][0pt]{\genfrac{}{}{}{}{\raisebox{#1}{$#2$}}{\raisebox{-#1}{$#3$}}}

\newcommand{\Fee}{\ensuremath{\mathrm{Fee}}}

\newcommand{\BitcoinFee}{\ensuremath{\mathrm{BitcoinFee}}}
\newcommand{\Bitcoins}{\ensuremath{\mathrm{Bitcoins}}}
\newcommand{\MoneyFee}{\ensuremath{\mathrm{MoneyFee}}}
\newcommand{\Money}{\ensuremath{\mathrm{Money}}}
\newcommand{\Intercept}{\ensuremath{\mathrm{Intercept}}}
\newcommand{\LogVol}{\ensuremath{\mathrm{LogVol}}}
\newcommand{\LinVol}{\ensuremath{\mathrm{LinVol}}}
\newcommand{\VolSmall}{\ensuremath{\mathrm{VolSmall}}}
\newcommand{\VolBig}{\ensuremath{\mathrm{VolBig}}}
\newcommand{\Tzero}{\ensuremath{\mathrm{T_{0}}}}
\newcommand{\Tone}{\ensuremath{\mathrm{T_{1}}}}
\newcommand{\Tholid}{\ensuremath{\mathrm{T_{holid}}}}
\newcommand{\Date}{\ensuremath{\mathrm{Date}}}
\newcommand{\AnomalousDays}{\ensuremath{\mathrm{AnomalousDays}}}
\newcommand{\EarlyAdopters}{\ensuremath{\mathrm{EarlyAdopters}}}
\newcommand{\AnomalousUsers}{\ensuremath{\mathrm{AnomalousUsers}}}
\newcommand{\Matchers}{\ensuremath{\mathrm{Matchers}}}
\newcommand{\Markus}{\ensuremath{\mathrm{Markus}}}
\newcommand{\Willy}{\ensuremath{\mathrm{Willy}}}

\newcommand{\Y}{\ensuremath{\mathrm{Y}}}

\newcommand{\Spread}{\ensuremath{\mathrm{Spread}}}

\makeatletter
\let\@fnsymbol\@arabic
\makeatother

\makeatletter
\def\@BTrule[#1]{%
  \ifx\longtable\undefined
    \let\@BTswitch\@BTnormal
  \else\ifx\hline\LT@hline
    \nobreak
    \let\@BTswitch\@BLTrule
  \else
     \let\@BTswitch\@BTnormal
  \fi\fi
  \global\@thisrulewidth=#1\relax
  \ifnum\@thisruleclass=\tw@\vskip\@aboverulesep\else
  \ifnum\@lastruleclass=\z@\vskip\@aboverulesep\else
  \ifnum\@lastruleclass=\@ne\vskip\doublerulesep\fi\fi\fi
  \@BTswitch}
\makeatother

\begin{document}

\maketitle
\thispagestyle{headings} \pagestyle{plain}
\addtocounter{footnote}{5}


\small{
\input{tr_abstract.tex}
}

\normalsize

\input{tr_introduction.tex}

\input{tr_background.tex}

\input{tr_identification.tex}

\input{tr_results.tex}

\input{tr_discussion.tex}

\subsection*{Acknowledgments}
{\small
We would like to thank
JT Hamrick, 
Marie Vasek,
the participants at the University of Innsbruck's research seminars in spring 2019,
Simone Giansante,
Andrea Canidio,
Roberto Renò,
Jiahua Xu,
Antonio Scala,
and the anonymous reviewers of WEIS 2019
for their comments on earlier versions of this work.
}

\printbibliography{}

\newpage

\input{tr_tables.tex}

\appendix

\renewcommand\thefigure{\thesection.\arabic{figure}} 
\renewcommand{\thetable}{\thesection.\arabic{table}}

\newpage
\input{tr_appendix_deltas.tex}

\newpage
\input{tr_appendix_fees.tex}

\newpage
\input{tr_appendix_learning.tex}

\newpage
\input{tr_appendix_individual_mkts.tex}

\newpage
\input{tr_appendix_datacleaning.tex}

\newpage
\input{tr_appendix_supplem.tex}

\end{document}

%% file: tr_abstract.tex
\begin{abstract}

We mine the leaked history of trades on Mt.\ Gox, the dominant Bitcoin exchange from 2011 to early 2014, to detect the triangular arbitrage activity conducted within the platform.
The availability of user identifiers per trade allows us to focus on the historical record of 440 investors, detected as arbitrageurs, and consequently to describe their trading behavior. 
%
We begin by showing that a considerable difference appears between arbitrageurs when indicators of their expertise are taken into account.
In particular, we distinguish between those who conducted arbitrage in a single or in multiple markets: using this element as a proxy for trade ability, we find that arbitrage actions performed by expert users are on average non profitable when transaction costs are accounted for, while skilled investors conduct arbitrage at a positive and statistically significant premium. 
%
Next, we show that specific trading strategies, such as splitting orders or conducting arbitrage non aggressively, are further indicators of expertise that increase the profitability of arbitrage.
Most importantly, we exploit within-user (across hours and markets) variation and document that expert users make profits on arbitrage by reacting quickly to plausible exogenous variations on the official exchange rates.
We present further evidence that such differences are chiefly due to a better ability of the latter in incorporating information, both on the transactions costs and on the exchange rates volatility, eventually resulting in a better timing choice at small time scale intervals. 
%
Our results support the hypothesis that arbitrageurs are few and sophisticated users.  
\end{abstract}

\textbf{JEL Classifications}: C58, D53, G11, G40

\textbf{Keywords}: Arbitrage, Bitcoin, Cryptocurrency Exchanges, Financial Econometrics, Behavior of Economic Agents

\newpage

%% file: tr_introduction.tex
\section{Introduction} \label{tr:introduction}

Arbitrage, the simultaneous purchase and sale of the same asset in two different markets for a risk-free profit, is a key concept in economics and finance. The concept is so important because the \emph{absence} of arbitrage opportunities is a necessary condition for market equilibrium (\cite{harrison1979}). Intuitively, whenever an arbitrage opportunity emerges, some arbitrageur will exploit it until the mechanism of supply and demand has eliminated the price difference. This `law of one price' makes the no-arbitrage principle a powerful solution concept in financial theory. It is a common foundation of the  Capital Asset Pricing Model (\cite{sharpe1964,lintner1965,mossin1966}), the arbitrage pricing theory (\cite{ross1976arbitrage}), the theory of option pricing (\cite{merton1973theory,black1973}), the efficient market hypothesis (\cite{fama1970}), and many other theories. 

In practice, arbitrage is never risk-free. Since purchase and sale are not executed in an atomic\footnote{Incidentally, this may change with decentralized exchanges on programmable cryptocurrency platforms.} transaction across markets, the arbitrageur bears the risk of incomplete execution or concurrent price changes. Moreover, the asset traded in both markets may not be exactly the same, and there may be political risk premia if the markets operate in different jurisdictions \citep{aliber1973}. 
These risks, in addition to other certain transaction costs, impose a lower bound on the price difference needed for profitable arbitrage. 
The orthodox economic response, in line with the efficient market hypothesis, is to imagine that many small arbitrageurs each take an infinitesimally small portion of the risk (and hence profit).
However, \citet{shleifer1997} challenge exactly this view in their landmark work on practical arbitrage in financial markets:
\begin{quote}
``[A]rbitrage is conducted by relatively few professional, highly specialized investors who combine their knowledge with resources of outside investors to take large positions.'' (p.~36)
\end{quote}
The authors support this claim by referring to the bounded rationality of many investors, 
``millions of little traders are typically not the ones who have the knowledge and information to engage in arbitrage.'' (p.~36)
While this is plausible and likely cross-checked by expert market participants, the evidence remains anecdotal. 
Most surprising to us is the fact that 20 years after these statements were published, we still could not find any academic paper that provides an empirical answer to the question `Who are the arbitrageurs?'\footnote{Of course, we simply may have missed the relevant source and are grateful for pointers from our readers.}, despite the topic being perceived as an issue of compelling relevance in current research~\citep{gromb2010limits}.
Indeed, in the best case scenario, a comprehensive answer to this question may not only reconcile economic theory with the reality on financial markets, but also refine the assumptions about arbitrageurs in theoretical studies that derive optimal trading strategies in the presence of arbitrageurs~\citep[and the works cited therein]{moallemi2012}, and ultimately contribute to understanding why persistent limitations to arbitrage emerge.

In this work we seek to provide a partial answer from a very singular market, namely the exchange market between convertible currency and cryptocurrency in the early years of Bitcoin. The choice of market and time is opportunistic. We mine a leaked dataset of individual and identified trades from Mt.\ Gox, a meanwhile defunct exchange that enjoyed a dominant market position before its collapse in early 2014. Focusing on the triangular arbitrage activity, we first quantify its magnitude within the Mt.\ Gox platform, and then we introduce proxy measures for the trading ability of the investors, to show that the expert users conduct more profitable actions, and are more responsive to exogenous shocks on the official exchange rate: a statistically significant relation exists between the investors' expertise and the profitability of arbitrage. Our results confirm the anecdotal evidence that arbitrageurs are few and sophisticated users.

The contributions of this work are manifold.
First, by investigating the dynamics of the Mt.\ Gox platform, we enrich the existing literature that focuses on the economic role of cryptocurrency exchanges in the Bitcoin ecosystem \citep[e.g.,][]{moore2018toit,griffin2020bitcoin}. 
Our analyses provide several insights on the Mt.\ Gox market structure and on its internal trading mechanisms, contributing in this sense more broadly to a better understanding of the microstructure of the cryptocurrency markets. 
Besides this, we provide a detailed estimation of the explicit transaction costs borne by the Mt.\ Gox users when trading within the platform (while authors previously mainly focused on the analysis of the fees paid for transactions on the Bitcoin network; see e.g. \cite{moser2015trends,dimitri2019transaction,easley2019mining}).
Further, to preprocess the leaked dataset we propose our own deduplication method, that grounds upon (and partially improves) existing methods introduced in the literature \citep{scaillet2017high,feder2018impact,gandal2018price}, and whose validity is documented extensively by comparisons with external sources of information.


Second, the availability of user-specific labels enables the design of novel methodologies to investigate arbitrage, based on the analysis of identified sequences of trades. 
Indeed, previous empirical research focuses mainly on assessing the validity of consolidated theories on arbitrage \citep[e.g.,][]{roll1980empirical,malkiel2003efficient}, or on detecting violations and anomalies arising in contrast with them \citep[e.g.,][]{lamont2003anomalies}; however, most of these works are based on aggregate information and do not tackle the topic from a bottom up approach.
Other studies do exploit transaction-level data, but lacking the user identifiers, and thus focus primarily on inferring information on the market \citep{lee1991inferring} or on the type of trader, rather than on the individuals features \citep{lee2000inferring}.
The work by \citet{wang2021cyclic}, closest to our investigation both in intentions and method, shows clearly that the cryptocurrency ecosystem unveiled unprecedented opportunities in quantitative finance analysis: the authors identify the cyclic arbitrages executed in Decentralized Exchanges (DEXs) within the Ethereum ecosystem. However, the pseudonimity guaranteed by the Ethereum protocol limits the possibility to trace users exactly.
In our framework, instead, each leg of all trades is labeled by a user identifier, allowing to trace the sequences of actions executed by the same user and to identify potential links among them.
Since in Mt.\ Gox the bitcoins could be traded within the same exchange against different fiat currencies, our method detects \textit{exactly} the actions of arbitrage as pairs of legs satisfying the textbook properties of arbitrage, that is, two legs (from different trades) executed by the same user, in different currency markets, and within a reasonably small neighborhood of time and volume. 
It is precisely in this sense that we intend the term `exactly': the matching legs are identified based on time delay and volume differences, which are known, and conditional on being executed by the same user, another condition which is known.
This crucial aspect gives us the unprecedented possibility to identify \textit{completely} the triangular arbitrage activity exploited within the platform, not as an aggregate but at the level of the individual arbitrage actions executed.
To the best of our knowledge, this work represents the first attempt to use this approach in the literature; thus, our work provides a contribution also by proposing new algorithms for quantitative finance who exploit micro level information and whose aim is to investigate the individual trading strategies.
Noteworthy, this method is also easily extendable to other contexts requiring a pattern identification through user identifiers.



Third, and most importantly, by identifying and describing the characteristics of the individual investors who conduct arbitrage in the presence of risk, our method provides empirical evidence that helps answering broader questions on the nature of arbitrage.
%
The closest papers 
we could find that investigate the user behavior in financial markets using user level information chiefly focus on the analysis of the individual traders choices, in order to profile investors in terms of risk attitude \citep[e.g.,][]{clark2005individual,kourtidis2011investors,de2019personality}.
However, no previous study investigated exhaustively the research questions that motivated our work.
This paper thus aims at filling, at least partially, an existing gap between the theoretical description of the arbitrage activity and the practical evidence from real markets.
Once the arbitrage activity is identified, we analyse and describe empirically, rather than anecdotally or using aggregate data, the individual trading patterns: we show that only a restricted group of users (N = 440) performs at least one arbitrage action, and an even smaller group of sophisticated users is responsible for the vast majority of trades; the users in such subset are mostly active in multiple currency markets, rather than in a single one, they conduct complex strategies (i.e., metaorders), and do not trade aggressively on the market, i.e. they prefer limit to market orders.
Most of all, though some of their actions yield losses, the arbitrage activity of such sophisticated investors is on average profitable, and they show greater ability in responding to variations (plausibly exogenous) of the official exchange rate, even after controlling for user fixed effects.
Crucially, we acknowledge instead that the arbitrage activity attributable to the non sophisticated users constitutes a small fraction of the total arbitrage activity identified, and on average it is non profitable when transaction costs are included.
This raises a relevant conceptual consideration: while a strict definition of arbitrage foresees the execution of such actions only for profit, arbitrageurs accept that they can incur losses, i.e., expected payoffs positive with probability smaller than one, even under fundamentally riskless conditions \citep{kondor2009risk}. 
However, in this context the non expert users systematically incur losses when the transaction costs are taken into account, and even if some of them are actually executing actions on average positive, thus conducting correctly arbitrage according to the textbook definition, their contribution is negligible in terms of numbers of trades. We thus conclude that arbitrage is carried mostly by few and specialized users.

We stress again the importance of the user identifiers for our analysis. Not only the methodology is based on algorithms that exploit them to capture pairs of legs likely forming arbitrage actions; their role is essential also as the devised identification strategy is based on the inclusion of user fixed effects that allow us to rule out the possibility that the differences in profitability arising when exchange rate variations take place might be influenced by unobservable user-specific ability.

In summary, in this paper we identify exactly the triangular arbitrage activity and the users who conduct it, and we provide the first form of empirical evidence to a relevant issue which up to now was only acknowledged anecdotally: \textit{who are the arbitrageurs?} The answer to this question is relevant in the literature and could help reducing the gap that currently exists between theoretical and practical arbitrage.
We point out that our findings may be contingent to this specific market, and that it is not trivial to extend such results to other financial markets, both for the uniqueness of the exchange platform itself and for the specific features that characterize the cryptocurrency market as a whole. Nonetheless, we believe that the general picture we describe reflects well also the features of practical arbitrage in a traditional financial market.

The remainder of the paper is organized as follows. In Section~\ref{tr:background} we provide some context on the cryptocurrency exchanges, and especially on the Mt.\ Gox platform. In Section~\ref{tr:identification} we describe the methodology implemented to identify the triangular arbitrage activity and to measure its profitability; we then report descriptive statistics on the users' trading patterns heterogeneity. In Section~\ref{tr:preliminary} we further explore such differences and provide preliminary evidence of the relationship between the user trade ability, captured by user-specific proxies of expertise, and the profitability of arbitrage. Next, in Section~\ref{tr:results} we investigate the responsiveness of arbitrageurs to plausible exogenous variations of the official exchange rates, and show that the expert traders react better to sharp variations (in terms of profitability of arbitrage). In Section~\ref{tr:discussion} we discuss our findings. Appendices~\ref{tr:appendix_deltas},~\ref{tr:appendix_fees},~\ref{tr:appendix_learning} provide additional robustness checks, 
while Appendix~\ref{tr:appendix_individual} outlines a description of the arbitrage activity on the major currency market (i.e., the EUR/USD market) taken individually. Appendices~\ref{tr:appendix_data_cleaning} and~\ref{tr:appendix_supplem} respectively report additional information on the cleaning procedures of the Mt.\ Gox leaked dataset, and supplemental figures and tables.

%% file: tr_background.tex
\section{Background} 
\label{tr:background}

\subsection{Arbitrage in cryptocurrency markets}

Arbitrage is a founding and unifying concept in financial economics\footnote{
In the words of \citet{ross1978simple}, ``it is surprising how much of what is central to modern finance is based solely on the arbitrage principles embodied in the basic valuation theorem'' (p.~471).
Many of the most influential works in finance ground on and contributed to the development and formalization of this fundamental concept. E.g., \cite{modigliani1958cost,black1973,merton1973theory,ross1976arbitrage,ross1978simple,cox1976valuation,cox1979option,harrison1979,harrison1981martingales,kreps1981arbitrage,delbaen1994general}.
}. 
It conveys a simple yet powerful message, with vast implications on the theory of asset pricing: assuming that agents are rational and prefer more to less, the no-arbitrage condition states that a portfolio yielding non negative payoffs must have a non negative cost. Excluding the case with zero prices, this statement is equivalent to saying that a unique vector of strictly positive state prices, defining unambiguously the price of all assets, indeed exists. When this condition is not met, then there must be a portfolio yielding non negative payoffs without requiring any initial investment - that is, a profitable and riskless investment, which is precisely the definition of arbitrage opportunity (\cite{ross1976risk,ross1978simple,varian1987arbitrage,dybvig1989arbitrage}).

A more recent stream of research reports however evidence that market anomalies arise in the form of persistent mispricings (\cite{harris1986price,froot1999stock,lamont2003can}), despite the presence of sophisticated investors seeking to exploit such opportunities: 
arbitrage entails risks and costs and de facto it is often limited.
E.g., in \citet{de1990noise} irrational noise traders can operate on an optimistic or pessimistic bias, creating a risk that prevents arbitrageurs from exploiting mispricings and causing price divergences from fundamental values even in the absence of fundamental risk; \citet{shleifer1997} show that when arbitrageurs operate in a principal agency relationship, principals may mistakenly evaluate the arbitrageurs on the basis of their performance and refuse to provide the required capital, especially in the extreme circumstances where additional capital would be most needed.
\citet{merton1987simple} points out that markets are not frictionless and investors need to account for additional costs (in capital requirements, entry, strategy implementation, non-istantaneous diffusion of information)\footnote{A survey covering the literature on the limits of arbitrage can be found in \citet{barberis2003survey} and in \citet{gromb2010limits}.}.

The nascent cryptocurrency market represents a promising area of research in this sense, as its innovative and unconventional market design provides an alternative standpoint to assess the validity of established methods and theory from traditional finance.
Bitcoin, the most prominent cryptocurrency in terms of market capitalization\footnote{Around 850 billion \$ at the time of writing (according to Coinmarketcap.com \url{https://bit.ly/3iXhZnj}). This, and all the following links, were accessed on 17 August 2021.}, is a decentralized system which records transfers between parties denominated in bitcoin (units of cryptocurrency) in a public ledger. By contrast, exchanges are centralized entities in the Bitcoin ecosystem that provide interfaces to conventional payment systems by allowing its users to trade units of cryptocurrency against fiat money (\cite{BCEM2015-JEP}). Typical exchanges manage and match orders in a private limit order book, and update their customers' account balances in cryptocurrency or fiat money when trades are executed. As a result, exchanges are the place where price formation occurs. Trades on exchanges are kept in a private ledger and have no effect on the public ledger unless users withdraw cryptocurrency from the exchange to a wallet under their own control (most exchanges publish aggregate information about prices and volume). 

We describe two strategies to exploit arbitrage that are relevant for the cryptocurrency markets and for our context: arbitrage across exchanges and within the same exchange. 
In the first case, intuitively, an arbitrageur observes a price difference between two exchanges, buys a bitcoin at the cheaper place, then transfers it to the more expensive place, where the bitcoin is sold for a profit. The transfer of bitcoins between exchanges would be observable in the public ledger and could be associated in time with differences in published prices, thereby generating evidence for arbitrage. However, bitcoin transactions are too slow and (at times) too costly for this strategy. Instead, arbitrageurs must maintain a stock of both bitcoins and fiat money in accounts at each exchange in order to react quickly to price differences. The funds can be balanced at a lower frequency and not necessarily correlated with observable price differences. Therefore, while arbitrage opportunities are measurable from published data, there is no way to identify arbitrage transactions or arbitrageurs from the public ledger, and in this regard cryptocurrency exchanges do not offer researchers any advantage over foreign exchange markets. This fact was already mentioned in the first attempt to study arbitrage in Bitcoin (\cite{petrov2013}). Interestingly, this term paper documents that market participants actively explored arbitrage opportunities in spring 2013\footnote{See also, e.g., \url{https://bit.ly/2GXhSer} and \url{https://bit.ly/2FwHNJk} for anecdotal evidence.} by pointing to two websites that track suitable price differences (see Figure~\ref{fig:cointhink}), and one open source software trading bot that exploits arbitrage opportunities. In the simplest case, this strategy of arbitrage requires an investor to open accounts at two exchanges; consequently, to identify arbitrage activity, information from the private ledgers of both the two exchanges involved is needed.

Besides this form of two-point arbitrage, the second relevant strategy is triangular arbitrage: most of the cryptocurrency exchanges offer the possibility to trade bitcoins (or other cryptoassets) against more than one fiat currency. 
%
%
Using bitcoin as a vehicle currency, investors can compare the implied relative price of traditional currencies to the official exchange rate and look for the presence of mispricings. 
We restrict our analysis to the second strategy, as this form of arbitrage has the advantage that information required to detect the arbitrage activity is entirely contained in the private ledger of a single exchange.


\subsection{Relation to Prior Work}

\begin{figure}[htbp]

	\begin{center}
	\begin{tikzpicture}[>=stealth,scale = 0.9,x=16mm,y=16mm]

	
	\draw [->] (10.8,0) coordinate (origin) --(20.5,0);
		\foreach \t in {11,12,13,14,15,16,17,18,19,20}
			\draw (\t,0)--++(0,-4pt)++(.5,0) node [below] {20\t};
	
	\draw [->] (origin)-- node [rotate=90,above=9mm]  {Bitcoin price in USD} ++(0,5.5);
		\foreach \y in {0,1,2,3,4}
		{	
			\draw (origin)++(0,1)++(0,\y) coordinate (base) --++(-4pt,0) node [left] {$10^\y$};
			\foreach \tick in {-0.699,-0.5229,-0.3979,-0.301,-0.2218,-0.1549,-0.0969,-0.0458}
				\draw (base)++(0,\tick) -- ++(-2pt,0);
		}
		
	
	\begin{scope}[blue,thick]
		\clip (origin) rectangle (20.185,5.5);
		\input{immagini/tikz/log-price}
	\end{scope}	
	
	
	\begin{scope}[y=9mm]
	\scriptsize

	\draw [|-|,brown,thick] (11.25,6.5) -- 
		node [below=2pt] {Mt.\ Gox dataset}
		(13.9100,6.5); 
	
	\draw [|-|,thick] (11.4385845,2.3) -- 
		node [below=1pt] {
		\cite{dong2015bitcoin}}
		(13.99,2.3); 
		
	\draw [|-|,thick] (11.75,5) -- 
		node [below=1pt] {
		\hspace{-1.3cm}
		\cite{smith2016analysis}}
		(14.08,5); 
		
	\draw [|-|,gray,thick] (12,3.2) -- 
		node [below=1pt] {
		\cite{badev2014bitcoin}}
		(14.534,3.2); 
	
	\draw [|-|,thick] (12,1.4) -- 
		node [below=1pt] {
		\cite{yu2018revisit}}
		(17.58,1.4); 
		
	\draw [|-|,thick] (13.085,0.6) -- 
		node [below=1pt] {
		\cite{pichl2017volatility}}
		(17.331,0.6); 
		
	\draw [|-|,gray,thick] (13.167,9.5) -- 
		node [below=2pt] {
		\cite{kruckeberg2020decentralized}}
		(18.333,9.5); 
	
	\draw [|-|,thick] (13.333,4.1) -- 
		node [below=1pt] {
		\cite{reynolds2018deviations}}
		(15.99,4.1); 
			
	\draw [|-|,thick] (14,6) -- 
		node [below=2pt] {\hspace{1.4cm}
		\cite{pieters2015bitcoin}} 
		(14.99,6); 
			
	\draw [|-|,gray,thick] (14.41666,7.5) -- 
		node [below=1pt] {\hspace{0.5cm}
		\cite{pieters2017financial}} 
		(15.5833,7.5); 
		
	\draw [|-|,thick] (14.32,5) -- 
		node [below=1pt] {
		\cite{nan2019bitcoin}}
		(18.74,5); 
		
	\draw [|-|,thick] (15,2.3) -- 
		node [below=1pt] {
		\cite{hirano2018analysis}}
		(18.13,2.3); 
			
	\draw [|-|,gray,thick] (15,8.5) -- 
		node [below=2pt] {\hspace{-1cm}
		\cite{bistarelli2019model}}
		(18.250,8.5); 
		
	\draw [|-|,gray,thick] (16,3.2) -- 
		node [below=2pt] {
		\hspace{-0.8cm} \cite{kroeger2017law}}
		(16.669,3.2); 
		
	\draw [|-|,thick] (17,6) -- 
		node [below=2pt] {\hspace{0.5cm}
		\cite{makarov2020trading}}
		(18.1666,6); 
	
	\draw [|-|,gray,thick] (17.918,7.8) -- 
		node [below=2pt] {\hspace{0.5cm}
		\cite{hattori2021relationship}}
		(18.670,7.8); 
		
	\draw [|-|,gray,thick] (17.917,1.4) -- 
		node [below=2pt] {\hspace{0.5cm}
		\cite{shynkevich2020bitcoin}}
		(18.750,1.4); 
		
	\draw [|-|,gray,thick] (18,3.8) -- 
		node [below=2pt] {
		\cite{lee2020pricing}}
		(19.250,3.8); 
		
	\draw [|-|,gray,thick] (18.250,6.85) -- 
		node [below=2pt]  {\hspace{1.2cm}
		\cite{hautsch2018limits}}
		(18.750,6.85); 
		
		
	\end{scope}
	\end{tikzpicture}
	\end{center}
	\caption{Related work in temporal and market context}
	\label{fig:timeline} 
	\floatfoot{\emph{Notes:}
	Bitcoin price in USD on log scale. 
	Black shaded works focus on triangular arbitrage. The gray ones study other forms of arbitrage within the Bitcoin ecosystem.}
\end{figure}
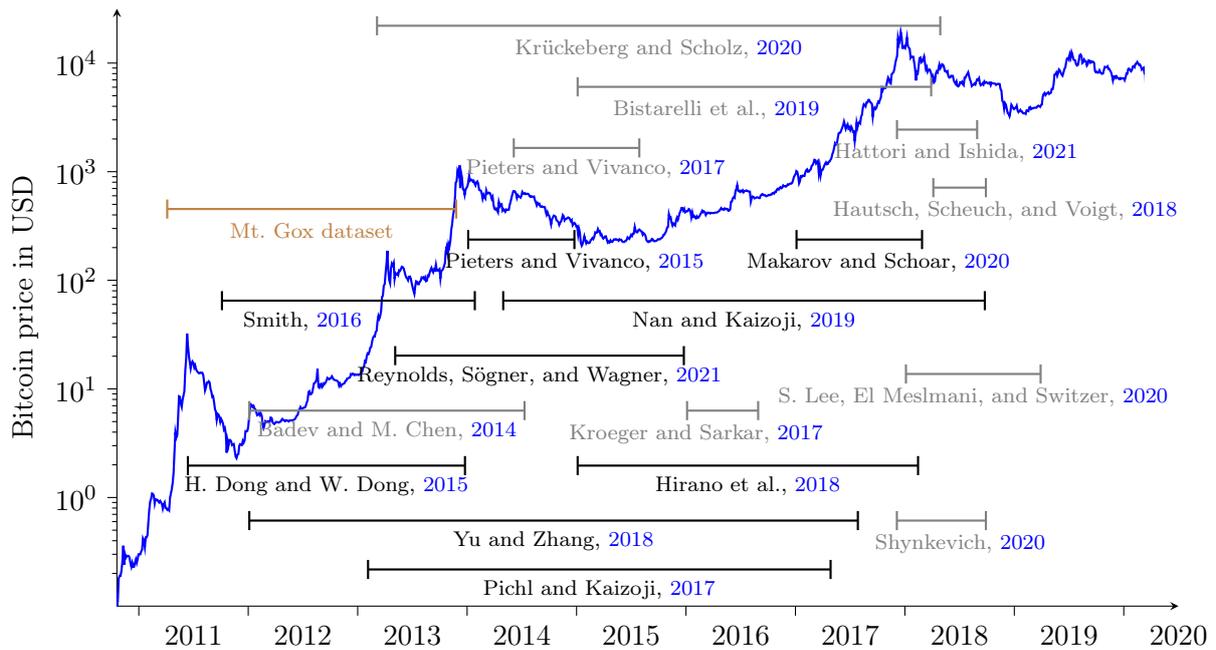

The body of prior works applying financial econometrics to time series data from cryptocurrency exchanges is vast and not easy to navigate\footnote{\cite[e.\,g.,][]{%
glaser2014bitcoin,
garcia2014digital, 
brandvold2015price,
yermack2015bitcoin,
cheung2015crypto,
ciaian2016economics,
athey2016bitcoin,
bouri2016return,
katsiampa2017volatility,
wheatley2018bitcoin,
dyhrberg2018investible}.}. 
When interpreting it, it is important to keep in mind that most studies use rather short and often non-overlapping samples. The maturing market for cryptocurrencies has exhibited extraordinary volatility as it transitioned through several epochs. Consequently, the time-series contain multiple structural breaks, which make it hard, if not impossible, to draw conclusions that generalize to the cryptocurrency as a whole. To illustrate this, we depict in Figure~\ref{fig:timeline} the sample periods of the works discussed in the following, along with the bitcoin price in USD on a logarithmic scale: this accounts best for the order of magnitude differences between epochs.
We restrict our review to focused studies of arbitrage opportunities and the exploration of market imperfections. The latter are relevant for our method because they inform us about transaction costs, which constrain the exploitability of apparent arbitrage opportunities.

Triangular arbitrage in the Bitcoin ecosystem has been widely investigated by several authors: the early years of Bitcoin trading, which are relevant for our work, are covered by \citet{dong2015bitcoin} and \citet{smith2016analysis}. The former test the bitcoin market for the presence of triangular arbitrage opportunities between the main cryptocurrency exchanges and the spot currency markets, and exploit price decomposition methods to study to what extent bitcoin behaves as a currency or as a financial asset. They find evidence of persistent price deviations, and conclude that the observed arbitrage stickiness can be explained only if users treat bitcoin as a financial asset rather than as a currency. The latter focuses on Mt.\ Gox aggregate data, and besides showing that shocks in that market do not affect rates in conventional venues, the author exploits the implied exchange rates in the market to conclude that bitcoin has gold-like (rather than currency-like) properties. They also suggest that there must exist a group of investors who enforce price convergence through their activity, given the degree of efficiency observed in the market.
\citet{pichl2017volatility} 
and \citet{reynolds2018deviations} 
study triangular arbitrage in a similar epoch, between 2013 and 2017: they both exploit data from  Bitcoincharts.com against daily currency spot rates, and find evidence of unexploited arbitrage opportunities. The former investigate triangular arbitrage without considering transaction costs and find unexploited opportunities especially in the Chinese market, while the latter report that in the bitcoin exchanges in exam the persistent mispricings arise only when Bitcoin is used as a vehicle currency, while no deviations from parity arise when considering the rate implied between traditional fiat currencies. 
\citet{pieters2015bitcoin} observe that, from January to December 2014, there are triangular arbitrage opportunities between different exchanges on the USD and EUR markets. As a noteworthy detail, they remark that the exchange rate for the Argentinian pesos (ARS) on Local Bitcoins, a peer-to-peer exchange, is closer to the ARS black market exchange rate than to the official one. 
Related to that, \citet{makarov2020trading} 
and \citet{yu2018revisit} 
report evidence of unexploited arbitrage opportunities and suggest that capital controls played an essential role in causing market frictions. %
\citet{hirano2018analysis} 
deploy machine learning techniques to test the Efficient Market Hypothesis in the Bitcoin market, showing that information inefficiencies arise in the form of triangular arbitrage opportunities, especially in minor currency markets.
Finally, \citet{nan2019bitcoin} 
analyse triangular arbitrage on the EUR/USD/BTC currency markets against the EUR/USD FX spot market using FX futures contracts to hedge risk, and show that arbitrage is a competitive strategy.

Beside this, also arbitrage on the same rate but across markets (two-point arbitrage) is well investigated. Empirical studies obtain similar conclusions on the presence of arbitrage opportunities and agree that price deviations, even in different time epochs, emerge and are persistent (\cite{badev2014bitcoin,pieters2017financial,kroeger2017law,kruckeberg2020decentralized}); evidence of mispricings across markets is found also in works proposing theoretical models on arbitrage which are fitted on empirical data (\cite{hautsch2018limits,bistarelli2019model}).
Other recent studies focus instead on the nascent futures market for Bitcoin: \citet{hattori2021relationship}, \citet{shynkevich2020bitcoin}, and \citet{lee2020pricing} obtain partially contrasting findings on the efficiency of such markets (specifically, the first two sources find evidence of efficiency in the markets; the disagreement with the tenor of most other literature can be attributed to the time window and the fact that the futures market operates in a single geographical area).
For completeness, we mention that
other studies investigate more broadly arbitrage in the cryptocurrency market (e.g., \cite{gandal2014competition,fischer2019statistical,leung2019constructing,crepelliere2020arbitrage,}).

In summary, based on heterogeneous methods and studying different periods in time with data of different frequency, the literature pretty consistently reports unexploited arbitrage opportunities in cryptocurrency markets. This does not imply that arbitrage does not happen, but might rather indicate that the costs and risks of arbitrageurs are under-estimated. Anecdotal evidence from forums, the existence of web-based arbitrage tools, and code repositories for trading bots indicate that arbitrage does happen (\cite{petrov2013}). 

\subsection{The Mt.\ Gox exchange and the leaked data set} 

All the above-reviewed studies have in common that they analyze aggregated price (and sometimes volume) time series. 
Our approach differs in that we use individual-level data from the internal ledger of a major exchange, Mt.\ Gox. The availability of such micro-information is of remarkable relevance: it allows to isolate trends specific to focused groups of traders, as e.g. the arbitrageurs.
 
Mt.\ Gox played a prominent role during the early years of Bitcoin: established in 2010, it was essentially 
the first cryptocurrency exchange and dominated market with around 80-90\% of total trading volume until late 2012. It was structured as an order-driven market based on a continuous two sided auction, and formally without any designated specialists.
The first competitors entered the market within a short time delay: Bitstamp and BTC-e in mid-2011, BTC China at the end of 2011. Other exchange services entered the market in the following months, and most of them were shut down after a brief period of activity \citep{moore2013beware,ceruleo2014bitcoin}.

Since the beginning of Spring 2013, a series of events gradually undermined the Mt.\ Gox credibility\footnote{
11 March 2013: Mt.\ Gox temporarily suspends bitcoin deposits after hard fork. \url{https://bit.ly/2GWPklj};\\ 
11 April 2013: Mt.\ Gox went down after unexpected increase in the trading activity. \url{https://bit.ly/3lEuSUW};
2 May 2013: Coinlab files a lawsuit against Mt.\ Gox \url{https://bit.ly/3duT755}; 
14 May 2013: the Department of Homeland Security issues a seizure warrant for an account owned by a Mt.\ Gox's U.S. subsidiary. \url{https://bit.ly/3k0yhwV}; 
5 August 2013: Mt.\ Gox announces significant losses due to crediting deposits. \url{https://bit.ly/317TA8n}.
}, 
and customers started experiencing delays when withdrawing fiat money\footnote{18 April 2013: users point out withdrawal delays. \url{https://bit.ly/3lHAyO7};
4 July 2013: Mt.\ Gox resumes the U.S. withdrawals halted on 20 June. \url{https://bit.ly/3doQxxH}, \url{https://bit.ly/3lEvPfY}. 
}. 
Consequently, the volume traded in Mt.\ Gox decreased significantly in the following months, and the bitcoins started to be traded at a large premium in Mt.\ Gox\footnote{See \url{https://bit.ly/2FshVy6}.}: in Spring 2013 the competitors of Mt.\ Gox had already gained a consistent share of the market, pushing Mt.\ Gox to just under 60\% in the summer of 2013 (see Figure~\ref{fig:history_exch} 
for the market share evolution over time.)  The exchange stopped withdraws at once on 7 February 2014, and filed for bankruptcy two weeks later. The former CEO was arrested after criminal charges of fraud and embezzlement in 2015, and found guilty of falsifying data in 2019. Exchange closure is a common phenomenon in the cryptocurrency space, and a source of concern for investors, as witnessed by the survival analysis of 80 exchanges in \citet{moore2018toit}.

\begin{figure}[tp]
    \centering
    \begin{subfigure}[t]{\textwidth}\vspace{-0.3cm}
        \hspace{-0.8cm}
        \includegraphics[width=1.05\textwidth]{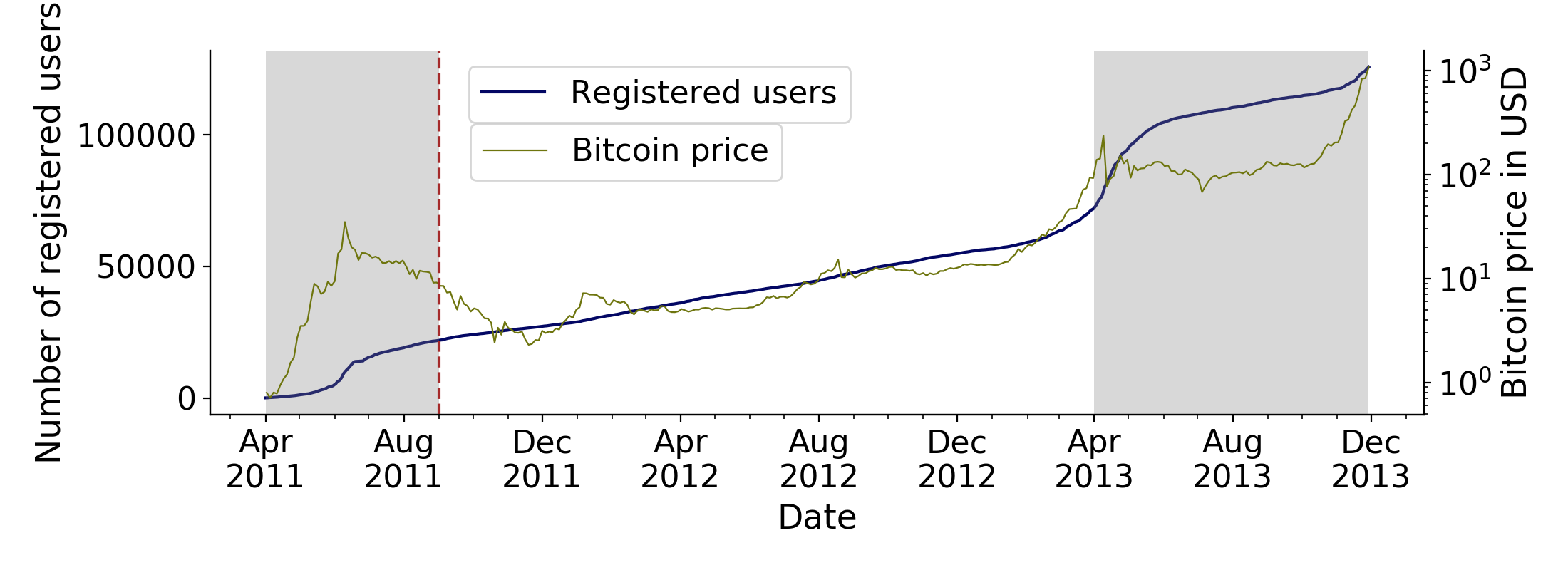}
		\caption{}
		\label{fig:TotUsrs}
    \end{subfigure}
    \begin{subfigure}[t]{\textwidth}\vspace{-0.3cm}
        \hspace{-0.5cm}
        \includegraphics[width=0.95\textwidth]{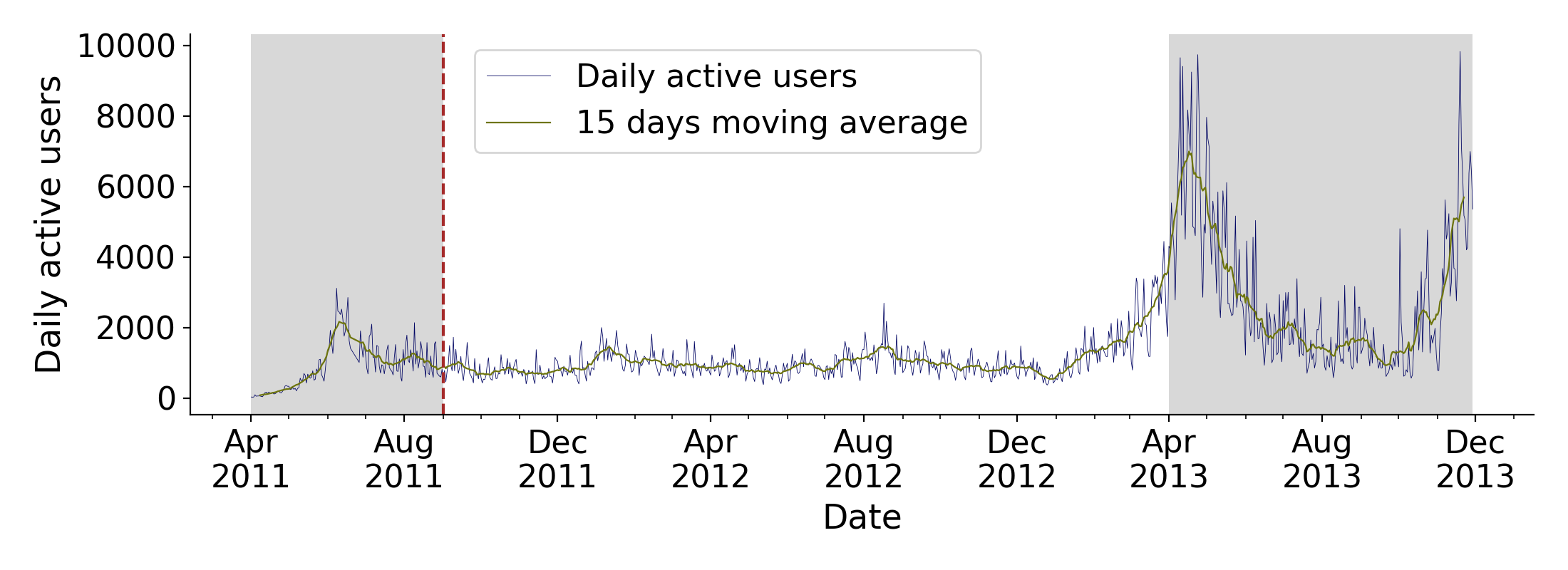}
		\caption{}
		\label{fig:ActUsrs} 
    \end{subfigure}
    \caption{Descriptive statistics of Mt.\ Gox users} \label{fig:usrs}
    \vspace{-0.5cm}
	\floatfoot{\emph{Notes:} Panel~(\subref{fig:TotUsrs}) shows the growth of registered users in relation to the bitcoin price (the latter is reported on a logarithmic scale). Panel~(\subref{fig:ActUsrs}) shows the number of daily active users. The brown vertical line indicates the date of introduction of the multi-currency trading; the gray shaded area represents the area excluded by the sample.}
\end{figure}

Our main dataset was leaked to the public in 2014 as a series of CSV files. 
They contain around 7.5 million trades executed in between 1 April 2011 and 30 November 2013; each trade is composed of two buy and sell legs, reported as separate rows; some information is trade-specific (trade identifier, date of execution, amount of bitcoins exchanged), while other variables are leg-specific (buy or sell type, user identifier, transaction costs paid); further information on the variables is provided in Appendix~\ref{tr:appendix_data_cleaning}. The vast majority (87.9\%) of trades are in USD, followed by EUR (7.7\%)\footnote{
From September 2011 on, users could trade bitcoins also in exchange for EUR, CAD, GBP, CHF, RUB, AUD, SEK, DKK, HKD, PLN, CNY, SGD, TBH, NZD, JPY (\url{https://bit.ly/314a5Cg}).}.
Figure~\ref{fig:usrs} visualizes selected indicators on how Mt.\ Gox's user base evolved over time, reaching a total of more than 125,000 at the end of 2013.
The plots outline intuitively that the peaks of interest towards the cryptoasset (Panel~\subref{fig:TotUsrs}) and of activity within in the market (Panel~\subref{fig:ActUsrs}) correspond with periods of exponential growth of the bitcoin price.

We highlight a relevant point for our analysis: whilst the dataset covers a longer period of time, we restrict our sample by excluding the trades executed after March 2013, due to the increasing difficulties reported by the users from April 2013 on in managing withdrawal operations and to the other facts documented above. The total users active from April 2011 to March 2013 are approximately 72,000, and the trades are around 5.5 million. 
Note also that the sample is further restricted, as the possibility to trade in currencies other than USD was introduced in September 2011. Thus the identified arbitrage actions fall in a time period ranging from September 2011 to March 2013.
Noteworthy, as Figure~\ref{fig:usrs} shows, the time window considered also coincides with an epoch of constant activity within the exchange platform, and with linear growth rate of the new registered users. 

The leaked data set has been widely analyzed by a number of prior works, which explore research questions that fall apart from arbitrage: e.g., the presence of metaorder executions (\cite{donier2015million}), unusual price jumps in the BTC/USD exchange rate (\cite{scaillet2017high}), the effects of distributed denial-of-service (DDoS) attacks on the trading activity (\cite{feder2018impact}), the impact of 
suspicious activity in the Mt.\ Gox exchange that likely engaged price manipulation (\cite{gandal2018price}, and \cite{chen2019market}, the latter trying to answer the same question through the lenses of network science), and herding behavior (\cite{haryanto2019disposition}).

In summary, the dataset is meanwhile well researched and largely accepted in the literature. This, along with our own comparisons to external aggregate information reported in Appendix~\ref{tr:appendix_data_cleaning}, support its validity and authenticity (and addresses the concerns that may arise to a cautious reader as the source of the dataset is anonymous and there is no way to utterly verify its correctness)\footnote{Comparisons refer to the data published by Bitcoincharts.com. In addition, as in \citet{scaillet2017high}, we match our data with an aggregated dataset published by Mt.\ Gox. The advantage is twofold: we further assess the validity of the leaked dataset, and we collect additional trade-specific information contained only in the latter dataset, i.e., which legs are aggressive (market) orders. Further details are discussed in Appendix~\ref{tr:appendix_data_cleaning}.
}.
Moreover, according to The Guardian\footnote{\url{https://bit.ly/2Iu77Rk}}, several members of the bitcoin community claimed to have found their own transactions in the dataset. Finally, certain facts established in the court case against the former CEO of Mt.\ Gox seem to plausibly explain patterns in the dataset\footnote{This statement is based on personal communication. The authors have not read the Japanese files.}.

We rest on the work of \citet{gandal2018price}, \citet{feder2018impact}, and \citet{scaillet2017high} to pre-process and clean the original leaked dataset. This stage chiefly consists in finding duplicate rows and in identifying (and correcting, when possible) misreported data.
The details for this procedure are described in Appendix~\ref{tr:appendix_data_cleaning}. 
It is worth noting that our aggregation technique differs from the above reference in that we aggregate the trades belonging to the same user occurring within the same second. 
Put it differently, we assume that such actions belong to the same executed order, in compliance with the operating principle of the Mt.\ Gox filling mechanism.\footnote{\url{https://bit.ly/33YCxaG}. 
} 
Order speed analyses on other cryptocurrency exchanges reveal that a one-second time scale is suitable to measure order execution delays\footnote{See \url{https://bit.ly/3iRJBdu}. Tests show that the time required to add limit orders and execute market orders are comparable, and that execution delays last on average from 10ms to 100/200ms, with skewed tails; a non-negligible amount of trades exists whose latency approaches the second level.}; traditional financial markets show much shorter latencies (see, e.g., \cite{budish2015high,hasbrouck2013low,kirilenko2015latency}).

Finally, a comment on research ethics and data privacy stands to reason. The internal ledger of Mt.\ Gox contains data that, in principle, can be linked to natural persons by matching it with other records. Moreover, the users appearing in this dataset had no expectation that their individual trades will become public. We therefore take utmost care that none of our analyses singles out users that have not been singled out in other work (which we always document with a proper citation). Specifically, we map all user identifiers in the dataset to a consecutive sequence of integers, preserving the order but not the numbers. Therefore, user identifiers in our figures should not be directly related to identifiers in the data source.  Moreover, we do not possess additional data which would allow linking records to natural persons, nor are we aware of a source where this data could be gathered. Therefore, we believe that the harm caused by our study is minimal while there are clear benefits in shedding light into a fundamental question in finance. Readers seeking to replicate the general methods described here are advised to make similar considerations before working with the data. The dataset can be provided upon request to ensure reproducibility.

%% file: immagini/tikz/log-price.tex
\draw
(10.636,-0.155)--
(10.644,-0.155)--
(10.652,-0.155)--
(10.66,-0.222)--
(10.668,-0.222)--
(10.677,-0.222)--
(10.685,-0.222)--
(10.693,-0.222)--
(10.701,-0.222)--
(10.71,0.176)--
(10.718,-0.222)--
(10.725999999999999,-0.222)--
(10.734,-0.222)--
(10.742,-0.222)--
(10.751,-0.222)--
(10.759,-0.222)--
(10.767,-0.222)--
(10.775,0.079)--
(10.784,0.0)--
(10.792,0.079)--
(10.8,0.0)--
(10.808,0.041)--
(10.816,0.255)--
(10.825,0.279)--
(10.833,0.279)--
(10.841,0.322)--
(10.849,0.398)--
(10.858,0.556)--
(10.866,0.38)--
(10.874,0.477)--
(10.882,0.447)--
(10.89,0.462)--
(10.899000000000001,0.462)--
(10.907,0.462)--
(10.915,0.447)--
(10.923,0.398)--
(10.932,0.362)--
(10.94,0.398)--
(10.948,0.362)--
(10.956,0.362)--
(10.964,0.398)--
(10.973,0.431)--
(10.981,0.398)--
(10.989,0.431)--
(10.997,0.477)--
(11.005,0.477)--
(11.014,0.477)--
(11.022,0.505)--
(11.03,0.519)--
(11.038,0.623)--
(11.047,0.602)--
(11.055,0.544)--
(11.063,0.633)--
(11.071,0.633)--
(11.079,0.653)--
(11.088,0.785)--
(11.096,0.892)--
(11.104,0.964)--
(11.112,0.987)--
(11.121,1.037)--
(11.129,1.033)--
(11.137,1.017)--
(11.145,0.944)--
(11.153,0.982)--
(11.162,0.982)--
(11.17,0.978)--
(11.178,0.964)--
(11.186,0.954)--
(11.195,0.968)--
(11.203,0.954)--
(11.211,0.94)--
(11.219,0.898)--
(11.227,0.934)--
(11.236,0.954)--
(11.244,0.903)--
(11.252,0.903)--
(11.26,0.886)--
(11.268,0.892)--
(11.277,0.881)--
(11.285,0.996)--
(11.293,1.037)--
(11.301,1.076)--
(11.31,1.14)--
(11.318,1.233)--
(11.326,1.425)--
(11.334,1.597)--
(11.342,1.551)--
(11.350999999999999,1.574)--
(11.359,1.733)--
(11.367,1.932)--
(11.375,1.928)--
(11.384,1.862)--
(11.392,1.833)--
(11.4,1.9)--
(11.408,1.939)--
(11.416,1.976)--
(11.425,2.178)--
(11.433,2.283)--
(11.441,2.51)--
(11.449,2.396)--
(11.458,2.301)--
(11.466,2.246)--
(11.474,2.187)--
(11.482,2.229)--
(11.49,2.253)--
(11.499,2.24)--
(11.507,2.197)--
(11.515,2.216)--
(11.523,2.181)--
(11.532,2.162)--
(11.54,2.154)--
(11.548,2.144)--
(11.556000000000001,2.144)--
(11.564,2.151)--
(11.573,2.149)--
(11.581,2.148)--
(11.589,2.126)--
(11.597,2.058)--
(11.605,1.985)--
(11.614,2.015)--
(11.622,2.057)--
(11.63,2.057)--
(11.638,2.07)--
(11.647,2.06)--
(11.655,2.0)--
(11.663,1.973)--
(11.671,1.927)--
(11.679,1.932)--
(11.688,1.877)--
(11.696,1.812)--
(11.704,1.792)--
(11.712,1.714)--
(11.721,1.77)--
(11.729,1.763)--
(11.737,1.74)--
(11.745,1.695)--
(11.753,1.722)--
(11.762,1.702)--
(11.77,1.68)--
(11.778,1.626)--
(11.786,1.64)--
(11.795,1.588)--
(11.803,1.413)--
(11.811,1.52)--
(11.818999999999999,1.477)--
(11.827,1.53)--
(11.836,1.529)--
(11.844,1.516)--
(11.852,1.483)--
(11.86,1.49)--
(11.868,1.493)--
(11.877,1.427)--
(11.885,1.375)--
(11.893,1.362)--
(11.901,1.394)--
(11.91,1.4)--
(11.918,1.477)--
(11.926,1.491)--
(11.934,1.486)--
(11.942,1.483)--
(11.951,1.522)--
(11.959,1.505)--
(11.967,1.526)--
(11.975,1.611)--
(11.984,1.609)--
(11.992,1.612)--
(12.0,1.65)--
(12.008,1.736)--
(12.016,1.82)--
(12.025,1.857)--
(12.033,1.851)--
(12.041,1.833)--
(12.049,1.843)--
(12.058,1.818)--
(12.066,1.811)--
(12.074,1.784)--
(12.082,1.757)--
(12.09,1.761)--
(12.099,1.777)--
(12.107,1.759)--
(12.115,1.774)--
(12.123,1.755)--
(12.132,1.676)--
(12.14,1.641)--
(12.148,1.666)--
(12.156,1.703)--
(12.164,1.697)--
(12.173,1.694)--
(12.181,1.702)--
(12.189,1.698)--
(12.197,1.697)--
(12.205,1.736)--
(12.214,1.732)--
(12.222,1.695)--
(12.23,1.682)--
(12.238,1.676)--
(12.247,1.686)--
(12.255,1.696)--
(12.263,1.695)--
(12.271,1.694)--
(12.279,1.69)--
(12.288,1.696)--
(12.296,1.698)--
(12.304,1.715)--
(12.312,1.724)--
(12.321,1.713)--
(12.329,1.707)--
(12.337,1.702)--
(12.345,1.712)--
(12.353,1.708)--
(12.362,1.708)--
(12.37,1.7)--
(12.378,1.708)--
(12.386,1.711)--
(12.395,1.71)--
(12.403,1.712)--
(12.411,1.713)--
(12.419,1.717)--
(12.427,1.722)--
(12.436,1.739)--
(12.444,1.754)--
(12.452,1.76)--
(12.46,1.797)--
(12.468,1.812)--
(12.477,1.833)--
(12.485,1.817)--
(12.493,1.818)--
(12.501,1.825)--
(12.51,1.827)--
(12.518,1.829)--
(12.526,1.844)--
(12.534,1.873)--
(12.542,1.892)--
(12.551,1.972)--
(12.559,1.979)--
(12.567,1.95)--
(12.575,1.952)--
(12.584,1.965)--
(12.592,2.005)--
(12.6,2.05)--
(12.608,2.055)--
(12.616,2.066)--
(12.625,2.09)--
(12.633,2.187)--
(12.641,2.019)--
(12.649000000000001,2.011)--
(12.658,2.041)--
(12.666,2.047)--
(12.674,2.015)--
(12.682,2.025)--
(12.69,2.052)--
(12.699,2.047)--
(12.707,2.06)--
(12.715,2.078)--
(12.723,2.101)--
(12.732,2.093)--
(12.74,2.088)--
(12.748,2.096)--
(12.756,2.1)--
(12.764,2.116)--
(12.773,2.1)--
(12.781,2.085)--
(12.789,2.083)--
(12.797,2.079)--
(12.805,2.077)--
(12.814,2.074)--
(12.822,2.063)--
(12.83,2.029)--
(12.838,2.048)--
(12.847,2.026)--
(12.855,2.04)--
(12.863,2.043)--
(12.871,2.048)--
(12.879,2.052)--
(12.888,2.073)--
(12.896,2.076)--
(12.904,2.096)--
(12.912,2.097)--
(12.921,2.102)--
(12.929,2.106)--
(12.937,2.136)--
(12.945,2.131)--
(12.953,2.14)--
(12.962,2.138)--
(12.97,2.127)--
(12.978,2.134)--
(12.986,2.129)--
(12.995,2.131)--
(13.003,2.133)--
(13.008,2.128)--
(13.016,2.132)--
(13.025,2.141)--
(13.033,2.156)--
(13.041,2.157)--
(13.049,2.191)--
(13.058,2.208)--
(13.066,2.255)--
(13.074,2.248)--
(13.082,2.296)--
(13.09,2.327)--
(13.099,2.322)--
(13.107,2.345)--
(13.115,2.38)--
(13.123,2.386)--
(13.132,2.438)--
(13.14,2.47)--
(13.148,2.485)--
(13.156,2.483)--
(13.164,2.523)--
(13.173,2.535)--
(13.181,2.623)--
(13.189,2.671)--
(13.197,2.647)--
(13.205,2.671)--
(13.214,2.717)--
(13.222,2.847)--
(13.23,2.857)--
(13.238,2.948)--
(13.247,2.966)--
(13.255,3.069)--
(13.263,3.153)--
(13.271,3.272)--
(13.279,3.097)--
(13.288,3.0)--
(13.296,2.973)--
(13.304,3.106)--
(13.312,3.154)--
(13.321,3.137)--
(13.329,3.16)--
(13.337,3.014)--
(13.345,3.064)--
(13.353,3.054)--
(13.362,3.063)--
(13.37,3.049)--
(13.378,3.091)--
(13.386,3.086)--
(13.395,3.103)--
(13.403,3.125)--
(13.411,3.122)--
(13.419,3.112)--
(13.427,3.084)--
(13.436,3.047)--
(13.444,3.026)--
(13.452,3.017)--
(13.46,3.0)--
(13.468,3.032)--
(13.477,3.035)--
(13.485,3.017)--
(13.493,2.976)--
(13.501,2.94)--
(13.51,2.903)--
(13.518,2.875)--
(13.526,2.944)--
(13.534,2.991)--
(13.542,2.987)--
(13.551,2.964)--
(13.559,2.964)--
(13.567,2.987)--
(13.575,2.994)--
(13.584,3.026)--
(13.592,3.021)--
(13.6,3.028)--
(13.608,3.012)--
(13.616,3.028)--
(13.625,2.992)--
(13.633,2.997)--
(13.641,3.043)--
(13.649000000000001,3.038)--
(13.658,3.071)--
(13.666,3.096)--
(13.674,3.114)--
(13.682,3.089)--
(13.69,3.073)--
(13.699,3.102)--
(13.707,3.094)--
(13.715,3.143)--
(13.723,3.09)--
(13.732,3.089)--
(13.74,3.096)--
(13.748,3.104)--
(13.756,3.016)--
(13.764,3.084)--
(13.773,3.094)--
(13.781,3.105)--
(13.789,3.131)--
(13.797,3.158)--
(13.805,3.27)--
(13.814,3.309)--
(13.822,3.254)--
(13.83,3.334)--
(13.838,3.307)--
(13.847,3.362)--
(13.855,3.464)--
(13.863,3.516)--
(13.871,3.594)--
(13.879,3.641)--
(13.888,3.738)--
(13.896,3.9)--
(13.904,3.911)--
(13.912,4.006)--
(13.921,3.98)--
(13.929,4.056)--
(13.937,3.849)--
(13.945,3.991)--
(13.953,3.947)--
(13.962,3.839)--
(13.97,3.836)--
(13.978,3.788)--
(13.986,3.832)--
(13.995,3.854)--
(14.003,3.865)--
(14.011,3.907)--
(14.019,3.962)--
(14.027,3.917)--
(14.036,3.926)--
(14.044,3.925)--
(14.052,3.909)--
(14.06,3.916)--
(14.068,3.892)--
(14.077,3.874)--
(14.085,3.903)--
(14.093,3.911)--
(14.101,3.893)--
(14.11,3.833)--
(14.118,3.829)--
(14.126,3.826)--
(14.134,3.799)--
(14.142,3.751)--
(14.151,3.786)--
(14.159,3.767)--
(14.167,3.754)--
(14.175,3.828)--
(14.184,3.801)--
(14.192,3.799)--
(14.2,3.807)--
(14.208,3.803)--
(14.216,3.786)--
(14.225,3.751)--
(14.233,3.768)--
(14.241,3.699)--
(14.249,3.66)--
(14.258,3.652)--
(14.266,3.661)--
(14.274000000000001,3.645)--
(14.282,3.627)--
(14.29,3.719)--
(14.299,3.683)--
(14.307,3.698)--
(14.315,3.701)--
(14.323,3.642)--
(14.332,3.655)--
(14.34,3.643)--
(14.348,3.631)--
(14.356,3.652)--
(14.364,3.646)--
(14.373,3.65)--
(14.381,3.65)--
(14.389,3.694)--
(14.397,3.723)--
(14.405,3.758)--
(14.414,3.793)--
(14.422,3.821)--
(14.43,3.821)--
(14.438,3.818)--
(14.447,3.796)--
(14.455,3.758)--
(14.463,3.784)--
(14.471,3.773)--
(14.479,3.77)--
(14.488,3.761)--
(14.496,3.778)--
(14.504,3.813)--
(14.512,3.8)--
(14.521,3.796)--
(14.529,3.802)--
(14.537,3.793)--
(14.545,3.796)--
(14.553,3.797)--
(14.562,3.793)--
(14.57,3.775)--
(14.578,3.768)--
(14.586,3.775)--
(14.595,3.77)--
(14.603,3.767)--
(14.611,3.771)--
(14.619,3.74)--
(14.627,3.719)--
(14.636,3.69)--
(14.644,3.712)--
(14.652,3.7)--
(14.66,3.705)--
(14.668,3.68)--
(14.677,3.676)--
(14.685,3.681)--
(14.693,3.675)--
(14.701,3.676)--
(14.71,3.673)--
(14.718,3.629)--
(14.725999999999999,3.603)--
(14.734,3.627)--
(14.742,3.602)--
(14.751,3.591)--
(14.759,3.555)--
(14.767,3.515)--
(14.775,3.561)--
(14.784,3.581)--
(14.792,3.596)--
(14.8,3.592)--
(14.808,3.586)--
(14.816,3.554)--
(14.825,3.545)--
(14.833,3.538)--
(14.841,3.511)--
(14.849,3.53)--
(14.858,3.54)--
(14.866,3.568)--
(14.874,3.601)--
(14.882,3.59)--
(14.89,3.552)--
(14.899000000000001,3.565)--
(14.907,3.565)--
(14.915,3.575)--
(14.923,3.579)--
(14.932,3.575)--
(14.94,3.56)--
(14.948,3.543)--
(14.956,3.554)--
(14.964,3.507)--
(14.973,3.521)--
(14.981,3.528)--
(14.989,3.518)--
(14.997,3.496)--
(15.005,3.498)--
(15.014,3.42)--
(15.022,3.474)--
(15.03,3.441)--
(15.038,3.346)--
(15.047,3.318)--
(15.055,3.334)--
(15.063,3.368)--
(15.071,3.405)--
(15.079,3.373)--
(15.088,3.341)--
(15.096,3.357)--
(15.104,3.347)--
(15.112,3.342)--
(15.121,3.347)--
(15.129,3.37)--
(15.137,3.376)--
(15.145,3.39)--
(15.153,3.38)--
(15.162,3.402)--
(15.17,3.441)--
(15.178,3.436)--
(15.186,3.439)--
(15.195,3.472)--
(15.203,3.451)--
(15.211,3.454)--
(15.219,3.418)--
(15.227,3.425)--
(15.236,3.397)--
(15.244,3.384)--
(15.252,3.392)--
(15.26,3.404)--
(15.268,3.405)--
(15.277,3.373)--
(15.285,3.348)--
(15.293,3.357)--
(15.301,3.348)--
(15.31,3.368)--
(15.318,3.355)--
(15.326,3.353)--
(15.334,3.364)--
(15.342,3.377)--
(15.350999999999999,3.375)--
(15.359,3.379)--
(15.367,3.373)--
(15.375,3.374)--
(15.384,3.364)--
(15.392,3.378)--
(15.4,3.372)--
(15.408,3.374)--
(15.416,3.359)--
(15.425,3.353)--
(15.433,3.352)--
(15.441,3.36)--
(15.449,3.361)--
(15.458,3.373)--
(15.466,3.395)--
(15.474,3.387)--
(15.482,3.381)--
(15.49,3.4)--
(15.499,3.419)--
(15.507,3.408)--
(15.515,3.429)--
(15.523,3.43)--
(15.532,3.492)--
(15.54,3.456)--
(15.548,3.442)--
(15.556000000000001,3.442)--
(15.564,3.461)--
(15.573,3.468)--
(15.581,3.46)--
(15.589,3.451)--
(15.597,3.449)--
(15.605,3.419)--
(15.614,3.433)--
(15.622,3.425)--
(15.63,3.409)--
(15.638,3.372)--
(15.647,3.358)--
(15.655,3.352)--
(15.663,3.361)--
(15.671,3.358)--
(15.679,3.363)--
(15.688,3.381)--
(15.696,3.378)--
(15.704,3.362)--
(15.712,3.359)--
(15.721,3.362)--
(15.729,3.363)--
(15.737,3.371)--
(15.745,3.379)--
(15.753,3.377)--
(15.762,3.377)--
(15.77,3.386)--
(15.778,3.39)--
(15.786,3.398)--
(15.795,3.419)--
(15.803,3.421)--
(15.811,3.438)--
(15.818999999999999,3.451)--
(15.827,3.484)--
(15.836,3.492)--
(15.844,3.607)--
(15.852,3.573)--
(15.86,3.58)--
(15.868,3.533)--
(15.877,3.504)--
(15.885,3.523)--
(15.893,3.514)--
(15.901,3.505)--
(15.91,3.554)--
(15.918,3.577)--
(15.926,3.558)--
(15.934,3.588)--
(15.942,3.62)--
(15.951,3.64)--
(15.959,3.666)--
(15.967,3.666)--
(15.975,3.642)--
(15.984,3.659)--
(15.992,3.626)--
(16.0,3.632)--
(16.008,3.636)--
(16.016,3.636)--
(16.025,3.656)--
(16.033,3.651)--
(16.041,3.633)--
(16.049,3.58)--
(16.058,3.624)--
(16.066,3.586)--
(16.074,3.593)--
(16.082,3.578)--
(16.09,3.571)--
(16.099,3.589)--
(16.107,3.574)--
(16.115,3.579)--
(16.123,3.591)--
(16.132,3.61)--
(16.14,3.623)--
(16.148,3.64)--
(16.156,3.627)--
(16.164,3.635)--
(16.173,3.626)--
(16.181,3.601)--
(16.189,3.615)--
(16.197,3.623)--
(16.205,3.618)--
(16.214,3.622)--
(16.222,3.615)--
(16.23,3.621)--
(16.238,3.62)--
(16.247,3.619)--
(16.255,3.619)--
(16.263,3.623)--
(16.271,3.623)--
(16.279,3.624)--
(16.288,3.627)--
(16.296,3.634)--
(16.304,3.639)--
(16.312,3.651)--
(16.321,3.665)--
(16.329,3.653)--
(16.337,3.656)--
(16.345,3.65)--
(16.353,3.662)--
(16.362,3.654)--
(16.37,3.659)--
(16.378,3.657)--
(16.386,3.64)--
(16.395,3.642)--
(16.403,3.652)--
(16.411,3.722)--
(16.419,3.726)--
(16.427,3.754)--
(16.436,3.766)--
(16.444,3.759)--
(16.452,3.823)--
(16.46,3.841)--
(16.468,3.876)--
(16.477,3.822)--
(16.485,3.823)--
(16.493,3.814)--
(16.501,3.825)--
(16.51,3.818)--
(16.518,3.829)--
(16.526,3.811)--
(16.534,3.822)--
(16.542,3.822)--
(16.551,3.828)--
(16.559,3.822)--
(16.567,3.821)--
(16.575,3.816)--
(16.584,3.816)--
(16.592,3.725)--
(16.6,3.758)--
(16.608,3.772)--
(16.616,3.768)--
(16.625,3.754)--
(16.633,3.757)--
(16.641,3.762)--
(16.649,3.765)--
(16.658,3.762)--
(16.666,3.757)--
(16.674,3.757)--
(16.682,3.781)--
(16.69,3.789)--
(16.699,3.794)--
(16.707,3.784)--
(16.715,3.783)--
(16.723,3.784)--
(16.732,3.774)--
(16.74,3.777)--
(16.748,3.781)--
(16.756,3.787)--
(16.764,3.783)--
(16.773,3.789)--
(16.781,3.789)--
(16.789,3.802)--
(16.797,3.805)--
(16.805,3.798)--
(16.814,3.815)--
(16.822,3.814)--
(16.83,3.836)--
(16.838,3.843)--
(16.847,3.835)--
(16.855,3.851)--
(16.863,3.858)--
(16.871,3.847)--
(16.879,3.852)--
(16.888,3.874)--
(16.896,3.867)--
(16.904,3.867)--
(16.912,3.862)--
(16.921,3.87)--
(16.929,3.883)--
(16.937,3.88)--
(16.945,3.886)--
(16.953,3.891)--
(16.962,3.89)--
(16.97,3.897)--
(16.978,3.922)--
(16.986,3.951)--
(16.995,3.969)--
(17.003,3.981)--
(17.008,4.006)--
(17.016,4.0)--
(17.025,3.959)--
(17.033,3.893)--
(17.041,3.914)--
(17.049,3.957)--
(17.058,3.952)--
(17.066,3.964)--
(17.074,3.961)--
(17.082,3.96)--
(17.09,3.993)--
(17.099,4.015)--
(17.107,4.022)--
(17.115,3.999)--
(17.123,4.001)--
(17.132,4.015)--
(17.14,4.022)--
(17.148,4.053)--
(17.156,4.061)--
(17.164,4.076)--
(17.173,4.11)--
(17.181,4.107)--
(17.189,4.076)--
(17.197,4.089)--
(17.205,4.099)--
(17.214,3.986)--
(17.222,4.046)--
(17.23,3.968)--
(17.238,4.017)--
(17.247,4.015)--
(17.255,4.033)--
(17.263,4.054)--
(17.271,4.073)--
(17.279,4.086)--
(17.288,4.068)--
(17.296,4.071)--
(17.304,4.092)--
(17.312,4.094)--
(17.321,4.11)--
(17.329,4.125)--
(17.337,4.161)--
(17.345,4.181)--
(17.353,4.218)--
(17.362,4.262)--
(17.37,4.25)--
(17.378,4.251)--
(17.386,4.303)--
(17.395,4.353)--
(17.403,4.358)--
(17.411,4.36)--
(17.419,4.382)--
(17.427,4.403)--
(17.436,4.429)--
(17.444,4.462)--
(17.452,4.432)--
(17.46,4.394)--
(17.468,4.415)--
(17.477,4.434)--
(17.485,4.398)--
(17.493,4.408)--
(17.501,4.384)--
(17.51,4.415)--
(17.518,4.398)--
(17.526,4.367)--
(17.534,4.369)--
(17.542,4.281)--
(17.551,4.355)--
(17.559,4.451)--
(17.567,4.409)--
(17.575,4.445)--
(17.584,4.457)--
(17.592,4.446)--
(17.6,4.507)--
(17.608,4.524)--
(17.616,4.588)--
(17.625,4.621)--
(17.633,4.613)--
(17.641,4.601)--
(17.649,4.637)--
(17.658,4.637)--
(17.666,4.661)--
(17.674,4.661)--
(17.682,4.642)--
(17.69,4.635)--
(17.699,4.623)--
(17.707,4.507)--
(17.715,4.566)--
(17.723,4.59)--
(17.732,4.577)--
(17.74,4.589)--
(17.748,4.62)--
(17.756,4.643)--
(17.764,4.635)--
(17.773,4.663)--
(17.781,4.683)--
(17.789,4.765)--
(17.797,4.748)--
(17.805,4.777)--
(17.814,4.771)--
(17.822,4.77)--
(17.83,4.788)--
(17.838,4.829)--
(17.847,4.867)--
(17.855,4.853)--
(17.863,4.818)--
(17.871,4.812)--
(17.879,4.896)--
(17.888,4.906)--
(17.896,4.916)--
(17.904,4.943)--
(17.912,4.996)--
(17.921,5.037)--
(17.929,5.064)--
(17.937,5.227)--
(17.945,5.178)--
(17.953,5.211)--
(17.962,5.285)--
(17.97,5.24)--
(17.978,5.139)--
(17.986,5.138)--
(17.995,5.153)--
(18.003,5.14)--
(18.011,5.177)--
(18.019,5.234)--
(18.027,5.159)--
(18.036,5.14)--
(18.044,5.132)--
(18.052,5.048)--
(18.06,5.061)--
(18.068,5.057)--
(18.077,5.058)--
(18.085,5.004)--
(18.093,4.947)--
(18.101,4.84)--
(18.11,4.916)--
(18.118,4.907)--
(18.126,4.976)--
(18.134,5.045)--
(18.142,5.051)--
(18.151,5.007)--
(18.159,5.014)--
(18.167,5.039)--
(18.175,5.061)--
(18.184,4.997)--
(18.192,4.943)--
(18.2,4.962)--
(18.208,4.919)--
(18.216,4.935)--
(18.225,4.941)--
(18.233,4.927)--
(18.241,4.9)--
(18.249,4.841)--
(18.258,4.871)--
(18.266,4.822)--
(18.274,4.832)--
(18.282,4.9)--
(18.29,4.923)--
(18.299,4.913)--
(18.307,4.951)--
(18.315,4.985)--
(18.323,4.951)--
(18.332,4.966)--
(18.34,4.989)--
(18.348,4.984)--
(18.356,4.969)--
(18.364,4.928)--
(18.373,4.928)--
(18.381,4.916)--
(18.389,4.924)--
(18.397,4.879)--
(18.405,4.866)--
(18.414,4.868)--
(18.422,4.883)--
(18.43,4.882)--
(18.438,4.882)--
(18.447,4.838)--
(18.455,4.822)--
(18.463,4.809)--
(18.471,4.83)--
(18.479,4.79)--
(18.488,4.785)--
(18.496,4.794)--
(18.504,4.821)--
(18.512,4.817)--
(18.521,4.828)--
(18.529,4.806)--
(18.537,4.796)--
(18.545,4.865)--
(18.553,4.865)--
(18.562,4.887)--
(18.57,4.9)--
(18.578,4.915)--
(18.586,4.881)--
(18.595,4.846)--
(18.603,4.827)--
(18.611,4.788)--
(18.619,4.796)--
(18.627,4.8)--
(18.636,4.812)--
(18.644,4.803)--
(18.652,4.829)--
(18.66,4.85)--
(18.668,4.846)--
(18.677,4.861)--
(18.685,4.813)--
(18.693,4.795)--
(18.701,4.801)--
(18.71,4.814)--
(18.718,4.802)--
(18.726,4.83)--
(18.734,4.818)--
(18.742,4.825)--
(18.751,4.82)--
(18.759,4.812)--
(18.767,4.818)--
(18.775,4.821)--
(18.784,4.796)--
(18.792,4.819)--
(18.8,4.814)--
(18.808,4.813)--
(18.816,4.812)--
(18.825,4.81)--
(18.833,4.799)--
(18.841,4.806)--
(18.849,4.809)--
(18.858,4.809)--
(18.866,4.807)--
(18.874,4.76)--
(18.882,4.746)--
(18.89,4.652)--
(18.899,4.639)--
(18.907,4.579)--
(18.915,4.631)--
(18.923000000000002,4.617)--
(18.932,4.573)--
(18.94,4.538)--
(18.948,4.532)--
(18.956,4.511)--
(18.964,4.551)--
(18.973,4.616)--
(18.981,4.602)--
(18.989,4.585)--
(18.997,4.576)--
(19.005,4.586)--
(19.014,4.586)--
(19.022,4.606)--
(19.03,4.562)--
(19.038,4.548)--
(19.047,4.559)--
(19.055,4.569)--
(19.063,4.555)--
(19.071,4.554)--
(19.079,4.538)--
(19.088,4.537)--
(19.096,4.539)--
(19.104,4.532)--
(19.112,4.564)--
(19.121,4.56)--
(19.129,4.557)--
(19.137,4.592)--
(19.145,4.596)--
(19.153,4.576)--
(19.162,4.584)--
(19.17,4.583)--
(19.178,4.588)--
(19.186,4.588)--
(19.195,4.589)--
(19.203,4.589)--
(19.211,4.602)--
(19.219,4.608)--
(19.227,4.603)--
(19.236,4.596)--
(19.244,4.614)--
(19.252,4.618)--
(19.26,4.691)--
(19.268,4.715)--
(19.277,4.725)--
(19.285,4.705)--
(19.293,4.716)--
(19.301,4.722)--
(19.31,4.731)--
(19.318,4.716)--
(19.326,4.724)--
(19.334,4.726)--
(19.342,4.761)--
(19.351,4.76)--
(19.359,4.803)--
(19.367,4.893)--
(19.375,4.897)--
(19.384,4.913)--
(19.392,4.882)--
(19.4,4.907)--
(19.408,4.941)--
(19.416,4.932)--
(19.425,4.91)--
(19.433,4.892)--
(19.441,4.883)--
(19.449,4.912)--
(19.458,4.947)--
(19.466,4.958)--
(19.474,5.009)--
(19.482,5.042)--
(19.49,5.047)--
(19.499,5.031)--
(19.507,5.079)--
(19.515,5.05)--
(19.523,5.1)--
(19.532,5.072)--
(19.54,5.036)--
(19.548000000000002,5.027)--
(19.556,5.025)--
(19.564,4.99)--
(19.573,4.977)--
(19.581,4.982)--
(19.589,5.022)--
(19.597,5.071)--
(19.605,5.079)--
(19.614,5.063)--
(19.622,5.001)--
(19.63,5.009)--
(19.638,5.032)--
(19.647,5.017)--
(19.655,5.015)--
(19.663,4.977)--
(19.671,4.99)--
(19.679,5.025)--
(19.688,5.021)--
(19.696,5.004)--
(19.704,5.016)--
(19.712,5.011)--
(19.721,5.012)--
(19.729,5.001)--
(19.737,4.926)--
(19.745,4.915)--
(19.753,4.92)--
(19.762,4.911)--
(19.77,4.914)--
(19.778,4.934)--
(19.786,4.918)--
(19.795,4.903)--
(19.803,4.901)--
(19.811,4.905)--
(19.819,4.938)--
(19.826999999999998,4.965)--
(19.836,4.961)--
(19.844,4.964)--
(19.852,4.971)--
(19.86,4.945)--
(19.868,4.945)--
(19.877,4.927)--
(19.885,4.913)--
(19.893,4.882)--
(19.901,4.839)--
(19.91,4.876)--
(19.918,4.878)--
(19.926,4.863)--
(19.934,4.878)--
(19.942,4.866)--
(19.951,4.857)--
(19.959,4.852)--
(19.967,4.862)--
(19.975,4.854)--
(19.984,4.86)--
(19.992,4.86)--
(20.0,4.859)--
(20.008,4.842)--
(20.016,4.866)--
(20.025,4.905)--
(20.033,4.904)--
(20.041,4.947)--
(20.049,4.949)--
(20.058,4.936)--
(20.066,4.924)--
(20.074,4.934)--
(20.082,4.968)--
(20.09,4.972)--
(20.099,4.962)--
(20.107,4.992)--
(20.115,4.994)--
(20.123,5.01)--
(20.132,4.997)--
(20.14,4.982)--
(20.148,4.985)--
(20.156,4.969)--
(20.164,4.94)--
(20.173,4.95)--
(20.181,4.957)--
(20.189,4.905)--
(20.197,4.9)--
(20.205,4.713)--
(20.214,4.729)--
(20.222,4.794)--
(20.23,4.813)--
(20.238,4.83)--
(20.247,4.77)--
(20.255,4.823)--
(20.263,4.837)--
(20.271,4.858)--
(20.279,4.837)--
(20.288,4.836)--
(20.296,4.852)--
(20.304,4.853)--
(20.312,4.853)--
(20.321,4.878)--
(20.329,4.89)--
(20.337,4.946)--
(20.345,4.949)--
(20.353,5.0)--
(20.362,4.942)--
(20.37,4.969)--
(20.378,4.972)--
(20.386,4.991)--
(20.395,4.962)--
(20.403,4.949)--
(20.411,4.981)--
(20.419,4.975)--
(20.427,4.985)--
(20.436,4.985)--
(20.444,4.99)--
(20.452,4.976)--
(20.46,4.974)--
(20.468,4.972)--
(20.477,4.968)--
(20.485,4.967)--
(20.493,4.954)--
(20.501,4.961)--
(20.51,4.958)--
(20.518,4.971)--
(20.526,4.966)--
(20.534,4.968)--
(20.542,4.963)--
(20.551,4.963)--
(20.559,4.973)--
(20.567,4.98)--
(20.575,5.043)--
(20.584,5.046)--
(20.592,5.044)--
(20.6,5.07)--
(20.608,5.071)--
(20.616,5.057)--
(20.625,5.071)--
(20.633,5.09)--
(20.641,5.074)--
(20.649,5.066)--
(20.658,5.059)--
(20.666,5.06)--
(20.674,5.076)--
(20.682,5.02)--
(20.69,5.016)--
(20.699,5.015)--
(20.707,5.014)--
(20.715,5.039)--
(20.723,5.045)--
(20.732,5.023)--
(20.74,5.029)--
(20.748,5.029)--
(20.756,5.026)--
(20.764,5.028)--
(20.773,5.028)--
(20.781,5.053)--
(20.789,5.058)--
(20.797,5.054)--
(20.805,5.07)--
(20.814,5.114)--
(20.822,5.115)--
(20.83,5.123)--
(20.838,5.14)--
(20.847,5.147)--
(20.855,5.193)--
(20.863,5.186)--
(20.871,5.212)--
(20.879,5.203)--
(20.888,5.25)--
(20.896,5.272)--
(20.904,5.283)--
(20.912,5.234)--
(20.921,5.295)--
(20.929,5.289)--
(20.937,5.287)--
(20.945,5.268)--
(20.953,5.274)--
(20.962,5.289)--
(20.97,5.365)--
(20.978,5.357)--
(20.986,5.375)--
(20.995,5.419)--
(21.003,5.46)--
(21.008,5.508)--
(21.016,5.532)--
(21.025,5.609)--
(21.033,5.551)--
(21.041,5.593)--
(21.049,5.554)--
(21.058,5.551)--
(21.066,5.507)--
(21.074,5.512)--
(21.082,5.535)--
(21.09,5.525)--
(21.099,5.568)--
(21.107,5.59)--
(21.115,5.652)--
(21.123,5.674)--
(21.132,5.692)--
(21.14,5.748)--
(21.148,5.733)--
(21.156,5.67)--
(21.164,5.654);

%% file: tr_identification.tex
\section{Identification and description of the arbitrage activity} 
\label{tr:identification}

\subsection{Detection of the arbitrage actions}
\label{sec:detection}

\begin{figure}[tp]
    
    \begin{tikzpicture}[>=stealth,scale=1, every node/.style={scale=0.8}, trim left=-0.5cm]
    
    \pgfmathsetmacro{\xmax}{15} 
    \pgfmathsetmacro{\ymax}{6}
    \pgfmathsetmacro{\XTQ}{13}
    \pgfmathsetmacro{\YTQ}{2}
    \pgfmathsetmacro{\DT}{1.5}
    \pgfmathsetmacro{\DQ}{1.3}
    
    \draw [->] (0,0) coordinate (origin) --(\xmax,0)node[anchor=north west] {Time};
    \draw [->] (origin)--(0,\ymax) node[anchor=south east] {Quantity};
    
    \foreach\Ta/\Qa\X/\Y in
    {1.8/5/$t_{1}$/$q_{1}$,6.5/3.5/$t_{2}$/$q_{2}$,\XTQ/\YTQ/$t_{3}$/$q_{3}$
    }
    {
    \draw[color=gray!60, fill=gray!5] (\Ta-\DT,\Qa-\DQ) rectangle (\Ta+\DT,\Qa+\DQ);
     
    \draw[ultra thin,color = gray!60, dashed] (\Ta,\Qa) -- (0,\Qa);
    
    \draw (origin)++(0,\Qa) coordinate (base) --++(-4pt,0) node [left] {\Y};
    
    \draw (origin)++(\Ta,0) coordinate (base) --++(0,-4pt) node [below] {\X};
     
    }
    
    \draw[snake=brace, raise snake = 4pt] (\XTQ,\YTQ + \DQ) -- (\XTQ + \DT,\YTQ + \DQ);
    \draw[snake=brace, raise snake = 4pt] (\XTQ + \DT,\YTQ + \DQ) -- (\XTQ + \DT,\YTQ);
    \draw[gray, dashed] (\XTQ,\YTQ) -- (\XTQ,\YTQ + \DQ);
    \draw[gray, dashed] (\XTQ,\YTQ) -- (\XTQ + \DT,\YTQ);
    \draw[above = 5] (\XTQ + \DT/2,\YTQ + \DQ) node {{\Large\textbf{$\Delta T$}}};
    \draw[right = 4] (\XTQ + \DT,\YTQ + \DQ/2) node {{\Large\textbf{$\Delta Q$}}};
    
    \foreach\T/\Q/\Act/\Cur/\Col in {1.8/5/Buy/USD/red!80!black,
    3/1.1/Sell/EUR/black,
    2.2/4.2/Buy/EUR/red!80!black,
    5.7/2.5/Sell/USD/green!80!black,
    6.5/3.5/Buy/GBP/green!80!black,
    6.9/4.6/Buy/USD/red!80!black,
    8.8/2.5/Buy/EUR/black,
    11.6/4/Buy/USD/black,
    12.1/1.3/Buy/GBP/red!80!black,
    \XTQ/\YTQ/Sell/GBP/red!80!black}
    {
    \filldraw[\Col,text = black] (\T,\Q) circle (2pt) --++(0,2.5pt) node[anchor = south] {\Act,\Cur};
    
    \draw (origin)++(0,\Q) coordinate (base) --++(-4pt,0);
    
    \draw (origin)++(\T,0) coordinate (base) --++(0,-4pt);
    
    \draw[ultra thin, dashed] (\T,\Q) -- (\T,0);
    }
    
    \draw[ultra thin, color = gray!60, dashed] (0,2.5) -- (5.7,2.5);
    
    \draw [|-|,color = green!40!black,thick] (6.5-\DT-0.1,2.5) -- 
		node [left=1pt] {$ \delta Q$}
		(6.5-\DT-0.1,3.5);
    
    \draw [|-|,color = green!40!black,thick] (5.7,3.5-\DQ-0.1) -- 
		node [below=2pt] {$ \delta T$}
		(6.5,3.5-\DQ-0.1);
	
    \end{tikzpicture}
    \caption{Algorithm to identify the arbitrage actions. Legs executed by user $i$. Only the green ones form a potential arbitrage action}
	\label{fig:algo_identification}
	\floatfoot{\emph{Notes:} each dot represents a leg executed by the illustrative user $i$, who buys and sells bitcoins (Buy,Sell) against three different currencies (USD,EUR,GBP), as a function of time on the x-axis and volume on the y-axis.
	The legs are compared only when they are in a small enough neighborhood of time and volume, defined by $ \Delta T $ and $ \Delta Q $. There are three candidate groups of legs: we exclude the legs in the neighborhood of [$t_{1},q_{1}$], as both are buy legs, and of [$t_{3},q_{3}$], since the investor trades against the same currency; only the green actions in the neighborhood of [$t_{2},q_{2}$] form a potential arbitrage action, which is characterized by the values $ \delta T$ and $ \delta Q $, respectively smaller or equal than $ \Delta T $ and $ \Delta Q $ and non-negative by construction. Note that the red dot in the neighborhood of [$t_{2},q_{2}$] is excluded as it collides with both the two other legs. Note also that a leg can form a single arbitrage action, and if there is more than a matching leg, the closest in time is chosen.}
\end{figure}
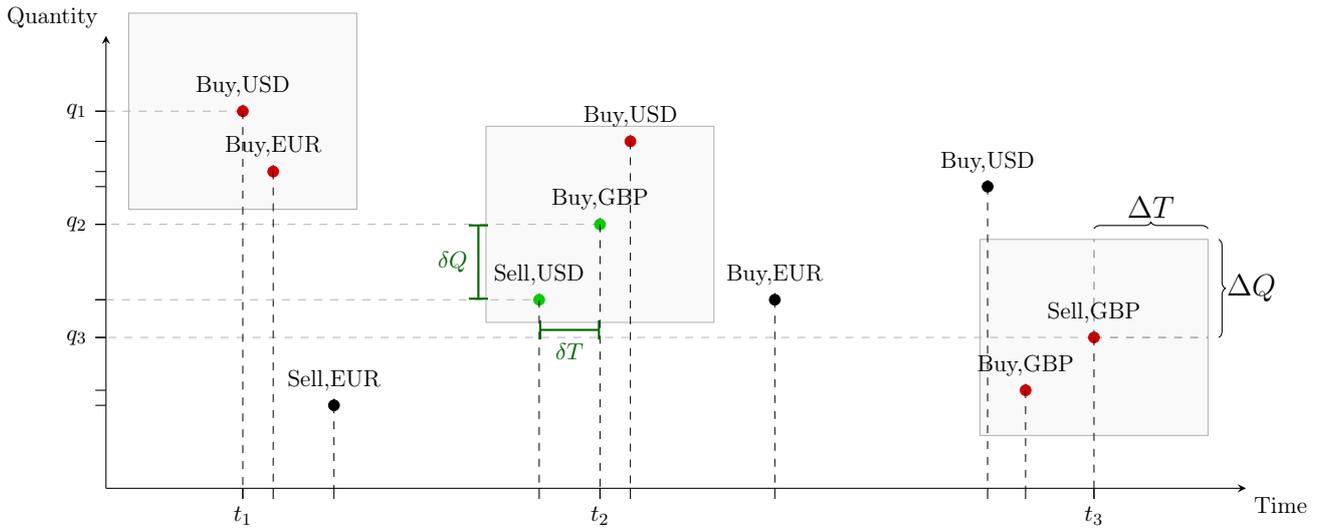

By definition, triangular arbitrage opportunities in currency markets arise when the exchange rate implied by the ratio of two fiat currencies quoted against a third vehicle currency (in our context, bitcoin) diverges from the official exchange rate.
Thus, an investor seeking to exploit such opportunities needs access to at least two currency markets quoted against the same third currency:
Mt.\ Gox users were granted the possibility to trade within the same platform in multiple fiat-to-bitcoin markets,\footnote{\url{https://bit.ly/2Fu1Rfp}}
and were entitled to have only one personal account at a time.\footnote{\url{https://bit.ly/3j0EkAl}.}
Bearing in mind that all the legs are labeled by individual identifiers, this setting is ideal to study triangular arbitrage at the micro (individual) level. We thus implement an algorithm to identify arbitrage actions, as described in Figure~\ref{fig:algo_identification}: the underlying idea is that it is possible to detect a potential arbitrage action in the form of a pair of (buy, sell) legs executed in two separate trades by the same investor using different currencies, and their identification is \textit{exact} --- though conditional on the choice of a boundary for the maximum time delay and volume difference between the pair of legs. That is, the triangular arbitrage activity can be observed completely within the private log of one exchange alone.

We focus on the investors that traded bitcoins against more than one fiat currency --- only 3,825, out of 71,808 total users; 
around 1,600,000 legs in the leaked dataset are attributable to them. 
A subgroup of 307 investors exchanged bitcoins for more than two fiat currencies, being involved in around 800,000 legs.
For each leg executed by this group of users we look for potential matches that form arbitrage actions in a neighborhood of time and volume $\pm \Delta T$ and $\pm \Delta Q$: we explore the Mt.\ Gox log searching for pairs of buy and sell legs that move a nearly equivalent amount of bitcoins, executed (almost) simultaneously by the same user in two separate trades, and exchanged for different fiat currencies --- hence in different fiat-to-bitcoin currency markets. 
We create a data set with all the potential\footnote{Even if these actions respect all the textbook properties of arbitrage, and it is hard to elaborate alternative trading strategies explaining such patterns, we cannot utterly exclude that part of them are false positives. We assume them to be true positives and relax from now on the term \textit{potential}; further details are given in Section~\ref{tr:discussion}.} triangular arbitrage actions, irrespective of the currency market they refer to.

\begin{figure}[tp]
	\centering
	\begin{subfigure}{0.475\textwidth}
		\includegraphics[width=\textwidth]{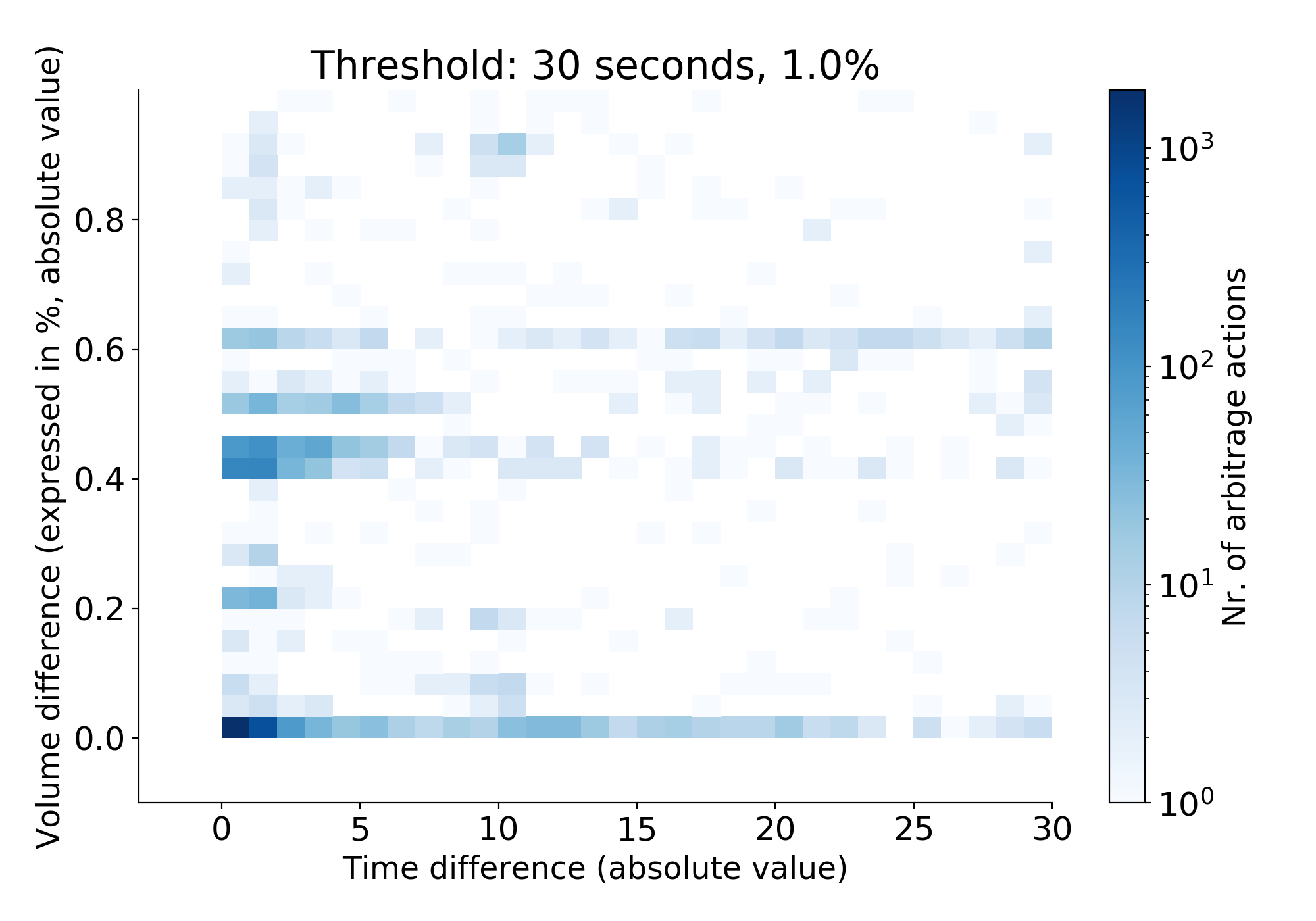}%
		\caption{$\Delta T$ = 30s, $\Delta Q$ = 1\%}
		\label{fig:t30q1}
	\end{subfigure}
	\begin{subfigure}{0.475\textwidth}
		\includegraphics[width=\textwidth]{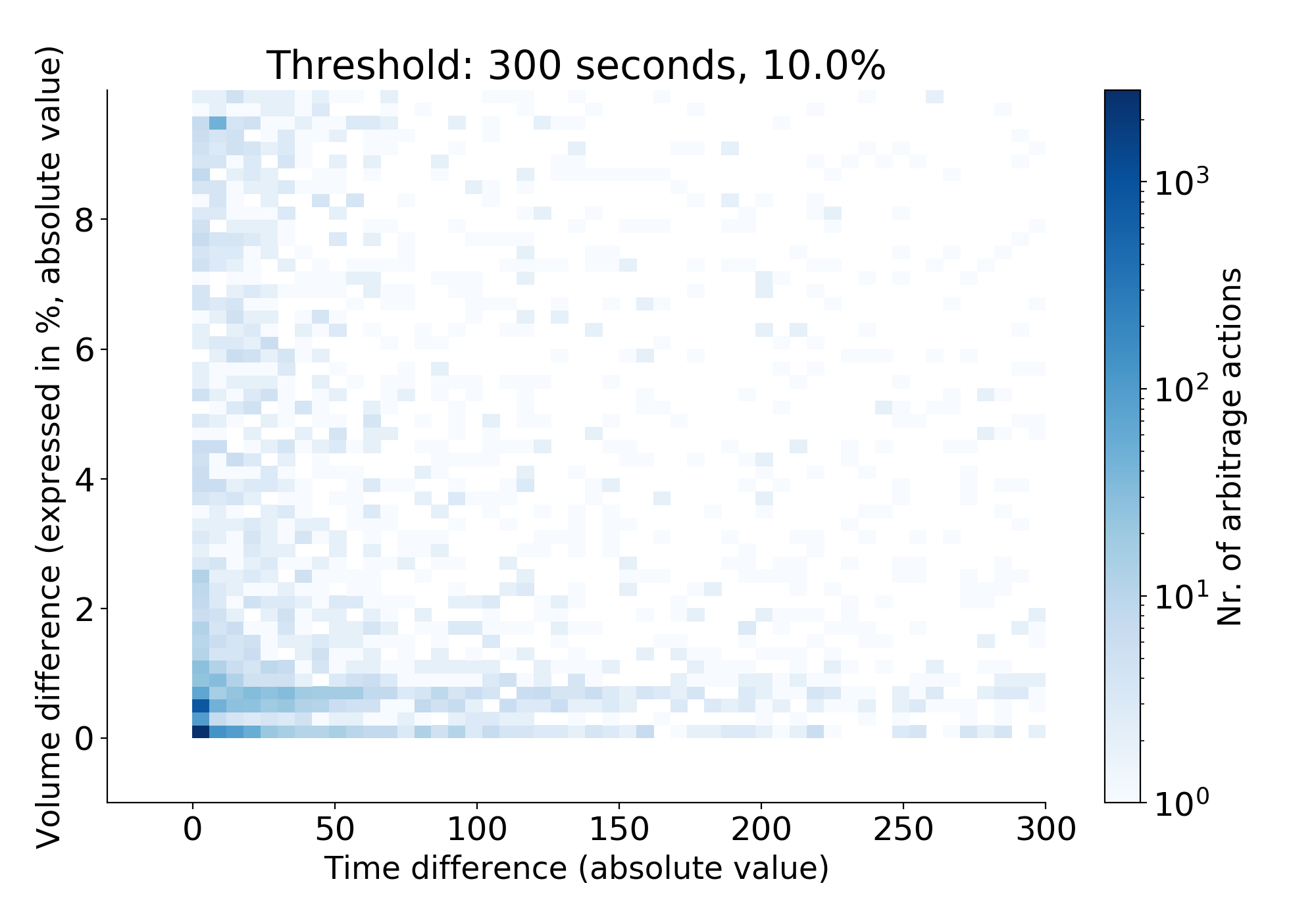}
		\caption{$\Delta T$ = 300s, $\Delta Q$ = 10\%}
		\label{fig:t300q10}
	\end{subfigure}
	\caption{Distribution of the arbitrage actions by $ \delta T$ and $ \delta Q$, given the boundaries $\Delta T $ and $\Delta Q $}
	\label{fig:arbact_distribution} 
	\floatfoot{\emph{Notes:} each arbitrage action is characterized by a $ \delta T$ and a $ \delta Q$ representing the distance in volume and time between the two legs composing such action. By construction, they are respectively smaller than $ \Delta T $ and $ \Delta Q $. Both panels report the number of actions in logarithmic scale, as a function of their $ \delta T$ (x-axis) and $ \delta Q$ (y-axis); Panel~\subref{fig:t30q1} focuses on the smaller interval [$ \Delta T $ = 30s, $ \Delta Q $ = 1\%], Panel~\subref{fig:t300q10} on [$ \Delta T $ = 300s, $ \Delta Q $ = 10\%]. }
\end{figure}

In Figure~\ref{fig:arbact_distribution} we illustrate how such actions distribute in the space delimited by $\pm \Delta T$ and $\pm \Delta Q$.
We present two different cases: 
in Panel~\subref{fig:t30q1}, where $\Delta T$ = 30 seconds and $\Delta Q$ = 1\%, the number of actions detected is N = 4,464; in Panel~\subref{fig:t300q10}, $\Delta T$ = 300s, $\Delta Q$ = 10\%, and N = 6,629.\footnote{Figure~\ref{fig:gradient} shows how the arbitrage actions increase for larger values of $\Delta T$ and $\Delta Q$. Table~\ref{tab:definitions} defines formally the parameters $\Delta T$ and $\Delta Q$, as well as the action-specific variables $ \delta T$ and $ \delta Q$.} 
Some interesting insights come along from this graphical analysis. 
In particular, as one can see, a large percentage of the trades is observed within small intervals, as the density of actions has a marked peak in the proximity of $ \delta T$ = 0s and $ \delta Q$ = 0\%.
This does not come unexpectedly, and it matches the textbook definition according to which arbitrage is performed through simultaneous actions involving almost equivalent securities. 
However, this has also interesting practical implications, as it suggests that likely such precision was achieved via automated trading.

While most of the arbitrage occurred within a tight interval $[\Delta T, \Delta Q]$, we hold a more conservative approach and scrutinize in the baseline analysis all the actions occurring within the intervals $[\Delta T = 300s, \Delta Q = 10\%]$, i.e. as depicted in Panel~\ref{fig:t300q10} (as a robustness check, in Appendix~\ref{tr:appendix_deltas} we provide the findings of our analyses obtained on different intervals of $[\Delta T, \Delta Q]$). 
The vast majority of the 6,629 identified arbitrage actions are performed in the GBP/USD and the EUR/USD markets (respectively 32.2\% and 29.1 \%), followed by the EUR/GBP (14.4\%), the AUD/USD (7.5\%), and the AUD/GBP (4.7\%) markets.

\subsection{Profitability of the arbitrage actions} 
\label{sec:profit}


Each arbitrage action, which is conducted on a specific fiat-to-fiat currency market, entails some profits (or losses) for the investor, depending on the spread between the exchange rate implied by the same action and the official rate.
We then measure the profitability of an arbitrage action as follows: 

{\vspace{-0.1cm}
\begin{equation*}
        Spread = \frac{ImpER - OffER}{OffER} \cdot 100,
\end{equation*} 
}
where $OffER$ is the hourly official rate\footnote{We use the hourly open prices of the official exchange rates published on \url{https://www.histdata.com/}
. This dataset lacks information for few minor currencies (CNY, THB, NOK, RUB): as a consequence, we could not calculate the associated profits for 35 arbitrage actions, which are excluded from the analyses where this data is required. E.g., user 5121X in Figure~\ref{fig:5121X} conducted 796 arbitrage actions, but we can calculate the profitability only for 782 of them.} and $ImpER$ is the implied one.
To compare them, by construction we use the direct quotation with the currency of the buy leg acting as the (fixed) foreign currency. That is, 

{\vspace{-0.1cm}
\begin{equation*}
        OffER = CUR_{B}toCUR_{S},
\end{equation*}
}
where $CUR_{B}$ is the fiat currency used to trade bitcoins on the buy side, and $CUR_{S}$ on the sell side, and 

{\vspace{-0.1cm}
\begin{equation*}
        ImpER = \myfrac[2pt]{Fiat_{S}}{BTC_{S}} \cdot \myfrac[2pt]{BTC_{B}}{Fiat_{B}}.
\end{equation*}
}

Noteworthy, the leaked log includes information on the explicit transaction costs incurred by the Mt.\ Gox users (i.e., the fees associated to each leg of all trades\footnote{Payable in bitcoins or in fiat currency, and sometimes partly in bitcoin and partly in fiat currency. Users could configure how to pay fees: see \url{https://bit.ly/34Wyb3h}.}).
Although additionally implicit costs may (and are likely to) exist, this feature of the dataset is especially important, as it allows us to account for the costs that directly affect the profitability of the arbitrage activity.
Thus, in the baseline investigation we adjust the actual profitability by incorporating the leg-specific fees in the prices paid to trade bitcoins, as described formally in Table~\ref{tab:definitions}, which provides a recap of the main variables introduced in this work.
However, as a robustness check, we consider two additional ways to account for the explicit fees (i.e., on a first scenario we exclude them, on another we estimate the fees a user would expect to pay given the official Mt.\ Gox schedule) which are discussed in Appendix~\ref{tr:appendix_fees}.

The average arbitrage action is worthy of a profit which is 0.42\% of the hourly official rate between the fiat currencies.
The average amount of bitcoins traded are equivalent to 52 USD (see Panel A of Table~\ref{tab:triangarb_all} and Table~\ref{tab:sumstats} for summary statistics).

\subsection{A preliminary inspection of the data}
\label{sec:indicators}

\begin{figure}[htbp]
	\centering
	\begin{subfigure}{0.475\textwidth}
	    \includegraphics[width=\textwidth]{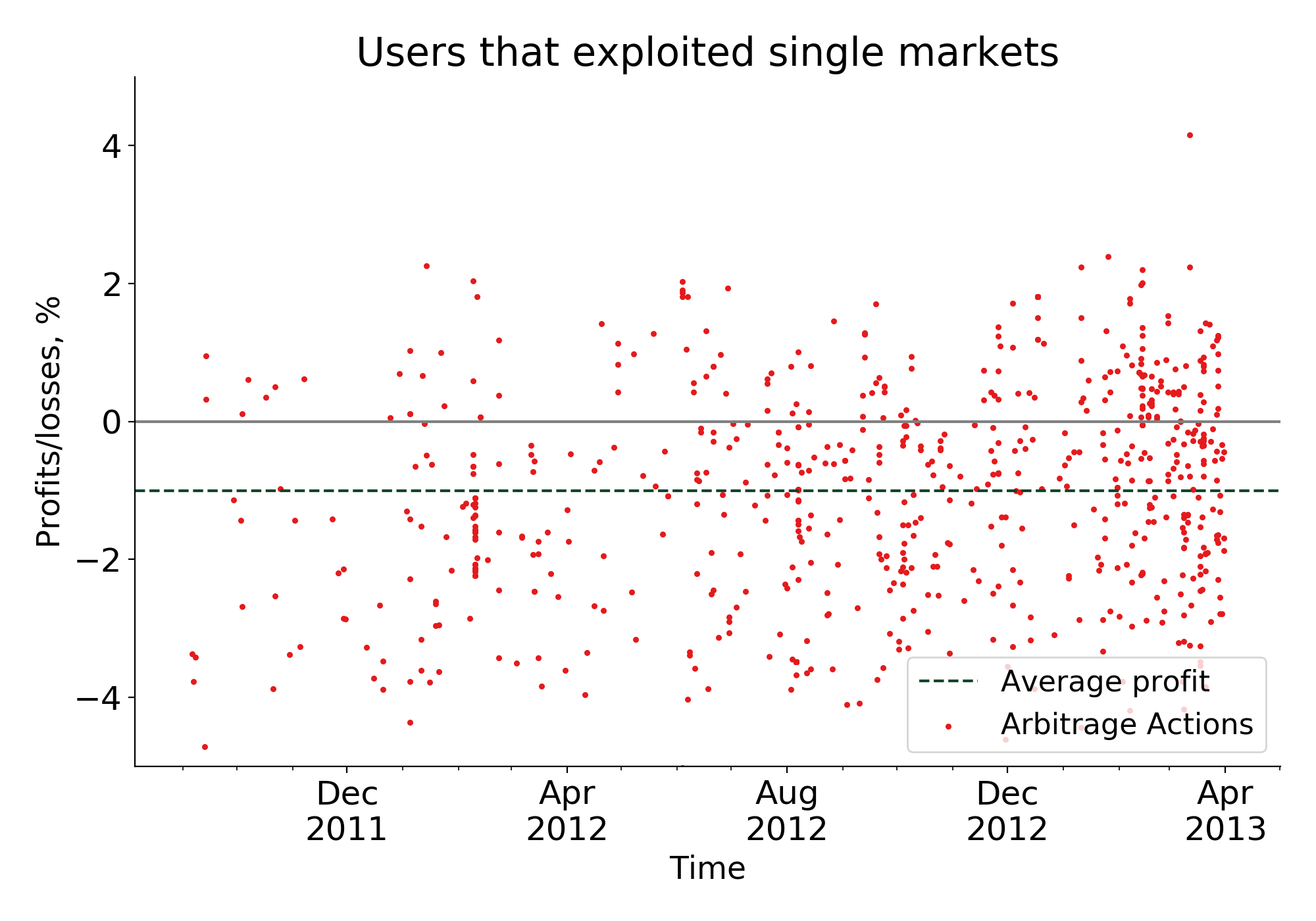}%
		\caption{}
		\label{fig:sing_profitsVStime}
	\end{subfigure}
	\begin{subfigure}{0.475\textwidth}
	    \includegraphics[width=\textwidth]{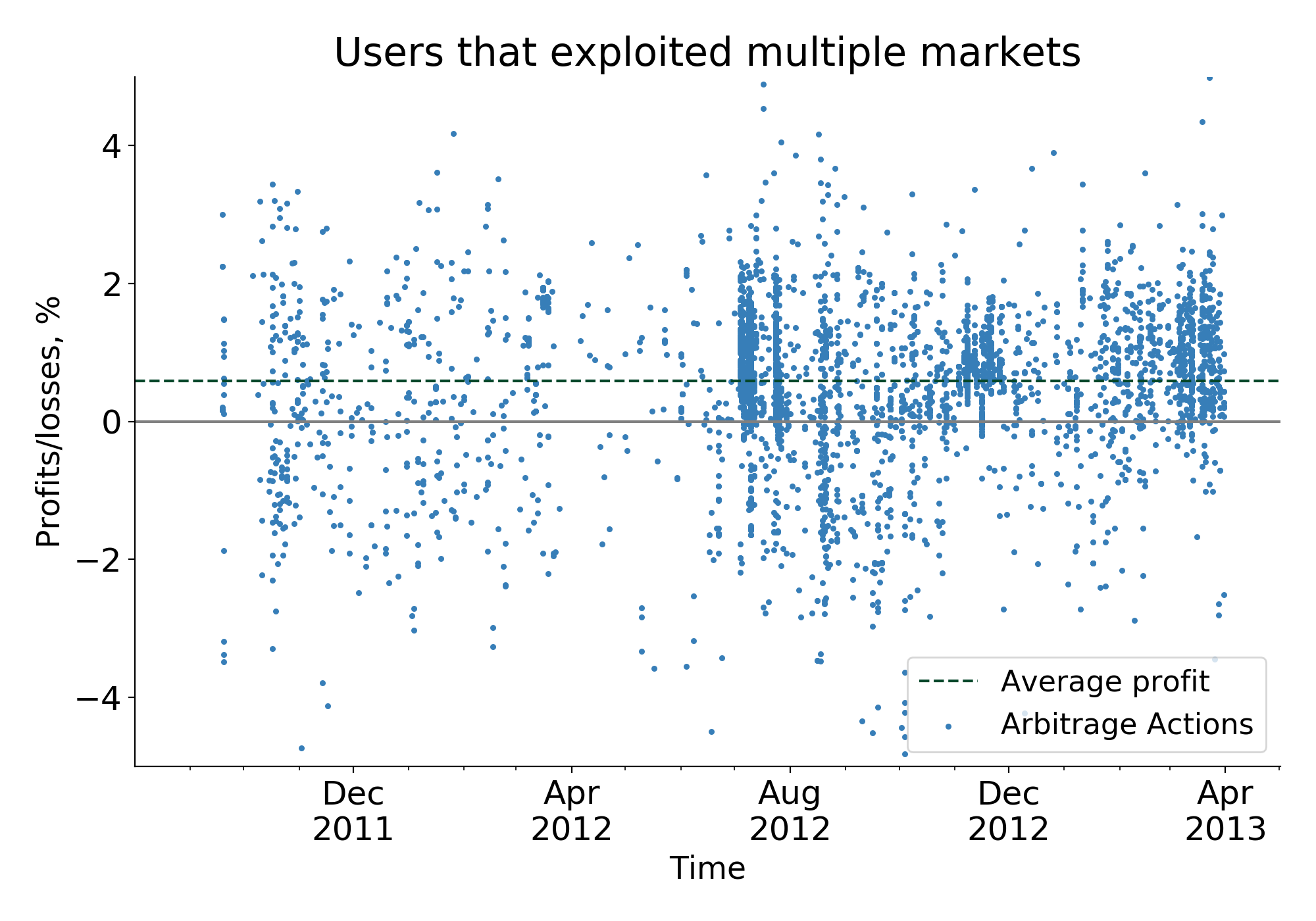}
		\caption{}
		\label{fig:mult_profitsVStime}
	\end{subfigure}	
	\caption{Profitability of the arbitrage actions. Users grouped by the number of currency markets exploited for arbitrage}
	\label{fig:profits_vs_time}
	\floatfoot{\emph{Notes:} the plot on the left reports the actions executed by arbitrageurs that exploited a single market. Viceversa, the plot on the right refers to the arbitrageurs who operated on multiple markets. The y-axis reports the profitability of the actions (including fees), depicted as dots, and the x-axis shows their evolution and deployment in time. Note that a negligible number of values may exceed the threshold [-5\%,5\%] on the y-axis. We do not show them (here and in the following plots) to focus on the area of interest.}
\end{figure}

The structure of micro-data we collect allows us to uncover a number of patterns regarding the behavior and nature of the arbitrageurs.
Notweorthy, in disagreement with theory, we note that arbitrageurs are \textit{few} --- the set of 6,629 identified arbitrage actions is executed by a total of 440 users (roughly 0.6\% of the total users).
Furthermore, the arbitrageurs' behavior seems to show a heterogeneous pattern.
First, a majority of 395 arbitrageurs explored the presence of opportunities on a single implied fiat-to-fiat currency market --- i.e., they exchanged bitcoins for exactly two fiat currencies\footnote{
All arbitrage actions involve two fiat currencies traded against bitcoins. Thus, arbitrage actions always refer to a specific fiat-to-fiat currency market. From now on we will imply this concept.
}. Others (N = 45) traded in multiple fiat-to-fiat markets, by exchanging bitcoins for at least three fiat currencies.
Remarkable differences appear when comparing the two groups: Figure~\ref{fig:profits_vs_time} reports the arbitrage activity of the users who exploited a single market --- Panel~(\subref{fig:sing_profitsVStime}) --- and multiple markets --- Panel~(\subref{fig:mult_profitsVStime}). Each dot is an arbitrage action whose x-coordinate is the time of execution and whose y-coordinate is the associated percentage profit/loss. Actions above the gray line are profitable, while actions that lie below determined losses for the users; a dashed line outlines the average profitability. The plots provide graphical evidence that the arbitrage actions executed by users in the latter group are on average more profitable and positive, while the ones in the former are on average negative. Differences in time distribution are evident too. 
Further evidence of such differences is provided by Panels B and C of Table~\ref{tab:triangarb_all}, that respectively refer to users who exploited single and multiple markets, and report additional relevant information specific to individual actions, such as the profitability with alternative measurements of the explicit transaction costs, the time delay or the volume difference between the buy and sell sides. They show that the actions executed by users who exploited single markets are on average non profitable, unless fees are excluded, while those conducted by users who exploited multiple markets are always on average profitable. The expected fees instead overestimate the real fees paid for both groups, and the differences between the actual fees paid and the expected fees are larger for the `Multiple' group.
Moreover, the actions in that group are more precise ($\delta T$ and $\delta Q$ are on average closer to zero) and, interestingly, are smaller in terms of moved volumes, both considering the amount of bitcoins and of fiat currency. 
A possible explanation, which we recall below, is that such users exploit more complex strategies and split orders to reduce the overall market impact.

Second, from Table~\ref{tab:triangarb_all} it emerges also that most of the arbitrage activity is conducted by the users who exploited multiple markets (N = 5,906 against N = 723). Indeed, the three most active users performed 
32.8\%, 12\%, and 10.4\% of the total actions, and all of them were active in multiple currency markets. Among those who executed arbitrage on a single currency market, only 11 users performed 10 or more actions, and the most active one performed just 27 actions. Table~\ref{tab:percentiles} provides further information on the number of actions executed by the two groups.

\begin{figure}[tp]
	\centering
	\begin{subfigure}{0.475\textwidth}
	    \includegraphics[width=\textwidth]{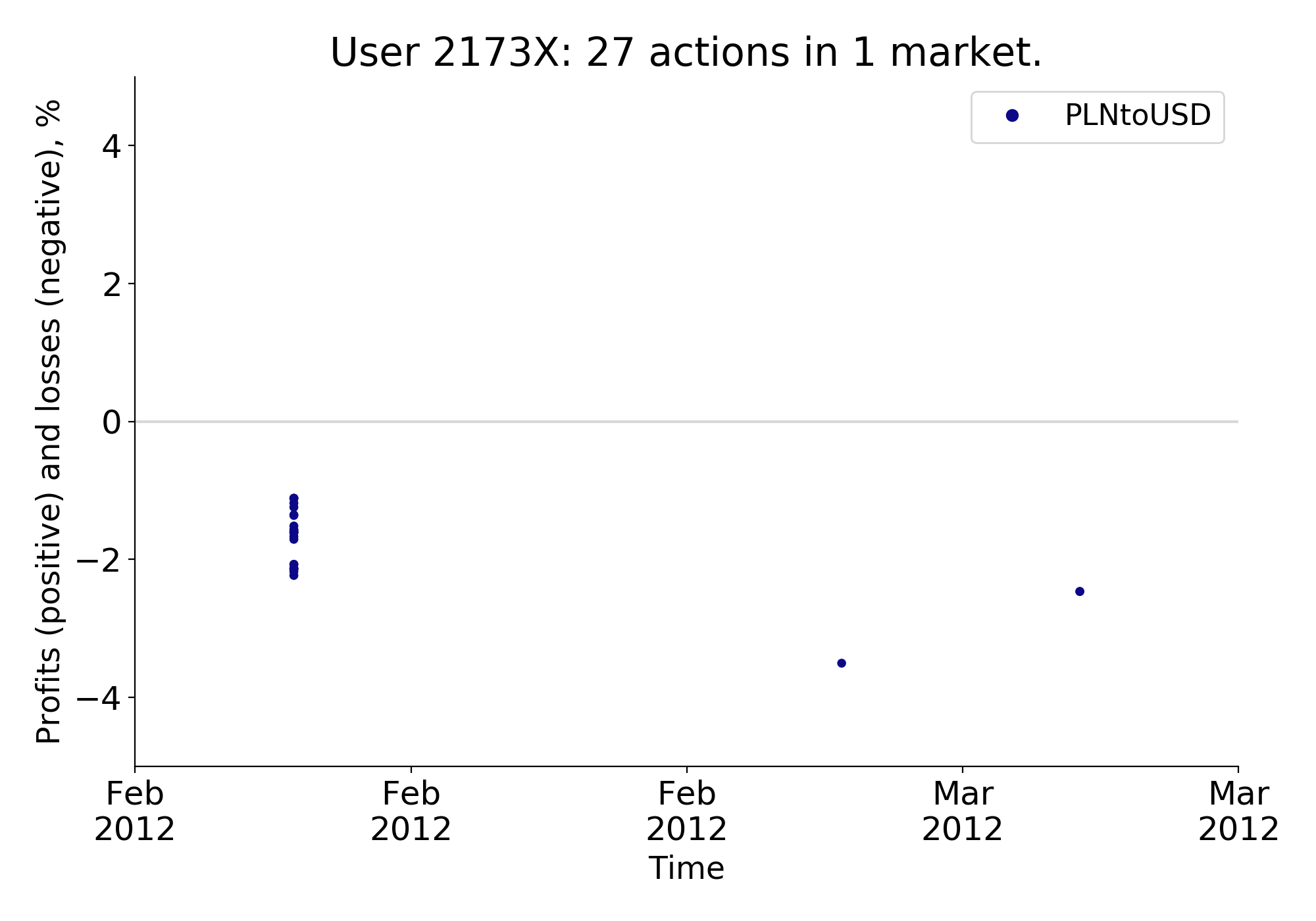}%
		\caption{}
		\label{fig:2173X}
	\end{subfigure}
	\begin{subfigure}{0.475\textwidth}
		\includegraphics[width=\textwidth]{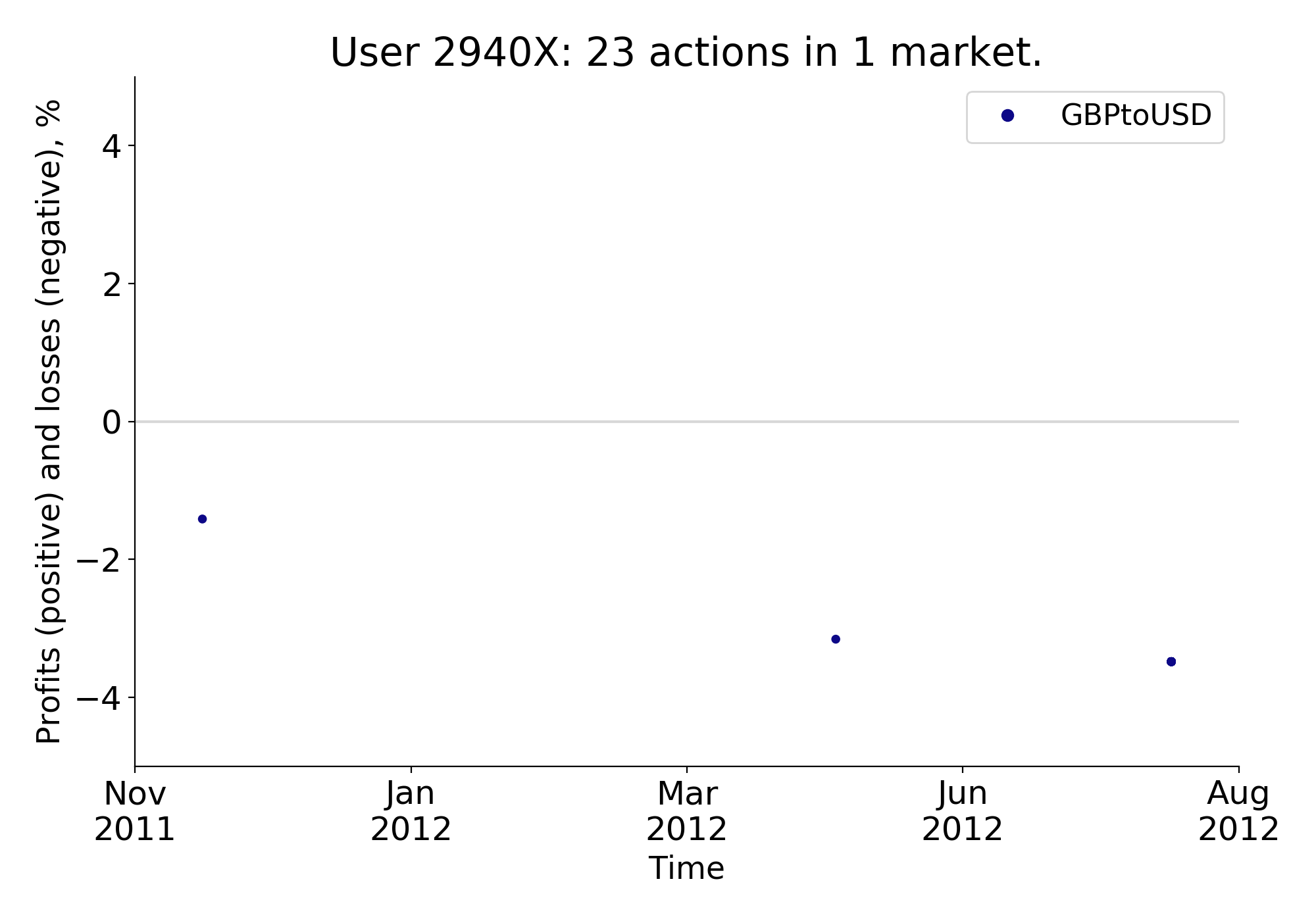}
		\caption{}
		\label{fig:2940X}
	\end{subfigure}
	\begin{subfigure}{0.475\textwidth}
		\includegraphics[width=\textwidth]{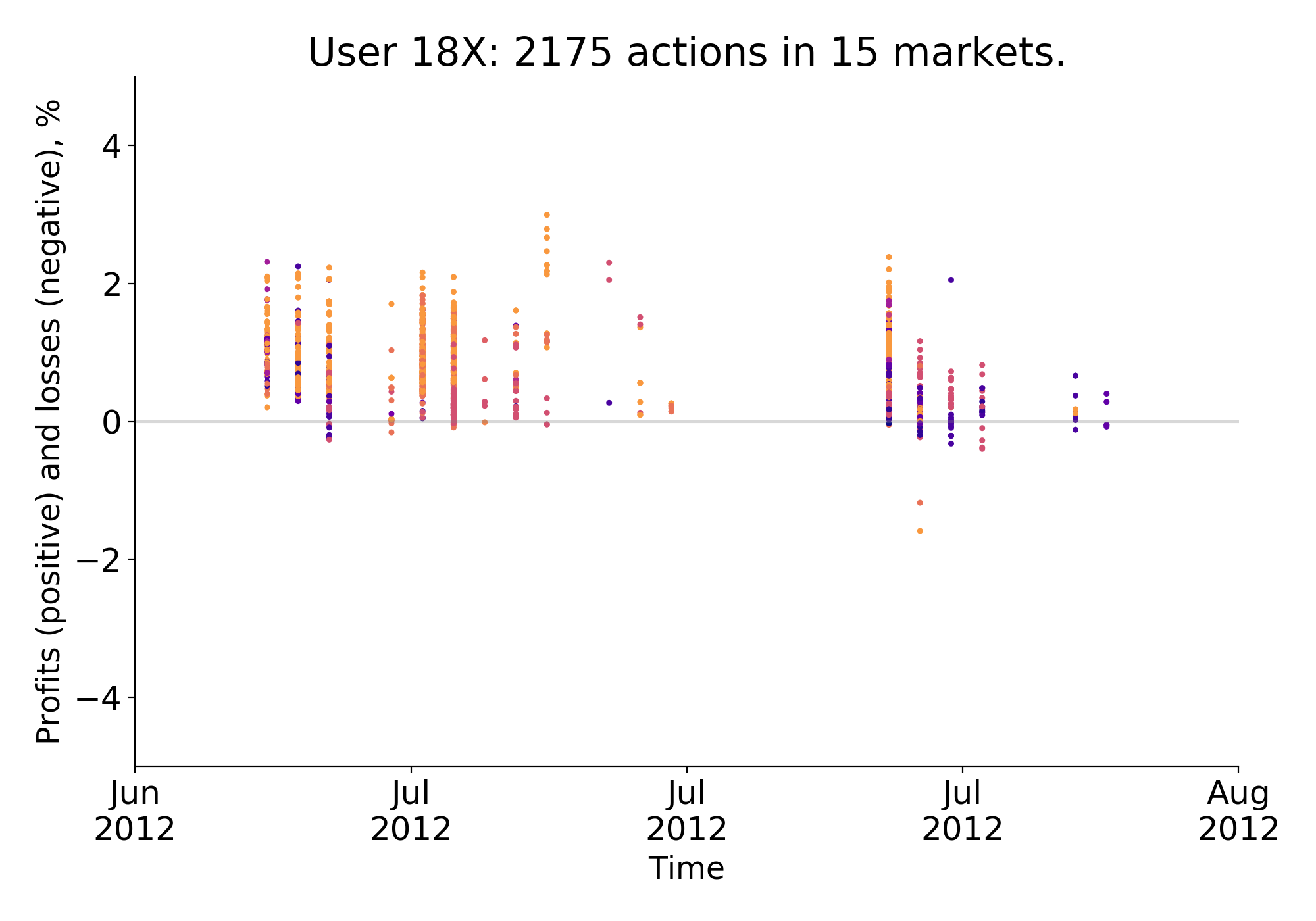}
		\caption{}
		\label{fig:18X}
	\end{subfigure}
	\begin{subfigure}{0.475\textwidth}
		\includegraphics[width=\textwidth]{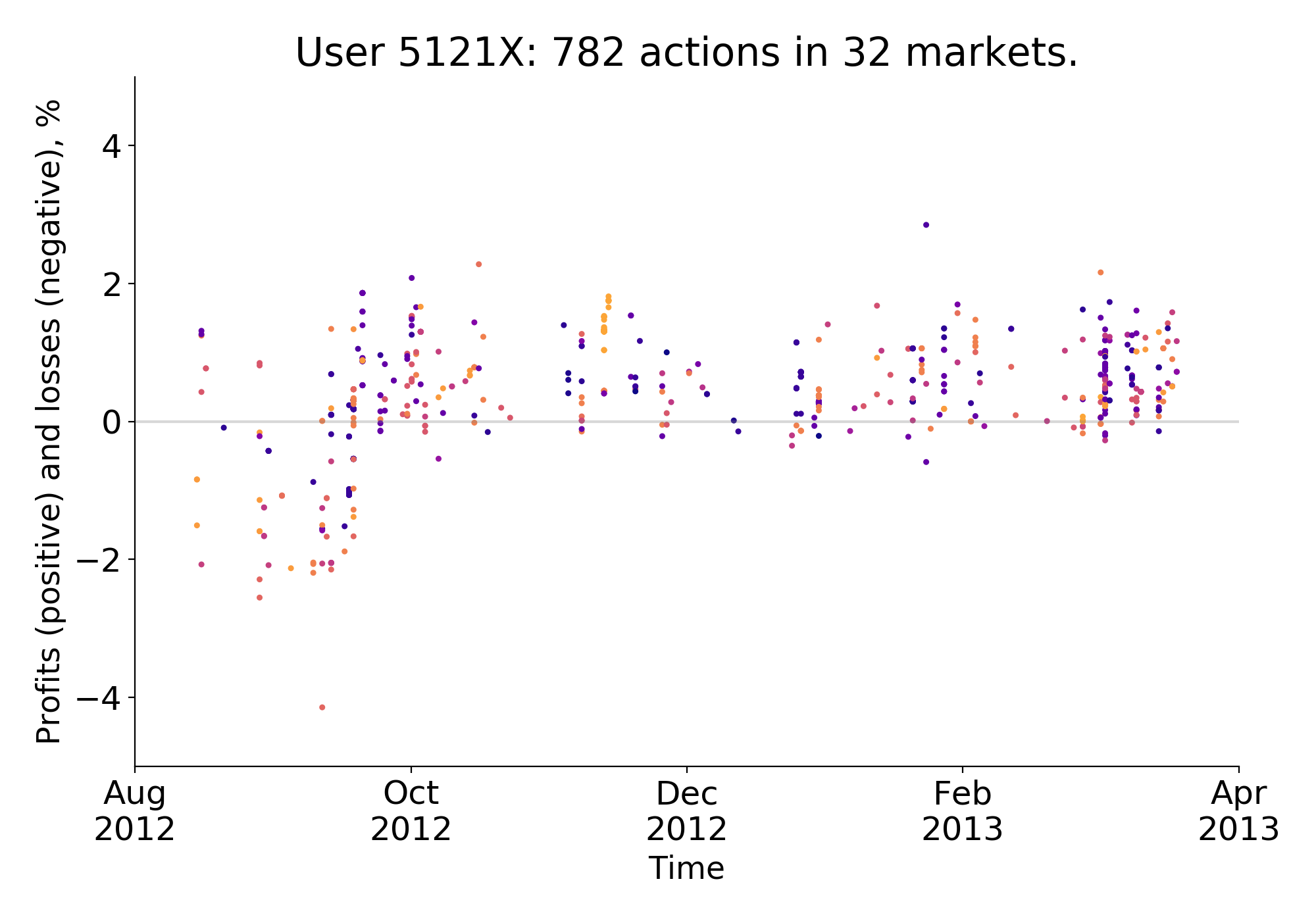}
		\caption{}
		\label{fig:5121X}
	\end{subfigure}
	\caption{Profitability and trading patterns across arbitrageurs}
	\label{fig:patterns} 
	\floatfoot{\emph{Notes:} the panels report the two most active users in a single currency market (above) and in multiple markets (below). The y-axis indicates the profitability of the actions, depicted as dots, and the x-axis shows their evolution and deployment in time. The different colors correspond to actions conducted in different currency markets. We do not report the legend for the two plots below as the number of markets is too high (15 and 37). We hide the last unit of each user identifier to preserve the anonymity. A negligible number of values may exceed the threshold [-5\%,5\%] on the y-axis.}
\end{figure}

Third, arbitrageurs that operate on multiple markets are also more acquainted with sophisticated algorithms, such as metaorders.
We follow \citet{donier2015million} and define as metaorder a group of at least 5 arbitrage actions executed by the same user in the same market (and in the same `buy/sell direction'), so that the delay between each sequential action never exceeds 60 seconds\footnote{
Note that it is not our main purpose to exactly identify metaorders; rather, we use this measure as an indicator to verify which users exploit these strategies more systematically. As a robustness check, we repeated the procedure by changing the minimum number of actions and the time delay, and the differences are negligible.
}.
As illustrated in Table~\ref{tab:metaorders}, only 13 arbitrageurs executed metaorders, which are typically composed of less than 10 actions --- each delayed of around 20 seconds --- and moved an average amount of bitcoins equivalent to few hundreds of dollars. Only 5 users performed more than 5 metaorders, and all of them exploited multiple currency markets and executed more than 100 arbitrage actions.
Fourth, arbitrageurs that operate on multiple markets are less likely to behave aggressively. We follow \citet{scaillet2017high} and define the aggressive bids and asks respectively as the buy or sell legs that initiate the market orders.
Thus an aggressive arbitrage action is an action with at least one aggressive leg. 
Table~\ref{tab:aggressive_orders} shows that aggressive orders have been used only by users who executed less than 30 actions; on average, they are not profitable.

Noteworthy, clustering the arbitrage actions executed by the same users unfolds interesting insights, and provides further evidence that such differences map into heterogeneous patterns of profitability of arbitrage.
In Figure \ref{fig:patterns}, for instance, we illustrate graphically the trading pattern of the most active users in a single currency market (Panels~\subref{fig:2173X} and~\subref{fig:2940X}) and in multiple markets (Panels~\subref{fig:18X} and~\subref{fig:5121X})\footnote{
For completeness, we provide the trading patterns of other traders in \ref{fig:patterns_supplem}.
Furthermore, Figure~\ref{fig:main_indicators} shows intuitively that a relationship between the variables introduced above and a profitable execution of the arbitrage activity indeed exists.}.
The dots indicate the profits/losses (y-axis) across time (x-axis) on arbitrage actions. 
Differences in profits are considerable.
While users in~(\subref{fig:2173X}) and~(\subref{fig:2940X}) systematically incur in losses when trading (as dots lie below the gray line), the others typically obtain profits by executing far more complex trading patterns. 
It is also worth remarking important differences between their strategies, e.g. when comparing users 18X and 5121X.
Trades performed by the first are concentrated in few weeks (around July 2012); its actions appear as consequential and related and are likely to being part of one or a sequence of metaorders. 
The trading pattern of the latter is steady and spans across a longer period of time. Both strategies are profitable, non-trivial and likely executed via algorithmic trading. 

\section{Trade ability and profitability of arbitrage}
\label{tr:preliminary}

In this paper we hypothesize that arbitrage profitability is a function of the user's trade ability.
As laid out in the previous section, our most preferred indicator for trade ability is arbitrage on multiple markets (which we complement with three other variables --- number of actions, execution of metaorders and of aggressive orders\footnote{To further explore the relationship between these variables, we perform a PCA, whose results are reported in Table~\ref{tab:pca}. Table~\ref{tab:pearson} shows instead the Pearson correlation across the main variables of 
our model.}).
Indeed, it is relatively simple to conduct arbitrage exploring opportunities on a single currency market. Evidence suggests that most of the users attempt to do arbitrage in this form (and non-systematically, i.e., in few and dispersed trials, on average non profitable). Few users explore instead more than one market looking for arbitrage opportunities.
This activity is in fact far from trivial and requires skills and expertise: users active in multiple markets must set up complex --- and likely automated --- strategies in order to handle funds in different fiat currencies and to correctly incorporate the increasing amount of disposable information on price variations (the potential number of markets to observe grows non-linearly with the number of currencies used). A further sign of expertise is that operating in multiple markets leads to risk reduction through diversification. 
Descriptive evidence provided above suggests also that users that make arbitrage through multiple markets obtain on average higher profits.
This conclusion is potentially threatened by two facts.
First, trade ability may not be fixed but may increase with trading.
Second, the correlation between trade ability and profits may be affected by an omitted variables bias.
In this section, we take these two aspects into our analysis.

\subsection{Learning-by-doing and trade ability}
\label{tr:learning}

\begin{figure}[t]
	\centering
	\includegraphics[width=\textwidth]{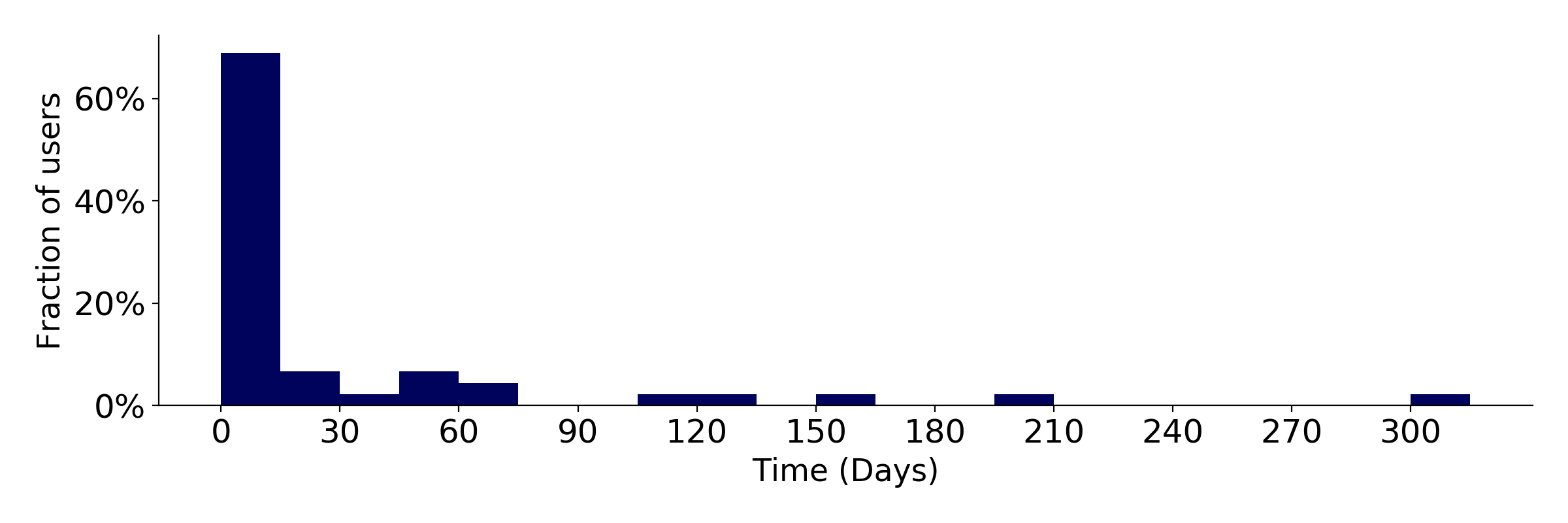}%
	\caption{Days passed between the first arbitrage action of a user and the first one in a new market}
	\label{fig:days_diff} 
	\floatfoot{\emph{Notes}: the graph refers only to the arbitrageurs active on multiple markets. For about 70\% of users only 0 to 14 days passed between the first arbitrage action and the first one in another currency market (first bin).
	}
\end{figure}

The validity of our analysis relies on the fact that arbitrage through trading on multiple markets is a sign of trade ability that a user holds before starting operating on Mt. Gox.
Thus our analysis fails to capture the link between expertise and profits if, for example, a user does arbitrage on a single market for long and only after a period of training the user starts arbitraging using other currencies. 
In Figure \ref{fig:days_diff} we show that such scenario is unlikely to hold in our context.
The plot illustrates the distribution of arbitrageurs active in multiple markets across days passed between the first arbitrage action of a user and the first one in a new market.
As one can see, the distribution is concentrated in the first bin which gathers arbitrageurs that operate on a new market within 14 days since its first arbitrage action.
This bin collects about 70\% of the arbitrageurs active in multiple markets.

In sum, for the vast majority of arbitrageurs the time passed between their first arbitrage action and the first one in a new currency market is short --- less than 15 days for 70\% of them and a month for about 80\%.
We therefore conclude that in our context investors sophistication unlikely changed considerably over time. 
Nonetheless, for robustness we also replicate our analysis excluding users if time delay between the first arbitrage action and the first one in a new market is large, i.e., we remove those that do not fall within the first bin (see Tables~\ref{tab:tab1_bins},~\ref{tab:tab1_alt_x_bins},~\ref{tab:tab2_bins} 
in Appendix~\ref{tr:appendix_learning}).

\subsection{Regression analysis}

The difference in profits recorded by users that operate on a single market and users that operate on multiple markets is likely to be biased.
For one thing, the latter group may invest a considerable larger amount of money on arbitrage than the former group of arbitrageurs.
As the expected profit from trading is larger one may expect that also effort does.
For another, profitability may stem from a specific feature of a market or on specific shocks that operate on a single time frame --- e.g., external events affecting the volatile Bitcoin ecosystem, sharp price variations and high volatility, and also internal structural changes within Mt. Gox.

A more rigorous way to investigate such difference in profit from arbitrage actions between the two groups of users is to estimate the following regression:
\begin{equation}\label{reg:1}
\Spread_{i,j,p,t} =  \beta_{0}  + \beta_{1} \text{Trade\,Ability}_{j} +  \beta_{2}\, \text{USD}_{i,j,p,t} + \theta_p + \phi_t + \varepsilon_{i,j,p,t},
\end{equation}
where $i$ indicates arbitrage actions, $j$ users, $p$ the pair of currencies identifying a dyad, and $t$ hours. 
Residuals, $\varepsilon_{i,j,p,t}$, are clustered at user-level to account for redundant information across actions made by the same user.

The outcome, $\Spread_{i,j,p,t}$, is the distance in price between the implied and the official exchange rate of an arbitrage action $i$ performed by a user $j$, on the hour $t$, using a dyad of currencies $p$. As described in Section~\ref{sec:profit}, it is reported as a percentage, and by construction the action is profitable when the former is larger than the latter.
The explanatory variable of interest in $\text{Trade\,Ability}_{j}$ is a variable conveying information on the expertise of the user $j$ who conducted the action $i$. 
The coefficient of interest is thus $\beta_{1}$, the conditional difference in profits between expert and non expert users (whose profits are captured by the constant, $\beta_0$).

Eq.~\ref{reg:1} also controls for the volume of the trades, expressed in dollars (and divided by 10,000). 
This variable is preferred to the volume of bitcoins traded as the the latter is subject to high price volatility in time. To construct this variable, prices of the actions not in USD are converted to allow comparisons across currency markets.
Most importantly, we include a set of currency pair (dyad) fixed effects, $\theta_p$, that allow us to compare arbitrage actions operated using the same couple of currencies.
We also introduce hourly time fixed effects, $\phi_t$. 
As we have explained in Section~\ref{tr:background}, Mt.\ Gox operated at the outset of the bitcoin uptake, was the first exchange platform with a significant relevance, and it was hit by several shocks.
Time fixed effects allow to absorb any potential shocks occurred on the market.
Besides this, by comparing arbitrage actions filled in the very same hour, we likely capture contingent conditions of the market strictly related to risk, such as liquidity, volatility, depth of the market, that otherwise would be hard to capture given the `two-leg' (and `two-currency') structure of the arbitrage actions considered in the analysis.\footnote{
We selected this time scale as a result of a trade-off between granularity and feasibility of the analyses (a smaller scale would be too demanding for a FE-based analysis).
}

In Table~\ref{tab:tab1_base} we present our estimations results where trade ability is proxied by the dummy variable $D(Currencies)$, equal to 0 if the user conducted arbitrage in a single currency market, and 1 if arbitrage is conducted in multiple markets.
Overall, these estimations are statistically significant and corroborate our hypothesis that sophisticated arbitrageurs trade on average at a positive premium, relative to less sophisticated users.
Namely, column (1) reports the estimate of the correlation between profitability and expertise;
in columns (2) and (3) we add separately time and dyad fixed effects;
in column (4) we add both fixed effects in the regression.
Some observations are relaxed when including the fixed effects, either because in some hours one single trade was executed, or because a trade is the only one executed in a minor market.

\begin{figure}[tp]
	\centering
	\begin{subfigure}{0.475\textwidth}
		\includegraphics[width=\textwidth]{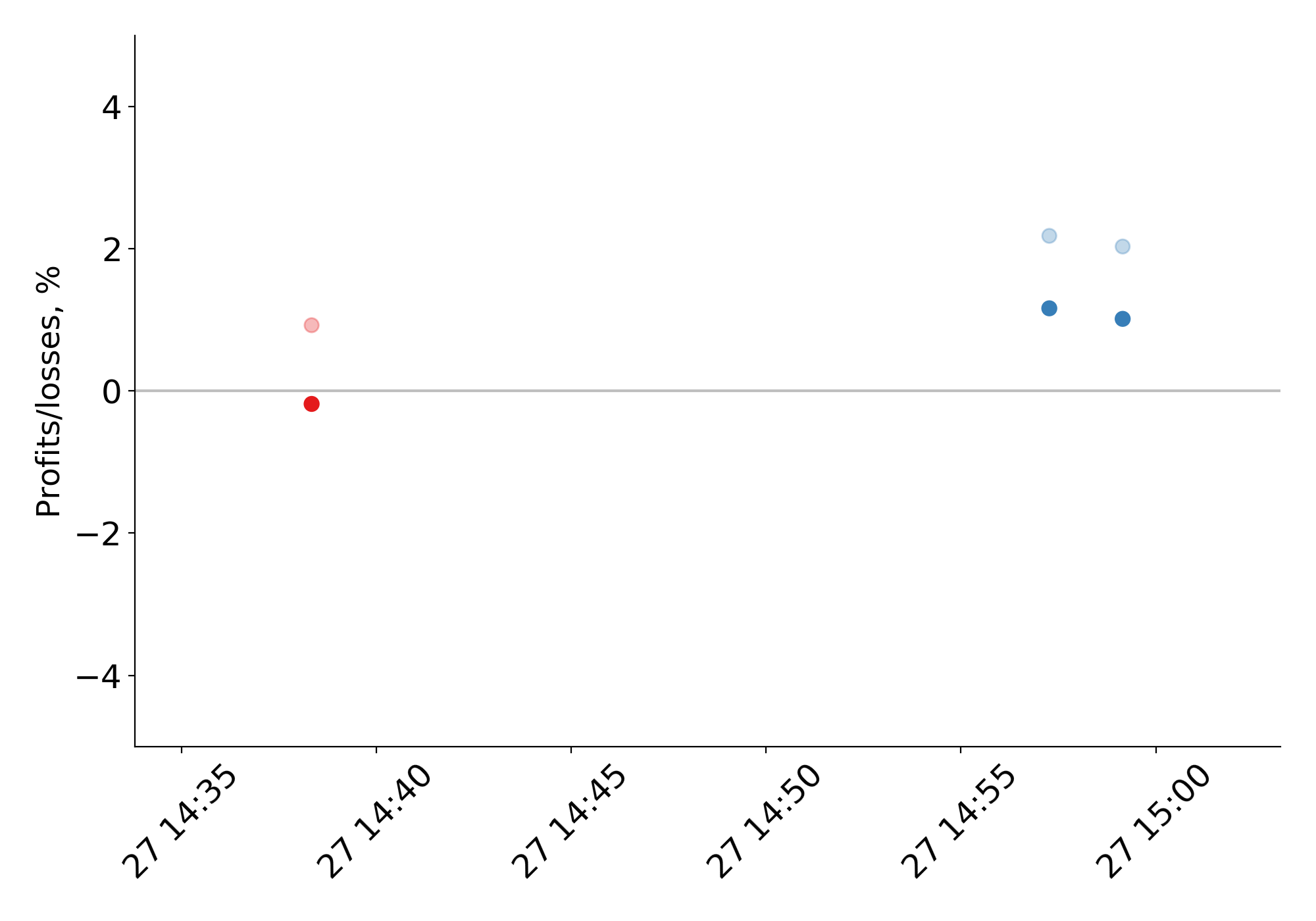}%
		\caption{Date: 27 October 2012, h.14}
		\label{fig:monetilla}
	\end{subfigure}
	\begin{subfigure}{0.475\textwidth}
		\includegraphics[width=\textwidth]{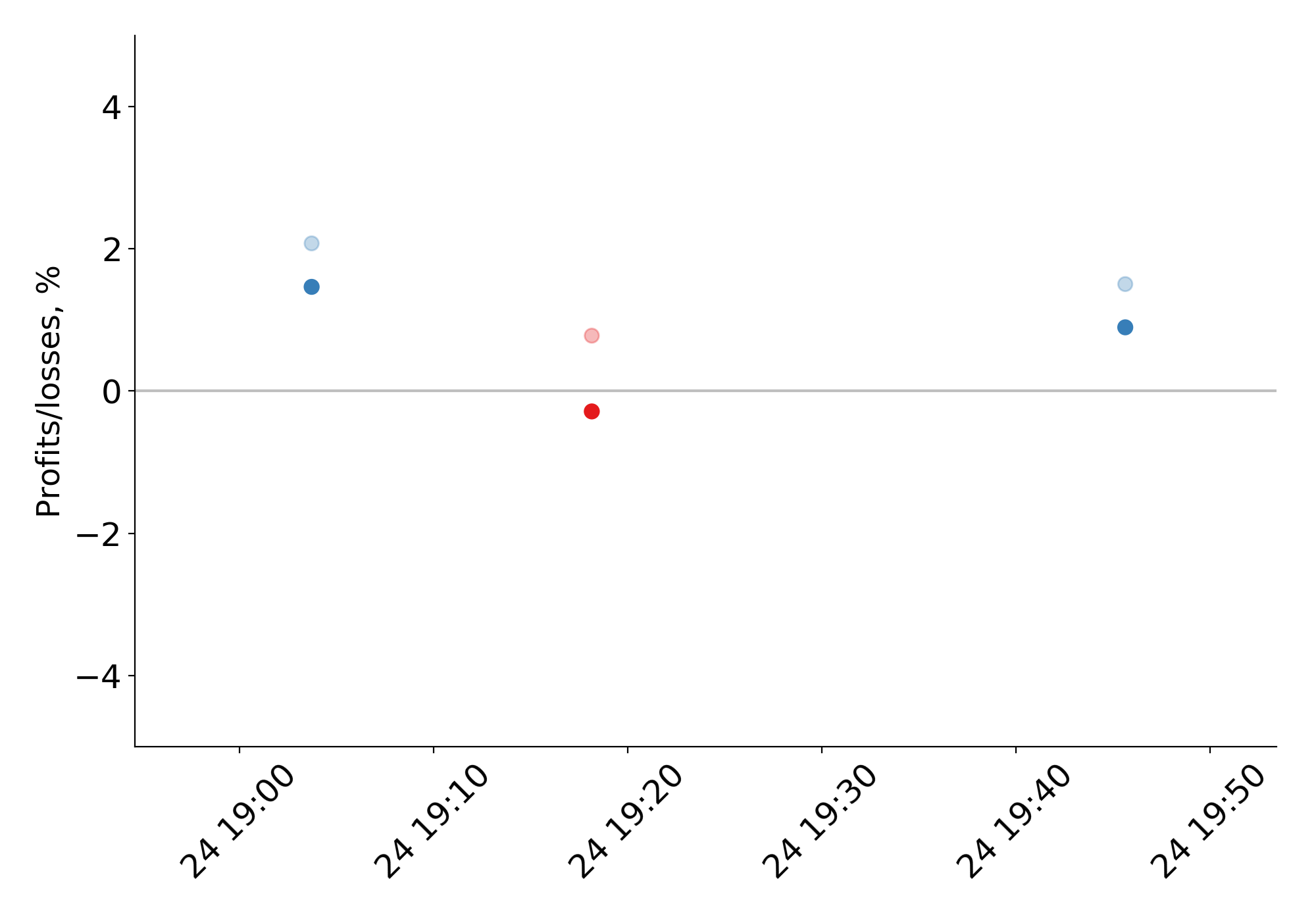}
		\caption{Date: 24 March 2013, h.19}
		\label{fig:monetillb}
	\end{subfigure}
	\caption{The `monetary illusion' effect}
	\label{fig:monetill} 
	\floatfoot{\emph{Notes:} the plots show the arbitrage activity in two different time windows. Each action is reported twice, once including and once excluding the transaction costs (the former is slightly transparent, in order to distinguish them). The y-axis reports the profits/losses, and the x-axis the date of execution. In both cases the non expert users (in red) conduct less profitable activity, and once the fees are included their actions yield losses.}
\end{figure}

The effect is also economically relevant.
Focusing on column (4), we find that the average sophisticated user traded at a premium of 1.292\%, relative to the unsophisticated arbitrageurs --- a difference which is slightly above a standard deviation in profitability. 
Finally, it is worth noting that the constant term is systematically negatively estimated across the four specifications, indicating that the average less expert arbitrageur in Mt. Gox incurred in a \textit{loss} from such activity. 
This result is of particular interest if compared to Table~\ref{tab:app_tab1fees}, where we repeat this analysis by computing the spreads with alternative measures of the transaction costs: the results show that the constant term is non-negative only when the transaction costs are not taken into account. Our interpretation is that the non sophisticated users do not account correctly for the costs of conducting the arbitrage strategy, in a mechanism akin to the one originating the monetary illusion phenomenon~\citep{shafir1997money}, and thus incur in unprofitable activity. Figure~\ref{fig:monetill} provides an example for two different illustrative time windows (27 October 2012 and 24 March 2013): each action is reported twice, once without the transaction costs (and slightly transparent) and once including the fees paid. Red dots represent the actions executed by investors who executed arbitrage in a single market, while the blue ones are executed by expert investors active in many markets. The x-axis is the time of execution of the action, the y-axis is the profit/loss. All these arbitrage actions are affected by the transaction costs, which reduce the yielded profits. However, in the case of the non expert users, not only the actions are in general less profitable, they become even unprofitable once the transaction costs are included, both when they are similar across users (Panel~\ref{fig:monetilla}), and when they vary across them (Panel~\ref{fig:monetillb}).

Table~\ref{tab:tab1_alt_x} employs alternative explanatory variables for trade ability described in Section \ref{sec:indicators}.
Column (4) of Table~\ref{tab:tab1_base} is reported in column (1) for easiness of comparison. 
Column (2) reports estimations of $\beta_1$ when trade ability is proxied by the logarithm of the number of currencies used; 
in (3) we explore the effect of the logarithmic number of arbitrage actions executed by the user on the profitability of the action. 
Both estimations are positive and statistically significant.
In column (4) trade ability is proxied by the dummy variable $D(Metaorder)$, which is equal to 1 for \textit{all} the actions conducted by users that executed at least one metaorder. 
Similarly, in column (5) we use the dummy variable $D(Aggressive)$ to indicate whether the user executed or not at least one aggressive action. 
We stress that both the latter two variables report user-specific and not action-specific information.
When trade ability is proxied by $D(Metaorder)$ the sophisticated users are more profitable, but the estimation is less precise; 
instead, we find that profit reduced significantly when the arbitrage activity is conducted by users who executed at least one aggressive arbitrage action.\footnote{To better interpret the magnitude of $\beta_1$ in Table~\ref{tab:sumstats} we report summary statistics for the variables employed.} %
Finally, column (6) shows the relationship between profits and the scores of the first component obtained with the principal component analysis. Also in this case, the $\beta_{1}$ coefficient is positive and statistically significant.
Specifically, we estimate that an arbitrageur with a trade ability score which is a standard deviation above the mean obtained a premium which is around half of the standard deviation in profits (i.e., $\frac{0.224 \times 2.83}{1.26}$).

%% file: tr_results.tex
\section{Trade ability and responsiveness in arbitrage}
\label{tr:results}


\begin{figure}[tp]
	\centering
	\begin{subfigure}{0.475\textwidth}
		\includegraphics[width=\textwidth]{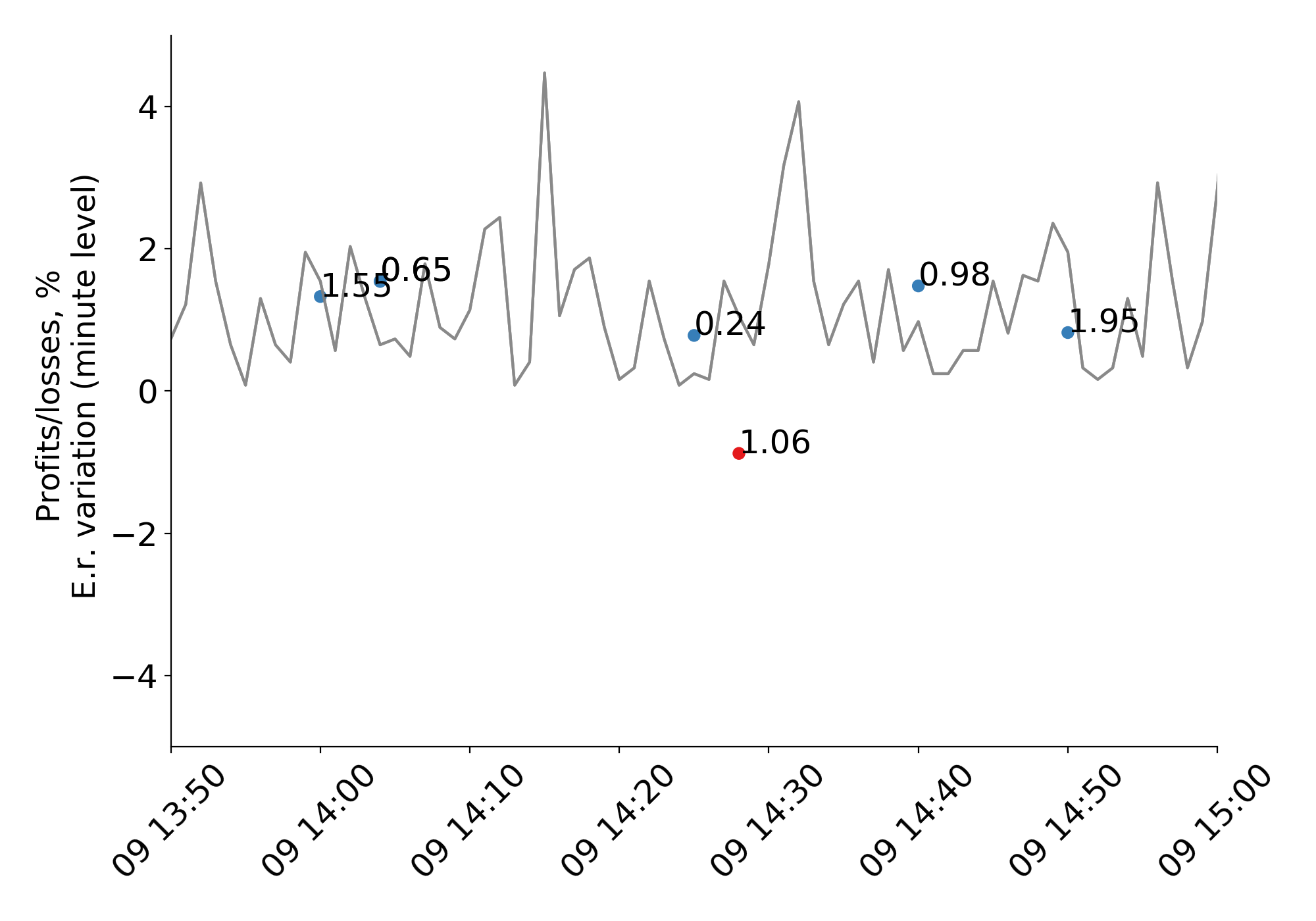}%
		\caption{Date: 9 July 2012, H. 14}
		\label{fig:within-c}
	\end{subfigure}
	\begin{subfigure}{0.475\textwidth}
		\includegraphics[width=\textwidth]{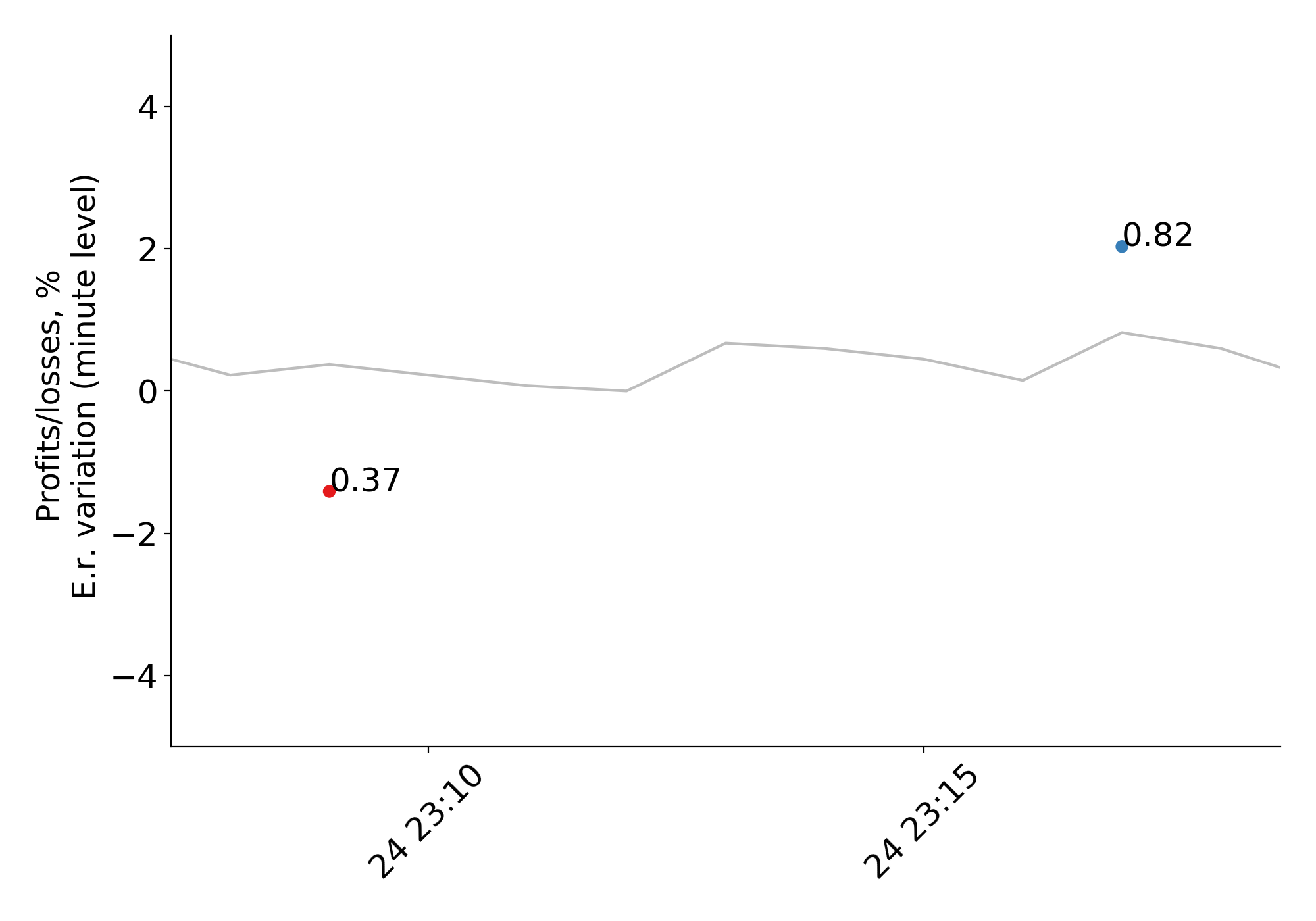}
		\caption{Date: 24 January 2013, H. 23}
		\label{fig:within-d}
	\end{subfigure}
	\begin{subfigure}{0.475\textwidth}
		\includegraphics[width=\textwidth]{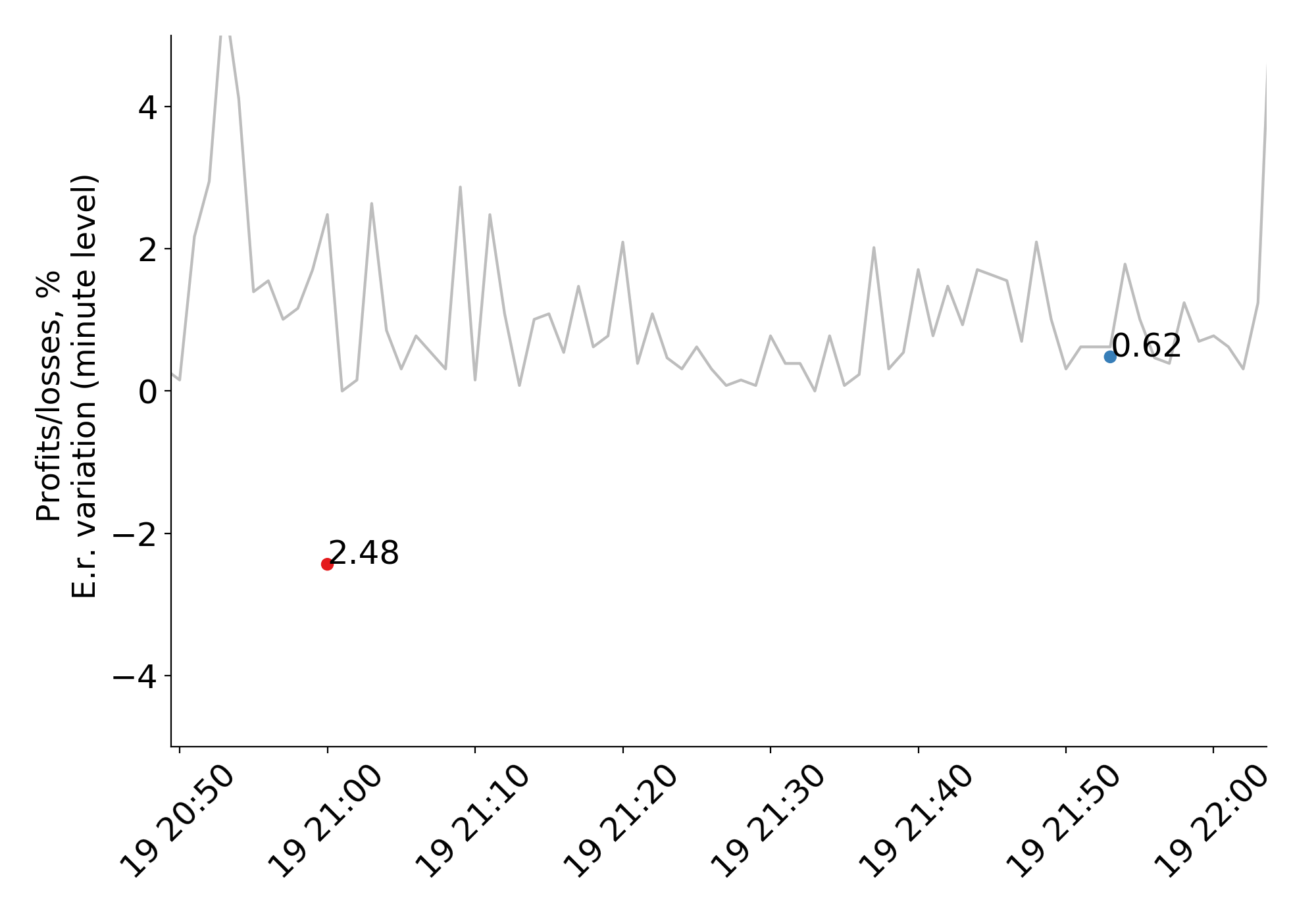}%
		\caption{Date: 19 March 2013, H. 21}
		\label{fig:within-c2}
	\end{subfigure}
	\begin{subfigure}{0.475\textwidth}
		\includegraphics[width=\textwidth]{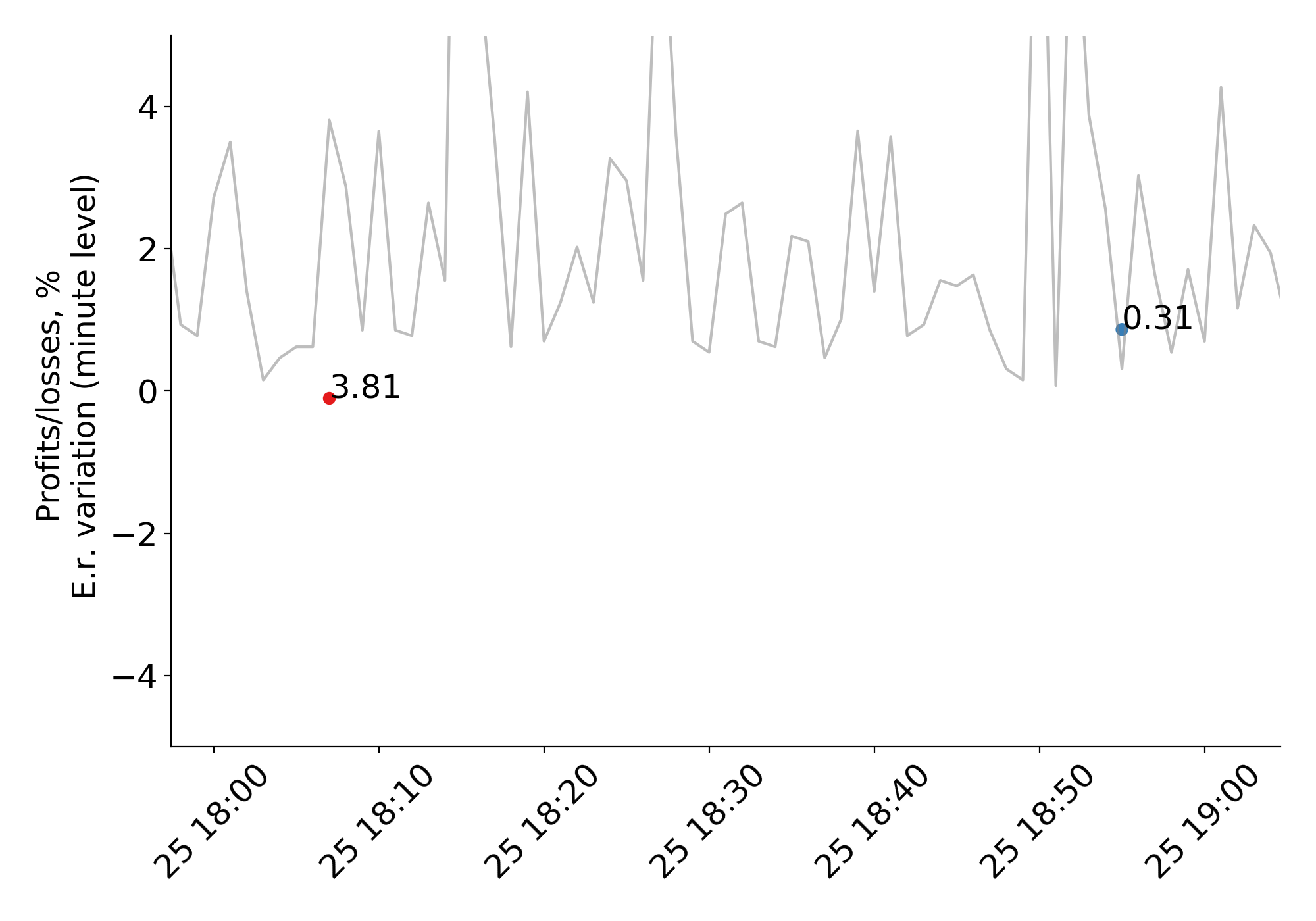}
		\caption{Date: 25 March 2013, H. 18}
		\label{fig:within-d2}
	\end{subfigure}
	\caption{Illustrative examples of within hours dynamics}
	\label{fig:minuterates} 
	\floatfoot{\emph{Notes:} each panel reports one plot for a different time window. The plots include only actions executed in the EUR/USD market. The actions are dots whose y-axis is the profit/loss, and the x-axis is the date of execution. Red ones are executed by non experts, blue ones by experts. The gray line is the percentage variation of the exchange rate with respect to the previous minute (multiplied by 100), and the labels report its value at the minute of execution of each action.}
\end{figure}

The evidence documented thus far suggests that expert users are more likely to make profits on arbitrage relative to non experts.
Why?
In this section we show that differences in profits stem from a better ability of the former in responding to quick fluctuations which makes arbitrage more (or less) profitable. 
In Figure~\ref{fig:minuterates} we provide a graphical intuition behind our argument. We draw the \textit{minute-level} percentage variation of the EUR/USD official rate in four different hour time windows (9 July 2012 h.14, 24 January 2013 h.23, 19 March 2013 h.21, 25 March 2013 h.18 respectively in Panels~\ref{fig:within-c}, \ref{fig:within-d}, \ref{fig:within-c2}, \ref{fig:within-d2}) and the actions placed during the same time frames.
%
Each dot in Figure~\ref{fig:minuterates} is an arbitrage action, whose y-axis is the profit/loss, and the x-axis is the time of execution. 
The value annotated in proximity of each dot is the specific value of the rate variation in that minute, and it is introduced for the ease of comparison. (Note that the value of the rate variation is multiplied by 100.)
The plots show that users react differently in the presence of exchange rate fluctuations, and are more or less able to turn such changes into their favor. Consider for instance Panel~\ref{fig:within-c2}: the expert user conducts arbitrage in a time window with smaller volatility, and its activity is more profitable. A similar pattern appears in Panel~\ref{fig:within-d2}, while in Panels~\ref{fig:within-c} and~\ref{fig:within-d} the differences in volatility are smaller. However, also in these cases, non expert users are less able to exploit in a profitable way the variation of the official rate. 
%

Indeed, a typical profitable situation in financial markets arises when unexpected deviations occur in fundamental values.
Because of structural frictions, adjustments across markets are not automatic, giving rise to opportunities in conducting arbitrage operations.
We exploit this fact in our analysis and reconstruct, from the hourly evolution of the official exchange rate in a market, the percent variation in the exchange rate with respect to the previous hour (see Table~\ref{tab:definitions}). 
Our variable $\Delta R_{p,t}$ takes a higher value when the official exchange rate, on a pair of currencies $p$, observed in the hour $t$, increases more relative to the previous hour.
It is therefore worth remarking that $\Delta R_{p,t}$ varies both across currency markets and time but not within.
%

The advantage in using this strategy is twofold.
First, the exploitation of these temporary opportunities is typically not obvious but requires expertise and/or the execution of automated orders.
Hence, when variation in the exchange rate is relatively prominent than usual times it is likely that expert users take advantage of them and make profits.
Second, as users who trade in Mt.\ Gox are small, their actions are unlikely to affect such deviations.
It is therefore plausible to assume that users are exchange rate \emph{takers} and deviations in the official exchange rate exogenous.

We employ this variable $\Delta R_{p,t}$ on the right-hand side of our regression and interact it with trade ability to test whether profits obtained by expert arbitrageurs are larger when fluctuations in the exchange rates are larger.
This is written as follows:
\begin{equation}\label{reg:2}
\Spread_{i,j,p,t} = \beta_{1} (\text{Trade\,Ability}_{j} \times \Delta R_{p,t}) + \beta_{2} \Delta R_{p,t} + \beta_3\,\text{USD}_{i,j,p,t} + \alpha_{j} + \theta_p + \phi_t + \varepsilon_{i,j,p,t}.
\end{equation}
As one can see, our main variable of interest in Eq. \ref{reg:2} is now time variant.
This allows us to employ a set of user fixed effects, $\alpha_{j}$, which permit to absorb any sort of heterogeneity that one may expect across users.
This includes education, financial literacy, and other unobservables that are likely to correlate with our measure of trade ability.
The inclusion of $\alpha_{j}$ also implies that our chief variation in the identification of $\beta_1$ is the variation across hours within a user.
$\beta_1$ can now be interpreted as the difference in profit, between expert and non expert users, following a 1 per cent increase in the rate of change of the official exchange rate.
$\beta_2$ captures the effect of a 1 per cent increase in the rate of change of the official exchange rate on the arbitrage profits made by non expert users.
These effects are additionally identified by including daily time fixed effects, $\phi_t$, and currency dyad fixed effects, $\theta_p$, and by controlling for the USD equivalent amount of bitcoin traded, $\text{USD}_{i,j,p,t}$.
Standard errors are clustered at user-level as above.


Table~\ref{tab:tab2_alt_x} reports estimations using different proxies of trade ability.
In columns (1-2) we use the variable $D(Currencies)$.
We then repeat the analyses by using alternative proxies, namely $Log(Currencies)$ (3-4), the log of the number of actions (5-6), the execution of metaorders (7-8) or aggressive actions (9-10), and finally the scores from the principal component analysis reported in Table~\ref{tab:pca} (11-12). 
Overall, we find that an increase in the rate of change of the official exchange rate generates a higher profit for arbitrage made by expert users, even when user fixed effects are included (columns 2, 4, 6, 8, 10, 12).
However, $\beta_{1}$ is not statistically significant in columns (2) and (6), perhaps due to the fact that the inclusion of user fixed effects is particularly demanding: 
as we showed in Table~\ref{tab:percentiles}, many of the users active in a single market executed just one action. 
This leads to the exclusion of a large number of observations from this group, making more difficult to obtain stable and statistically significant results.
Finally, the result for column (10) --- relative to the aggressiveness of the actions --- goes against our expectations, but the coefficient is not statistically significant.
%
%
In support of this analysis, we report additional robustness checks in Appendices~\ref{tr:appendix_deltas}, ~\ref{tr:appendix_fees}, and~\ref{tr:appendix_learning} that further validate our results.
 
In summary, these findings indicate that sophisticated investors are able to take into account and exploit in their favor quick changes in the official exchange rate better than non expert users, and that this ability leads to higher profits. 
For example, looking at the effect reported in column~(12) of Table~\ref{tab:tab2_alt_x}, we estimate that an arbitrageur with a trade ability score which is a standard deviation above the mean obtained a profit which is 1.347\% following a 1 per cent increase in the rate of change of the official exchange rate (i.e., $0.476 \times 2.83$); note that this premium accounts for more that a standard deviation of the dependent variable.
Our interpretation, indeed, is that the expert arbitrageurs are more able with respect to the others to react to price deviations, and thus their activity is also more profitable.

In conclusion, thus, we hypothesize that differences in profits stem from the experts' ability in incorporating information in their strategies (perhaps using APIs, and/or automated trading algorithms); conversely, as low sophisticated arbitrageurs do not incorporate this information correctly, their actions are negatively affected by variations in the official exchange rates.
In operative terms, this ability 
in better capturing information results in a greater expertise at the moment of choosing \textit{when} conducting the arbitrage actions: the timing of execution at the micro scale is a salient element which determines a crucial difference between a profitable and a non profitable action. 
Thus, ultimately, the differences between the two groups result in a better choice, time-wise, that yields higher profits to the expert users with respect to the non expert ones.

%% file: tr_discussion.tex
\section{Discussion} \label{tr:discussion}

The findings in the behavioral finance literature challenge the conventional economic interpretation of theoretical arbitrage that would foresee, in the presence of risk, the intervention of many small traders with homogeneous expectations, not subject to capital constraints, and risk-neutral towards a small enough exposure on the market.
Practitioners are well aware that this description is far from reality: arbitrageurs are few, sophisticated, and specialized traders. 
However, the evidence of this statement still remains anecdotal, and no empirical study exists to describe who are the arbitrageurs.
In this work we investigate the Mt.\ Gox leaked dataset,
in order to provide empirical evidence that some of its customers indeed conducted triangular arbitrage activity, as well as an explanation to the observed dynamics. By these means, we try and answer to some open questions on the nature of arbitrage, and on the users who conduct it.

First, we identify the set of potential arbitrage  actions, consisting of about 6,600 pairs of buy and sell legs, providing evidence that arbitrage was indeed conducted within the Mt.\ Gox platform.
Coherently with the theory, in the vast majority of cases these pairs are almost simultaneous and involve the same or nearly equivalent security (in terms of volumes of bitcoins moved).
The most interesting results concern the characteristics of the investors involved in arbitrage. Out of more than 70,000 users that traded through the Mt.\ Gox platform in the time window considered, only less than 4,000 traded bitcoins against more than one currency, thus greatly reducing the set of potential arbitrageurs, and just 440 of them are responsible for at least one detected arbitrage action. Major differences across users arise even within this group, if we account for indicators approximating the user expertise. In particular, we focus on the distinction between the arbitrageurs who performed arbitrage only on a single currency market and those who conducted it on multiple currency markets, assuming that it is a good proxy of a trader's expertise: all the members of the former group performed few actions, while the latter, even though the minority, are responsible for the vast majority of actions. 
They often follow complex trading patterns, including strategies that entail splitting orders into smaller ones to reduce market impact, and operate taking into account the penalizing effect that an excessive aggressiveness would have on their trades. 
Indeed, our findings are consistent with a scenario in which many uninformed traders move prices far from the fundamental value, and the arbitrageurs absorb the demand shocks by providing liquidity on the market.

We thus devise a model to quantify the effect of the ability on the profitability of the arbitrage actions: the results show that the actions executed by skilled users are significantly more profitable, while those performed by the non skilled ones are on average non profitable when transaction costs are included. Graphic support shows that the distribution in time of the arbitrage actions tends to be different across the two groups. Most of all, the differential effect between non expert and expert users in the presence of movements of the official exchange rate is positive and statistically significant, even after including user fixed effects: the latter are more responsive to exogenous shocks on the official exchange rate.
We hypothesize that these differences across the two groups of users are chiefly due to a better ability of the latter in incorporating the available information within the trading strategies, perhaps by exploiting automated algorithms and dedicated APIs. Indeed, the sophisticated users take into account the penalising effect of the transaction costs in a more systematic way, and they show greater ability in exploiting movements of the official exchange rates. Ultimately, this translates to a better choice in terms of timing execution of the arbitrage activity also at a small scale level.

In summary, our findings clearly support the thesis that arbitrageurs are few: we identify $ N = 440 $ users who conducted at least one arbitrage action over a total of $ N = 71,808 $ Mt.\ Gox users.
Furthermore, the non skilled arbitrageurs executed only a small fraction of the total arbitrage actions ($ N = 723 $ against $ N = 5,906 $), despite being a larger group ($ N = 395$ against $ N = 45$), and on average such activity is non profitable. 
%
%
On the contrary, skilled arbitrageurs execute many arbitrage actions, they accept to incur potential losses on individual actions, but overall their trading strategy systematically yields profits.
As the essential property of arbitrage is the \textit{advantageous} exploitation of the mispricing of an asset, our findings support the claim that arbitrage is performed by a small number of professional investors.

%
%

External validity, however, might be a concern, even more so in the light of the peculiarities of the Bitcoin and Mt.\ Gox ecosystem. For instance, according to the model described by \citet{shleifer1997}, in the conventional financial markets the arbitrageurs exploit their knowledge operating on someone else's funds, thus taking large positions in an agency relationship, where the interplay of capital and risk plays a prominent role. In the Mt.\ Gox market, the average arbitrage action is small (less than a hundred of dollars): we hypothesize that in this context the agency relationship does not take place, and the arbitrageurs are not operating on behalf of someone else. Rather, they are likely investing their own - more limited - funds. This difference could be explained by the fact that we are considering a niche market at its early stages: at the time, the bitcoin market was relatively unknown to the major financial investors. An alternative explanation for the small dimension of the average trade is that expert users implemented also complex strategies involving splitting orders.

The recent findings by \citet{wang2021cyclic} partially mitigate the potential concerns on external validity. 
The authors study cyclic arbitrage across decentralized exchanges (DEXs) for the second most prominent cryptocurrency after Bitcoin, that is, Ethereum. 
This framework shows relevant differences with respect to the one we analyze: the Ethereum protocol makes intensively use of smart contracts, and the investigation was conducted on a very different time epoch, i.e. from May 2020 to January 2021. They observe almost 300,000 arbitrage cycles, another remarkable difference (and a sign that the cryptocurrency market has matured with time).
%
Most interestingly for our purposes, they notice relevant differences between two groups of users, those who exploit private smart contracts and those who conduct arbitrage using public protocols, the former experimenting a much higher success rate (62\% against 28.4\%). Creating private smart contracts is complex and requires deep knowledge of the Ethereum ecosystem, in order not to incur large losses: this finding, obtained for a different cryptocurrency and in a very different epoch, is strictly related to our study and consistent with it, thus further supporting the validity of our investigation.

Some other limitations to this work stand out: a valid objection is that all these actions could be false positives. It is impossible to prove undoubtedly that all of them are arbitrage actions, unless a direct proof is provided by the Mt.\ Gox users (we thus encourage the Mt. Gox arbitrageurs among the readers to contact the authors, and share their comments). However, some major elements emerge in support of our hypothesis. First, it is hard to find alternative explanations to the existence of several pairs of trades executed according to patterns so specific such as those we described throughout the paper. Second, we explicitly consider several variants and controls in our regressions, in order to rule out alternative hypotheses. Third, the fact that many actions - predominantly profitable - are ascribable to few large users is in our opinion another meaningful factor.

Another limitation is that we implicitly assume that we are providing an `upper bound' to the total triangular arbitrage activity: assuming that the identified actions are true positives, then our algorithm ideally detects all triangular arbitrage activity. However, this is likely not true: it is possible that some arbitrage actions have a more complex structure than just being composed of two buy/sell legs (e.g., they might involve many cycles and/or several currencies), and that the aggregation method described in Section~\ref{tr:background} to emulate the Mt.\ Gox internal matching mechanism is not thoroughly accurate. Thus, even though we believe our procedure is a good approximation, we acknowledge that we might be estimating imperfectly the total number of arbitrage actions. 
Furthermore, we select the arbitrage actions in a reasonably small neighborhood of time and volume, but we do not ground our choice on a theoretical support; rather, we base it on empirical evidence. For this reason, we adopt a conservative scenario with larger boundaries, and we report additional estimations both on a less conservative scenario and on a case with even larger thresholds in Appendix~\ref{tr:appendix_deltas}. 
We do not investigate the effect of the arbitrage activity in enforcing the `law of one price': an interesting development of the current work would be to inspect whether, and to what extent, the arbitrage activity had an effect on the market efficiency.
%

Finally, the analysis focuses on a single exchange platform: on the one side it is an advantage, as all information required to detect arbitrage activity is entirely contained in the private ledger. On the other side, triangular arbitrage only aligns prices in one market, whereas an essential function of arbitrage is its function of `information carrier' across markets. So, while the evidence might be stronger for triangular arbitrage within the same market, this result may miss generality.

\newpage

%% file: tr_tables.tex
\section*{Tables}
\label{tr:tables}

{\small
\begin{xltabular}{\textwidth}{lX}
    \caption{Definitions}
    \label{tab:definitions} \\
    \toprule
    Variable & Description/formalization \\
    \midrule
    \makecell[l]{\\ Arbitrage \\ Action} & Action composed of two legs executed by the same user in different trades using different currencies. The time delay and volume difference cannot exceed a threshold $[\Delta T, \Delta Q]$. In Each arbitrage action is obtained by merging a buy and sell leg.
    \begin{equation}
        Arbitrage Action = (Leg_{Buy},Leg_{Sell})
    \end{equation}
    \\
    $ \Delta T$ & Maximum time delay allowed (e.g., 300 seconds in the baseline analysis)
    \\[1ex]
    $ \Delta Q$ & Maximum volume difference allowed (e.g., 10\% in the baseline analysis)
    \\[4ex]
    $ \delta T$ & Time delay between $ Leg_{B} $  and $ Leg_{S} $, expressed in seconds. By definition smaller or equal to $ \Delta T$:
    \begin{equation}
         \delta T = |T_{B} - T_{S}| \leq  \Delta T
    \end{equation}
    \\
    $ \delta Q$ &  Volume difference (Bitcoins traded) between $ Leg_{B} $  and $ Leg_{S} $, expressed as a percentage. 
    By definition smaller or equal to $ \Delta Q$:
    \begin{equation}
        \delta Q = \frac{|Vol_{B} - Vol_{S}|}{(Vol_{B} + Vol_{S}) / 2}\cdot 100 \leq  \Delta Q
    \end{equation}
    \\[10ex]
    \makecell[l]{Official \\ exchange rate} &  By convention, each arbitrage action is compared to the official exchange rate in the following way: 
    \begin{equation}
        OffER = CUR_{B}toCUR_{S}
    \end{equation}
    that is, if the buy leg of an arbitrage action is performed in EUR and the Sell one is in USD, then we consider the official exchange rate EURtoUSD. If the Buy side is in USD, and the Sell one in EUR, then it is compared to the e.r. USDtoEUR.
    \\[23ex]
    $\Delta R$ & Hourly variation of the official exchange rate expressed as a percentage:
    \begin{equation}
        \Delta R = \frac{|OffER_{t_{1}} - OffER_{t_{0}}|}{OffER_{t_{0}}} \cdot 100
    \end{equation}
    \\
    Dyad & Pair of currencies that defines the fiat-to-fiat currency market to which the arbitrage action belongs. E.g., the dyad (EUR,USD) refers to actions whose currencies are $CUR_{B}$ = EUR and $CUR_{S}$ = USD or viceversa (as they `refer' to the same currency market).
    \\[5ex]
    \makecell[l]{\\ Implied \\ exchange rate}  &  The implied exchange rate is calculated by comparing the price of bitcoins in the two legs. The latter row includes fees.
    \begin{equation}
        ImpER =
        \begin{cases}
            \myfrac[5pt]{Fiat_{S}}{BTC_{S}}\cdot\myfrac[5pt]{BTC_{B}}{Fiat_{B}} \text{\hspace{26ex} without fees} \\[4ex]
            \myfrac[5pt]{Fiat_{S}-Fee_{f,S}}{BTC_{S} + Fee_{b,S}} \cdot \myfrac[5pt]{BTC_{B} + Fee_{f,B}}{Fiat_{B}-Fee_{b,B}} \text{ \hspace{6ex}  with fees}
        \end{cases}
    \end{equation}
    The pedices B and S refer to the buy and sell side; $f$ and $b$ indicate respectively if the term $Fee$ is denominated in fiat or in bitcoins.
    \\ [34ex]
    Profit (Spread) &  Spread between the implied and the official rate divided by the official rate, expressed as a percentage. By construction, profits arise when ImpER $>$ OffER.
    \begin{equation}
        Spread = \frac{ImpER - OffER}{OffER} \cdot 100
    \end{equation} 
    %
    \\  [1ex]
    Metaorder & Metaorders are identified as sequences of at least 5 arbitrage actions executed by the same user, in the same market, and such that the time passed between each action is less than one minute. \emph{Note}: we partly follow the methodology described in \citet{donier2015million}, with some differences: the authors consider a larger time delta (one hour) bewteen each action, and contrary to them we use an arbitrary parameter (N=5) to define the minimum length of a metaorder. While we do not provide the results here, we varied the two thresholds and noticed that the differences are negligible for our purposes, and this setting reduces the false positives classification. 
    \\ [30ex]
    Aggressive & Arbitrage action composed by at least one aggressive leg (that is, a leg that initiated a market order). \\ [5ex]
    Equiv. \$ & Value of a trade expressed in dollars. We use this variable to indicate the value of a trade since the bitcoin value is not stable in time.\\
    \bottomrule
\end{xltabular}
}

\begin{table}[H]
    \centering
    \begin{tabular}{lrrrrrrr}
    \toprule
    \multicolumn{8}{c}{\textbf{Panel A: all arbitrage actions (N = 6629)}}\\
    \midrule
         &      Mean &     St.D. &     Min &       25\% &       50\% &       75\% &       Max \\
    \midrule
    Profits, fees, \% &      0.42 &      1.26 &  -11.35 &      0.075 &      0.621 &      1.096 &     18.16 \\
     P., exp. fees, \% &      0.28 &      1.22 &   -7.46 &     -0.191 &      0.375 &      0.982 &     18.24 \\
    P., no fees, \% &      1.05 &      1.21 &   -6.40 &      0.490 &      1.110 &      1.696 &     19.60 \\
    Bitcoins &      4.12 &     12.56 &    0.00 &      0.039 &      0.807 &      3.261 &    334.14 \\
    'Equiv. \$' &     52.54 &    169.63 &    0.00 &      0.359 &      7.400 &     41.424 &   4666.66 \\
    $ \delta T $ (s) &     29.04 &     59.09 &    0 &      0 &      1 &     24 &    300 \\
    $ \delta Q $ (\%) &      1.30 &      2.46 &    0.00 &      0.000 &      0.215 &      0.863 &      9.99 \\
    \midrule
    \multicolumn{8}{c}{\textbf{Panel B: actions of users who exploited single markets (N = 723)}}\\
    \midrule
     Profits, fees, \% &     -1.00 &      1.96 &  -11.35 &     -2.191 &     -0.933 &      0.118 &     18.16 \\
      P., exp. fees, \%p &     -0.95 &      1.91 &   -7.46 &     -2.161 &     -0.891 &      0.172 &     18.24 \\
     P., no fees, \% &      0.11 &      1.91 &   -6.40 &     -1.135 &      0.134 &      1.275 &     19.60 \\
    Bitcoins &      7.89 &     21.16 &    0.00 &      0.253 &      2.000 &      7.472 &    288.35 \\
    'Equiv. \$' &    118.06 &    340.25 &    0.00 &      4.014 &     27.299 &     95.708 &   4666.66 \\
    $ \delta T $ (s) &     59.95 &     68.21 &    0 &     13 &     34 &     86 &    297 \\
    $ \delta Q $ (\%) &      1.04 &      1.77 &    0.00 &      0.461 &      0.602 &      0.602 &      9.82 \\
    \midrule
    \multicolumn{8}{c}{\textbf{Panel C: actions of users who exploited multiple markets (N = 5906)}}\\
    \midrule
   Profits, fees, \% &      0.59 &      1.02 &   -7.40 &      0.208 &      0.688 &      1.128 &     10.13 \\
      P., exp. fees, \%p &      0.42 &      1.02 &   -7.34 &     -0.009 &      0.448 &      1.019 &     10.15 \\
     P., no fees, \% &      1.16 &      1.04 &   -6.28 &      0.577 &      1.178 &      1.719 &     10.79 \\
    Bitcoins &      3.66 &     10.97 &    0.00 &      0.030 &      0.606 &      2.995 &    334.14 \\
    'Equiv. \$' &     44.52 &    132.48 &    0.00 &      0.318 &      5.767 &     35.087 &   3862.71 \\
    $ \delta T $ (s) &     25.26 &     56.74 &    0 &      0 &      1 &     16 &    300 \\
    $ \delta Q $ (\%) &      1.34 &      2.53 &    0.00 &      0.000 &      0.000 &      0.928 &      9.99 \\
    \bottomrule
    \end{tabular}
    \caption{Descriptive statistics of the arbitrage actions} 
	\label{tab:triangarb_all}
	\floatfoot{\emph{Notes:} actions identified at $\Delta T$ = 300s and $\Delta Q$ = 10\%. Panel A describes the main features of all the arbitrage actions, while Panel B reports the statistics for the subset of actions (N = 723) executed by users that performed arbitrage in a single currency market. Panel C refers to those executed by investors active in multiple markets (N = 5,906).}
\end{table}

\begin{table}[H]
    \centering
    \begin{tabular}{lrrrrrrrrr}
    \toprule
    {} &  Mean &  Std &  Min &  25\% &  50\% &  75\% &  90\% &  95\% &   Max \\
    \midrule
    Group \textit{Single} (N = 395)  &     1 &    2 &    1 &    1 &    1 &    2 &    2 &    5 &    27 \\
    Group \textit{Multiple} (N = 45) &   131 &  366 &    2 &    4 &   11 &   28 &  392 &  690 &  2175 \\
    \bottomrule
    \end{tabular}
    \caption{Statistics on the number of actions executed by the arbitrageurs}
	\label{tab:percentiles}
	\floatfoot{\emph{Notes:} we split the users in two groups, that is, those who performed arbitrage on a \textit{Single} and on \textit{Multiple} markets. The statistics describe the mean, standard deviation, minimum, maximum, and percentiles of the number of actions performed by the two subgroups of users. Note that, by construction, the users in the group \textit{Multiple} performed at least two arbitrage actions; thus, they are involved in at least four trades. Similarly, users in the group \textit{Single} conducted at least two trades.
	}
\end{table}

\begin{table}[H]
    \centering
    {\small
    \begin{tabular}{lcccccc}
    \toprule
    {} &  Percentage &  \makecell[c]{Number of \\ metaorders} &  \makecell[c]{Avg. \\ length} &  \makecell[c]{Avg. \\ time delay} &  \makecell[c]{Avg. \\ Bitcoins} &  \makecell[c]{Avg. \\ Equiv. dollars} \\
    \midrule
    18X   &       54.07 &               91 &        12.92 &            13.33 &          52.54 &               369.38 \\
    1245X &       80.00 &                2 &         6.00 &            23.83 &           7.43 &                97.73 \\
    1964X &       44.10 &               11 &         7.82 &            26.73 &          35.81 &               178.17 \\
    2173X &       18.52 &                1 &         5.00 &            14.00 &          40.00 &               234.55 \\
    2286X &       35.71 &                1 &         5.00 &            26.25 &           5.00 &               297.55 \\
    2717X &        3.45 &                1 &         5.00 &            47.00 &           0.59 &                 6.45 \\
    2940X &       91.30 &                1 &        21.00 &            17.75 &           2.36 &                25.80 \\
    3174X &       63.28 &               40 &        10.60 &            29.22 &          30.81 &               346.89 \\
    4156X &       29.00 &                7 &         8.29 &            28.46 &           9.97 &                70.54 \\
    4325X &       22.73 &                1 &         5.00 &            29.50 &          55.00 &              1118.86 \\
    4901X &       56.06 &                1 &        37.00 &            11.36 &          16.55 &               162.88 \\
    5121X &       29.40 &               26 &         9.00 &            15.51 &           1.32 &                35.86 \\
    6688X &       20.97 &                2 &         6.50 &            20.07 &           7.36 &               242.74 \\
    \bottomrule
    \end{tabular}
    }
    \caption{Arbitrage actions executed via metaorders, descriptive statistics}
    \floatfoot{\emph{Notes:} for each user (rows), we identify the sequences of actions with the characteristics of metaorders. Only the 13 users reported here performed metaorders. \textit{Percentage} indicates the number of actions that are part of metaorders over the total number of arbitrage actions executed by the user; the second column represents the number of metaorders identified. The other columns describe average values on the metaorders executed by each user and respectively report the average number of actions that compose a metaorder, the average time delay between the actions in the same metaorder, the mean volume of a metaorder expressed in dollars and in bitcoins. User identifiers are anonymized.}
    \label{tab:metaorders}
\end{table}

\begin{table}[H]
    \centering
    \begin{tabular}{lrrrrrrr}
    \toprule
     &   Mean &  St.d. &    Min &    25\% &   50\% &   75\% &     Max \\
    \midrule
    Arbitrage actions (N)&  6.572 &  8.223 &  1.000 &  1.000 &  2.00 &  11.0 &  28.000 \\
    Spread  (\%)   & -1.106 &  1.434 & -5.354 & -2.13 & -0.975 &  -0.058 &   2.243 \\
    $Currencies_{d}$ (dummy)   &  0.278 &  0.449 &  0.000 &  0.000 &  0.00 &   1.0 &   1.000 \\
    \bottomrule
    \end{tabular}
    \caption{Descriptive statistics of the aggressive arbitrage actions (N = 313)}
    \floatfoot{\emph{Notes:} out of N = 6,629 arbitrage actions, just N = 313 are aggressive actions, that is, arbitrage actions in which at least one of the two legs of the arbitrage action is an aggressive order. They are executed by users who performed few arbitrage actions ($1^{st}$ row: 6.57 on average, and maximum 28); on average they are not profitable ($2^{nd}$ row), and they are executed primarily by users active only on single markets ($3^{rd}$ row).}
    \label{tab:aggressive_orders}
\end{table}

\begin{table}[H]
    \centering
    \caption{Principal Component analysis}
    \label{tab:pca}
    \begin{tabular}{lrrrrr}
        \toprule
        {} &  $D(Currencies)$ &  $Log(Actions)$ &  $D(Metaorder)$ &  $D(Aggressive)$ &  Expl. variance, \% \\
        \midrule
        PC1 &              0.54 &             0.63 &             0.53 &             -0.15 &               53.38 \\
        \bottomrule
    \end{tabular}
    \floatfoot{\emph{Notes:} the sample is the set of users (N = 440). We consider the four main indicators that we exploit in Section~\ref{tr:results}: the dummy variable that classifies users who exploited single or multiple markets, $D(Currencies)$; the logarithm of the actions executed, $Log(Actions)$; whether the action is part of a metaorder ($D(Metaorder)$) or aggressive ($D(Aggressive)$). Values are standardized by constructing their z-score. 
    The table shows the loadings of all the variables (columns) for the first principal components (row). The last column reports the explained variance of the component.}
\end{table}

\begin{table}[H]
    \centering
    \caption{Relationship between trade ability and profits} 
    \label{tab:tab1_base}
    \begin{tabular}{l*{4}{c}}
\toprule
Dep. var.:        & \multicolumn{4}{c}{Spread (with fees)} \\
\cmidrule{2-5}
            &\multicolumn{1}{c}{(1)}         &\multicolumn{1}{c}{(2)}         &\multicolumn{1}{c}{(3)}         &\multicolumn{1}{c}{(4)}         \\
\midrule
D(Currencies)&      1.6180\sym{***}&      1.5791\sym{***}&      1.2421\sym{***}&      1.2917\sym{***}\\
            &    (0.1900)         &    (0.1943)         &    (0.1584)         &    (0.1659)         \\
[1em]
Equiv. \$   &      3.4652\sym{**} &      2.6151\sym{*}  &      0.6556         &      0.1912         \\
            &    (1.7532)         &    (1.5862)         &    (1.2465)         &    (1.1280)         \\
[1em]
Constant    &     -1.0420\sym{***}&     -0.9985\sym{***}&     -0.6141\sym{***}&     -0.6506\sym{***}\\
            &    (0.1593)         &    (0.1775)         &    (0.1500)         &    (0.1604)         \\
\midrule
Time FE     &           N         &           N         &           Y         &           Y         \\
Dyad FE     &           N         &           Y         &           N         &           Y         \\
N           &        6594         &        6582         &        5307         &        5284         \\
R-squared   &        0.16         &        0.20         &        0.68         &        0.69         \\
    \bottomrule
    \end{tabular}
    \floatfoot{\emph{Notes}: the Table reports OLS estimates of the relationship between the dependent variable \textit{Spread}, that captures the profitability of an arbitrage action, and the variable D(Currencies), which is a proxy of the user trade ability, equal to 1 if the user conducted arbitrage in multiple markets, and 0 otherwise. We consider four different specifications of the model: (1) without including fixed effects, (2) with dyad fixed effects, (3) with time fixed effects, (4) with both. All columns include an additional control for the amount of volume traded, expressed in USD (and divided by 10,000). We report only the overall $R^2$. Errors are clustered at the user-level to account for intra-class correlation. Standard errors are reported in parentheses.  \sym{*} \(p<0.1\), \sym{**} \(p<0.05\), \sym{***} \(p<0.01\)}
\end{table}

\begin{table}[H]
    \centering
    \caption{Relationship between trade ability and profits, alternative proxies} 
    \label{tab:tab1_alt_x}
    \begin{tabular}{l*{6}{c}}
    \toprule
Dep. var.:        & \multicolumn{6}{c}{Spread (with fees)} \\
\cmidrule{2-7}
&\multicolumn{1}{c}{(1)}         &\multicolumn{1}{c}{(2)}         &\multicolumn{1}{c}{(3)}         &\multicolumn{1}{c}{(4)}                &\multicolumn{1}{c}{(5)}    &\multicolumn{1}{c}{(6)}  
                \\
    \midrule
D(Currencies)&      1.2917\sym{***}&                     &                     &                     &                     &                     \\
            &    (0.1659)         &                     &                     &                     &                     &                     \\
Log(Currencies)&                     &      0.9326\sym{**} &                     &                     &                     &                     \\
            &                     &    (0.4439)         &                     &                     &                     &                     \\
Log(Actions)&                     &                     &      0.3165\sym{***}&                     &                     &                     \\
            &                     &                     &    (0.0627)         &                     &                     &                     \\
D(Metaorder)&                     &                     &                     &      0.2877         &                     &                     \\
            &                     &                     &                     &    (0.1914)         &                     &                     \\
D(Aggressive)&                     &                     &                     &                     &     -1.5280\sym{***}&                     \\
            &                     &                     &                     &                     &    (0.1796)         &                     \\
PC1         &                     &                     &                     &                     &                     &      0.2242\sym{***}\\
            &                     &                     &                     &                     &                     &    (0.0466)         \\
Equiv. \$   &      0.1912         &      0.1525         &      1.2060         &      0.0733         &      0.3375         &      0.8175         \\
            &    (1.1280)         &    (1.2621)         &    (1.3191)         &    (1.3384)         &    (1.1244)         &    (1.3500)         \\
[1em]
Constant    &     -0.6506\sym{***}&     -1.0453         &     -1.4603\sym{***}&      0.3359\sym{**} &      0.6113\sym{***}&     -1.0717\sym{***}\\
            &    (0.1604)         &    (0.7807)         &    (0.4143)         &    (0.1614)         &    (0.0424)         &    (0.3593)         \\
\hline
Time FE     &           Y         &           Y         &           Y         &           Y         &           Y         &           Y         \\
Dyad FE     &           Y         &           Y         &           Y         &           Y         &           Y         &           Y         \\
N           &        5284         &        5284         &        5284         &        5284         &        5284         &        5284         \\
R-squared   &        0.69         &        0.69         &        0.72         &        0.67         &        0.69         &        0.70         \\
    \bottomrule
    \end{tabular}
    \floatfoot{\emph{Notes}: the Table reports OLS estimates of the relationship between the dependent variable \textit{Spread} and alternative proxies of the user trade ability: (1) D(Currencies) provides a baseline reference by repeating column (4) of Table~\ref{tab:tab1_base}; (2) Log(Currencies) is the logarithm of the number of currency markets exploited by the user; (3) Log(Actions) is the logarithm of the number of arbitrage actions executed by the user; (4) and (5), D(Metaorder) and D(Aggressive), are respectively dummy variables that indicate whether the user conducted metaorders or aggressive actions. (6) PC1 is the score of each variable obtained by performing a PC analysis as explained in Table~\ref{tab:pca}. All columns include time and dyad fixed effects, as well as an additional control for the amount of volume traded, expressed in USD (and divided by 10,000). We report only the overall $R^2$. Errors are clustered at the user-level to account for intra-class correlation.  Standard errors are reported in parentheses.  \sym{*} \(p<0.1\), \sym{**} \(p<0.05\), \sym{***} \(p<0.01\)}
\end{table}

\begin{landscape}
\begin{table}[H]
    \centering
    \vspace{-0.3cm}
    \caption{Responsiveness to official rate variations}
    \label{tab:tab2_alt_x}
    {\footnotesize
\begin{tabular}{l*{12}{c}}
\toprule
Dep. var.:        & \multicolumn{12}{c}{Spread (with fees)} \\
\cmidrule{2-13}
            &\multicolumn{1}{c}{(1)}         &\multicolumn{1}{c}{(2)}         &\multicolumn{1}{c}{(3)}         &\multicolumn{1}{c}{(4)}         &\multicolumn{1}{c}{(5)}         &\multicolumn{1}{c}{(6)}         &\multicolumn{1}{c}{(7)}         &\multicolumn{1}{c}{(8)}         &\multicolumn{1}{c}{(9)}         &\multicolumn{1}{c}{(10)}         &\multicolumn{1}{c}{(11)}         &\multicolumn{1}{c}{(12)}         \\
\midrule
$ \Delta R \times $D(Currencies)  &       6.905\sym{***}&       1.442         &                     &                     &                     &                     &                     &                     &                     &                     &                     &                     \\[-0.3em]
            &     (1.139)         &     (5.314)         &                     &                     &                     &                     &                     &                     &                     &                     &                     &                     \\[-0.3em]
$ \Delta R \times $Log(Currencies) &                     &                     &       3.213\sym{**} &       1.703\sym{**} &                     &                     &                     &                     &                     &                     &                     &                     \\[-0.3em]
            &                     &                     &     (1.528)         &     (0.723)         &                     &                     &                     &                     &                     &                     &                     &                     \\[-0.3em]
$ \Delta R \times $Log(Actions)&                     &                     &                     &                     &       1.427\sym{***}&       0.150         &                     &                     &                     &                     &                     &                     \\[-0.3em]
            &                     &                     &                     &                     &     (0.202)         &     (0.409)         &                     &                     &                     &                     &                     &                     \\[-0.3em]
$ \Delta R \times $D(Metaorder)&                     &                     &                     &                     &                     &                     &       3.693\sym{**} &       2.228\sym{***}&                     &                     &                     &                     \\[-0.3em]
            &                     &                     &                     &                     &                     &                     &     (1.600)         &     (0.821)         &                     &                     &                     &                     \\[-0.3em]
$ \Delta R \times $D(Aggressive)&                     &                     &                     &                     &                     &                     &                     &                     &      -6.786\sym{***}&       3.593         &                     &                     \\[-0.3em]
            &                     &                     &                     &                     &                     &                     &                     &                     &     (1.595)         &     (3.461)         &                     &                     \\[-0.3em]
$ \Delta R \times $PC1    &                     &                     &                     &                     &                     &                     &                     &                     &                     &                     &       1.048\sym{***}&       0.476\sym{**} \\[-0.3em]
            &                     &                     &                     &                     &                     &                     &                     &                     &                     &                     &     (0.159)         &     (0.211)        \\
$ \Delta R $     &      -5.441\sym{***}&      -0.742         &      -4.744\sym{**} &      -2.426         &      -7.060\sym{***}&      -0.258         &      -2.341\sym{*}  &      -1.045         &       0.168         &       0.617         &      -5.791\sym{***}&      -2.810\sym{*}  \\
            &     (1.196)         &     (5.279)         &     (2.125)         &     (1.590)         &     (1.320)         &     (2.423)         &     (1.241)         &     (0.854)         &     (0.526)         &     (0.956)         &     (1.187)         &     (1.546)         \\
Equiv. \$   &       1.150         &      -0.732         &       0.732         &      -0.861         &       2.128         &      -0.751         &       0.766         &      -0.804         &       0.824         &      -0.741         &       2.225         &      -0.812         \\
            &     (1.700)         &     (0.833)         &     (1.762)         &     (0.878)         &     (1.815)         &     (0.827)         &     (1.647)         &     (0.871)         &     (1.896)         &     (0.838)         &     (1.800)         &     (0.866)         \\
\hline
User FE     &           N         &           Y         &           N         &           Y         &           N         &           Y         &           N         &           Y         &           N         &           Y         &           N         &           Y         \\
Time FE     &           N         &           Y         &           N         &           Y         &           N         &           Y         &           N         &           Y         &           N         &           Y         &           N         &           Y        \\
Dyad FE     &           N         &           Y         &           N         &           Y         &           N         &           Y         &           N         &           Y         &           N         &           Y         &           N         &           Y         \\
N           &        6594         &        5142         &        6594         &        5142         &        6594         &        5142         &        6594         &        5142         &        6594         &        5142         &        6594         &        5142         \\
R-squared   &        0.05         &        0.75         &        0.02         &        0.75         &        0.07         &        0.75         &        0.02         &        0.75         &        0.02         &        0.75         &        0.06         &        0.75         \\
\bottomrule
\end{tabular}}
\floatfoot{\scriptsize \emph{Notes}: the Table describes the responsiveness to variations of the official exchange rate for the main proxies of trade ability, and their effect on profits. It reports OLS estimates for 12 different specifications, each including an interaction term between the official rate variation and a proxy of trade ability: D(Currencies) in (1-2), Log(Currencies) in (3-4), Log(Actions) in (5-6), D(Metaorder) in (7-8), D(Aggressive) in (9-10), PC1 in (11-12).
The first column of each alternative proxy is without fixed effects, while the second includes time, dyad and user fixed effects. All columns include a control for the amount of volume traded, expressed in USD (and divided by 10,000), and for the official rate variation. We report only the overall $R^2$. Errors are clustered at the user-level to account for intra-class correlation.  Standard errors are reported in parentheses.  \sym{*} \(p<0.1\), \sym{**} \(p<0.05\), \sym{***} \(p<0.01\)}
\end{table}
\end{landscape}

%% file: tr_appendix_deltas.tex
\section{Additional robustness checks on $\Delta T$ and $\Delta Q$}
\label{tr:appendix_deltas}

\setcounter{figure}{0}
\setcounter{table}{0}


As described in Section~\ref{tr:identification}, the algorithm implemented to identify the arbitrage actions compares the legs executed by the same user in a small neighborhood of time and volume, defined by the parameters $ \Delta T $ and $ \Delta Q $. 
In this Appendix we show that the results of our estimations do not vary significantly when we consider larger or smaller sets of identified arbitrage actions --- i.e., if we modify the boundaries for the time delay and volume difference by varying $\Delta T$ and $\Delta Q$.

All the results in the following always include dyad and hourly fixed effects (and, when allowed by the model, user fixed effects).
Table~\ref{tab:app_tab1} repeats the analysis in Table~\ref{tab:tab1_base} by imposing $\Delta T = 30s$ and $\Delta Q = 1\%$ in columns (1-2-3), and $\Delta T = 600s$ and $\Delta Q = 20\%$ in columns (4-5-6). The dependent variable is with fees (1-4), without fees (2-5), and with expected fees (3-6) --- further details on this are discussed in Appendix~\ref{tr:appendix_fees}. The results hold whatever the specification. Interestingly, the $R^2$ is higher for smaller $\Delta T$ and $\Delta Q$, letting hypothesize that larger boundaries lead to the inclusion of a higher amount of false positives.
Table~\ref{tab:app_tab1_alt_xdeltas} shows that similar considerations apply also when considering alternative specifications of the expertise.

Interesting findings are reported in Table~\ref{tab:app_tab2deltas}, that replicates Table~\ref{tab:tab2_alt_x} for $\Delta T = 30s$ and $\Delta Q = 1\%$ in columns (1) to (6), and $\Delta T = 600s$ and $\Delta Q = 20\%$ in (7) to (12). The results strengthen the intuition discussed above that larger boundaries reduce the statistical power of the model. Indeed, though the overall pattern is confirmed, the results for $\Delta T = 600s$ and $\Delta Q = 20\%$ are often not statistically significant and the $R^2$ is smaller. Noteworthy, instead, columns (1) to (6) provide even better results with respect to those reported in the main analysis (with $\Delta T = 300s$ and $\Delta Q = 10\%$), both considering the $\beta_1$ coefficients, and the $R^2$.

These results are especially important, as they demonstrate that the findings described in the main body of the paper are not circumscribed to a specific parametrization of $\Delta T$ and $\Delta Q$.
The boundaries $\Delta T = 300s$ and $\Delta Q = 10\%$ appear then to be reasonable, though their choice is based solely on empirical evidence. While it seems plausible that smaller intervals lead to higher statistical precision at the cost of excluding some true positives from the sample, larger values of $\Delta T$ and $\Delta Q $ lead to the inclusion of additional points in the sample, at the cost of a weaker statistical power and likely of a larger fraction of false positives. 
In addition, we point out that we could incur in a self-selection bias by further restricting the boundaries on time and volume. Indeed, as shown in Section~\ref{tr:identification}, less skilled users tend to execute actions with larger $\delta T$ and $\delta Q$. We thus prefer an intermediate, more conservative approach, and focus on $\Delta T = 300s$ and $\Delta Q = 10\%$ in the main analysis. 


\bigskip
\bigskip

\begin{table}[H]
    \centering
    \caption{Relationship between trade ability and profits. Robustness check with alternative $\Delta T$, $\Delta Q$}
    \label{tab:app_tab1}
    \begin{tabular}{l*{6}{c}}
\toprule
           Thresholds: &\multicolumn{3}{c}{30s, 1\%}                                     &\multicolumn{3}{c}{600s, 20\%}                                   \\
           Dep. var.: 
           &\multicolumn{1}{c}{With}   &\multicolumn{1}{c}{No}&\multicolumn{1}{c}{Exp.}
          &\multicolumn{1}{c}{With }   &\multicolumn{1}{c}{No}&\multicolumn{1}{c}{Exp.}\\ \cmidrule(lr){2-4}\cmidrule(lr){5-7}
            &\multicolumn{1}{c}{(1)}         &\multicolumn{1}{c}{(2)}         &\multicolumn{1}{c}{(3)}         &\multicolumn{1}{c}{(4)}         &\multicolumn{1}{c}{(5)}         &\multicolumn{1}{c}{(6)}         \\
\midrule
D(Currencies) &      1.2287\sym{**} &      0.8634\sym{**} &      1.1033\sym{**} &      1.2304\sym{***}&      0.8557\sym{***}&      1.1568\sym{***}\\
            &    (0.4830)         &    (0.3784)         &    (0.4219)         &    (0.1349)         &    (0.1221)         &    (0.1376)         \\
[1em]
Equiv. \$   &      0.6671         &      0.3140         &      0.4420         &     -0.3636         &     -1.0275         &     -0.5332         \\
            &    (0.8899)         &    (0.7671)         &    (0.7590)         &    (0.8931)         &    (0.8572)         &    (0.8843)         \\
[1em]
Constant    &     -0.5669         &      0.2005         &     -0.6963\sym{*}  &     -0.6018\sym{***}&      0.3510\sym{***}&     -0.6794\sym{***}\\
            &    (0.4602)         &    (0.3622)         &    (0.4028)         &    (0.1274)         &    (0.1153)         &    (0.1270)         \\
\midrule
Time FE     &           Y         &           Y         &           Y         &           Y         &           Y         &           Y         \\
Dyad FE     &           Y         &           Y         &           Y         &           Y         &           Y         &           Y         \\
N           &        3935         &        3935         &        3935         &        6554         &        6554         &        6554         \\
R-squared   &        0.71         &        0.76         &        0.75         &        0.64         &        0.66         &        0.65         \\
\bottomrule
    \multicolumn{7}{l}{\footnotesize Standard errors in parentheses. \sym{*} \(p<0.1\), \sym{**} \(p<0.05\), \sym{***} \(p<0.01\)}\\
    \multicolumn{7}{l}{\footnotesize Errors are clustered at user-level to account for intra-class correlation.}\\
    \end{tabular}
\end{table}

\begin{landscape}
\begin{table}[]
    \centering
    \caption{Relationship between trade ability and profits, alternative proxies.  Robustness check with alternative $\Delta T$, $\Delta Q$}
    \label{tab:app_tab1_alt_xdeltas}
    {\small
    \begin{tabular}{l*{12}{c}}
\toprule
Dep. var.:        & \multicolumn{12}{c}{Spread (with fees)} \\
\cmidrule{2-13}            
Thresholds: &\multicolumn{5}{c}{30s, 1\%}                                                                                 &\multicolumn{7}{c}{600s, 20\%}                                                                                                                           \\\cmidrule(lr){2-6}\cmidrule(lr){7-13}
            &\multicolumn{1}{c}{(1)}         &\multicolumn{1}{c}{(2)}         &\multicolumn{1}{c}{(3)}         &\multicolumn{1}{c}{(4)}         &\multicolumn{1}{c}{(5)}         &\multicolumn{1}{c}{(6)}         &\multicolumn{1}{c}{(7)}         &\multicolumn{1}{c}{(8)}         &\multicolumn{1}{c}{(9)}         &\multicolumn{1}{c}{(10)}         &\multicolumn{1}{c}{(11)}         &\multicolumn{1}{c}{(12)}         \\
\hline
D(Currencies)&      1.2287\sym{**} &                     &                     &                     &                     &                     &      1.2304\sym{***}&                     &                     &                     &                     &                     \\
            &    (0.4830)         &                     &                     &                     &                     &                     &    (0.1349)         &                     &                     &                     &                     &                     \\
Log(Currencies)&                     &      1.6369\sym{*}  &                     &                     &                     &                     &                     &      0.6337\sym{*}  &                     &                     &                     &                     \\
            &                     &    (0.9536)         &                     &                     &                     &                     &                     &    (0.3271)         &                     &                     &                     &                     \\
Log(Actions)&                     &                     &      0.4534\sym{***}&                     &                     &                     &                     &                     &      0.2703\sym{***}&                     &                     &                     \\
            &                     &                     &    (0.1032)         &                     &                     &                     &                     &                     &    (0.0331)         &                     &                     &                     \\
D(Metaorder)&                     &                     &                     &      0.8736\sym{***}&                     &                     &                     &                     &                     &      0.8423\sym{***}&                     &                     \\
            &                     &                     &                     &    (0.2414)         &                     &                     &                     &                     &                     &    (0.1460)         &                     &                     \\
D(Aggressive)&                     &                     &                     &                     &     -0.9744\sym{***}&                     &                     &                     &                     &                     &     -1.4934\sym{***}&                     \\
            &                     &                     &                     &                     &    (0.3440)         &                     &                     &                     &                     &                     &    (0.1726)         &                  \\
PC1         &                     &                     &                     &                     &                     &      0.4722\sym{**} &                     &                     &                     &                     &                     &      0.1880\sym{***}\\
            &                     &                     &                     &                     &                     &    (0.2107)         &                     &                     &                     &                     &                     &    (0.0166)         \\
Equiv. \$   &      0.6671         &      1.0319         &      0.4933         &      0.8404         &      0.8652         &      0.8447         &     -0.3636         &     -0.5168         &      1.1817         &      1.1333         &      0.3792         &      1.4998         \\
            &    (0.8899)         &    (0.8542)         &    (0.4411)         &    (0.8684)         &    (0.8575)         &    (0.7163)         &    (0.8931)         &    (1.1656)         &    (1.0516)         &    (1.4024)         &    (1.0271)         &    (1.2314)         \\
[1em]
Constant    &     -0.5669         &     -2.3328         &     -2.3788\sym{***}&     -0.2127         &      0.6279\sym{***}&     -2.0841\sym{*}  &     -0.6018\sym{***}&     -0.5195         &     -1.1929\sym{***}&     -0.1444         &      0.5949\sym{***}&     -0.8669\sym{***}\\
            &    (0.4602)         &    (1.7354)         &    (0.6900)         &    (0.2425)         &    (0.0536)         &    (1.2259)         &    (0.1274)         &    (0.5656)         &    (0.2230)         &    (0.1219)         &    (0.0397)         &    (0.1420)         \\
\hline
Time FE     &           Y         &           Y         &           Y         &           Y         &           Y         &           Y         &           Y         &           Y         &           Y         &           Y         &           Y         &           Y         \\
Dyad FE     &           Y         &           Y         &           Y         &           Y         &           Y         &           Y         &           Y         &           Y         &           Y         &           Y         &           Y         &           Y         \\
N           &        3935         &        3935         &        3935         &        3935         &        3935         &        3935         &        6554         &        6554         &        6554         &        6554         &        6554         &        6554         \\
R-squared   &        0.71         &        0.74         &        0.78         &        0.71         &        0.71         &        0.75         &        0.64         &        0.63         &        0.66         &        0.63         &        0.64         &        0.66         \\
\bottomrule
    \multicolumn{13}{l}{Standard errors in parentheses. \sym{*} \(p<0.1\), \sym{**} \(p<0.05\), \sym{***} \(p<0.01\)}\\
    \multicolumn{13}{l}{Errors are clustered at user-level to account for intra-class correlation.}\\
    \end{tabular}}
\end{table}
\end{landscape}

\begin{landscape}
\begin{table}[H]
    \centering
    \caption{Responsiveness to official rate variations. Robustness check with alternative $\Delta T$, $\Delta Q$}
    \label{tab:app_tab2deltas}
    {\footnotesize
    \begin{tabular}{l*{12}{c}}
\toprule
Dep. var.:        & \multicolumn{12}{c}{Spread (with fees)} \\
\cmidrule{2-13}
Thresholds: &\multicolumn{6}{c}{30s, 1\%}              &\multicolumn{6}{c}{600s, 20\%}             \\
            \cmidrule(lr){2-7}\cmidrule(lr){8-13}
            &\multicolumn{1}{c}{(1)}         &\multicolumn{1}{c}{(2)}         &\multicolumn{1}{c}{(3)}         &\multicolumn{1}{c}{(4)}         &\multicolumn{1}{c}{(5)}         &\multicolumn{1}{c}{(6)}         &\multicolumn{1}{c}{(7)}         &\multicolumn{1}{c}{(8)}         &\multicolumn{1}{c}{(9)}         &\multicolumn{1}{c}{(10)}         &\multicolumn{1}{c}{(11)}         &\multicolumn{1}{c}{(12)}         \\
\midrule
$ \Delta R \times $D(Currencies)&       1.420         &                     &                     &                     &                     &                     &      -0.094         &                     &                     &                     &                     &                     \\[-0.2em]
            &     (1.108)         &                     &                     &                     &                     &                     &     (3.143)         &                     &                     &                     &                     &                     \\[-0.2em]
$ \Delta R \times $Log(Currencies)&                     &       3.049\sym{*}  &                     &                     &                     &                     &                     &       0.275         &                     &                     &                     &                     \\[-0.2em]
            &                     &     (1.546)         &                     &                     &                     &                     &                     &     (0.980)         &                     &                     &                     &                     \\[-0.2em]
$ \Delta R \times $Log(Actions)&                     &                     &       0.593\sym{***}&                     &                     &                     &                     &                     &       0.040         &                     &                     &                     \\[-0.2em]
            &                     &                     &     (0.173)         &                     &                     &                     &                     &                     &     (0.356)         &                     &                     &                     \\[-0.2em]
$ \Delta R \times $D(Metaorder)&                     &                     &                     &       2.404\sym{*}  &                     &                     &                     &                     &                     &       2.024\sym{**} &                     &                     \\[-0.2em]
            &                     &                     &                     &     (1.429)         &                     &                     &                     &                     &                     &     (0.814)         &                     &                     \\[-0.2em]
$ \Delta R \times $D(Aggressive)&                     &                     &                     &                     &      -0.608         &                     &                     &                     &                     &                     &       1.823         &                     \\[-0.2em]
            &                     &                     &                     &                     &     (1.324)         &                     &                     &                     &                     &                     &     (2.420)         &                     \\[-0.2em]
$ \Delta R \times $PC1    &                     &                     &                     &                     &                     &       0.704\sym{***}&                     &                     &                     &                     &                     &       0.317         \\[-0.2em]
            &                     &                     &                     &                     &                     &     (0.245)         &                     &                     &                     &                     &                     &     (0.204)         \\
$ \Delta R $     &       0.000         &      -4.226         &      -2.547         &      -0.693         &       1.433         &      -2.561\sym{*}  &       0.390         &      -0.188         &       0.035         &      -1.394         &       0.246         &      -2.160         \\
            &         (.)         &     (3.074)         &     (1.524)         &     (1.024)         &     (1.121)         &     (1.430)         &     (3.226)         &     (1.782)         &     (2.410)         &     (0.965)         &     (0.667)         &     (1.694)         \\
Equiv. \$   &       0.012         &       0.002         &      -0.021         &       0.009         &       0.011         &      -0.008         &      -0.935         &      -0.956         &      -0.939         &      -1.037         &      -0.927         &      -1.009         \\
            &     (0.583)         &     (0.575)         &     (0.586)         &     (0.584)         &     (0.583)         &     (0.586)         &     (0.838)         &     (0.825)         &     (0.825)         &     (0.881)         &     (0.833)         &     (0.861)         \\
\midrule
User FE     &           Y         &           Y         &           Y         &           Y         &           Y         &           Y         &           Y         &           Y         &           Y         &           Y         &           Y         &           Y         \\
Time FE     &           Y         &           Y         &           Y         &           Y         &           Y         &           Y         &           Y         &           Y         &           Y         &           Y         &           Y         &           Y         \\
Dyad FE     &           Y         &           Y         &           Y         &           Y         &           Y         &           Y         &           Y         &           Y         &           Y         &           Y         &           Y         &           Y         \\
N           &        3923         &        3923         &        3923         &        3923         &        3923         &        3923         &        6322         &        6322         &        6322         &        6322         &        6322         &        6322         \\
R-squared   &        0.81         &        0.81         &        0.81         &        0.81         &        0.81         &        0.81         &        0.71         &        0.71         &        0.71         &        0.71         &        0.71         &        0.71         \\
\bottomrule
    \multicolumn{13}{l}{Standard errors in parentheses. \sym{*} \(p<0.1\), \sym{**} \(p<0.05\), \sym{***} \(p<0.01\)}\\
    \multicolumn{13}{l}{Errors are clustered at user-level to account for intra-class correlation.}\\
    \end{tabular}}
\end{table}
\end{landscape}

%% file: tr_appendix_fees.tex
\section{Fees scheme \& robustness checks on alternative measures of transaction costs}
\label{tr:appendix_fees}

\setcounter{figure}{0}
\setcounter{table}{0}

\begin{table}[H]
    \centering
    \begin{tabular}{ccccccccc}
    \toprule
    &     &  & \multicolumn{2}{c}{Without fees}  &  \multicolumn{2}{c}{With fees} &  \multicolumn{2}{c}{Expected fees} \\
    \cmidrule(r{0.1cm}){4-5} \cmidrule(l{0.1cm}){6-7}
    \cmidrule(l{0.1cm}){8-9}
     $ Leg_{Buy} $ &    $ Leg_{Sell} $   &     official e.r.  & implied e.r.   &   P/L & implied e.r.   &   P/L & implied e.r.   &   P/L   \\    
    \midrule
     EUR   &   USD    &   1.400     &   1.405    &    profit  &   1.402    &    profit &   1.399    &    \textit{\textbf{loss}}  \\
     EUR   &    USD   &   1.400     &   1.395    &   loss  &   1.393    &   loss    &   1.393    &   loss    \\
     USD   &    EUR   &   0.714     &   0.717    &    profit  &   0.715    &    profit  &   0.715    &    profit  \\
     USD   &    EUR   &   0.714     &   0.716    &    profit   &   0.713    &    \textit{\textbf{loss}}   &   0.715    &    profit    \\
     \bottomrule
    \end{tabular}
    \caption{Profitability of the arbitrage actions, illustrative example on the EUR, USD currencies}  
    \floatfoot{\emph{Notes:} by construction, the official and implied rates are in EURtoUSD in rows 1 and 2, and USDtoEUR in rows 3 and 4. Actions are profitable when the implied rate is higher than the official one. Fees reduce the margins for profit and worsen losses. The implied rate may differ when using the expected fees and the fees reported in the dataset (e.g., the bold entries of row 1 and 4).}
    \label{tab:profitability}
\end{table}

Our main analyses take into account the explicit transaction costs borne by the users to trade bitcoins within the Mt.\ Gox exchange platform. In this Appendix we discuss two alternative ways to consider them; as shown with an illustrative example in Table~\ref{tab:profitability}, fees change when estimated in different ways.
The first alternative approach is to examine the case without fees at all, to assess to what extent the users include this information in their trading strategies. 
The second is more complex and entails an estimation of the fees a user would expect to pay. Indeed, according to the official fee schedule posted by Mt.\ Gox, the magnitude of the fees paid should depend on the volume traded by the user in the last month --- i.e., most active users in the recent past pay relatively lesser. However, in some cases divergences exist between the fees paid by the users and the amount they should pay based on the fee schedule. 
We thus devise and fit a fee model, which we discuss below, to estimate the expected transaction costs for each leg, given the user's trading history at the time of such trade (note that the model estimation is performed over the full Mt.\ Gox dataset, i.e. it takes into account all trades included in the Mt.\ Gox dataset, from April 2011 to November 2013).
Then, as a robustness check, we repeat the main estimations reported in Tables~\ref{tab:tab1_base},~\ref{tab:tab1_alt_x},~\ref{tab:tab2_alt_x} for the two alternative fee measurements.

\begin{table}[h]
	\centering
	\caption{Mt.\ Gox posted fee schedule} 
	\label{tab:feeschedule}
	{\small
	\begin{tabular}{p{4cm} p{1.5cm}  p{4cm} p{1.5cm}}
		\toprule
		Volume (BTCs) &  Fees & Volume (BTCs) & Fees\\
		\midrule
		$ 0 $ to $ < 100 $ &  $ 0.60\% $ &   $ 10000 $ to $ < 25000 $ &  $ 0.30\%  $   \\
		$ 100 $ to $ < 200 $ & $ 0.55\%  $&   $ 25000 $ to $ < 50000 $ &  $ 0.29\% $   \\
		$ 200 $ to $ < 500  $&$  0.53\%  $ &  $ 50000 $ to $ < 100000 $ &  $ 0.28\% $ \\
		$ 500 $ to $ < 1000 $ & $ 0.50\% $ &  $  100000 $ to $ < 250000 $ & $ 0.27\% $  \\
		$ 1000 $ to $ < 2000 $ & $ 0.46\% $ &   $ 250000 $ to $ < 500000 $ &  $ 0.26\% $      \\
		$ 2000 $ to $ < 5000 $ & $ 0.43\% $ &   $ > 500000 $ &  0.25\%      \\
		$ 5000 $ to $ < 10000 $ & $ 0.40\% $ &   - &     -    \\
		\bottomrule
	\end{tabular}
	\floatfoot{\emph{Notes:} discounts are based on the user's trading volume over the last 720 hours.}
	}
\end{table}

\begin{figure}[tp]
    \caption{Empirical fee schedule} 
	\label{fig:fees} 
	\begin{center}
		\includegraphics[width=\textwidth]{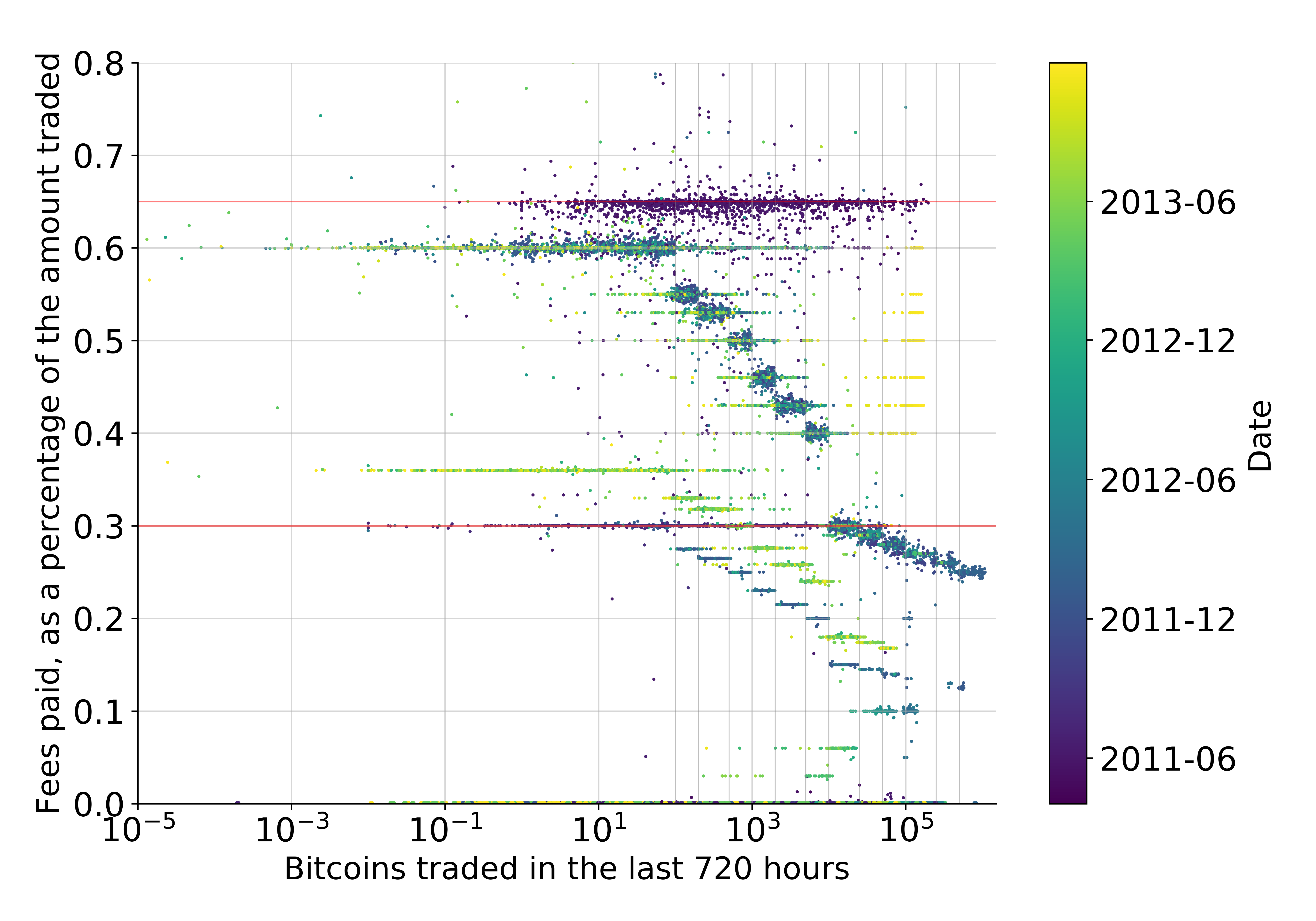}
	\floatfoot{\emph{Notes:} fees paid per leg, expressed as the percentage of the value of the trade (y-axis). According to Mt.\ Gox's schedule, fees depend on the amount of bitcoins traded by the user in the last 720 hours (on the x-axis). Color indicates the time of the trade. The lines of the grid help to graphically delimit the volume bands corresponding to different discount brackets. Two horizontal red lines help to identify the two time windows when fees were fixed: $0.65\%$ from 1 April 2011 to 23 June 2011, and $0.3\%$ from 24 June 2011 to 18 August 2011. Many legs pay no fees, and a limited number of legs pays fees as low as $0.1$ percent or less. In addition, in some particular circumstances (e.g., Easter and Christmas holidays in 2011 and 2012), fees were halved, thus explaining some of the values that do not correspond to the official fee schedule. The plot shows a random sample of $N = 100,000$ legs. Note that a small fraction of trades (around 15,000 over the total of 7.5 million trades) have fees equal or above 1\% and are likely to correspond to misreported data. We explicitly decided to plot only the legs between 0\% and 0.8\%.}
	\end{center}
\end{figure}

Table~\ref{tab:feeschedule} shows the fee schedule published by Mt.\ Gox; we could confirm that the scheme was valid at least from 16 October 2011 until mid-February 2013\footnote{\url{https://bit.ly/374zXBW}, \url{https://bit.ly/2GYVcdF}, and \url{https://bit.ly/2IrPRMp}}. Discounts were based on the volume of bitcoins traded by the user over the last 720 hours\footnote{\url{https://bit.ly/2SWwJIx}} (that is, 30 days).

We first compare the posted and the real fee schedules. In the Mt.\ Gox dataset, transaction costs are reported in two entries: excluding minor exceptions, bitcoin buyers were charged with fees in bitcoin, while sellers with fees in fiat money\footnote{\url{https://bit.ly/34Wyb3h}}. For each leg of every trade, we compute the actual fees paid as 
\begin{equation}
    \Fee = \frac{\BitcoinFee}{\Bitcoins}+ \frac{\MoneyFee}{\Money},
\end{equation}
where `$\BitcoinFee$' represents the fees paid on the amount of bitcoins traded, while `$\MoneyFee$' represents the fees paid on the amount of fiat money traded.

\begin{figure}[tp]
    \centering
    \caption{Patterns of waived fees} 
    \label{fig:nofees}
    \begin{subfigure}[t]{1.07\textwidth}
        \hspace{-0.2cm}
        \includegraphics[width=\textwidth]{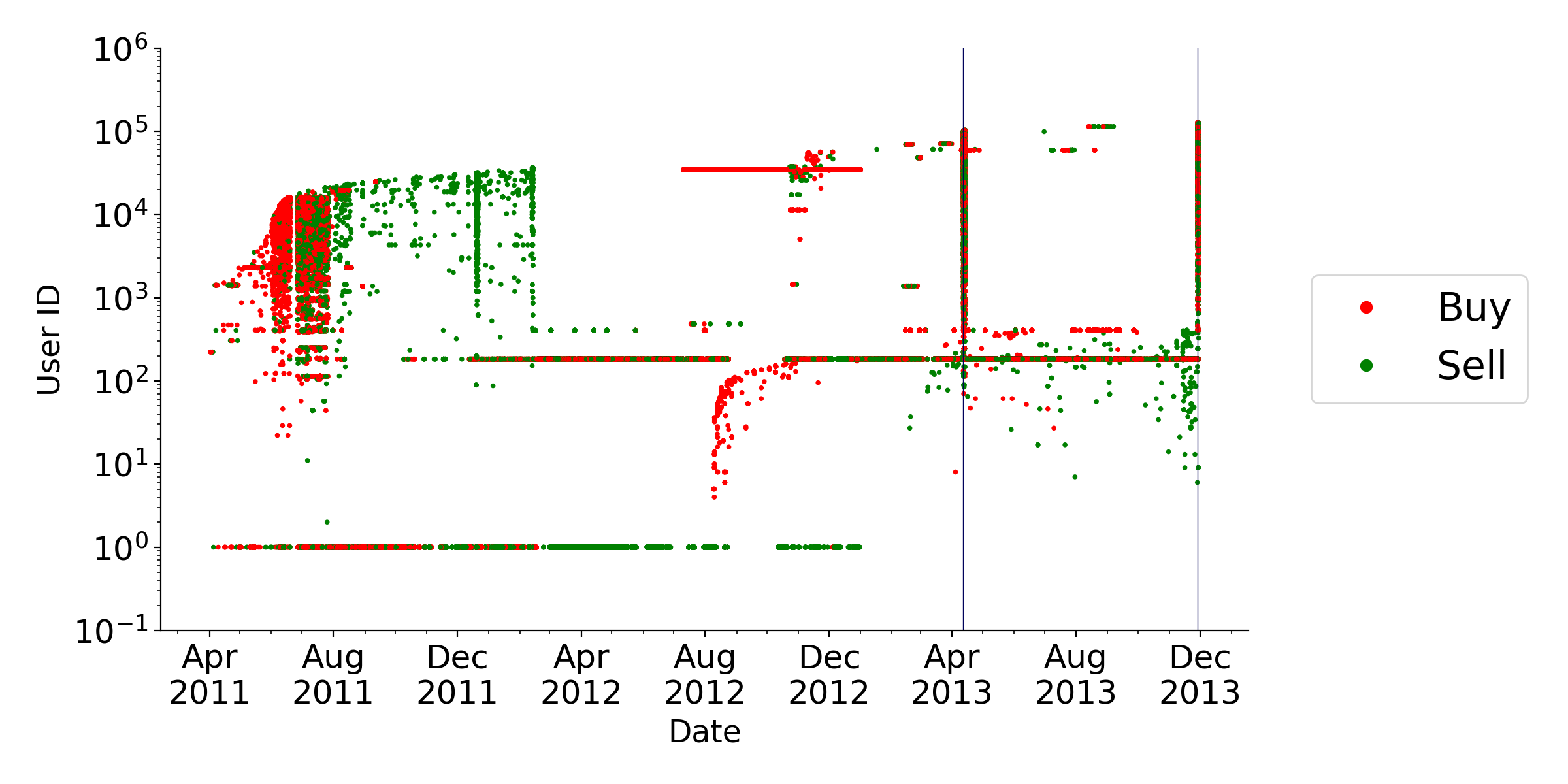}
		\caption{}
		\label{fig:feezero}
    \end{subfigure}
    \begin{subfigure}[t]{\textwidth}
        \centering
        \includegraphics[width=0.95\textwidth]{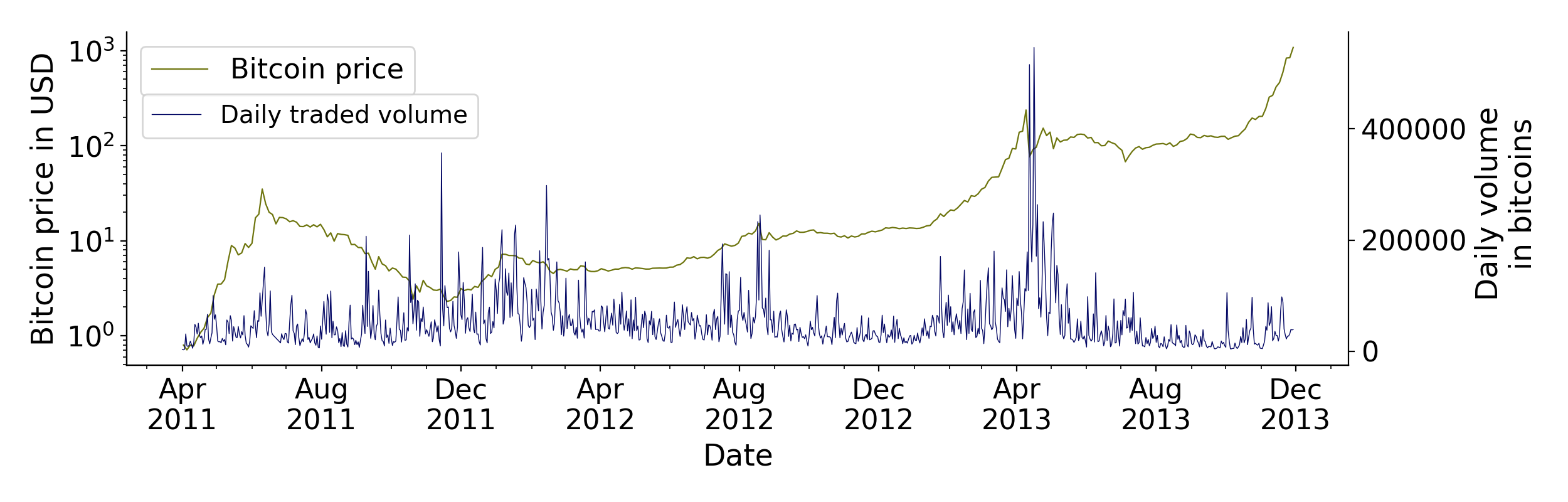}
		\caption{}
		\label{fig:volume_and_prices}
    \end{subfigure}
    \floatfoot{\emph{Notes}: Panel~(\subref{fig:feezero}) reports the distribution of the legs whose associated fees are zero, as a function of  time and user ID (the latter on a logarithmic scale, to focus on low IDs). Buy (red) and sell (green) orders differ by color. Many interesting patterns emerge: first, two users with low ID did not pay fees over extended time periods; moreover, they account for $ \sim 1,000,000 $ zero-fee legs over the total of $ \sim 1,700,000 $ zero-fee legs. Second, during some days (especially on 19, 20, and 21 December 2011; 12, 13, and 14 April 2013; 28 and 29 November 2013) there were anomalous increases in trade legs with zero fees. 
    Possible explanations include special events, such as a temporary downtime of Mt.\ Gox on 11 April 2013, and the exchange rate surpassing 1,000 \$/BTC for the first time on 27 November 2013. In both cases, the number of zero-fee trade legs increases shortly afterwards.   
    Panel~(\subref{fig:volume_and_prices}) depicts the daily bitcoin volume traded and the exchange rate in USD (the latter on a logarithmic scale). Observe from both panels that many users with low IDs seem to have \textit{coordinately} executed buy orders from August to around November 2012, and then exclusively sell orders in the the days preceding the Bitcoin price peak. The plot shows a random sample of $N= 200,000$ legs.}
\end{figure}

Figure~\ref{fig:fees} shows a sample ($N = 100,000$) of the empirical fees, focusing on the relationship between the actual transaction costs and the past volume traded; each dot represents the fees paid on a leg of a trade. By comparing the posted and the actual schedules, we note that most of the data points fall into the expected volume bands, although deviations exist: first, from 0.40\% to 0.20\%, many points follow a pattern that cannot be explained by the posted schedule; second, a non-negligible number of dots falls below the threshold of 0.20\%, suggesting the existence of privileged users, and a subset of legs is completely exempted from any kind of fee:  Figure~\ref{fig:nofees} reports additional information on them.

For this reason, instead of reverse-engineering the posted schedule, we take an empirical approach and fit a simple model that predicts the fees a user would have to pay given his trading history. 
The fee model is specified as:
\begin{align}
\begin{split}
\Fee_{i} = \beta_{0} &+ \beta_{1} \cdot \LogVol_{i} +  \beta_{2} \cdot \VolSmall_{i} +  \beta_{3} \cdot \VolBig_{i} + \beta_{4} \cdot \LogVol_{i} \cdot \VolSmall_{i} \\& + \beta_{5} \cdot \LogVol_{i} \cdot   \VolBig_{i} + \beta_{6} \cdot \Tzero_{i}  +  \beta_{7} \cdot \Tone_{i}  +  \beta_{8} \cdot \Tholid_{i} + \epsilon_{i}.
\\[4ex]
\end{split}
\end{align}
The independent variables have the following meaning:
\begin{itemize}
	\item $\LogVol$, the natural logarithm of the volume traded in the last 720 hours by the user who submitted the leg associated to the fee;
	\item $ \VolSmall $, a dummy variable equal to 1 if the volume traded in the last 720 hours is between 100 and 10,000 bitcoins;
	\item $ \VolBig $, a dummy variable equal to 1 if the volume traded in the last 720 hours exceeds 10,000 bitcoins. As it can be seen from Figure~\ref{fig:fees}, discount factors follow a linear trend with different slopes below and above the 10,000 bitcoins threshold. This is the reason why we introduced this dummy variable and the previous one, as well as their interaction terms with the $ \LogVol $ variable;
	\item $ \Tzero $, a dummy variable for trades executed between 1 April 2011 and 23 June 2011;
	\item $ \Tone $, a dummy variable for trades executed between 24 June 2011 and 18 August 2011;
	\item $ \Tholid $, a dummy variable for trades executed on `special days': from 26 December 2011 to 1 January 2012, from 2 to 7 April 2012; on 9 and 10 November 2012\footnote{respectively, Christmas holidays in 2011, Easter holidays in 2012 and first Bitcoin Friday Sale day. \\ \url{https://bit.ly/37317sY}}.
\end{itemize}

{\renewcommand{\arraystretch}{1}
\begin{table}[t!]
	\begin{center}
	\caption{Fee model for non-zero fees, coefficients fitted with OLS}
	\label{tab:OLSlog}
		\begin{tabular}{lccccc}
			\toprule
			Dependent variable:&\multicolumn{5}{l}{Fee as a percentage of the amount traded}  \\[1ex]
			& \multicolumn{5}{c}{Specification}\\
			\cmidrule{2-6}
			&\multicolumn{1}{c}{(1)}&\multicolumn{1}{c}{(2)}&\multicolumn{1}{c}{(3)}&\multicolumn{1}{c}{(4)}&\multicolumn{1}{c}{(5)}\\
			\midrule
			$ \Intercept $       & 0.638   & 0.640   &  0.557   & 0.559   & 0.561    \\
			& (0.0001)    & (0.0001)    & (0.0000)    & (0.0001)    & (0.0001)     \\
			$ \LogVol $        & -0.030  & -0.030  &       & -0.001  & -0.001   \\
			& (0.0000)    & (0.0000)    &               & (0.0000)    & (0.0000)     \\
			$ \VolSmall $     &      &      &   -0.107    & 0.150   & 0.152   \\
			&             &       &  (0.0001)    & (0.0002)    & (0.0002)     \\
			$ \VolBig $     &      &       & -0.284      & -0.201  & -0.212   \\
			&         &         &        (0.0001) & (0.0005)    & (0.0004)     \\
			$ \LogVol \ast \VolSmall$  &         &        &      & -0.037  & -0.037   \\
			&             &       &         & (0.0000)    & (0.0000)     \\
			$ \LogVol \ast \VolBig$    &       &     &       & -0.007  & -0.006   \\
			&             &        &         & (0.0001)    & (0.0001)     \\
			$ \Tzero $         &       & 0.164   &          &          & 0.158    \\
			&             & (0.0002)    &             &           & (0.0001)     \\
			$ \Tone $          &         & -0.162  &          &         & -0.170   \\
			&          & (0.0001)    &              &              & (0.0001)     \\
			$ \Tholid $       &        & -0.187  &            &            & -0.191   \\
			&              & (0.0002)    &          &          & (0.0002)     \\
			\hline
			$ R^{2} $     & 0.516     & 0.610   &   0.533      & 0.598     & 0.694   \\
			Obs.            & 13083547 & 13083547 & 13083547 & 13083547 & 13083547  \\
			\bottomrule
		\end{tabular}
	\end{center}
	\floatfoot{\emph{Notes:}  all variables are significant at the 0.1\% level. This is due to the high number of observations; however, we emphasize that these results are intended not so much to find significant effects as to predict fees. 
	Observations consist of legs whose fees are positive and their value is below 1\%.}
\end{table}
}

Table~\ref{tab:OLSlog} reports the estimated coefficients and goodness-of-fit indicators. In each specification, the constant term approximates the non-discounted official fee of 0.6\%, and the response variable is negatively correlated with an increase of the volume traded in the past 720 hours; as expected, when included in the model, $ \Tzero $ and $ \Tone $ respectively increase and decrease the constant term by a factor of around 0.16\%, while \Tholid has an even stronger negative effect (around -0.19\%). Finally, again in accordance with our expectations, both $\beta_{4}$ and $\beta_{5}$ are negative; the first one is bigger in absolute terms, thus indicating a steeper variation of discounts given the same variation in volume (as can be seen in Figure~\ref{fig:fees}). We chose a logarithmic model because the break points for the fee schedule in Table~\ref{tab:feeschedule} follow a logarithmic trend, and thus it was straightforward to consider the estimation on logarithmic volumes. To strengthen the results, we explored an alternative model using linear volumes; we report the results in Table~\ref{tab:OLSlin}.

The comparison of the two models shows that the overall pattern is similar, but in the logarithmic models the coefficients can be interpreted with less decimal digits, and the parameters' variations of signs are less frequent; in addition to that, in linear models (1) to (3) the intercept is smaller than expected (0.45\%) and the interaction terms $\beta_{4}$ and $ \beta_{5}$ are close to zero, in contrast with what we would expect. Regarding the explanatory power, the logarithmic model (1) outperforms the same model with linear volume ($ R^{2} $ = 0.516 versus $ R^{2} $ = 0.103), as well as the estimations (2) and (3). Model (4) and (5) have comparable explanatory power, but also in that case the logarithmic model has higher $R^{2}$.

{\renewcommand{\arraystretch}{1}
	\begin{table}
		\caption{Fee model for non-zero fees, coefficients fitted with OLS (linear alternative, not used)}
		\label{tab:OLSlin}
		\begin{center}
		    \begin{tabular}{lccccc}
			        \toprule
			        Dependent variable:&\multicolumn{5}{l}{Fee as a percentage of the amount traded}  \\[1ex]
			        & \multicolumn{5}{c}{Specification}\\
			        \cmidrule{2-6}
			        &\multicolumn{1}{c}{(1)}&\multicolumn{1}{c}{(2)}&\multicolumn{1}{c}{(3)}&\multicolumn{1}{c}{(4)}&\multicolumn{1}{c}{(5)}\\
			        \midrule
					$ \Intercept $        & 0.454   & 0.456   & 0.469   & 0.559   & 0.560    \\
					& (0.0000)    & (0.0000)    & (0.0000)    & (0.0001)    & (0.0001)     \\
					$ \LinVol $         & -0.000  & -0.000  &               & -0.000  & -0.000   \\
					& (0.0000)    & (0.0000)    &               & (0.0000)    & (0.0000)     \\
					$ \VolSmall $       &      &            &               & -0.071  & -0.069   \\
					&               &               &               & (0.0001)    & (0.0001)     \\
					$ \VolBig $       &          &             &          & -0.280  & -0.281   \\
					&               &               &               & (0.0001)    & (0.0001)     \\
					$ \LinVol \ast \VolSmall$ &            &       & -0.000  & 0.000   & 0.000    \\
					&           &              & (0.0000)    & (0.0000)    & (0.0000)     \\
					$ \LinVol \ast \VolBig$   &        &       & -0.000 & 0.000   & 0.000    \\
					&               &               & (0.0000)    & (0.0000)    & (0.0000)     \\
					$ \Tzero $         &      & 0.169   &             &          & 0.158    \\
					&               & (0.0002)    &               &               & (0.0001)     \\
					$ \Tone $           &             & -0.149  &       &             & -0.171   \\
					&               & (0.0002)    &              &              & (0.0001)     \\
					$ \Tholid $         &       & -0.202  &             &       & -0.192   \\
					&               & (0.0003)    &               &               & (0.0002)     \\
					\hline
					$ R^{2} $     & 0.103   & 0.197       & 0.144       & 0.585     & 0.682   \\
					Obs.            & 13083547 & 13083547 & 13083547 & 13083547 & 13083547  \\
					\bottomrule
			\end{tabular}
		\end{center}
	\floatfoot{\emph{Notes:} all variables are significant at 0.1\% level. This is due to the high number of observations; however, we emphasize that these results are intended not so much to find significant effects as to predict fees.
	Observations consist of legs whose fees are positive and their value is below 1\%.}
	\end{table}
}

{\renewcommand{\arraystretch}{1}
	\begin{table}
		\caption{Fee model for zero fees, coefficients fitted with the logistic regression}
		\label{tab:logitfees}
		\begin{center}
				\begin{tabular}{lccccc}
					\toprule
					Dependent variable:&\multicolumn{5}{l}{Dichotomous variable accounting for the presence of fees}  \\[1ex]
			        & \multicolumn{5}{c}{Specification}\\
			        \cmidrule{2-6}
			        &\multicolumn{1}{c}{(1)}&\multicolumn{1}{c}{(2)}&\multicolumn{1}{c}{(3)}&\multicolumn{1}{c}{(4)}&\multicolumn{1}{c}{(5)}\\
			        \midrule
					$ \Intercept $   & 5.104   & 4.429   & 1.670   & 3.246   & 3.257   \\
					& (0.0033)    & (0.0046)    & (0.0019)    & (0.0034)    & (0.0073)     \\
					$ \LogVol  $   & -0.378       & -0.314       &      &             & -0.050   \\
					& (0.0003)    & (0.0004)    &               &               & (0.0007)     \\
					$ \Bitcoins $    &             & 0.011   &           &        & 0.005    \\
					&               & (0.0001)    &               &               & (0.0001)     \\
					$ \Date $   &               & -0.001  & 0.001   &               & 0.002    \\
					&               & (0.0000)    & (0.0000)    &               & (0.0000)     \\
					$\AnomalousDays $        &         &         & -3.402  &         & -6.459   \\
					&               &               & (0.0034)    &               & (0.0060)     \\
					$\EarlyAdopters$ &        &           &         & -0.897       & -1.575   \\
					&               &               &               & (0.0037)    & (0.0055)     \\
					$\AnomalousUsers$     &      &      &            & -5.556       & -6.941  \\
					&               &               &               & (0.0044)    & (0.0065)     \\
					$ \Matchers $ &        &         &             & 0.929        & 0.799    \\
					&               &               &               & (0.0052)    & (0.0062)     \\
					$\Markus$      &        &             &          & -1.434       & -1.468   \\
					&               &               &               & (0.0156)    & (0.0159)     \\
					$\Willy$       &      &            &             & -0.418       & -0.675   \\
					&               &               &               & (0.0189)    & (0.0306)     \\
					\hline
					$ pseudo-R^{2} $       & 0.157        & 0.228        & 0.109        & 0.472        & 0.676         \\
					Obs.        & 14875192 & 14875192 & 14875192 & 14875192 & 14875192  \\
					\bottomrule
			\end{tabular}
		\end{center}
	\floatfoot{\emph{Notes:}  all variables are significant at 0.1\% level. This is due to the high number of observations; however, we emphasize that these results are intended not so much to find significant effects as to predict fees.}
	\end{table}
}

Even though not directly comparable, in the work of~\citet{kim2017transaction} work on the cost advantage of Bitcoin over cross-border ATM transactions the models for Bitcoin cross-border transaction costs achieve R-squared values in the order of 0.5 for the time period 2014-2015. Thus, it seems that cryptocurrency fees can be explained with linear regressions at this order of magnitude.

Finally, we propose a logit model (Table \ref{tab:logitfees}) to estimate the probability that a leg pay zero fees given user-specific and time-related variables: here the binary dependent variable is 1 if the trader paid some fee, and 0 otherwise. The model is specified as follows: 

\begin{equation}
\begin{split}
log\bigg( \frac{\Fee_{i}}{1-\Fee_{i}}\bigg) = \beta_{0} + \beta_{1} \cdot \LogVol_{i} +  \beta_{2} \cdot \Bitcoins_{i} + \beta_{3} \cdot \Date_{i} + \\
+ \beta_{4} \cdot \AnomalousDays_{i} + \beta_{5} \cdot \EarlyAdopters_{i} + \beta_{6} \cdot \AnomalousUsers_{i} + \\ 
+ \beta_{7} \cdot \Matchers_{i}+  \beta_{8} \cdot \Markus_{i}+  \beta_{9} \cdot \Willy_{i} + \epsilon_{i}
\end{split}
\end{equation}

and the independent variables have the following meaning:
\begin{itemize}
	\item $\Bitcoins$: the amount of bitcoins traded;
	\item $ \Date $: date of the trade execution;
	\item $\EarlyAdopters$: dummy variable equal to 1 for the first $ 16,000 $ user IDs in sequential order; 
	\item $\AnomalousDays $: dummy variable equal to 1 for the days with an anomalous presence of zero fees trades;
	\item $\AnomalousUsers$: dummy variable equal to 1 for some users who performed an unusually high number of actions with zero fees;
	\item $ \Matchers $: dummy variable equal to 1 for the legs whose complementary leg was executed by one of the aforementioned `Anomalous Users'; 
	\item $ \Willy $: dummy variable for the User ID associated to \textit{Willy};
	\item $ \Markus $: dummy variable for the \textit{Markus}' User ID not associated to manipulations (see Appendix~\ref{tr:appendix_data_cleaning} for the details on the users called by the literature \textit{Markus} and \textit{Willy});
\end{itemize}

Models (1) to (3) have very low pseudo-$R^{2}$; the largest part of the explanatory power of the model is associated with the variables that contain information on the Users IDs.
Coefficients $\beta_{1}$, $\beta_{2}$, and $\beta_{3}$ are very small, suggesting that their effect is significant but limited. As expected, instead, the variables $\AnomalousDays $ and $\AnomalousUsers $ are associated with negative and high coefficients. The probability that fees are paid decreases for the users defined as early adopters. Unexpectedly, the variable $\Matchers $ is associated to higher probability that the order might have paid a fee. 

In the following we repeat the main estimations in Tables~\ref{tab:tab1_base}, ~\ref{tab:tab1_alt_x}, ~\ref{tab:tab2_alt_x} for the case in which profits are measured without including the fees, and with the fees a user would expect to pay given his transaction history (Tables~\ref{tab:app_tab1fees},~\ref{tab:app_tab1_alt_xfees},~\ref{tab:app_tab2fees}). Results are consistent with those reported for the main estimations.

\begin{landscape}
\begin{table}[H]
    \centering
    \caption{Relationship between trade ability and profits. Robustness check with alternative measures of profits (no fees, expected fees)} 
    \label{tab:app_tab1fees}
    \begin{tabular}{l*{8}{c}}
\toprule
       Dep. var.:   &\multicolumn{4}{c}{Spread (Without fees)}                                                       &\multicolumn{4}{c}{Spread (Expected fees)}                                                      \\\cmidrule(lr){2-5}\cmidrule(lr){6-9}
            &\multicolumn{1}{c}{(1)}         &\multicolumn{1}{c}{(2)}         &\multicolumn{1}{c}{(3)}         &\multicolumn{1}{c}{(4)}         &\multicolumn{1}{c}{(5)}         &\multicolumn{1}{c}{(6)}         &\multicolumn{1}{c}{(7)}         &\multicolumn{1}{c}{(8)}         \\
\midrule
D(Currencies)&      1.0687\sym{***}&      1.1651\sym{***}&      0.8727\sym{***}&      0.9316\sym{***}&      1.3977\sym{***}&      1.4436\sym{***}&      1.1583\sym{***}&      1.2160\sym{***}\\
            &    (0.2204)         &    (0.1880)         &    (0.1416)         &    (0.1486)         &    (0.1845)         &    (0.1878)         &    (0.1643)         &    (0.1702)         \\
[1em]
Equiv. \$   &      2.3429         &      1.3837         &      0.0384         &     -0.4376         &      3.7186\sym{**} &      2.6949\sym{*}  &      0.4920         &      0.0296         \\
            &    (1.6352)         &    (1.3343)         &    (1.1974)         &    (1.1337)         &    (1.8186)         &    (1.6110)         &    (1.2224)         &    (1.0973)         \\
[1em]
Constant    &      0.0823         &      0.0048         &      0.2756\sym{**} &      0.2283         &     -0.9895\sym{***}&     -1.0211\sym{***}&     -0.7191\sym{***}&     -0.7642\sym{***}\\
            &    (0.1582)         &    (0.1640)         &    (0.1350)         &    (0.1428)         &    (0.1582)         &    (0.1676)         &    (0.1540)         &    (0.1616)         \\
\midrule
Time FE     &           N         &           N         &           Y         &           Y         &           N         &           N         &           Y         &           Y         \\
Dyad FE     &           N         &           Y         &           N         &           Y         &           N         &           Y         &           N         &           Y         \\
N           &        6594         &        6582         &        5307         &        5284         &        6594         &        6582         &        5307         &        5284         \\
R-squared   &        0.07         &        0.12         &        0.70         &        0.71         &        0.12         &        0.17         &        0.70         &        0.70         \\
    \bottomrule
    \multicolumn{9}{l}{\footnotesize Standard errors in parentheses.  \sym{*} \(p<0.1\), \sym{**} \(p<0.05\), \sym{***} \(p<0.01\) }\\
    \multicolumn{9}{l}{\footnotesize Errors are clustered at user-level to account for intra-class correlation.}\\
    \end{tabular}
\end{table}
\end{landscape}

\begin{landscape}
\begin{table}[H]
    \centering
    \caption{Relationship between trade ability and profits, alternative proxies. Robustness check with alternative measures of profits (no fees, expected fees)}
    \label{tab:app_tab1_alt_xfees}
    {\footnotesize
\begin{tabular}{l*{12}{c}}
\toprule
   Dep. var.:     &\multicolumn{6}{c}{Spread (Without fees)}                                                                             &\multicolumn{6}{c}{Spread (Expected fees)}                                                                            \\\cmidrule(lr){2-7}\cmidrule(lr){8-13}
            &\multicolumn{1}{c}{(1)}         &\multicolumn{1}{c}{(2)}         &\multicolumn{1}{c}{(3)}         &\multicolumn{1}{c}{(4)}         &\multicolumn{1}{c}{(5)}         &\multicolumn{1}{c}{(6)}         &\multicolumn{1}{c}{(7)}         &\multicolumn{1}{c}{(8)}         &\multicolumn{1}{c}{(9)}         &\multicolumn{1}{c}{(10)}        &\multicolumn{1}{c}{(11)}        &\multicolumn{1}{c}{(12)}         \\
\midrule
D(Currencies)&      0.9316\sym{***}&                     &                     &                     &                     &                     &      1.2160\sym{***}&                     &                     &                     &                     &                     \\
            &    (0.1486)         &                     &                     &                     &                     &                     &    (0.1702)         &                     &                     &                     &                     &                     \\
Log(Currencies)&                     &      0.6898\sym{**} &                     &                     &                     &                     &                     &      0.7413\sym{**} &                     &                     &                     &                     \\
            &                     &    (0.2885)         &                     &                     &                     &                     &                     &    (0.3382)         &                     &                     &                     &                     \\
Log(Actions)&                     &                     &      0.2176\sym{***}&                     &                     &                     &                     &                     &      0.2658\sym{***}&                     &                     &                     \\
            &                     &                     &    (0.0438)         &                     &                     &                     &                     &                     &    (0.0401)         &                     &                     &                     \\
D(Metaorder)&                     &                     &                     &      0.1007         &                     &                     &                     &                     &                     &      0.1775         &                     &                     \\
            &                     &                     &                     &    (0.0799)         &                     &                     &                     &                     &                     &    (0.1606)         &                     &                     \\
D(Aggressive)&                     &                     &                     &                     &     -1.1194\sym{***}&                     &                     &                     &                     &                     &     -1.4486\sym{***}&                     \\
            &                     &                     &                     &                     &    (0.1858)         &                     &                     &                     &                     &                     &    (0.1921)         &                     \\
PC1         &                     &                     &                     &                     &                     &      0.1472\sym{***}&                     &                     &                     &                     &                     &      0.1875\sym{***}\\
            &                     &                     &                     &                     &                     &    (0.0303)         &                     &                     &                     &                     &                     &    (0.0347)         \\
Equiv. \$   &     -0.4376         &     -0.4620         &      0.2527         &     -0.5631         &     -0.3278         &     -0.0410         &      0.0296         &     -0.0344         &      0.8589         &     -0.1167         &      0.1698         &      0.5297         \\
            &    (1.1337)         &    (1.2172)         &    (1.1696)         &    (1.2793)         &    (1.1218)         &    (1.2314)         &    (1.0973)         &    (1.2304)         &    (1.2157)         &    (1.2940)         &    (1.0901)         &    (1.2704)         \\ 
[1em]
Constant    &      0.2283         &     -0.0860         &     -0.2878         &      1.0234\sym{***}&      1.1390\sym{***}&      0.0300         &     -0.7642\sym{***}&     -0.9003         &     -1.3211\sym{***}&      0.2375\sym{*}  &      0.4241\sym{***}&     -0.9893\sym{***}\\
            &    (0.1428)         &    (0.5085)         &    (0.2903)         &    (0.0717)         &    (0.0271)         &    (0.2368)         &    (0.1616)         &    (0.5945)         &    (0.2649)         &    (0.1388)         &    (0.0293)         &    (0.2656)         \\
\hline
Time FE     &           Y         &           Y         &           Y         &           Y         &           Y         &           Y         &           Y         &           Y         &           Y         &           Y         &           Y         &           Y         \\
Dyad FE     &           Y         &           Y         &           Y         &           Y         &           Y         &           Y         &           Y         &           Y         &           Y         &           Y         &           Y         &           Y         \\
N           &        5284         &        5284         &        5284         &        5284         &        5284         &        5284         &        5284         &        5284         &        5284         &        5284         &        5284         &        5284         \\
R-squared   &        0.71         &        0.71         &        0.72         &        0.70         &        0.71         &        0.71         &        0.70         &        0.70         &        0.72         &        0.69         &        0.70         &        0.71         \\
\bottomrule
\multicolumn{13}{l}{Standard errors in parentheses. \sym{*} \(p<0.1\), \sym{**} \(p<0.05\), \sym{***} \(p<0.01\)}\\
    \multicolumn{13}{l}{Errors are clustered at user-level to account for intra-class correlation.}\\
\end{tabular}
}
\end{table}
\end{landscape}

\begin{landscape}
\begin{table}[H]
    \centering
    \caption{Responsiveness to official rate variations. Robustness check with alternative measures of profits (no fees, expected fees)}
    \label{tab:app_tab2fees}
    {\footnotesize
    \begin{tabular}{l*{12}{c}}
\toprule
           Dep. var.: &\multicolumn{6}{c}{Spread (Without fees)}               &\multicolumn{6}{c}{Spread (Expected fees)}             \\
            \cmidrule(lr){2-7}\cmidrule(lr){8-13}
            &\multicolumn{1}{c}{(1)}         &\multicolumn{1}{c}{(2)}         &\multicolumn{1}{c}{(3)}         &\multicolumn{1}{c}{(4)}         &\multicolumn{1}{c}{(5)}         &\multicolumn{1}{c}{(6)}         &\multicolumn{1}{c}{(7)}         &\multicolumn{1}{c}{(8)}         &\multicolumn{1}{c}{(9)}         &\multicolumn{1}{c}{(10)}         &\multicolumn{1}{c}{(11)}         &\multicolumn{1}{c}{(12)}         \\
\midrule
$ \Delta R \times $D(Currencies)&       1.176         &                     &                     &                     &                     &                     &       1.266         &                     &                     &                     &                     &                     \\
            &     (5.242)         &                     &                     &                     &                     &                     &     (5.226)         &                     &                     &                     &                     &                     \\
$ \Delta R \times $Log(Currencies)&                     &       1.880\sym{***}&                     &                     &                     &                     &                     &       1.767\sym{**} &                     &                     &                     &                     \\
            &                     &     (0.709)         &                     &                     &                     &                     &                     &     (0.707)         &                     &                     &                     &                     \\
$ \Delta R \times $Log(Actions)&                     &                     &       0.162         &                     &                     &                     &                     &                     &       0.157         &                     &                     &                     \\
            &                     &                     &     (0.408)         &                     &                     &                     &                     &                     &     (0.406)         &                     &                     &                     \\
$ \Delta R \times $D(Metaorder)&                     &                     &                     &       2.463\sym{***}&                     &                     &                     &                     &                     &       2.323\sym{***}&                     &                     \\
            &                     &                     &                     &     (0.854)         &                     &                     &                     &                     &                     &     (0.824)         &                     &                     \\
$ \Delta R \times $D(Aggressive)&                     &                     &                     &                     &       3.416         &                     &                     &                     &                     &                     &       3.346         &                     \\
            &                     &                     &                     &                     &     (3.466)         &                     &                     &                     &                     &                     &     (3.404)         &                     \\
$ \Delta R \times $PC1    &                     &                     &                     &                     &                     &       0.523\sym{**} &                     &                     &                     &                     &                     &       0.496\sym{**} \\
            &                     &                     &                     &                     &                     &     (0.216)         &                     &                     &                     &                     &                     &     (0.212)         \\
$\Delta R $     &      -0.496         &      -2.769\sym{*}  &      -0.355         &      -1.248         &       0.602         &      -3.182\sym{**} &      -0.607         &      -2.583\sym{*}  &      -0.343         &      -1.160         &       0.582         &      -2.998\sym{*}  \\
            &     (5.212)         &     (1.533)         &     (2.422)         &     (0.846)         &     (0.964)         &     (1.555)         &     (5.193)         &     (1.548)         &     (2.391)         &     (0.839)         &     (0.956)         &     (1.535)         \\
[1em]
Equiv. \$   &      -0.730         &      -0.872         &      -0.751         &      -0.810         &      -0.738         &      -0.818         &      -0.705         &      -0.839         &      -0.725         &      -0.780         &      -0.713         &      -0.788         \\
            &     (0.838)         &     (0.889)         &     (0.832)         &     (0.881)         &     (0.843)         &     (0.875)         &     (0.826)         &     (0.872)         &     (0.820)         &     (0.865)         &     (0.830)         &     (0.860)         \\
\midrule
User FE     &           Y         &           Y         &           Y         &           Y         &           Y         &           Y         &           Y         &           Y         &           Y         &           Y         &           Y         &           Y         \\
Time FE     &           Y         &           Y         &           Y         &           Y         &           Y         &           Y         &           Y         &           Y         &           Y         &           Y         &           Y         &           Y         \\
Dyad FE     &           Y         &           Y         &           Y         &           Y         &           Y         &           Y         &           Y         &           Y         &           Y         &           Y         &           Y         &           Y         \\
N           &        5142         &        5142         &        5142         &        5142         &        5142         &        5142         &        5142         &        5142         &        5142         &        5142         &        5142         &        5142         \\
R-squared   &        0.75         &        0.75         &        0.75         &        0.75         &        0.75         &        0.75         &        0.75         &        0.75         &        0.75         &        0.75         &        0.75         &        0.75         \\
\bottomrule
    \multicolumn{13}{l}{Standard errors in parentheses. \sym{*} \(p<0.1\), \sym{**} \(p<0.05\), \sym{***} \(p<0.01\)}\\
    \multicolumn{13}{l}{Errors are clustered at user-level to account for intra-class correlation.}\\
    \end{tabular}
    }
\end{table}
\end{landscape}

%% file: tr_appendix_learning.tex
\section{Robustness checks on learning-by-doing}
\label{tr:appendix_learning}

\setcounter{figure}{0}
\setcounter{table}{0}

Our analysis is based on the assumption that trade ability is an innate characteristic for the investors, i.e. it does not increase significantly over time through trading. In Section \ref{tr:preliminary} we address the concern that users might instead learn by trading and we show that this is unlikely in our context. As a robustness check, we repeat here the regressions reported in Tables~\ref{tab:tab1_base},~\ref{tab:tab1_alt_x},~\ref{tab:tab2_alt_x} by excluding the users active in multiple markets if the time passed between their first arbitrage action and the first one in a new currency market is large (i.e., more than 14 days). The results are shown in Tables~\ref{tab:tab1_bins},~\ref{tab:tab1_alt_x_bins},~\ref{tab:tab2_bins}, and their interpretation is in accordance with the one reported for the main estimations.

\vspace{1.5cm}

\begin{table}[H]
    \centering
    \caption{Relationship between trade ability and profits. Robustness check on learning-by-doing} 
    \label{tab:tab1_bins}
    \begin{tabular}{l*{4}{c}}
\toprule
Dep. var.:          &\multicolumn{4}{c}{Spread (with fees)} \\
\cmidrule{2-5}
            &\multicolumn{1}{c}{(1)}         &\multicolumn{1}{c}{(2)}         &\multicolumn{1}{c}{(3)}         &\multicolumn{1}{c}{(4)}         \\
\midrule
D(Currencies)&      1.6040\sym{***}&      1.6340\sym{***}&      0.8320\sym{***}&      0.8280\sym{***}\\
            &    (0.2099)         &    (0.2031)         &    (0.2391)         &    (0.2850)         \\
[1em]
Equiv. \$   &      2.4045         &      1.7055         &      0.1145         &      0.5385         \\
            &    (1.6007)         &    (1.3292)         &    (1.1578)         &    (0.9215)         \\
[1em]
Constant    &     -1.0294\sym{***}&     -1.0457\sym{***}&     -0.2440         &     -0.2314         \\
            &    (0.1605)         &    (0.1973)         &    (0.2202)         &    (0.2666)         \\
\hline
Time FE     &           N         &           N         &           Y         &           Y         \\
Dyad FE     &           N         &           Y         &           N         &           Y         \\
N           &        4817         &        4805         &        4053         &        4032         \\
R-squared   &        0.21         &        0.29         &        0.75         &        0.76         \\
    \bottomrule
\multicolumn{5}{l}{\footnotesize Standard errors in parentheses \sym{*} \(p<0.1\), \sym{**} \(p<0.05\), \sym{***} \(p<0.01\)}\\
\multicolumn{5}{l}{\footnotesize Errors are clustered at user-level to account for intra-class correlation.}\\
\end{tabular}
\end{table}

\begin{table}[H]
    \centering
    \caption{Relationship between trade ability and profits, alternative proxies. Robustness check on learning-by-doing}
    \label{tab:tab1_alt_x_bins}
    \begin{tabular}{l*{6}{c}}
\toprule
Dep. var.:          &\multicolumn{6}{c}{Spread (with fees)} \\
\cmidrule{2-7}
    &\multicolumn{1}{c}{(1)}         &\multicolumn{1}{c}{(2)}         &\multicolumn{1}{c}{(3)}         &\multicolumn{1}{c}{(4)}         &\multicolumn{1}{c}{(5)} 
    &\multicolumn{1}{c}{(6)}\\
\midrule
D(Currencies)&      0.8280\sym{***}&                     &                     &                     &                     &                     \\
            &    (0.2850)         &                     &                     &                     &                     &                     \\
[1em]
Log(Currencies)&                     &      1.5511\sym{**} &                     &                     &                     &                     \\
            &                     &    (0.6296)         &                     &                     &                     &                     \\
[1em]
Log(Actions)&                     &                     &      0.3741\sym{***}&                     &                     &                     \\
            &                     &                     &    (0.1056)         &                     &                     &                     \\
[1em]
D(Metaorder)&                     &                     &                     &      1.1473\sym{***}&                     &                     \\
            &                     &                     &                     &    (0.2191)         &                     &                     \\
[1em]
D(Aggressive)&                     &                     &                     &                     &     -1.1139\sym{***}&                     \\
            &                     &                     &                     &                     &    (0.2103)         &                     \\
[1em]
PC1         &                     &                     &                     &                     &                     &      0.2457\sym{***}\\
            &                     &                     &                     &                     &                     &    (0.0832)         \\
[1em]
Equiv. \$   &      0.5385         &      0.4517         &      0.0911         &      0.5386         &      0.7176         &      0.3940         \\
            &    (0.9215)         &    (0.8186)         &    (0.7401)         &    (0.9046)         &    (0.8718)         &    (0.8288)         \\
[1em]
Constant    &     -0.2314         &     -2.3370\sym{*}  &     -1.9417\sym{***}&     -0.5109\sym{**} &      0.5793\sym{***}&     -1.3833\sym{**} \\
            &    (0.2666)         &    (1.1859)         &    (0.7161)         &    (0.2200)         &    (0.0526)         &    (0.6772)         \\
\hline
Time FE     &           Y         &           Y         &           Y         &           Y         &           Y         &           Y         \\
Dyad FE     &           Y         &           Y         &           Y         &           Y         &           Y         &           Y         \\
N           &        4032         &        4032         &        4032         &        4032         &        4032         &        4032         \\
R-squared   &        0.76         &        0.79         &        0.81         &        0.76         &        0.76         &        0.78         \\
\bottomrule
\multicolumn{7}{l}{\footnotesize Standard errors in parentheses \sym{*} \(p<0.1\), \sym{**} \(p<0.05\), \sym{***} \(p<0.01\)}\\
\multicolumn{7}{l}{\footnotesize Errors are clustered at user-level to account for intra-class correlation.}\\
\end{tabular}
\end{table}

\begin{table}[H]
    \centering
    \caption{Responsiveness to official rate variations. Robustness check on learning-by-doing}
    \label{tab:tab2_bins}
    \begin{tabular}{l*{6}{c}}
\toprule
Dep. var.:          &\multicolumn{6}{c}{Spread (with fees)} \\
\cmidrule{2-7}
            &\multicolumn{1}{c}{(1)}         &\multicolumn{1}{c}{(2)}         &\multicolumn{1}{c}{(3)}         &\multicolumn{1}{c}{(4)}         &\multicolumn{1}{c}{(5)}         &\multicolumn{1}{c}{(6)}         \\
\midrule
$ \Delta R \times $D(Currencies)&      16.616\sym{**} &                     &                     &                     &                     &                     \\
            &     (6.460)         &                     &                     &                     &                     &                     \\
[1em]
$ \Delta R \times $Log(Currencies)&                     &       3.048         &                     &                     &                     &                     \\
            &                     &     (2.129)         &                     &                     &                     &                     \\
[1em]
$ \Delta R \times $Log(Actions)&                     &                     &       1.143\sym{**} &                     &                     &                     \\
            &                     &                     &     (0.504)         &                     &                     &                     \\
[1em]
$ \Delta R \times $D(Metaorder)&                     &                     &                     &       3.997         &                     &                     \\
            &                     &                     &                     &     (4.080)         &                     &                     \\
[1em]
$ \Delta R \times $D(Aggressive)&                     &                     &                     &                     &      -1.333         &                     \\
            &                     &                     &                     &                     &     (7.365)         &                     \\
[1em]
$ \Delta R \times $PC1    &                     &                     &                     &                     &                     &       0.991         \\
            &                     &                     &                     &                     &                     &     (0.598)         \\
[1em]
$\Delta R $     &     -15.963\sym{**} &      -5.242         &      -7.116\sym{*}  &      -3.198         &       0.668         &      -7.463         \\
            &     (6.042)         &     (4.754)         &     (3.605)         &     (3.883)         &     (1.580)         &     (4.947)         \\
[1em]
Equiv. \$   &      -0.024         &      -0.051         &      -0.114         &      -0.035         &      -0.035         &      -0.073         \\
            &     (0.640)         &     (0.639)         &     (0.646)         &     (0.647)         &     (0.647)         &     (0.645)         \\
\hline
User FE     &           Y         &           Y         &           Y         &           Y         &           Y         &           Y         \\
Time FE     &           Y         &           Y         &           Y         &           Y         &           Y         &           Y         \\
Dyad FE     &           Y         &           Y         &           Y         &           Y         &           Y         &           Y         \\
N           &        3957         &        3957         &        3957         &        3957         &        3957         &        3957         \\
R-squared   &        0.84         &        0.84         &        0.84         &        0.84         &        0.84         &        0.84         \\
    \bottomrule
\multicolumn{5}{l}{\footnotesize Standard errors in parentheses \sym{*} \(p<0.1\), \sym{**} \(p<0.05\), \sym{***} \(p<0.01\)}\\
\multicolumn{5}{l}{\footnotesize Errors are clustered at user-level to account for intra-class correlation.}\\
\end{tabular}
\end{table}

%% file: tr_appendix_individual_mkts.tex
\section{Analysis of individual fiat-to-fiat markets}
\label{tr:appendix_individual}

\setcounter{figure}{0}
\setcounter{table}{0}

\begin{table}[H]
    \centering
    \caption{Profitability of the arbitrage actions, illustrative example on \textit{individual} markets}
    \label{tab:profitability_2}
    \begin{tabular}{ccccccc}
    \toprule
    &     &  & \multicolumn{2}{c}{Without fees}  &  \multicolumn{2}{c}{With fees} \\
    \cmidrule(r{0.1cm}){4-5} \cmidrule(l{0.1cm}){6-7}
     buy leg &    sell leg   &     official e.r.  & implied e.r.   &   profit/loss & implied e.r.   &   profit/loss   \\    
    \midrule
     EUR   &   USD    &   1.40     &   1.405    &    profit  &   1.402    &    profit  \\
     EUR   &    USD   &   1.40     &   1.395    &   loss  &   1.390    &   loss    \\ 
     USD   &    EUR   &   1.40     &   1.395    &    profit  &   1.399    &    profit   \\
     USD   &    EUR   &   1.40     &   1.405    &    loss   &   1.410    &    loss    \\ 
     \bottomrule
    \end{tabular}
    \floatfoot{\emph{Notes:} to compare graphically the actions executed in the same currency market, we now construct the profits (spread between implied and official rates) keeping the official rate \textit{fixed}. That is, we compare all actions belonging to a given dyad to the same official rate. Here, for example, we consider the dyad (EUR,USD) and we express implied rates in EURtoUSD, independently of which currency appears in the buy side. When the implied rate is higher than the official one, it is profitable to buy bitcoins in EUR and sell them for USD, that is, to implicitly buy USD using EUR; viceversa when the implied rate is lower. Fees reduce the margins for profit and worsen losses.}
\end{table}

\begin{figure}[htbp]
	\centering
	\caption{Comparison between the implied and the official exchange rate for the detected arbitrage actions (left), and the multi-currency trades (right)}
	\label{fig:absolute}
	\begin{subfigure}{0.475\textwidth}
		\includegraphics[width=\textwidth]{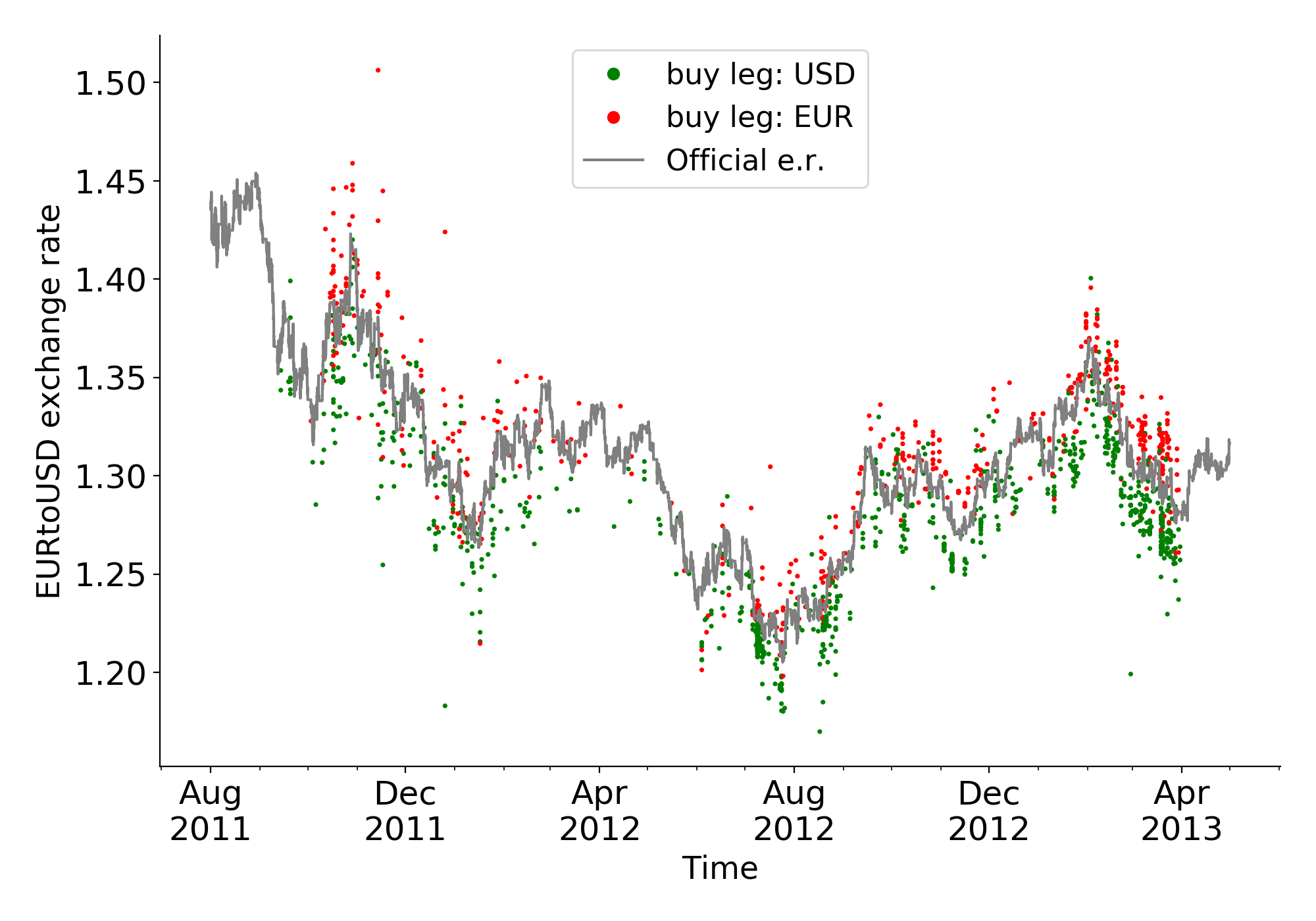}%
		\caption{{\small All arbitrage actions}}
		\label{fig:abs_EURUSD}
	\end{subfigure}
	\begin{subfigure}{0.475\textwidth}
		\includegraphics[width=\textwidth]{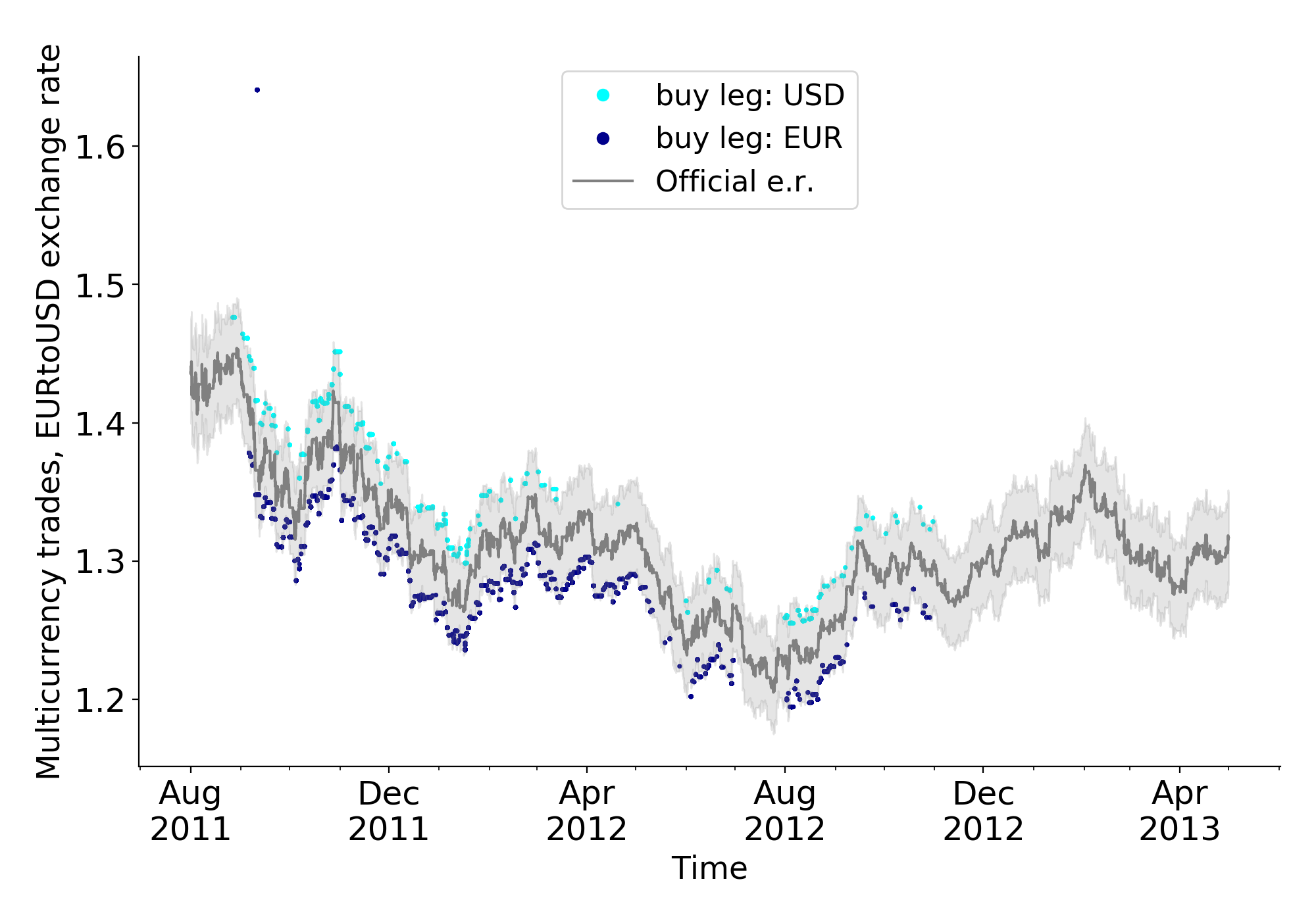}
		\caption{{\small Multi-currency trades, EUR/USD market}}
		\label{fig:tibanne_EURUSD}
	\end{subfigure}
	\floatfoot{\emph{Notes:} values are reported in absolute terms; the gray line represents the official exchange rate. Note that, while in Panel~(\subref{fig:abs_EURUSD}) the trades are distributed coherently with respect to a profit-based logic, the multi-currency trades (Panel~(\subref{fig:tibanne_EURUSD})) are always non-profitable and `anchor' the Bitcoin price across different markets.}
\end{figure}

Figure~\ref{fig:absolute} focuses on the EUR/USD market; Panel~\ref{fig:abs_EURUSD} reports the official exchange rate (gray line) along with all the N = 1,931 triangular actions performed in that market, depicted as dots whose color is red when the bitcoins are bought in EUR and sold in USD - that is, actions where users implicitly buy USD using EUR - and vice versa for the green ones. We will refer to them respectively as the `EUR buy' trades and the `USD buy' trades.
The implied rates in the plots are computed excluding the transaction costs. We underline that the purpose of arbitrage is to \textit{profit} from an asset mispricing across different markets: as shown in Table~\ref{tab:profitability_2}, the points above (below) the line are profitable if the investor buys bitcoins in EUR (USD) and sells for USD (EUR). 

Interestingly, we do not observe persistent price deviations with respect to the official exchange rate, nor we observe arbitrage activity taking the form of an organic response to mispricings from multiple individuals independently trading across currency markets in a univocal direction (determined by the sign of the price deviation), pushing prices towards equilibrium. Rather, we do observe that the implied exchange rate closely tracks the official one and that actions seemingly exploit opportunities arising due to price volatility between the two of them.
We interpret such dynamics as the consequence of many uninformed traders who push prices for a short time far from fundamental values, and the arbitrageurs as the liquidity providers that profit from these demand shocks by exploiting such mispricings. This interpretation is consistent with the results reported in the main body of the paper, with the limited aggressiveness of the expert users on the markets, and with the model of arbitrageurs as liquidity providers described in \citet{gromb2010limits}.

We also believe that the arbitrage opportunities were likely limited by the existence of a feature called multi-currency trading, which was introduced in September 2011 along with the possibility to trade in fiat currencies other than USD. 
Such feature facilitated the bitcoin trades against minor currencies by allowing the execution of orders involving different currencies in the buy and sell leg, for an additional fee of 2.5\%\footnote{Excluding the fact that the buy and the sell legs are executed in different currencies, these are in all respects normal trades: the two legs are labeled by the same trade ID and performed by different users. To conduct such trades, Mt.\ Gox acted as an intermediary via a virtual internal account. See \url{https://bit.ly/3jXtTPe}. For the purposes of our analysis, it is also interesting to notice that only around 1\% of the arbitrage actions are composed by legs that are part of multi-currency trades: arbitrageurs do not use this feature, as it entails additional fees.}.
It is very likely that this mechanism anchored the movements of the bitcoin price in minor markets to the main one, BTCtoUSD. 
Panel~\ref{fig:tibanne_EURUSD} compares the official exchange rate to the implied rate for the trades in EUR and USD that exploited the multi-currency feature. For reasons detailed in Appendix~\ref{tr:appendix_data_cleaning} and related to how Mt.\ Gox internally transcribed the trades, we provide information on this mechanism only with respect to the time period ranging from August 2011 to October 2012 excluding July 2012. These trades are systematically unprofitable. (If we account for and remove the 2.5\% fee, the dots converge steadily to the official exchange rate: the gray line shows the official EURtoUSD exchange rate, and the gray band delimits the $ \pm 2.5\% $ deviation area.)

\begin{figure}[tp]
	\centering
	\caption{Difference between the implied and the official exchange rate as a percentage of the exchange rate for the EURtoUSD market}
	\label{fig:relative}
	\begin{subfigure}{\textwidth}
		\includegraphics[width=0.5\textwidth]{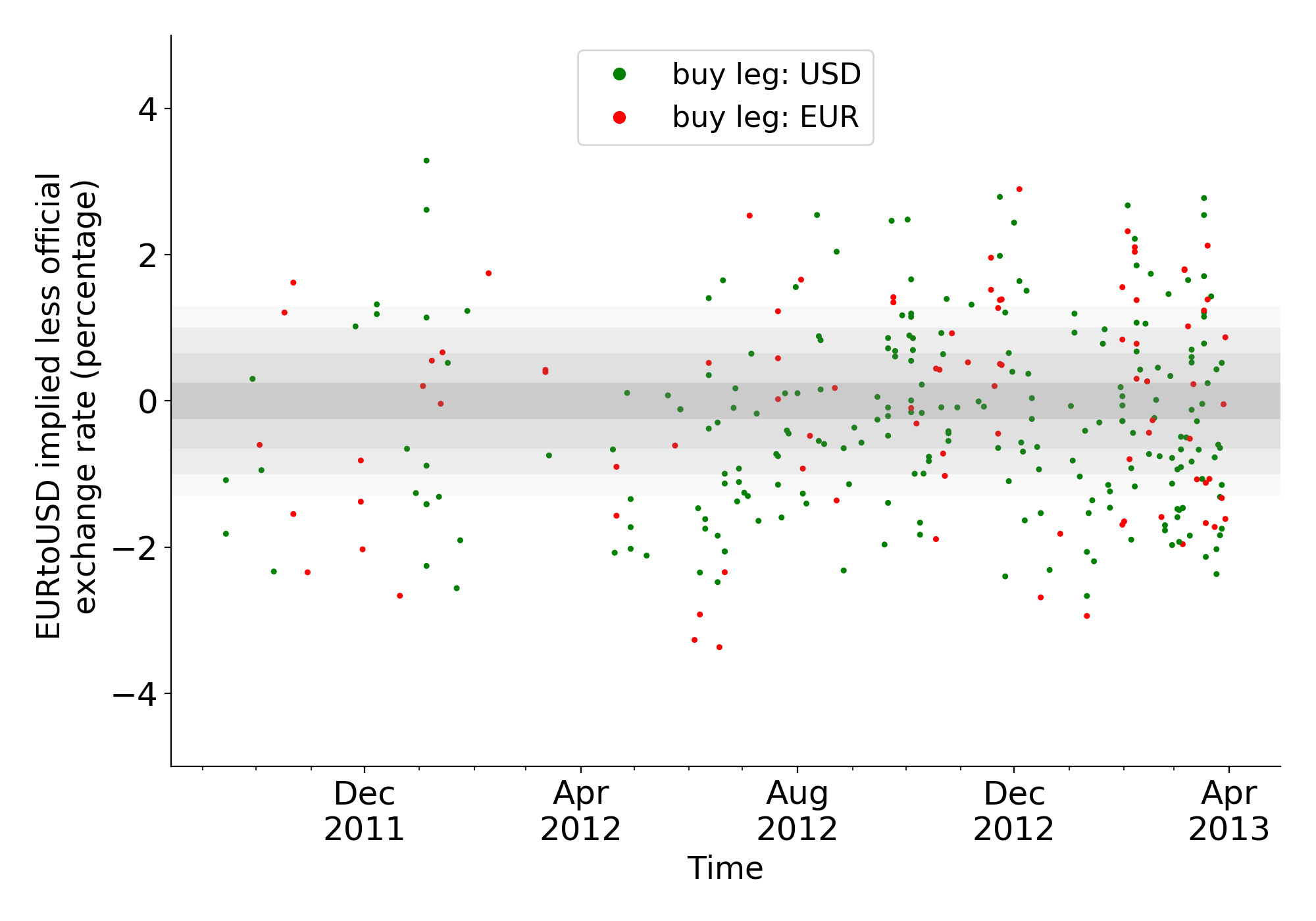}%
		\includegraphics[width=0.5\textwidth]{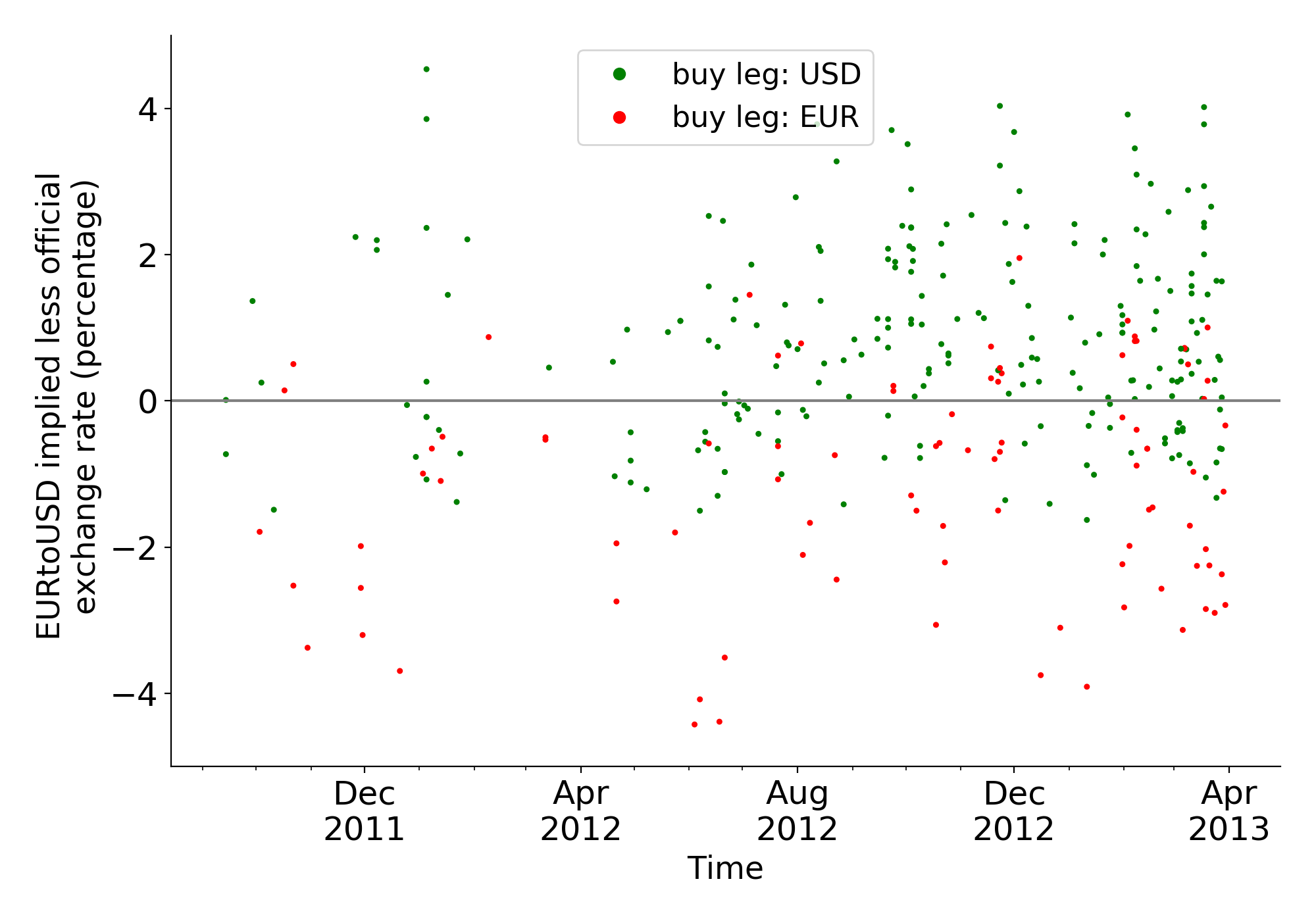}
		\caption{{\small Few actions, without fees (left) \& including fees (right)}}
		\label{fig:few}
	\end{subfigure}
	\centering
	\begin{subfigure}{\textwidth}
		\includegraphics[width=0.5\textwidth]{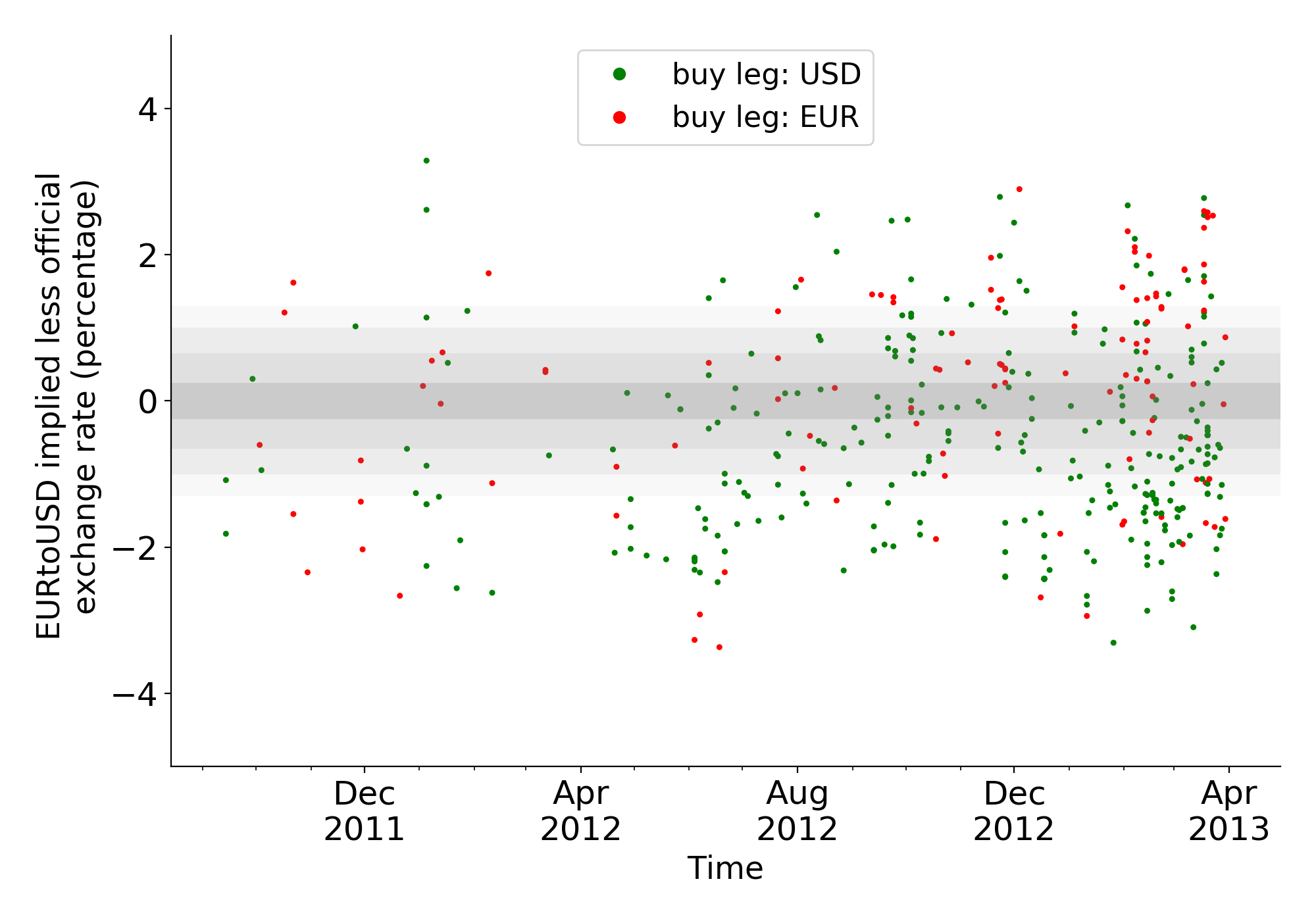}%
		\includegraphics[width=0.5\textwidth]{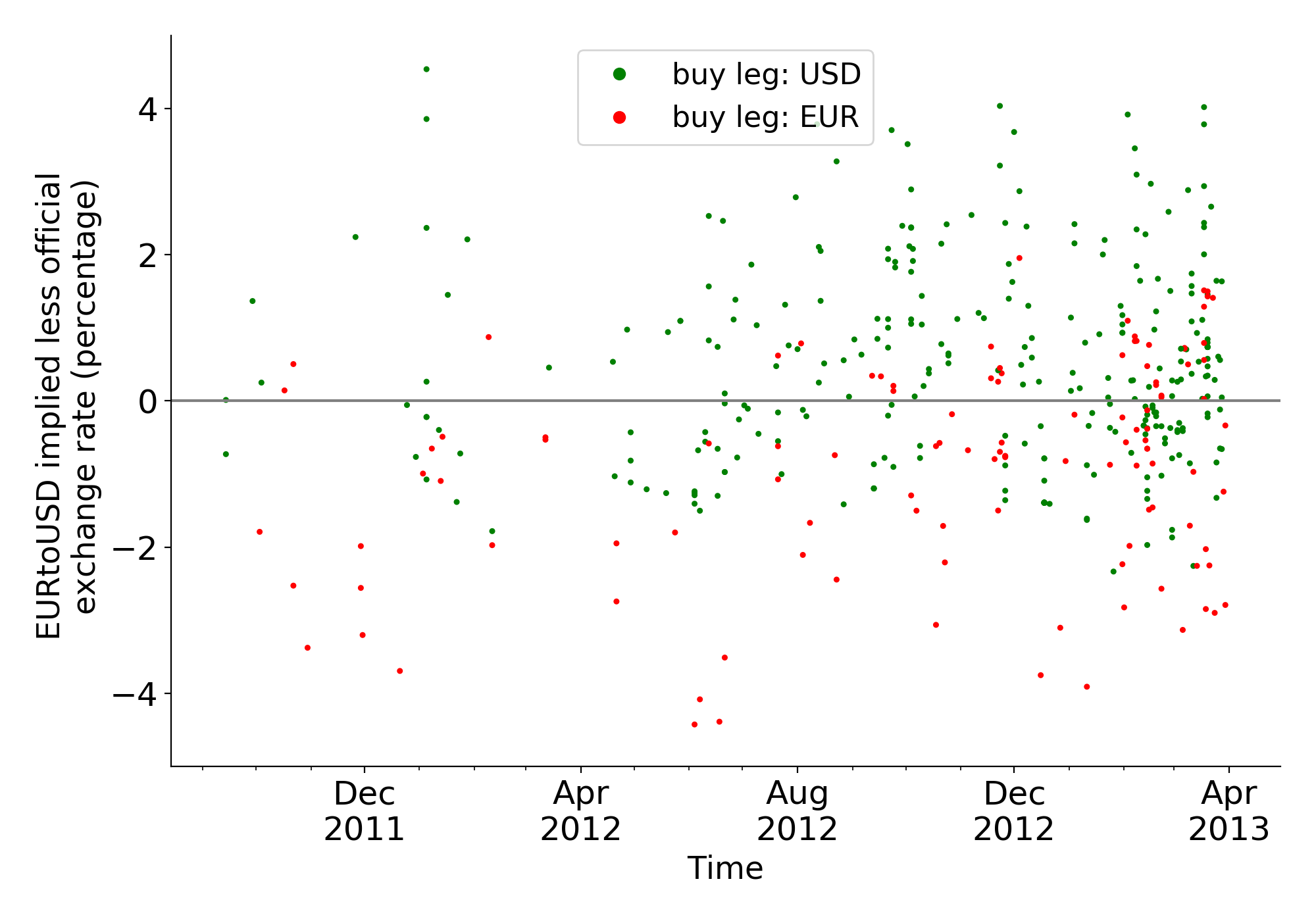}
		\caption{{\small Single market, without fees (left) \& including fees (right)}}
		\label{fig:two}
	\end{subfigure}
	\begin{subfigure}{\textwidth}
		\includegraphics[width=0.5\textwidth]{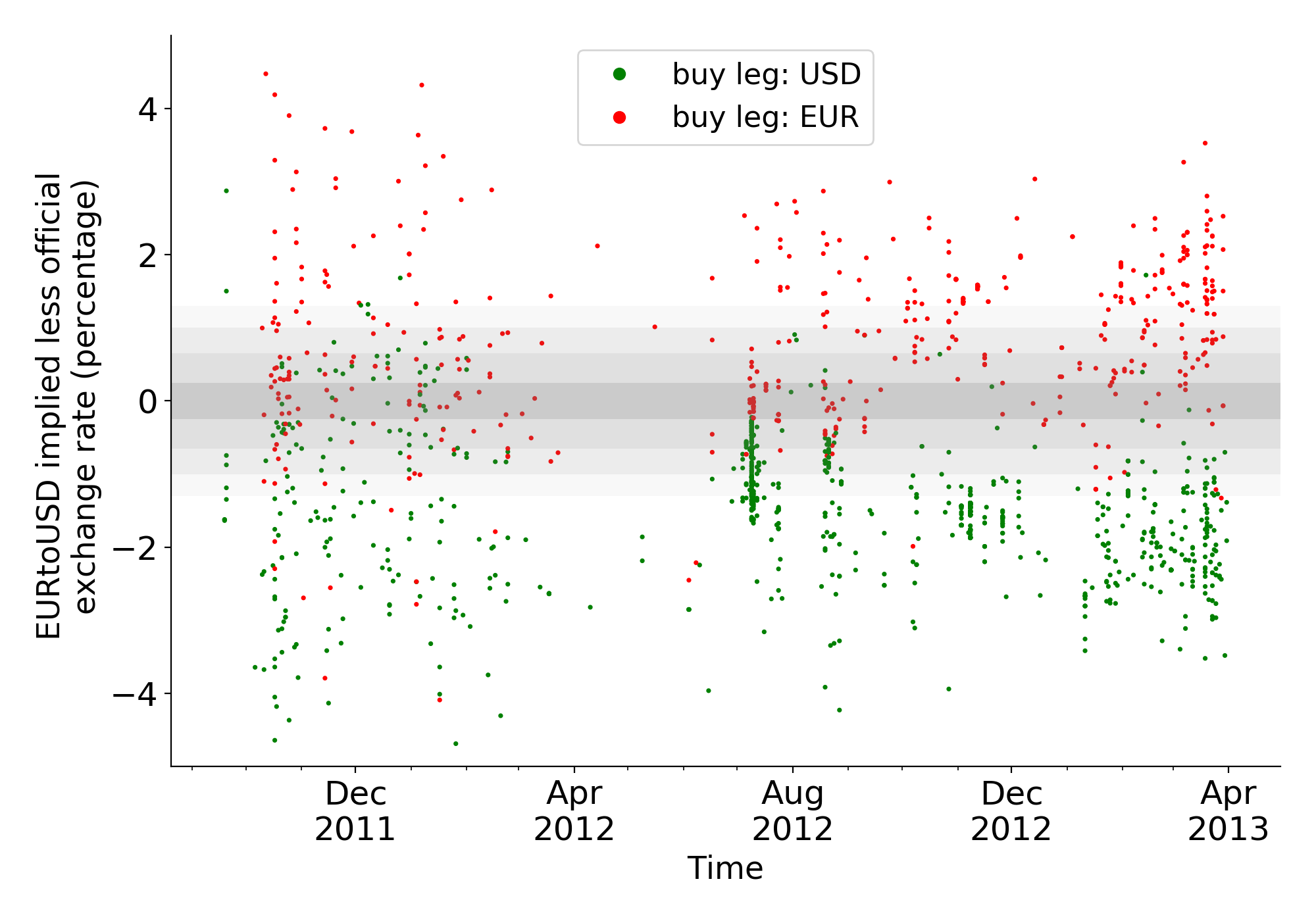}
		\includegraphics[width=0.5\textwidth]{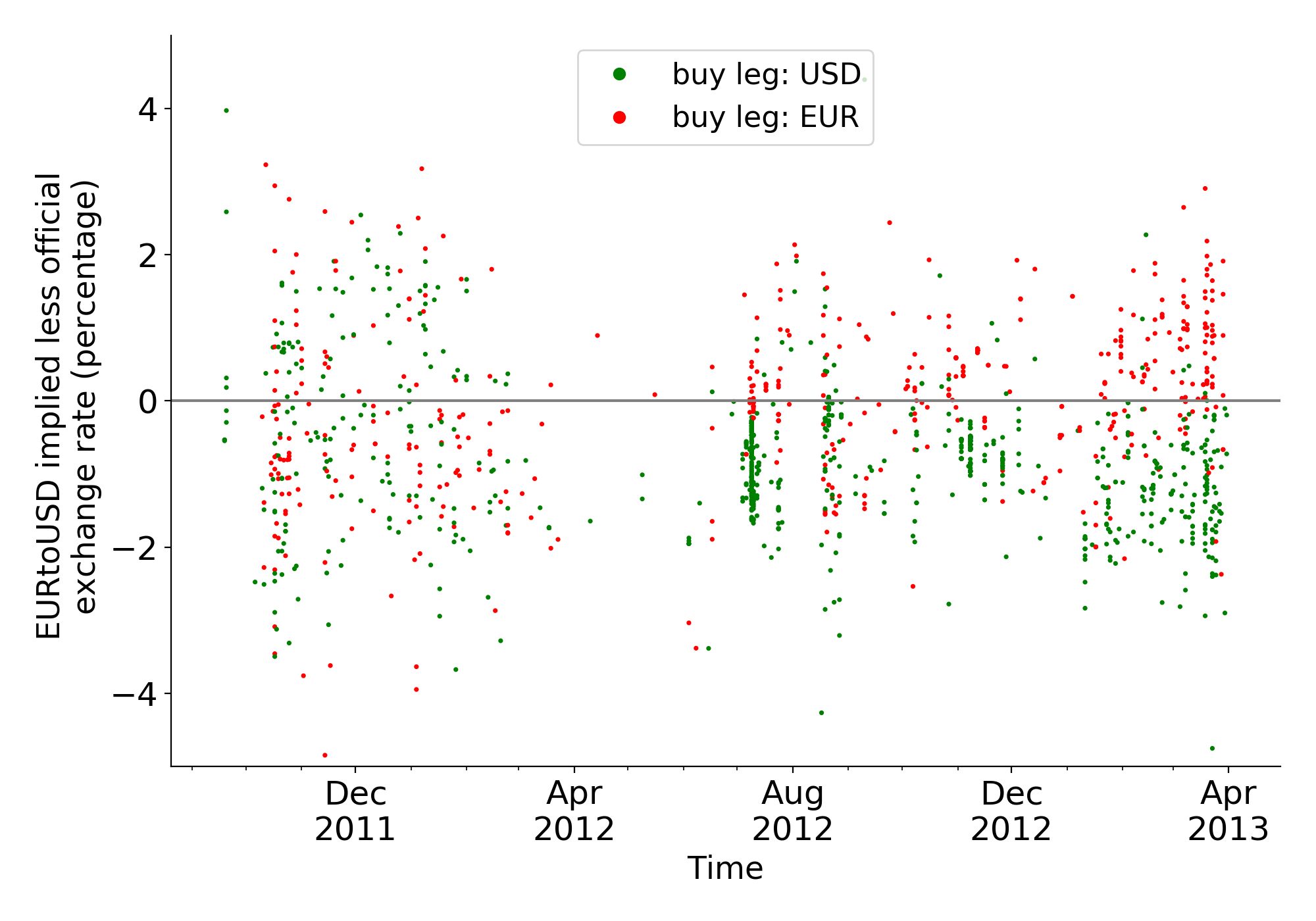}
		\caption{{\small Multiple market, without fees (left) \& including fees (right)}}
		\label{fig:manycurr}
	\end{subfigure}
	\floatfoot{\emph{Notes:} Panel~(\subref{fig:few}) reports only the actions executed by the users that made less than 10 arbitrage actions, Panel~(\subref{fig:two}) those executed by the users active in a single market, and Panel~(\subref{fig:manycurr}) those executed by users active in multiple markets. On the right, fees are included; on the left, fees are excluded. The gray shaded bands represent the area in which transaction costs might exceed the potential profits.}
\end{figure}

Further insights on the dynamics of the arbitrage activity result when we decompose the actions reported in Panel~\ref{fig:abs_EURUSD} per groups of users.
We thus report in Figure~\ref{fig:relative} the spread between the implied rate and the official rate as a percentage of the latter, and we consider three cases: we show the actions executed by users that made less than 10 arbitrage actions in Panel~\ref{fig:few}, and those made by users active in a single market in Panel~\ref{fig:two}. In Panel~\ref{fig:manycurr} we consider instead only the actions ascribable to the users active in multiple markets.
On the right side we include the transaction costs, while on the left side we exclude them; the gray shaded band approximates the area in which the transaction costs might\footnote{As discussed in Appendix~\ref{tr:appendix_fees}, the fees paid depend on the users' trading history, thus are specific to each individual and vary across users and in time.} exceed the potential gains originated by the spread between the implied and the official rate. The differences between the first two cases and the third are evident; in the former two, the points are more randomly dispersed around the zero, especially when the transaction costs are included, while the latter shows a pattern that distinguishes the EUR buy trades, mainly above the zero, and the USD buy trades, mainly below.
Second, in the latter scenario trades are less dispersed - they are more concentrated in shorter time periods - and seemingly part of complex automated strategies (as we discussed in Section~\ref{tr:identification}). 
Additional information on other currency markets is reported in Appendix~\ref{tr:appendix_supplem}.

%% file: tr_appendix_datacleaning.tex
\section{Dataset cleaning procedures} 
\label{tr:appendix_data_cleaning}

\setcounter{figure}{0}
\setcounter{table}{0}


The goal of this Appendix is to provide a reference documentation for the Mt.\ Gox leaked dataset. Indeed, the reported variables do not come with a documentation explaining their meaning, which is thus derived through comparisons with previous research conducted over it.
	
The first part of the polishing procedures is devoted to identify and remove the redundant duplicate rows contained in the leaked dataset. After a short description of the Mt.\ Gox leaked files, we follow the methodology described in the literature by~\citet{feder2018impact}, ~\citet{gandal2018price} and~\citet{scaillet2017high}: for reasons explained below, we slightly deviate from their approaches and implement our own deduplication methods. We discuss which files of the leaked dataset can be safely discarded without any loss of information and we control for the presence of duplicate rows in the remaining files. Second, we further remove the rows with misreported entries. In some cases the correct values can be retrieved: thus, when possible, we chose to extrapolate the plausibly correct values. Finally, we conduct additional sanity checks on the quality of the data: we verify the correctness of the filtered data by comparison to external sources of information, that is, a dataset made public by Mt.\ Gox, and aggregated data from the website Bitcoincharts.com, a benchmark for academic studies on cryptocurrency markets that collects information at the trade level on the main cryptocurrency exchange platforms.

\subsection{Description of the leaked log files}

The analysis is conducted over a total of 62 .CSV monthly files that cover a time period ranging from April 2011 to November 2013. 
The files, once merged, amount to 22,175,247 rows per 19 columns. Overall, they share a common structure: the data are reported as a sequence of trades, each identified by a \textit{trade ID}, and each composed of two rows corresponding to a buy and a sell legs. Most importantly, each leg is associated to a \textit{user ID}. A trade can be schematized as follows:

\begin{center}
\begin{tabular}{ccccccccc}
\toprule
    Date & Trade ID & User ID & Type & Bitcoins &  Money & Currency & Fees\\
    \midrule
   T &  0 & 1 & buy  & Amount$_{BTC}$ & Amount$_{CUR}$ & CUR & F1 \\
   T & 0 & 2 & sell & Amount$_{BTC}$ & Amount$_{CUR}$ & CUR & F2 \\
    \bottomrule
\end{tabular}  
\end{center}

Along with the standard matching mechanism, in September 2011 Mt.\ Gox introduced the multi-currency trading, a feature allowing investors to match orders even if executed by other users using a different fiat currency (at the cost of an additional fee). This matching mechanism, slightly more complex than the single currency trade, requires the administrator of the exchange platform to act as an intermediary between parties. These trades represent a marginal fraction of the whole log of trades.
	
The single leaked files largely overlap: several legs are duplicates and the data are not reported in a homogeneous format. Some patterns can anyway be identified. First, for each month it is possible to identify one primary file that contains all the relevant information, while all the remaining ones are redundant subsets. Second, the whole body of files can be grouped in 3 macro sets, each sharing the same structure; the differences across them mainly concern the way multi-currency trades are transcribed, and the columns stored. Thus, files are divided and analyzed in three blocks, sharing a similar pattern:
	
\begin{enumerate}
	\item files related to \textbf{April 2011}. Two different files are reported for April 2011, both composed of 15 columns: \textit{Trade\_Id, Date, User\_Id, Japan, Type, Currency, Bitcoins, Money, Money\_Rate, Money\_JPY, Money\_Fee, Money\_Fee\_Rate, Money\_Fee\_JPY, Bitcoin\_Fee, Bitcoin\_Fee\_JPY};
	no multi-currency trades are present.
	
	\item files from \textbf{May 2011 to October 2012}: for every month there is only one .CSV file, each with the same columns of the April 2011 files. Multi-currency trades are implemented as follows:
	
	\begin{center}
	\begin{tabular}{ccccc}
	\toprule
	    User ID & Type & Bitcoins &  Money & Currency \\
	    \midrule
	   1 & buy  & A$_{BTC}$ & A$_{CUR1}$ & CUR1 \\
	   TIBANNE\_LIMITED\_HK  & sell  & A$_{BTC}$ & A$_{CUR1}$ & CUR1 \\
	   2 & sell & A$_{BTC}$ & A$_{CUR2}$ & CUR2 \\
	   TIBANNE\_LIMITED\_HK  & buy  & A$_{BTC}$ & A$_{CUR2}$ & CUR2 \\
	    \bottomrule
	\end{tabular}  
	\end{center}
	
	we will refer to this implementation method as the `Tibanne' one;
	
	\item files from \textbf{November 2012 to November 2013}. Two files per month are available, one denominated 'Coinlab' and one denominated 'mtgox\_japan'. In addition, for February, March, and April 2013, also weekly datasets (with a similar nomenclature) are included. All of them are composed of 19 columns: the 15 aforementioned, plus \textit{User, User\_Id\_Hash, User\_Country, User\_State}; the `Coinlab' and weekly files are subsets of the original `mtgox\_japan' ones. The multi-currency trades are stored in the database as follows:
	
	\begin{center}
	\begin{tabular}{ccccc}
	\toprule
	    User ID & Type & Bitcoins &  Money & Currency \\
	    \midrule
	   3 & buy  & A$_{BTC}$ & A$_{CUR1}$ & CUR1 \\
	   THK  & sell  & A$_{BTC}$ & A$_{CUR1}$ & CUR1 \\
	   3 & buy  & A$_{BTC}$ & A$_{CUR1}$ & CUR1 \\
	   THK  & sell  & A$_{BTC}$ & A$_{CUR1}$ & CUR1 \\
	    \bottomrule
	\end{tabular}  
	\end{center}
	
	we will refer to this implementation method as the `THK' one. Note that the trades are misreported, as already noticed by previous researchers: each secondary leg - the `sell side' - copies the values of the first one, thus reporting twice the buy side of the trade. The sell side of the `THK' multi-currency trades is never reported. 
	
	\item The file related to \textbf{July 2012} represents an exception to this classification: it shares the same structure and properties of the second group, but the  multi-currency trades follow the `THK' scheme.
	
	Note that in all the files the `Tibanne' and `THK' methods are mutually exclusive.
	
\end{enumerate}

The 19 columns can be interpreted in the following way:

\begin{itemize}
	\item \textit{Trade\_Id}: identifier of the trade. Each trade identifier appears once associated to the buy leg and once to the sell leg composing the same trade; the identifiers are reported as a simple sequential increasing number until 19 June 2011, while from 26 June 2011 on, as suggested in~\cite{scaillet2017high}, they correspond to the concatenation of a POSIX timestamp and a microsecond timestamp\footnote{On 19 June 2011, Mt.\ Gox's website went down for several days after a security breach. See \url{https://bit.ly/3dtzNFf}.};
	
	\item \textit{Date}: date of execution of the trade; time is reported in format YYYY-MM-DD hh:mm:ss;
	
	\item \textit{Type}: value that determines whether the row represents a buy or a sell leg. The object of exchange is an amount of bitcoins; thus the buyer purchases bitcoins denominated in fiat currency, and viceversa for the seller;
	
	\item \textit{User\_Id}: parameter that identifies the user who performed the buy or sell action;
	
	\item \textit{Japan}: the meaning of this variable is not clear. It can take two values: 'JP' and 'NJP'; it could be a parameter informative on the geographical origin of the trade execution, or it might indicate a special category of users (or trades);
	
	\item \textit{Currency}: fiat currency used in the trade. In total, 17 different currencies are used: US Dollar, Euro, British Pound, Polish Zloty, Australian Dollar, Japanese Yen, Canadian Dollar, Swedish Krone, Swiss Franc, Russian Ruble, Chinese Yuan, New Zealand Dollar, Singapore Dollar, Hong Kong Dollar, Danish Krone, Norwegian Krone, Thai Bat.
	
	\item \textit{Bitcoins}: amount of bitcoins exchanged;
	
	\item \textit{Money}: quantity of fiat currency traded;
	
	\item \textit{Money\_JPY}: quantity of fiat currency traded, expressed in Japanese Yen;
	
	\item \textit{Money\_Rate}: exchange rate used to convert the value in the field `Money' into Japanese Yen;
	
	\item \textit{Money\_Fee} and \textit{Bitcoin\_Fee}: fees paid to perform the trade. Normally, fees were paid in fiat money by the seller and in bitcoins by the buyer;
	
	\item \textit{Money\_Fee\_JPY} and \textit{Bitcoin\_Fee\_JPY}: fees paid to perform the trade, expressed in Japanese Yen;
	
	\item \textit{Money\_Fee\_Rate}: exchange rate used to convert the value in the fields `Money\_Fee' and `Bitcoin\_Fee' into Japanese Yen; 
	
	\item \textit{User}: user identifier expressed in hexadecimal base;
	
	\item \textit{User\_Id\_Hash}: hashed representation of the user identifier;
	
	\item \textit{User\_Country}: National geographic location of the user;
	
	\item \textit{User\_State}: Regional geographic location of the user.
	
\end{itemize}

\subsection{Deduplication methods - comparisons with the literature}

Members of the Mt.\ Gox users community were among the first to explore the leaked dataset. Supposedly they wanted to prove misbehavior of the exchange in the events that lead to its bankruptcy on 28 February 2014. The volunteers analyzed the structure of the dataset, tried to identify potential malicious users, and pointed out key issues to keep in mind\footnote{See, for example: 
\url{https://bit.ly/34WhQvo}, 
\url{https://bit.ly/2FtFtCJ}, 
\url{https://bit.ly/3iWtBXB}, 
\url{https://bit.ly/3dt0gD4}, and 
\url{https://bit.ly/3lQi8e9}.}.
Prior to carrying out this analysis, we followed their example, and replicated the steps adopted by Gandal et al. and Feder et al., in order to remove duplicates from the dataset. However, we slightly deviate from their method in a way explained and justified in the following.

Previous works use two related methods to detect duplicates. The first one (method \textit{Conservative}) detects rows as duplicates if the following entries are equal: \textit{user id, timestamp, buy/sell action, amount in BTC, amount in Yen}. The other method detects rows as duplicates if the following entries are equal: \textit{user id, timestamp, buy/sell action, amount in BTC}. The latter is more aggressive because it removes a higher number of rows, hence previous works refer to it as \textit{Aggressive}.

However, these approaches also treat as duplicates unwanted legs, and do not take into account the likely presence of metaorders in the leaked dataset. Consider the case in which a user performs two exactly equivalent trades at the same moment, with the only difference that the complementary leg is executed by different trading partners: both deduplication methods mentioned above remove one of the two exactly equivalent legs. Thus, these methods reduce the dataset more than desirable.

To prevent this behavior, we slightly changed the deduplication \textit{Aggressive} method, by adding the \textit{trade id} value to the set of variables used to detect duplicates as in~\cite{scaillet2017high} (method \textit{TradeId}). As a result, rows are detected as duplicates if the following entries are equal: \textit{trade id, user id, timestamp, buy/sell action, amount in BTC}. 
To compare the results, we also implemented another deduplication technique (method \textit{Pairs}), based on the \textit{Aggressive} method, but the legs of a trade are not treated independently: rows are considered as duplicates \textit{only if} both legs are duplicates.

To clarify the differences among the different methods, a series of example trades and the resulting deduplications are shown in the following.

\textbf{Original sample.} Table~\ref{tab:dedOrig} shows the original table. It also corresponds to the deduplication results of method \textbf{TradeId}, meaning that in that specific case the \textit{TradeId} method does not remove any duplicate. Consider the following example: rows 937 and 939 show two equal legs, having the same values for \textit{user id, timestamp, buy/sell action, amount in BTC, amount in Yen}; however, since the \textit{trade id} marks them as distinct, they are not treated as duplicates.

\begin{table}
	\centering
	\caption{Original sample and result of \textit{TradeId} deduplication method}
	\label{tab:dedOrig}
	\begin{tabular}{lrlrlrr}
		\toprule
		{} &  Trade\_Id &                 Date &  User\_Id &  Type &  Bitcoins &   Money\_JPY \\
		\midrule
		\textbf{930} &     35837 &  11-04-04 14:23 &     2824 &   buy &      10.0 &  586.89 \\
		\textbf{931} &     35837 &  11-04-04 14:23 &      388 &  sell &      10.0 &  586.89 \\
		\textbf{932} &     35838 &  11-04-04 14:23 &     3111 &   buy &      10.0 &  578.42 \\
		\textbf{933} &     35838 &  11-04-04 14:23 &      388 &  sell &      10.0 &  578.42 \\
		\textbf{934} &     35839 &  11-04-04 14:23 &     2824 &   buy &      10.0 &  570.20 \\
		\textbf{935} &     35839 &  11-04-04 14:23 &      388 &  sell &      10.0 &  570.20 \\
		\textbf{936} &     35840 &  11-04-04 14:23 &     3111 &   buy &      10.0 &  570.00\\
		\textbf{937} &     35840 &  11-04-04 14:23 &      388 &  sell &      10.0 &  570.00 \\
		\textbf{938} &     35841 &  11-04-04 14:23 &     1000 &   buy &      10.0 &  570.00 \\
		\textbf{939} &     35841 &  11-04-04 14:23 &      388 &  sell &      10.0 &  570.00 \\
		\bottomrule
	\end{tabular}
\end{table}

\begin{table}
	\centering
	\caption{Result of \textit{Conservative} deduplication technique}
	\label{tab:dedCons}
	\begin{tabular}{lrlrlrr}
		\toprule
		{} &  Trade\_Id &                 Date &  User\_Id &  Type &  Bitcoins &   Money\_JPY \\
		\midrule
		\textbf{930} &     35837 &  11-04-04 14:23 &     2824 &   buy &      10.0 &  586.89 \\
		\textbf{931} &     35837 &  11-04-04 14:23 &      388 &  sell &      10.0 &  586.89 \\
		\textbf{932} &     35838 &  11-04-04 14:23 &     3111 &   buy &      10.0 &  578.42 \\
		\textbf{933} &     35838 &  11-04-04 14:23 &      388 &  sell &      10.0 &  578.42 \\
		\textbf{934} &     35839 &  11-04-04 14:23 &     2824 &   buy &      10.0 &  570.20 \\
		\textbf{935} &     35839 &  11-04-04 14:23 &      388 &  sell &      10.0 &  570.20 \\
		\textbf{936} &     35840 &  11-04-04 14:23 &     3111 &   buy &      10.0 &  570.00 \\
		\textbf{937} &     35840 &  11-04-04 14:23 &      388 &  sell &      10.0 &  570.00 \\
		\bottomrule
	\end{tabular}
\end{table}

\textbf{Conservative.} Here, instead, row 939 is considered a duplicate of row 937. To maintain the dataset coherent, both rows 939 \textit{and} 938 are removed. Row 935 is not a duplicate of row 933 because of the difference in the value of `Money\_JPY'. Table~\ref{tab:dedCons} shows the result.

\textbf{Aggressive.} In every trade User 388 is seller at the same date and quantity. Thus, independently on the partner, all are considered as duplicates of the trade at rows 930, 931. Results are shown in Table~\ref{tab:dedAggr}.

\begin{table}
	\centering
	\caption{Result of \textit{Aggressive} deduplication method}
	\label{tab:dedAggr}
	\begin{tabular}{lrlrlrr}
		\toprule
		{} &  Trade\_Id &                 Date &  User\_Id &  Type &  Bitcoins &   Money\_JPY \\
		\midrule
		\textbf{930} &     35837 &  11-04-04 14:23 &     2824 &   buy &      10.0 &  586.89 \\
		\textbf{931} &     35837 &  11-04-04 14:23 &      388 &  sell &      10.0 &  586.89 \\
		\bottomrule
	\end{tabular}
\end{table}

\textbf{Pairs.} Here, instead, we remove only trades where \textit{both} legs of a trade are duplicates according to the criterion \textit{user id, timestamp, buy/sell action, amount in BTC}. Note that Trade Id is \textit{not} considered to detect duplicates. As depicted in Table~\ref{tab:dedPairs}, pairs 934, 935 and 936, 937 are removed, while pair 938, 939 is kept, given the presence of a different user w.r.t. previous trades in the \textit{buy} side.

\begin{table}[htbp]
	\centering
	\caption{Result of \textit{Pairs} deduplication method}
	\label{tab:dedPairs}
	\begin{tabular}{lrlrlrr}
		\toprule
		{} &  Trade\_Id &                 Date &  User\_Id &  Type &  Bitcoins &   Money\_JPY \\
		\midrule
		\textbf{930} &     35837 &  11-04-04 14:23 &     2824 &   buy &      10.0 &  586.89 \\
		\textbf{931} &     35837 &  11-04-04 14:23 &      388 &  sell &      10.0 &  586.89 \\
		\textbf{932} &     35838 &  11-04-04 14:23 &     3111 &   buy &      10.0 &  578.42 \\
		\textbf{933} &     35838 &  11-04-04 14:23 &      388 &  sell &      10.0 &  578.42 \\
		\textbf{938} &     35841 &  11-04-04 14:23 &     1000 &   buy &      10.0 &  570.00 \\
		\textbf{939} &     35841 &  11-04-04 14:23 &      388 &  sell &      10.0 &  570.00 \\
		\bottomrule
	\end{tabular}
\end{table}

We follow the approach introduced in~\cite{scaillet2017high} and we conduct the analyses using the \textit{TradeId} deduplication method. As a robustness check, we also repeat them using the \textit{Pairs} method; results are coherent.

\subsection{Identification of the redundant files}

We verify that part of the files are a redundant copy. The leaked dataset includes two different versions for April 2011: the file `2011-04.csv', and the file `2011-04\_mtgox\_japan.csv'. The two differ by 41 rows, and only with respect to the field `User\_Id': all these legs are reported as executed by user 634 in the first, while this identifier was hidden (User\_Id = `DELETED') in the latter. 
All the `Coinlab' files from November 2012 to November 2013 contain only duplicated rows when compared to the corresponding monthly `mtgox\_japan' ones. Instead, a set of rows in the weekly `mtgox\_japan' files apparently result as non-duplicates, all of them being performed by user 634.  However, upon closer inspection, it emerges that also these rows are duplicated, if taking into account posthumous corrections to the fields `Money' and `User\_Id': in each of these instances, the user identifier is systematically changed from 698630 to 634, and the values in the `Money' field, seemingly random, are corrected to market values. Consistently with the literature, we assume the rows with user ID 698630 as the original ones. 
These trades were likely not supported by a real transfer of funds from an account to another, althought virtually executed: thus, as noted by~\cite{gandal2018price}, `no legitimate Mt.\ Gox customer received the currency Markus supposedly paid to acquire these claimed coins'\footnote{For further details, see also the online supplementary material \url{https://bit.ly/3jZDvcp} of~\cite{gandal2018price}.}. However, the parties involved in these trades with user 698630 were supposedly not aware of the underlying illicit activity; for them, the trade had normally occurred\footnote{One can verify that these trades had a market impact, see e.g. bitcoincharts.com \url{https://bit.ly/318Ql0B}.}
. For the purpose of our analysis, it is correct to consider these trades as if they had been executed at market prices. 

Once merged, after discarding the redundant files, the dataset is composed of 16,748,681 rows and 19 columns. 
Using the two deduplication techniques described before (Pairs and TradeId), we control for the presence of further duplicate trades in the remaining files. 
Note that, in this step, the latter methodology does not reduce further the dataset.

\subsection{Dataset sanity checks} 
\label{sec:SanityChecks}

\begin{figure}[htbp]
	\centering
	\includegraphics[width=0.8\textwidth]{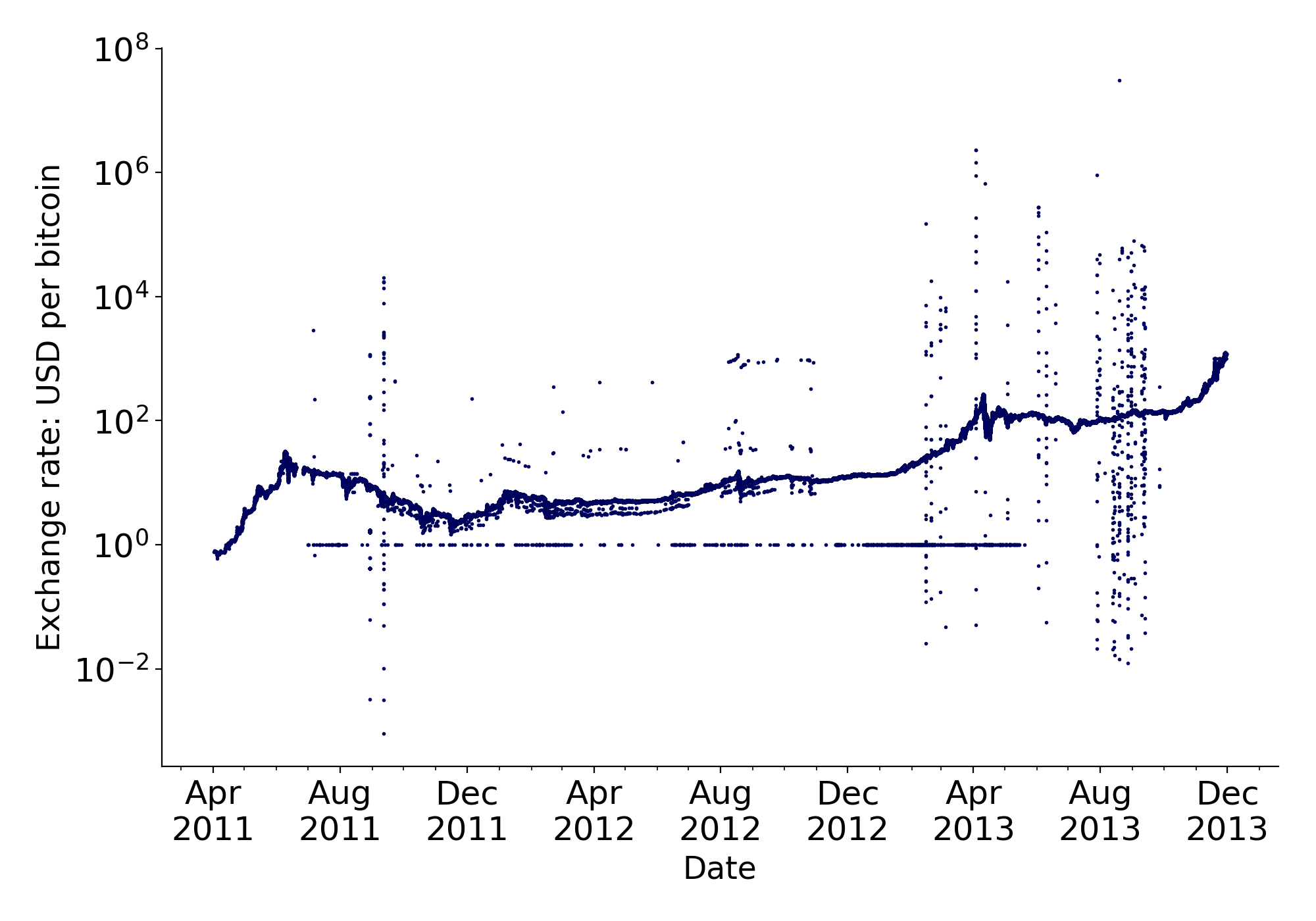}
	\caption{Exchange rate of a sample of USD trades in the merged dataset before the polishing procedures}
	\label{fig:exchRateOrig}
\end{figure}

\begin{figure}[htbp]
	\centering
	\begin{subfigure}[t]{0.45\textwidth}
		\centering
		\includegraphics[width=\textwidth]{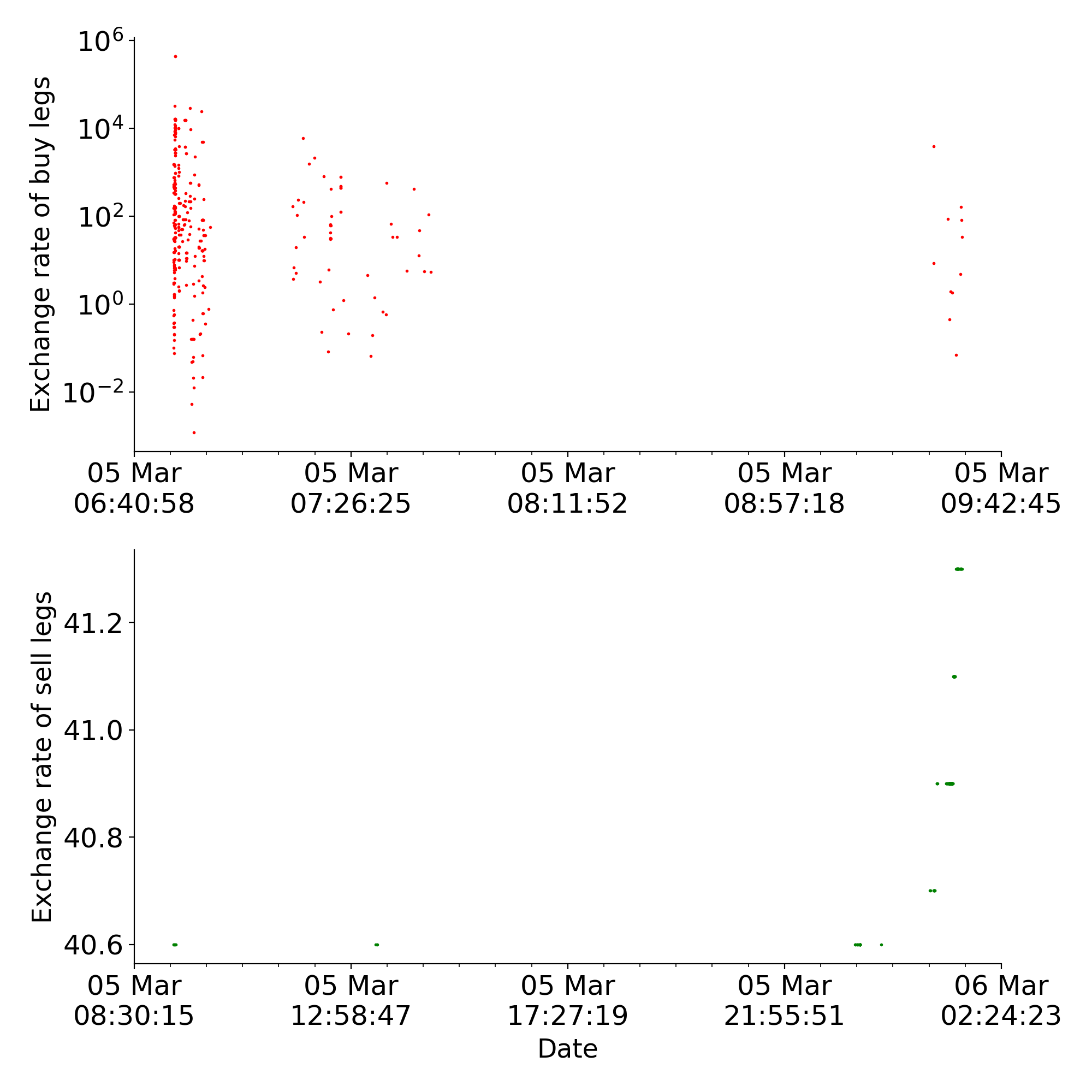}
		\caption{}
		\label{fig:698630Mar}
	\end{subfigure}
	\begin{subfigure}[t]{0.45\textwidth}
		\centering
		\includegraphics[width=\textwidth]{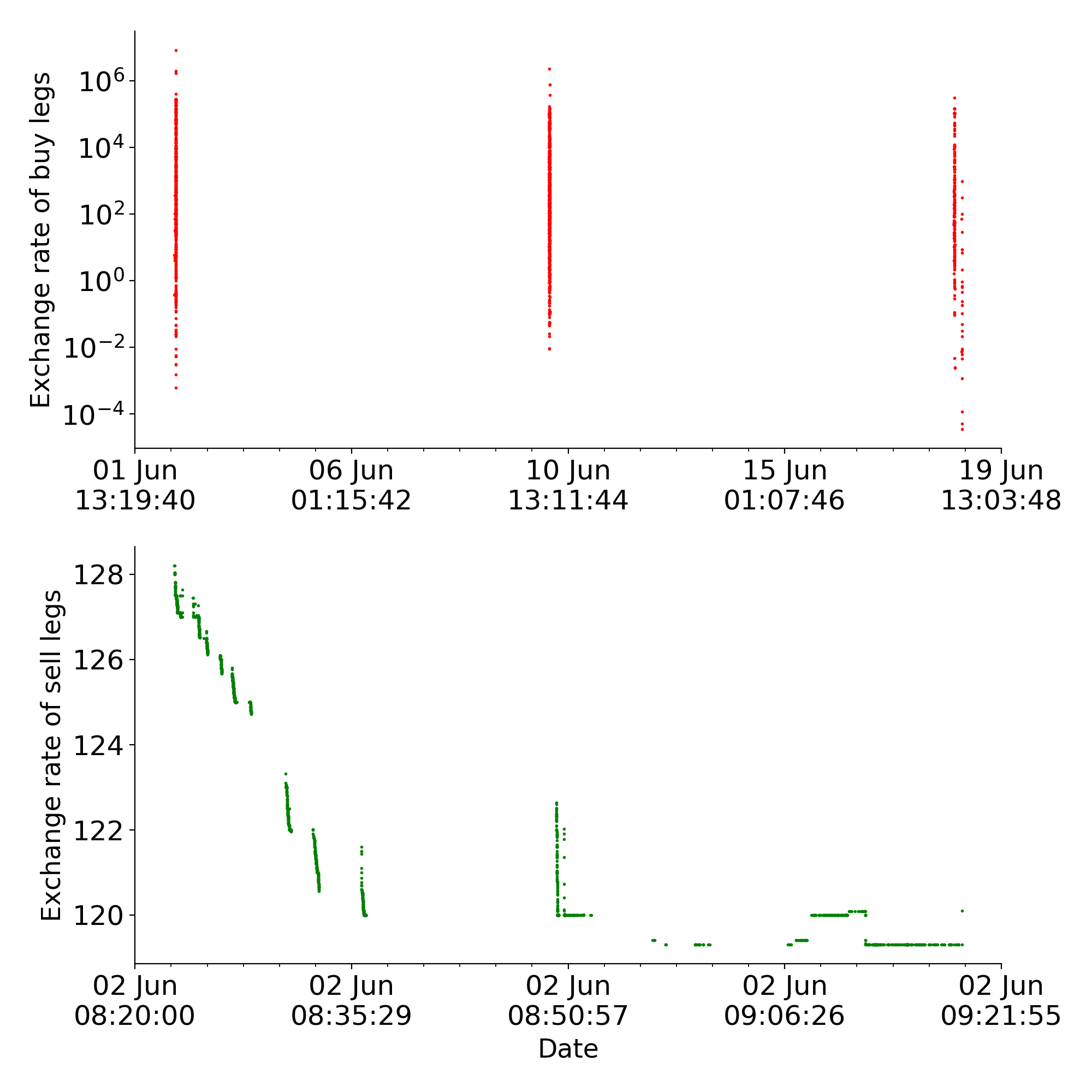}
		\caption{}
		\label{fig:698630Jun} 
	\end{subfigure}
	\caption{Exchange rate of trades executed by user 698630 in March (left) and June (right) 2013}
	\floatfoot{\emph{Notes:} the choice of the months is illustrative. Buy trades (upper panels) have random exchange rates, while  exchange rates of the sell trades (lower panels) are at the market prices.}
	\label{fig:698630}
\end{figure}

Figure~\ref{fig:exchRateOrig} shows the computed exchange rates for a sample of trades executed in US dollars in the leaked dataset after merging the single CSV files. Many values appear to be not related to market values. Given the relevance of this variable for our analysis, and to guarantee high quality standards for the data used, we perform additional sanity checks built on approaches used in the literature to further polish the dataset. The reported values refer to the TradeId deduplcation method: 
\begin{itemize}
	\item Last day's trades are removed, due to the presence of several inconsistencies;
	
	\item 115,275 self-trades are removed\footnote{The figures reported are comparable with those found in~\cite{scaillet2017high}
	. The difference is due to the fact that, in their analysis, they discard all non USD trades.};

	\item rows where `TIBANNE\_LIMITED\_HK' 
	and `THK' 
	act as intermediaries in the multi-currency trades are removed;
			
	\item As noted in~\cite{scaillet2017high}, around 136,000 trades (92,000 in USD) 
	`have a systematic data error whereby the fiat amount is the same in the primary and the secondary currency, and thus incorrect by a factor corresponding to the exchange rate between the two currencies'. All these are multi-currencies trades of the `Tibanne' type. As we will explain later, we do not remove these legs as the correct data can be retrieved.
	Additional 336,813 legs, all secondary legs belonging to multi-currency trades based on the `THK' method, are missing. Since it is not possible to recover the data for these missing legs, we remove from the dataset that we use for the analyses both the buy and the sell side. 
	However, for completeness, we also construct a version of the dataset that includes the primary legs of these trades.
	
	\item 31 trades with `DELETED' user identifier or the field `Bitcoins' = 0 are removed; 
	
	\item We verify that the dates are expressed in the same time zone (UTC);
	
	\item following~\cite{gandal2018price} and anonymous contributions from the Mt.\ Gox users community, we identify two key users:
	\begin{itemize}
		\item The literature refers to the account 698630 as \textbf{Markus}, and suspects pertain to the legality of its actions. It never paid fees, and all the trades where it appears as a buyer have seemingly random values in the `Money' field (supposedly because the virtual trades were not backed by real transactions: as a consequence, the trading log mechanism would interpret incorrectly the void entry and would fill it by simply copying and pasting the last `Money' value previously transcribed in the log). Figure~\ref{fig:698630} shows its trade pattern, distinguishing between buy and sell legs.
		Notice that this account is strictly linked to the one with user id = 634, as described before; user 634, active only in the first months of the datasets, seems to perform licit trades.
		
		\item The nickname \textbf{Willy} was given to 49 different accounts, bots sharing the same trade pattern and controlled by the same entity\footnote{\url{https://bit.ly/34VCsUx}.}: their `User\_Country' field is always marked as `??'; they only perform buy actions; these accounts activate one by one, once the previous has spent a definite round amount of dollars (usually \$2.5 millions).
		
		While Markus is active from February 2013 to September 2013, Willy's bots become active a few hours after the last Markus' trade, suggesting they are plausibly linked.			
	\end{itemize}
	
	It is likely that the same person owned all these accounts (634,698630, and the 49 bots).
	
	\item The field `User\_Id' for Markus trades performed by the account `698630' is changed to 635 (after controlling that no user has user identifier = 635), so that the accounts 698630 and 634 are sequentially linked, but they can be treated as a single user; the field `User\_Id' for Willy's bots is changed to 1000000, a number which is not owned by any other user. 
	
	\item  We add a column where we map all the user identifiers in the dataset to a consecutive sequence of integers, preserving the order but not the numbers. Throughout the analysis we will use this mapping to guarantee user's anonymity;
	
	\item We add a column for the multi-currency trades; the entries take the following values:
	
	\begin{enumerate}
		\setcounter{enumi}{-1}
		\item Standard trades;
		\item Tibanne multi-currency trades;
		\item THK multi-currency trades.
	\end{enumerate}
\end{itemize}

\subsection{Comparisons with other sources of information}

\subsubsection{The public Mt.\ Gox dataset}

To further ensure the validity of the data, we compare the leaked log to a dataset made public by Mt.\ Gox and used by~\cite{scaillet2017high} in their work. This dataset contains the history of trades executed in Mt.\ Gox, from its birth on 17 July 2010 to 17 December 2013, for a total of 8,605,998 trades. The main difference with the leaked dataset is that trades are not reported as a couple of legs, but as a single trade, thus they do not contain information on the users who performed the buy and sell actions; however, the trade identifiers follow the same scheme used in the leaked dataset, allowing to merge the two datasets.

Besides general fields already included in the leaked log (such as the trade identifier, the fiat money used, the volume traded, the exchange rate, the date), the public dataset contains additional information on the trades, specifying the typology of the order (possible values are `limit'; `market'; `limit,mixed\_currency'; `market,mixed\_currency') and which party initiated the trade (`bid' or `ask'). 
While there is no way to fully verify the correctness of these datasets, the comparison with a second source of information provides a solid robustness check. 
	
We first verify that for each trade \textit{the amount of bitcoins traded correspond exactly} in the two datasets. The values related to the money traded, instead, show some differences for around 310,000 legs. 
Data in the public dataset are reported as exchange rates, thus the following comparisons concern exchange rates. 
First, we find out that trades executed before 12 September 2013 in SEK and JPY are misreported in the public dataset by a factor of $ 10^{2} $, explaining the divergences for around 32,000 
legs. Second, around 135,000 
multi-currency `Tibanne' trades are wrongly transcribed in the leaked log, as the two legs report the same fiat amount; the comparison with the prices in the public dataset allows to correct them. Around 55,000 
trades are transactions that involve small amounts of bitcoins (less than $10^{-3}$), and in 8,000 
the difference between exchange rates is smaller than 1\% - suggesting for both those groups that the differences are due to different roundings.
Among the remaining `unexplained' legs (around 80,000
), the vast majority is related to trades executed by user 635 (around 23,000 
trades) and few other users: the first four users alone account for slightly less than 50,000 legs. Upon closer inspection, these errors seem to be caused by the same mechanism explained above for the Markus buy trades: the `Money' field, when the value is void, is apparently filled with the entry reported in the previous trade. E.g., an account performed 
more than 2,500 nearly consecutive trades in day 2011-09-11, the vast majority of which with random `Money' values.

Although we are aware that these steps do not fully ensure the correctness of the dataset, we perform as many sanity checks as possible to guarantee its validation. In addition, we consider the harm of a possible error stemming from the choice of correcting the values using the public dataset to be minimal, both for the purpose of this work and in the aim of providing a general, clean dataset usable in future analyses.

\begin{figure}[htbp]
	\centering
	\includegraphics[width=0.8\textwidth]{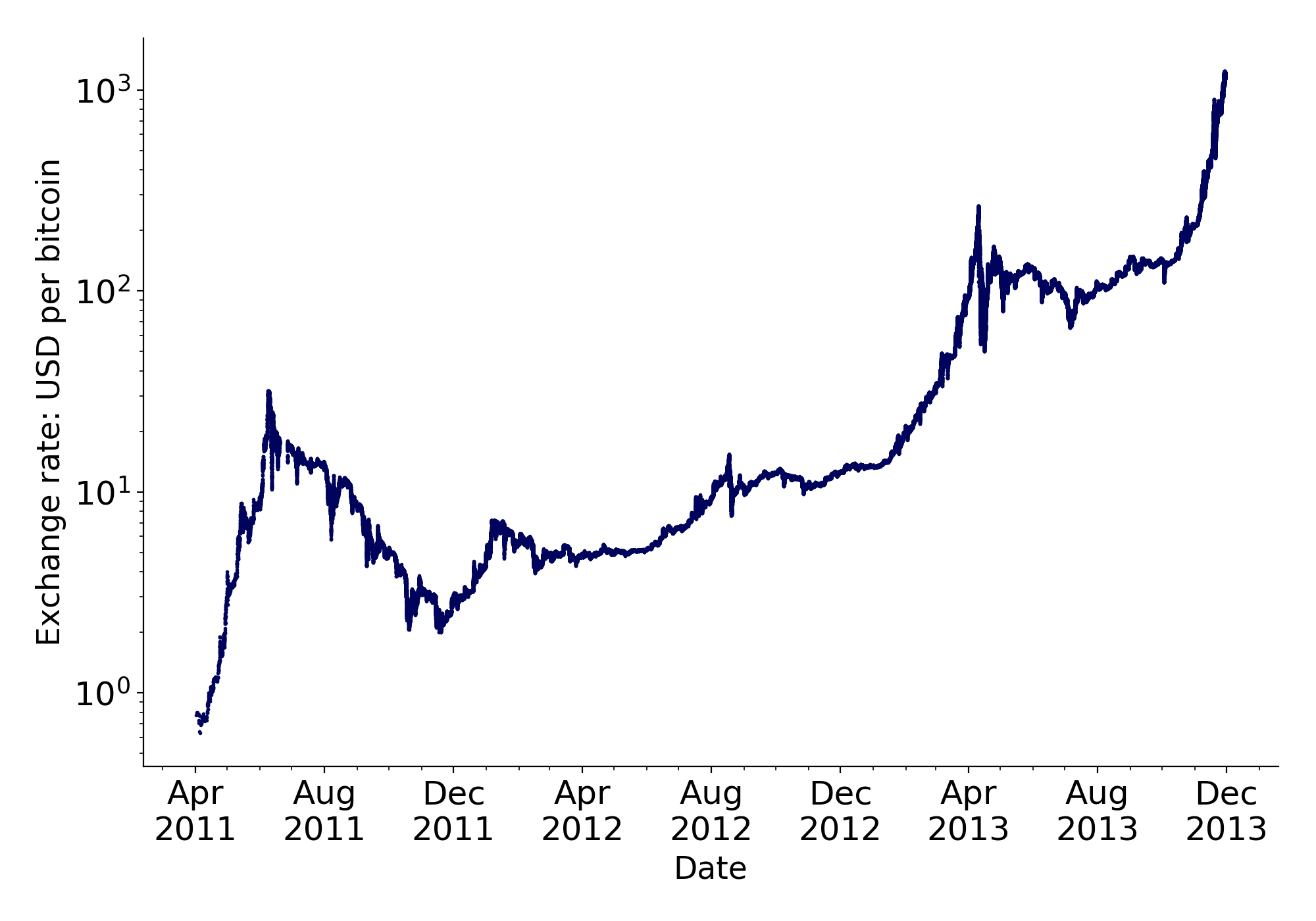}
	\caption{Exchange rate of a sample of USD trades in the merged dataset after the polishing procedures and the comparison with the public dataset}
	\label{fig:exchRateAfter}
\end{figure}

\begin{figure}[htbp]
	\centering
	\begin{subfigure}{0.48\textwidth}
		\centering
		\includegraphics[width=\textwidth]{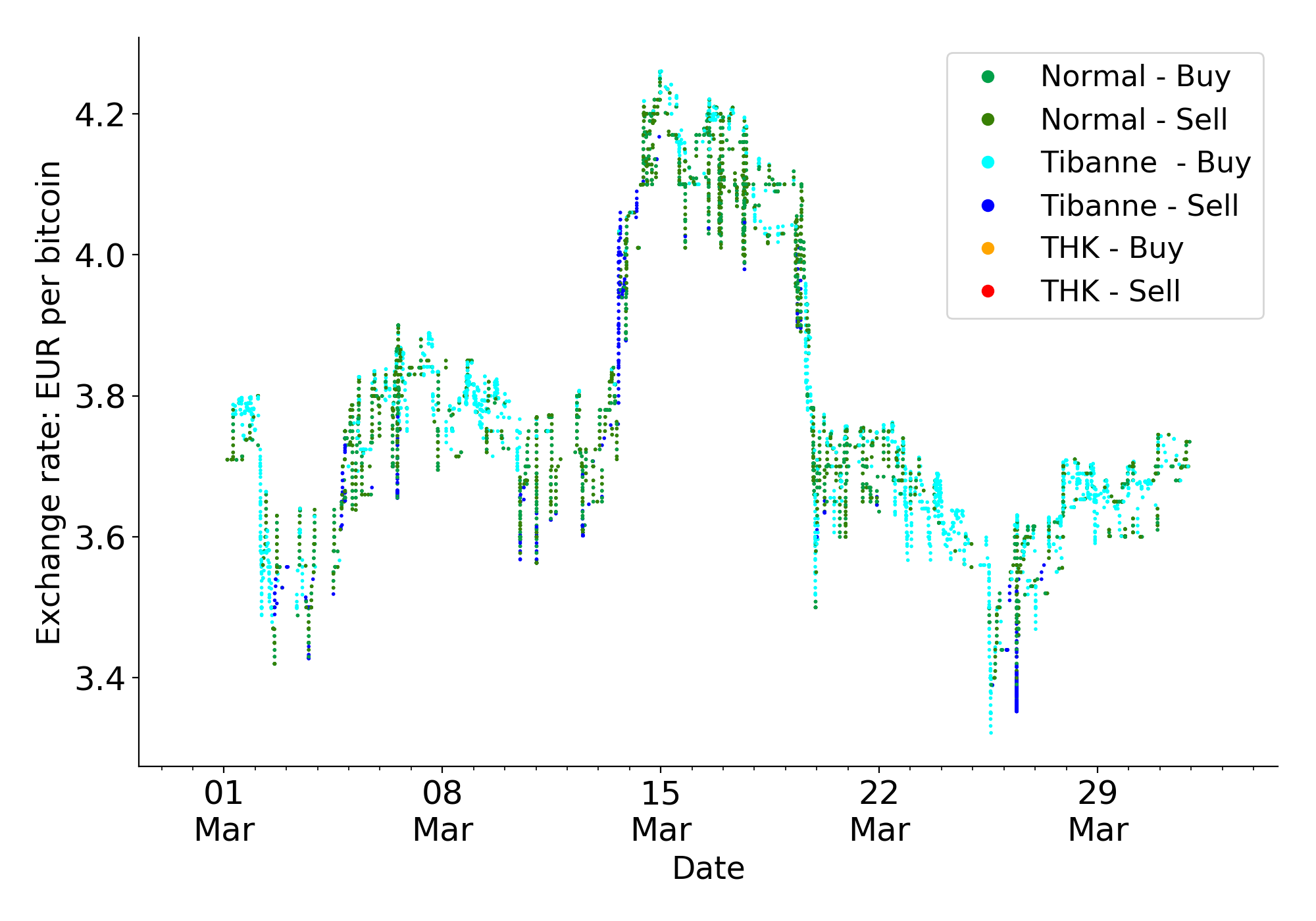}
		\caption{}
		\label{fig:AfterEUR12-03}
	\end{subfigure}
	\begin{subfigure}{0.48\textwidth}
		\centering
		\includegraphics[width=\textwidth]{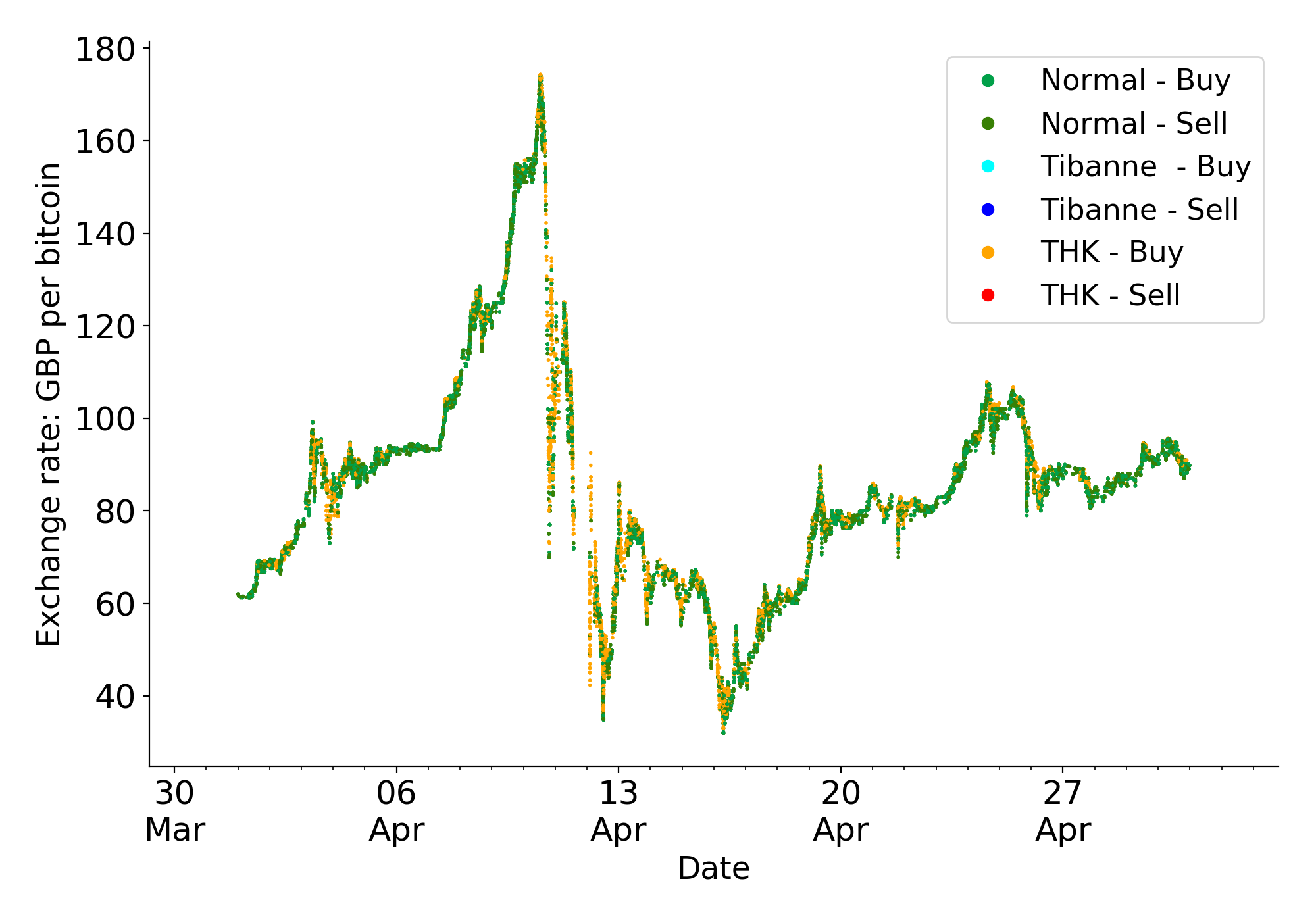}
		\caption{}
		\label{fig:AfterGBP13-04}
	\end{subfigure}
	\caption{Exchange rate of a sample of trades in EUR (March 2012) and in GBP (April 2013)}
	\floatfoot{\emph{Notes:} the choice of the reported months is illustrative}
	\label{fig:monthER}
\end{figure}

Figure~\ref{fig:exchRateAfter} shows the results for the USD market after the cleaning procedures. Figure~\ref{fig:monthER} shows the exchange rates for other markets (EUR,GBP) focusing on the typology of trades performed (wheher multi-currency or single-currency: we plot the monthly BTCtoEUR exchange rate for March 2012 on the left, and the BTCtoGBP e.r. for April 2013 on the right.

\subsubsection{volumes comparison with Bitcoincharts data}

\begin{figure}[htbp]
	\centering
	\begin{subfigure}{0.48\textwidth}
		\centering
		\includegraphics[width=\textwidth]{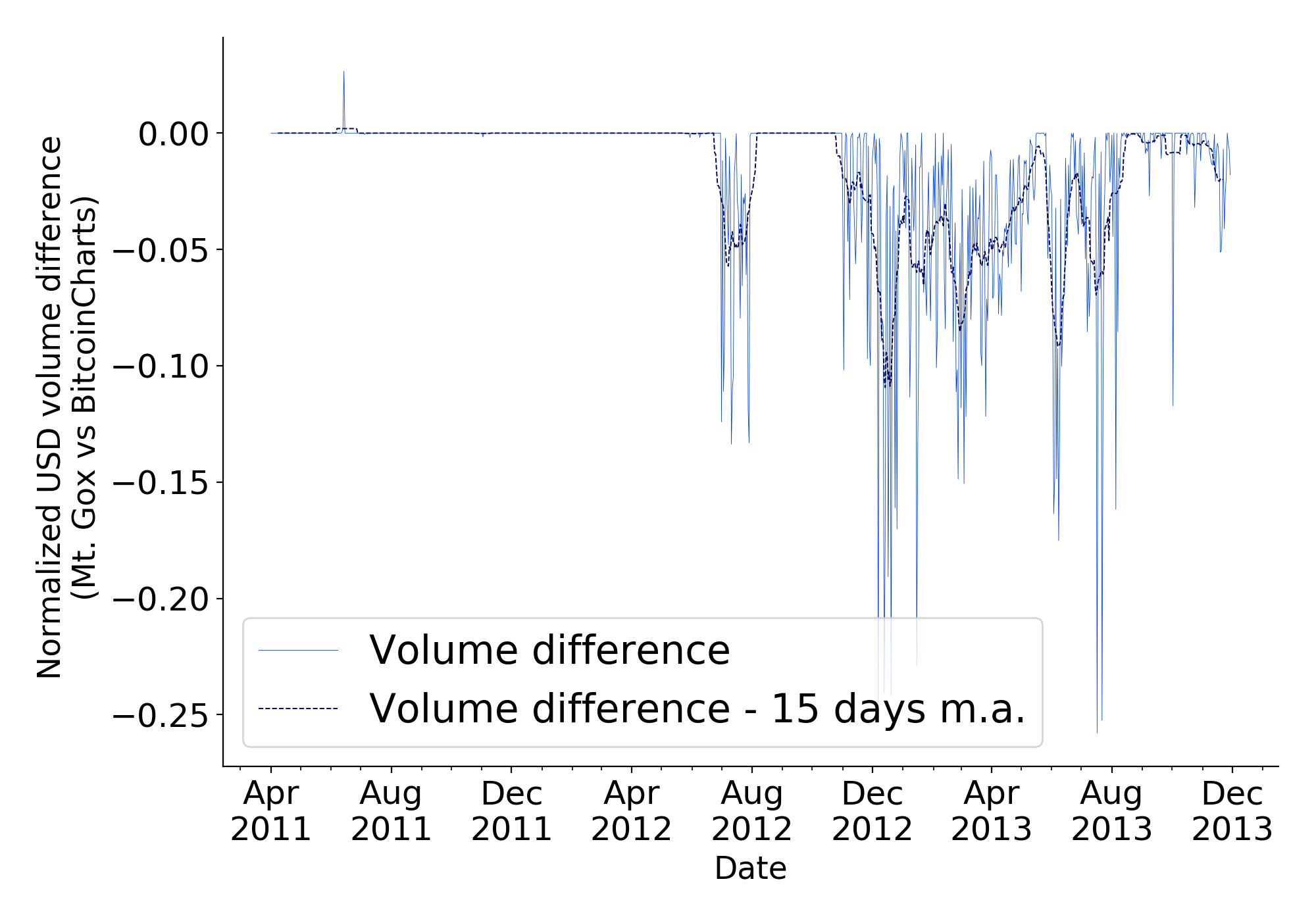}
		\caption{}
		\label{fig:ConcatTHK}
	\end{subfigure}
	\begin{subfigure}{0.48\textwidth}
		\centering
		\includegraphics[width=\textwidth]{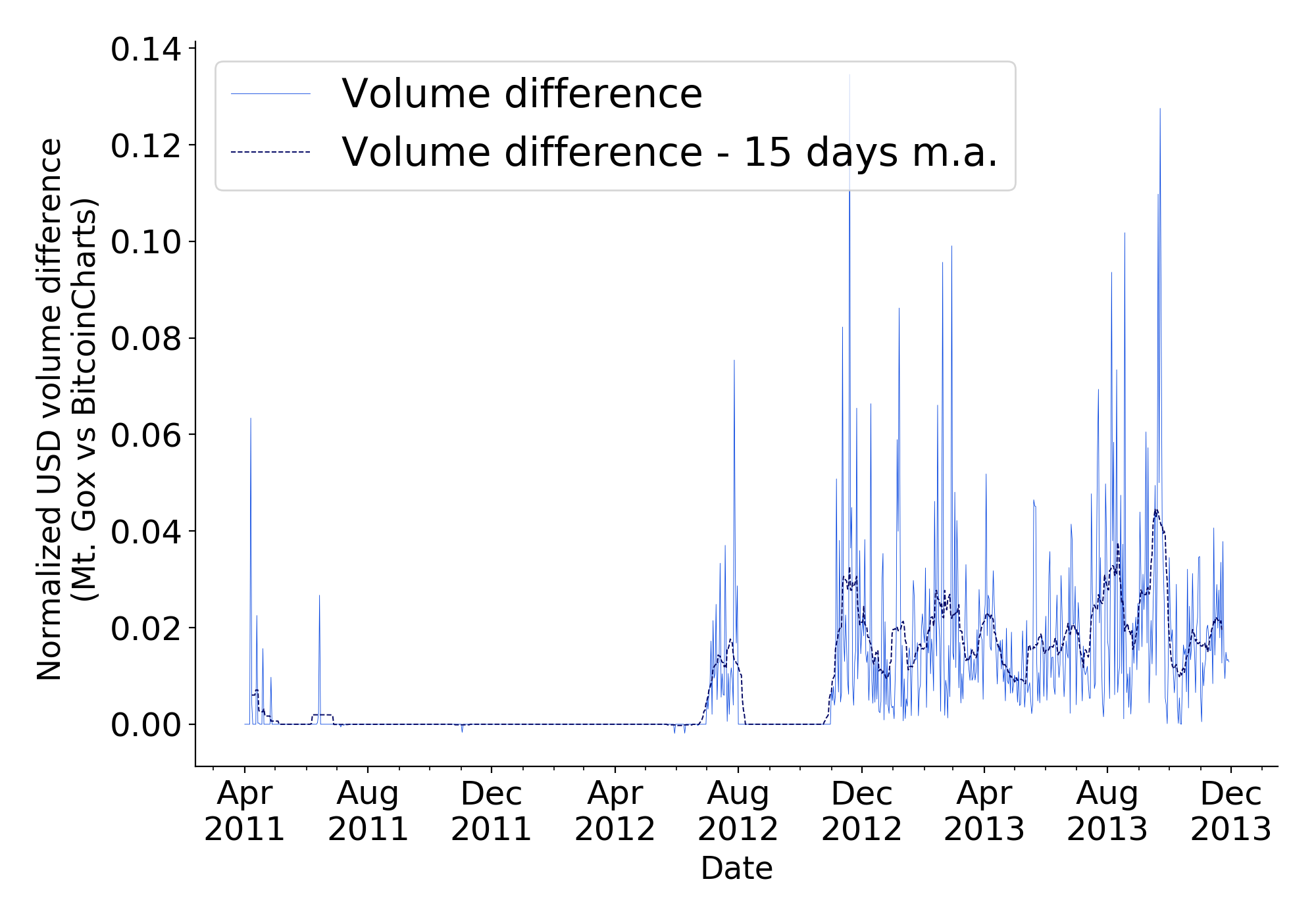}
		\caption{}
		\label{fig:datamtgoxTHK}
	\end{subfigure}
	\caption{Volumes comparison with Bitcoincharts.com data \textit{before} performing the dataset sanity checks, using the deduplication method `TradeID'}
	\floatfoot{\emph{Notes:} Figure~\ref{fig:ConcatTHK} reports the differences between the daily volumes in the leaked dataset and the Bitcoincharts data; Figure~\ref{fig:datamtgoxTHK} reports the `upper bound' of the USD traded volumes that we would observe if all the missing secondary THK legs were in USD.}
	\label{fig:btchartsTHK}
\end{figure}

To get a form of external validation, we compare the daily USD volumes in the leaked dataset with the data made public by Bitcoincharts.com\footnote{See \url{https://bit.ly/2H52exr}.}. This website provides reliable data and is a benchmark in most of the literature on Bitcoin exchange platforms. All the following plots compare USD-denominated volumes and represent the difference of volumes obtained from the leaked dataset against the bitcoincharts volumes, and then normalized on the volumes of the Mt.\ Gox's log. The lines represent the daily differences and their 15 days centered moving average.
Figure~\ref{fig:ConcatTHK} shows the difference between the bitcoincharts daily volumes and the leaked dataset, after the deduplication procedure (using the method 'Trade Id' \textit{and} including the primary legs of the THK trades), but before the cleaning procedure reported in Section \ref{sec:SanityChecks}.

The volumes differ only in the months were the `THK' multi-currency method is implemented. As we argued before, the secondary legs of such trades are missing. It is likely that some of them were in USD, plausibly explaining the missing volumes. Thus, Figure~\ref{fig:ConcatTHK} can be interpreted \textit{as if} no THK secondary leg were in USD; that is, as a `lower bound' for USD-denominated volumes traded. Viceversa, Figure~\ref{fig:datamtgoxTHK} plots the volume differences one would observe if all the secondary legs - of the THK trades whose first leg is not in USD - were in USD. This can be considered as an `upper bound' to USD volumes traded. 
The plots suggest that this intuition is correct and that the difference between the Mt.\ Gox dataset and the Bitcoincharts data is due to the missing multi-currency THK secondary legs.

\begin{figure}[htbp]
	\centering
	\begin{subfigure}{0.48\textwidth}
		\centering
		\includegraphics[width=\textwidth]{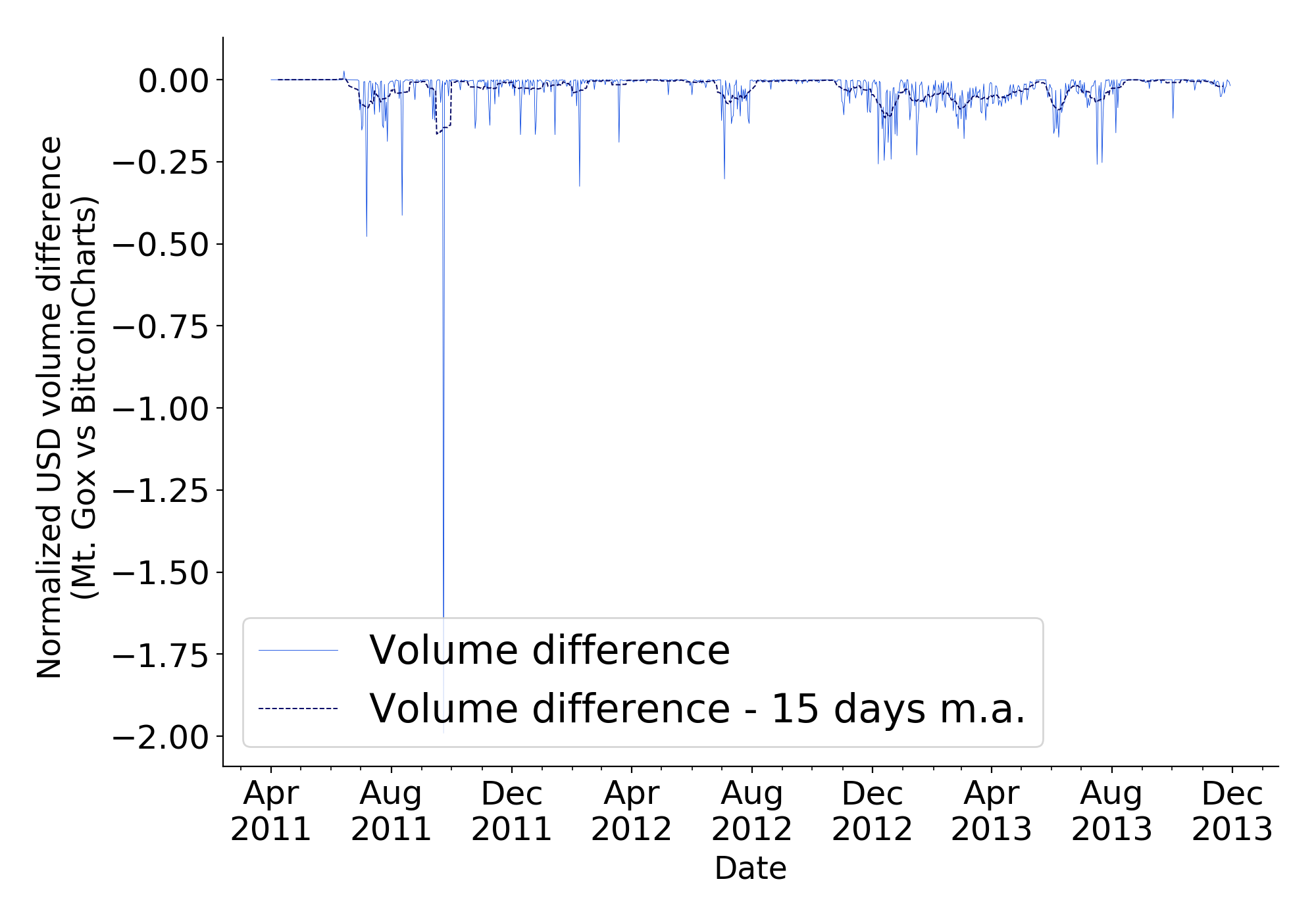}
		\caption{}
		\label{fig:FinalTidTHK}
	\end{subfigure}
	\begin{subfigure}{0.48\textwidth}
		\centering
		\includegraphics[width=\textwidth]{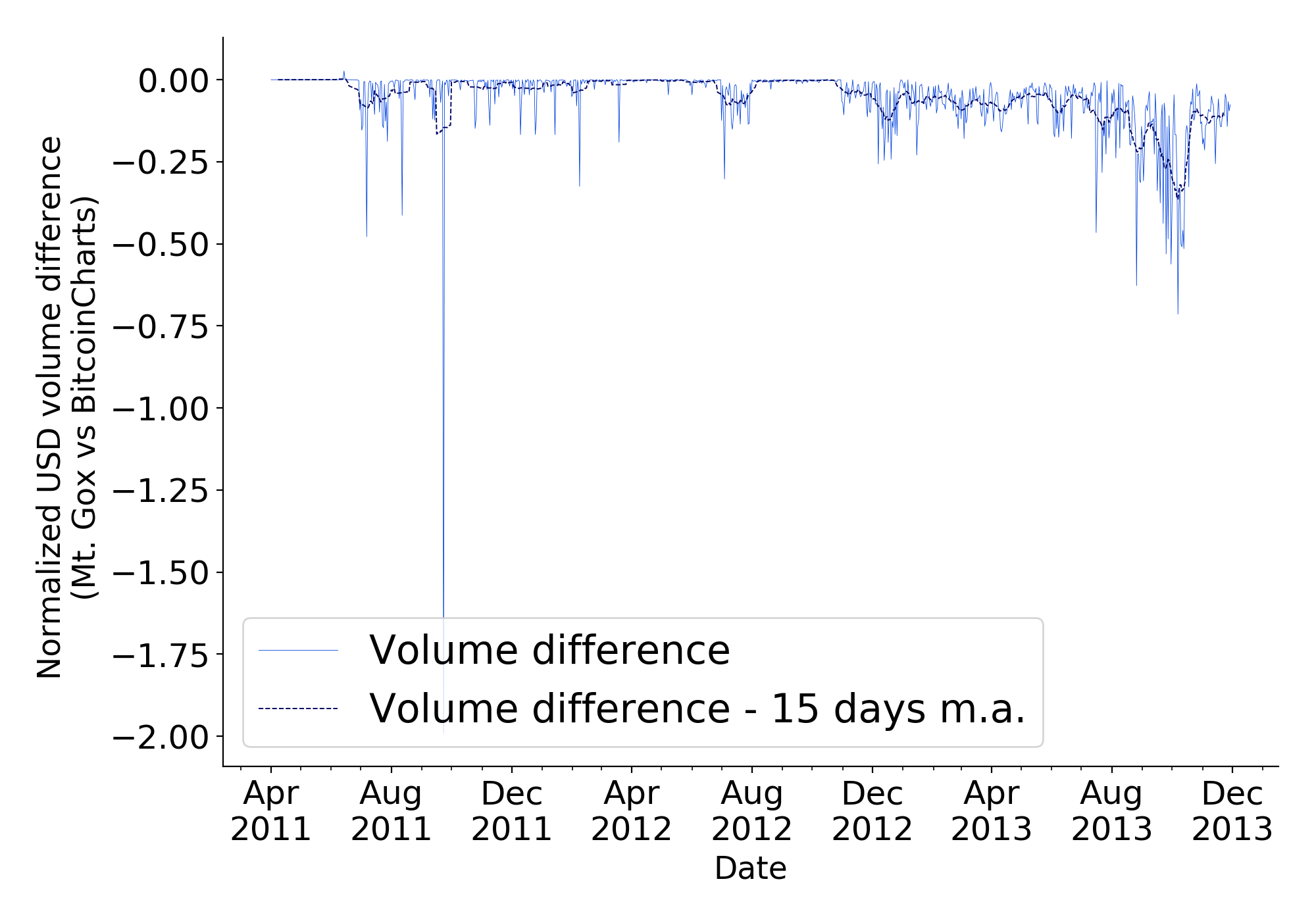}
		\caption{}
		\label{fig:FinalTidNoTHK}
	\end{subfigure}
	\caption{Volumes comparison \textit{after} performing the dataset sanity checks}
	\floatfoot{\emph{Notes:} comparison with Bitcoincharts.com, using the deduplication method `TradeID'. Figure \ref{fig:FinalTidTHK} reports the difference when THK trades are included; Figure \ref{fig:FinalTidNoTHK} without THK trades.}
	\label{fig:FinalBtcharts}
\end{figure}

Finally, in Figure~\ref{fig:FinalBtcharts} we compare how the volume differences change when including (left) and excluding (right) the `THK' trades, after implementing the sanity checks in Section \ref{sec:SanityChecks}.

\subsubsection{Comparison with other results in the literature}

We identify the trading activity of the users Markus and Willy, and we report the aggregated amount of bitcoins and fiat currency that they exchanged according to our deduplication approaches. We compare these figures with the findings in~\cite{gandal2018price}. The results slightly change depending on the deduplication method used, but are consistent with those found previously in the literature:

\begin{table}[ht]
	\begin{minipage}[b]{0.45\linewidth}
		\begin{tabular}{l}
			\vspace{0.3cm}
			\textit{Pairs without THK}: \\
			\vspace{0.3cm}
			bitcoins bought by Markus: 302,928.392 \\
			\vspace{0.3cm}
			bitcoins sold by Markus: 35,535.441 \\
			\vspace{0.3cm}
			Dollars spent by Willy: 100,739,634.260 \\
			\vspace{1cm}
			bitcoins bought by Willy: 237,689.192\\
			\vspace{0.3cm}
			\textit{TradeId without THK}: \\   
			\vspace{0.3cm}
			bitcoins bought by Markus: 306,339.336 \\
			\vspace{0.3cm}
			bitcoins sold by Markus: 35,867.176 \\
			\vspace{0.3cm}
			Dollars spent by Willy: 102,710,284.814 \\
			\vspace{0.3cm}
			bitcoins bought by Willy: 242,794.122\\
		\end{tabular}
	\end{minipage}
	\hspace{0.5cm}
	\begin{minipage}[b]{0.45\linewidth}
		\begin{tabular}{l}
		\vspace{0.3cm}
		\textit{Pairs with THK}:\\
		\vspace{0.3cm}
		bitcoins bought by Markus: 330,994.764 \\
		\vspace{0.3cm}
		bitcoins sold by Markus: 35,535.441 \\
		\vspace{0.3cm}
		Dollars spent by Willy: 108,247,263.292 \\
		\vspace{1cm}
		bitcoins bought by Willy: 261,437.290\\
		\vspace{0.3cm}
		\textit{TradeId with THK}:\\   
		\vspace{0.3cm}
		bitcoins bought by Markus: 335,897.702 \\
		\vspace{0.3cm}
		bitcoins sold by Markus: 35,867.176 \\
		\vspace{0.3cm}
		Dollars spent by Willy: 110,376,397.492 \\
		\vspace{0.3cm}
		bitcoins bought by Willy: 266,984.786$^{a}$\\
		\end{tabular}
	\end{minipage}
\vspace{0.2cm}

$^{a}${\small \textit{If we include the 1,147.948 bitcoins traded by Willy in the last day (removed from the dataset), we obtain exactly the amount found by Gandal et al.: 268,132.734 bitcoins}.}
\end{table}	

To conclude, we proved that the results of our deduplication methods are consistent with (and partially develop) those implemented in previous analyses.
Several checks are performed to ensure the quality of our dataset, and show that our procedure provides a high quality dataset which is consistent with external sources of information largely accepted in the literature. Further, we merge the leaked dataset with public information on the Mt.\ Gox trades and we provide it in order to ensure reproducibility. We rest our analyses on the \textit{TradeId} methodology, and we exclude the THK trades whose secondary leg is missing.

%% file: tr_appendix_supplem.tex
\section{Supplemental Figures and Tables} 
\label{tr:appendix_supplem}

\setcounter{figure}{0}
\setcounter{table}{0}

\begin{figure}[H]
	\begin{center}
		\includegraphics[width=0.5\textwidth]{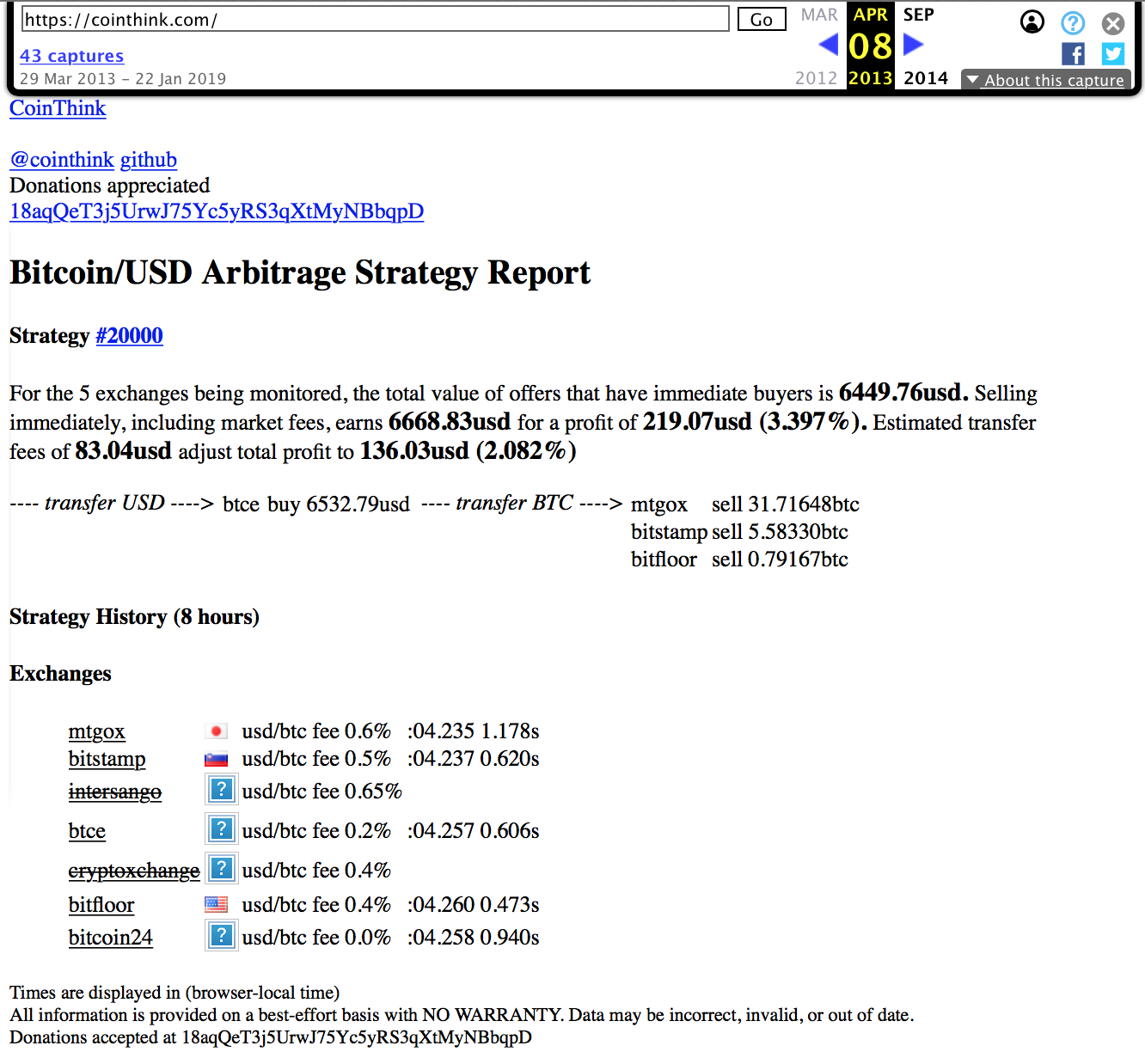}
	\end{center}
	\caption{Example of an online arbitrage tool (captured by the Internet Archive on 8 April 2013)}
	\label{fig:cointhink}
\end{figure}

\begin{figure}
    \includegraphics[width=\textwidth]{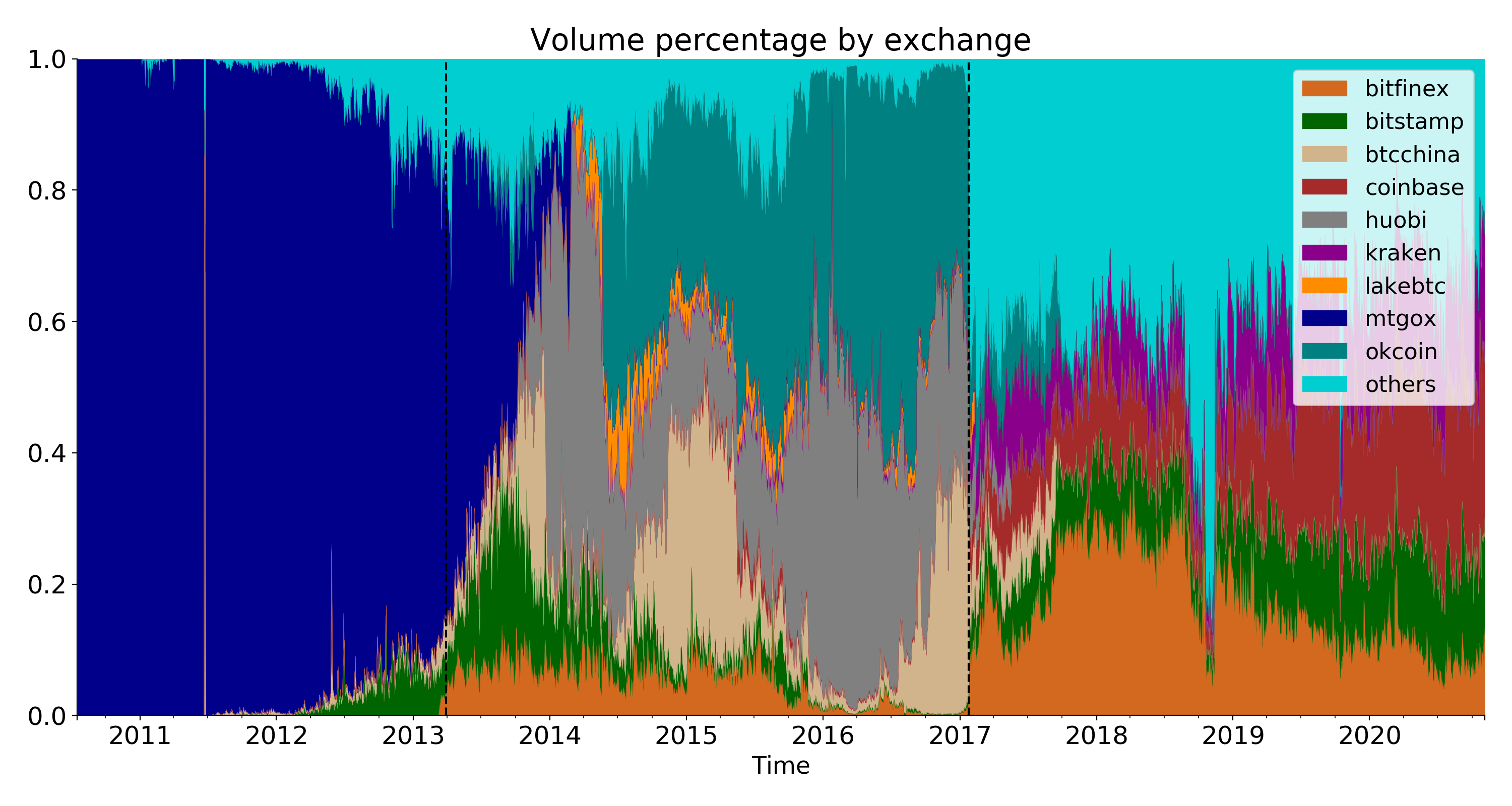}
    \caption{Evolution in time of the volumes traded in the main cryptocurrency exchanges}
    \label{fig:history_exch} 
    \floatfoot{\emph{Notes:} evolution of the daily volume percentage traded in the main exchanges. It is possible to roughly identify three different phases with distinctive patterns. The first one coincides with the epoch dominated by Mt.\ Gox, from mid 2010 to 2013; the second, from 2014 to 2017, corresponds to a time window mostly dominated by Chinese exchanges; the last and current phase is more heterogeneous, and few but competing exchanges share the market and operate against multiple fiat currencies (and cryptocurrencies). Data from \url{https://data.bitcoinity.org/} }
\end{figure}

\begin{table}[H]
    \centering
    \caption{Summary statistics}
    \label{tab:sumstats}
    \begin{tabular}{lrrrrrrr}
    \toprule
    ($ N = 6,629$) &   Mean &   St.D. &    Min &  25\% &  50\% &   75\% &      Max \\
    \midrule
    Profits, fees, \%  &   0.42 &    1.26 & -11.35 &  0.08 &  0.62 &   1.10 &    18.16 \\
    'Equiv. \$'        &  52.54 &  169.63 &   0.00 &  0.36 &  7.40 &  41.42 &  4666.66 \\
    $ \Delta R $ (abs) &   0.06 &    0.08 &   0.00 &  0.01 &  0.04 &   0.08 &     1.16 \\
    D(Currencies)      &   0.89 &    0.31 &   0.00 &  1.00 &  1.00 &   1.00 &     1.00 \\
    Log(Currencies)    &   1.64 &    0.51 &   0.69 &  1.10 &  1.61 &   1.95 &     2.48 \\
    Log(Actions)       &   5.94 &    2.14 &   0.00 &  5.30 &  6.54 &   7.68 &     7.68 \\
    D(Metaorder)       &   0.67 &    0.47 &   0.00 &  0.00 &  1.00 &   1.00 &     1.00 \\
    D(Aggressive)      &   0.06 &    0.23 &   0.00 &  0.00 &  0.00 &   0.00 &     1.00 \\
    PC1                &   6.57 &    2.83 &  -0.77 &  5.03 &  8.16 &   8.83 &     8.83 \\
    \bottomrule
    \end{tabular}
    \floatfoot{\emph{Notes:} this table summarizes the main variables employed in the regression analyses. The sample is the number of arbitrage actions (N = 6,629). The profits are expressed as a percentage of the spread between the implied and the official rate, as explained in Table \ref{tab:definitions}; the `Equiv. \$' term is the value of the trade expressed in USD dollars. The remaining rows refer to alternative measures of the investors trading ability. D(Currencies) is equal to 1 if the user exploited multiple currency markets to conduct arbitrage. It is a dummy variable, as well as D(Metaorder) and D(aggressive), respectively equal to 1 if the action was conducted by a user who executed metaorders or aggressive orders. Log(Currencies) is the logarithm of the currency markets exploited by the investor who conducted the arbitrage action, while Log(Actions) is the logarithm of the number of actions they executed. PC1 is the scores of the arbitrage action, obtained by performing a principal component analysis as described in Table \ref{tab:pca}.}
\end{table}

\begin{table}[H]
    \centering
    \caption{Pearson correlation for the main variables used in the model}
	\label{tab:pearson}
	{\footnotesize
	\begin{tabular}{lccccccccc}
    \toprule
    {} & $ D(Cur) $ & $ Log(Cur) $ & $ Log(Act) $ & $ D(Met) $ & $ D(Agg) $ &   PC1 & Profits & Eq. \$ &  $ \Delta R $ \\
    \midrule
    $ D(Currencies) $   &                 1 &                0.65 &              0.8 &             0.41 &             -0.44 &  0.76 &            0.39 &     -0.14 &         -0.10 \\
    $ Log(Currencies) $ &                   &                   1 &             0.74 &             0.63 &             -0.31 &  0.79 &            0.22 &     -0.18 &         -0.08 \\
    $ Log(Actions) $    &                   &                     &                1 &             0.57 &             -0.53 &   0.9 &            0.43 &     -0.18 &         -0.11 \\
    $ D(Metaorder) $    &                   &                     &                  &                1 &             -0.35 &  0.86 &            0.24 &     -0.17 &         -0.04 \\
    $ D(Aggressive) $   &                   &                     &                  &                  &                 1 & -0.52 &           -0.28 &      0.13 &          0.01 \\
    PC1                 &                   &                     &                  &                  &                   &     1 &            0.39 &      -0.2 &         -0.09 \\
    Profits,fees,\%     &                   &                     &                  &                  &                   &       &               1 &     -0.01 &         -0.02 \\
    Equiv. \$           &                   &                     &                  &                  &                   &       &                 &         1 &          0.04 \\
    $ \Delta R $        &                   &                     &                  &                  &                   &       &                 &           &          1.00 \\
    \bottomrule
    \end{tabular}
    }
	\floatfoot{{\small\emph{Notes}: The correlation is constructed on the sample of the actions (N = 6,629).}}
\end{table}

\begin{figure}[H]
	\begin{center}
		\includegraphics[width=0.8\textwidth]{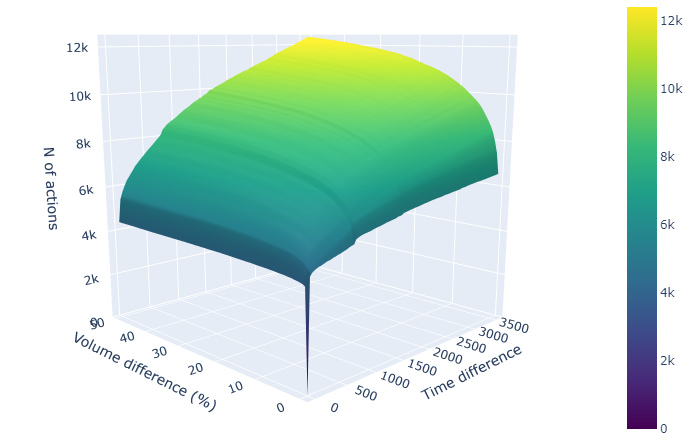}
	\end{center}
	\caption{Growth of the triangular arbitrage actions as a function of $\Delta T$ and $\Delta Q$}
	\label{fig:gradient}
\end{figure}

\begin{figure}[tp]
	\centering
	\caption{Trading patterns of the $ 3^{rd} $ to the $ 8^{th} $ most active users in multiple markets
	}
	\label{fig:patterns_supplem}
	\begin{subfigure}{0.475\textwidth}
		\includegraphics[width=\textwidth]{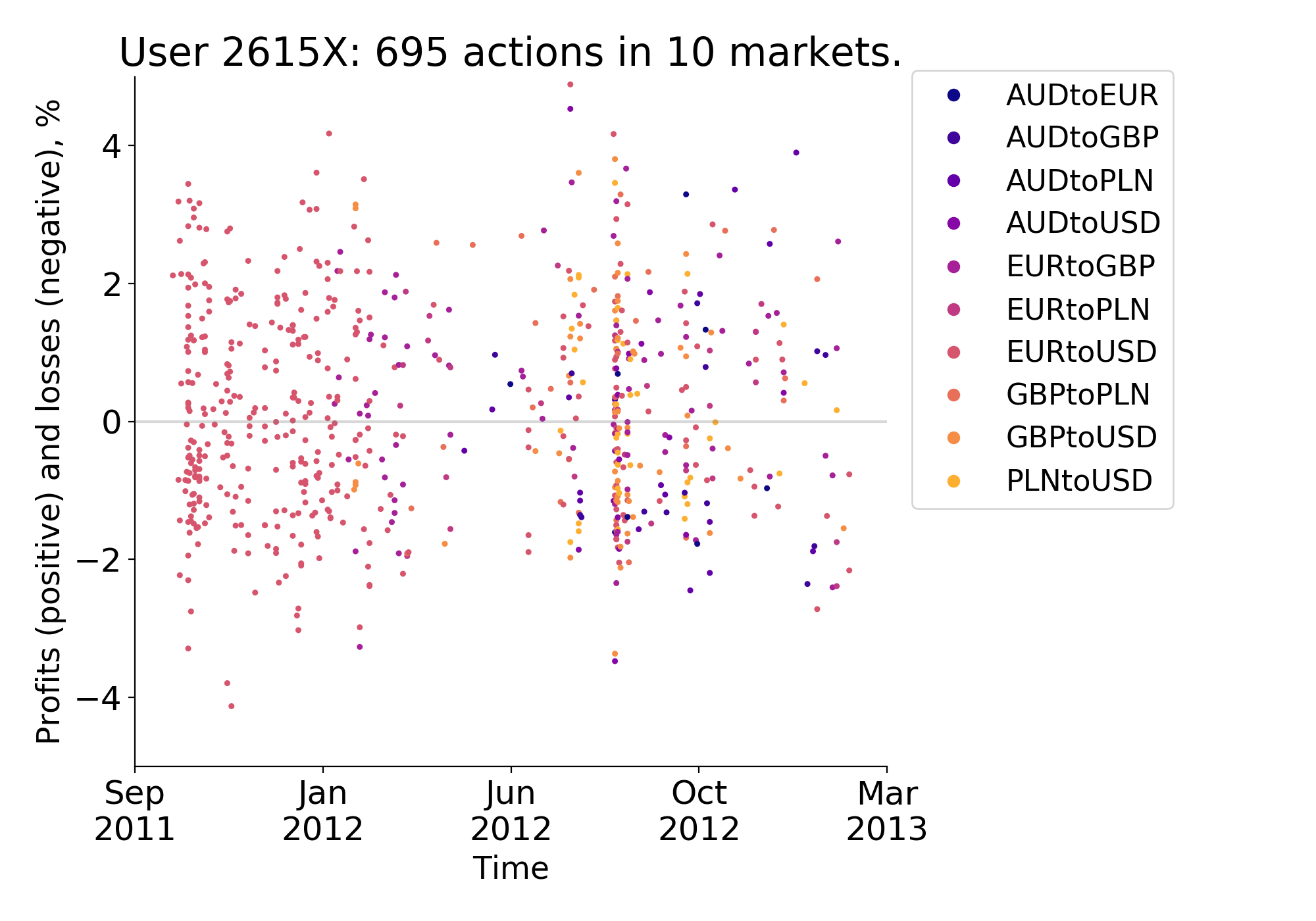}%
		\caption{}
		\label{fig:2615X}
	\end{subfigure}
	\begin{subfigure}{0.475\textwidth}
		\includegraphics[width=\textwidth]{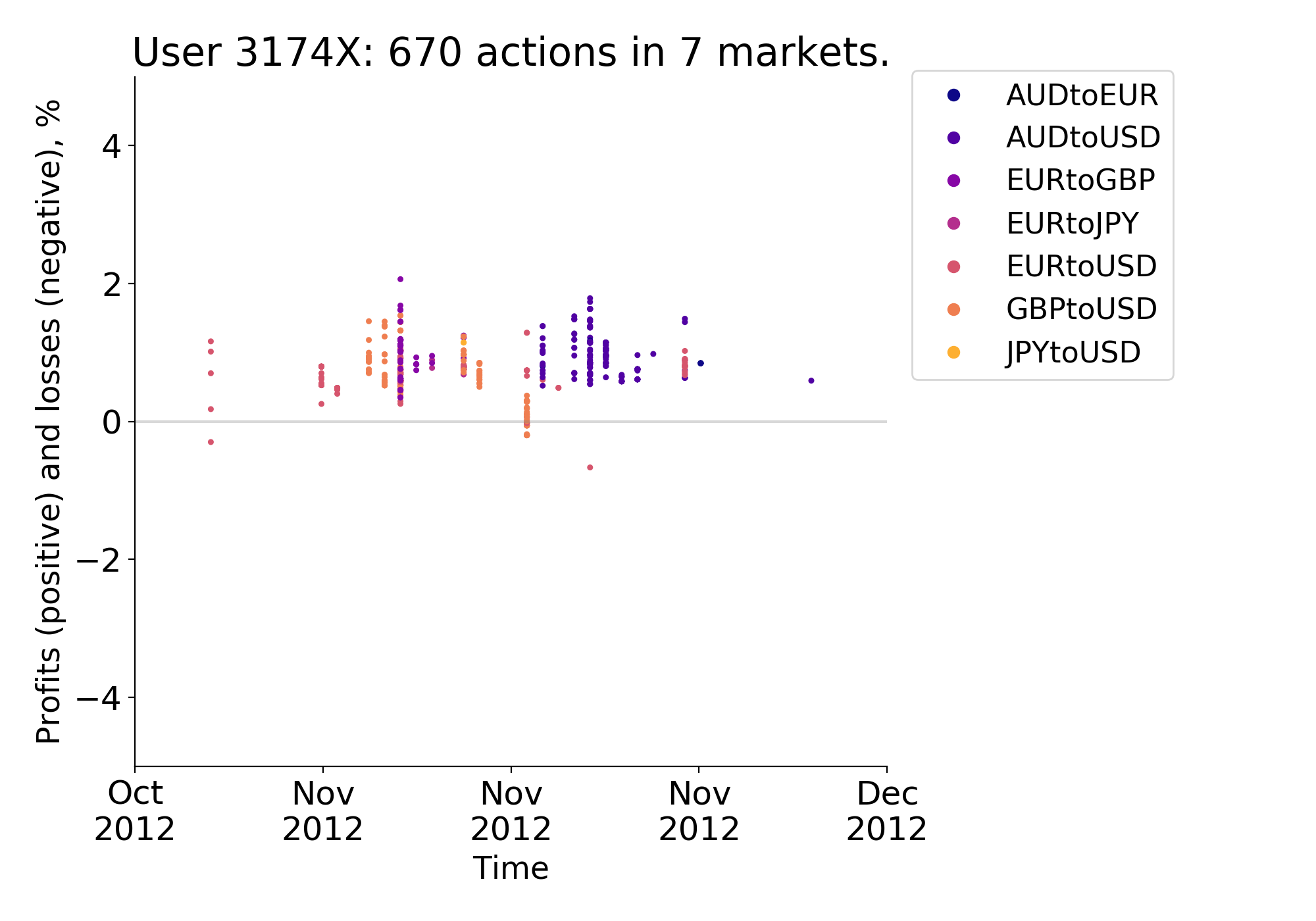}
		\caption{}
		\label{fig:3174X}
	\end{subfigure}
	\begin{subfigure}{0.475\textwidth}
		\includegraphics[width=\textwidth]{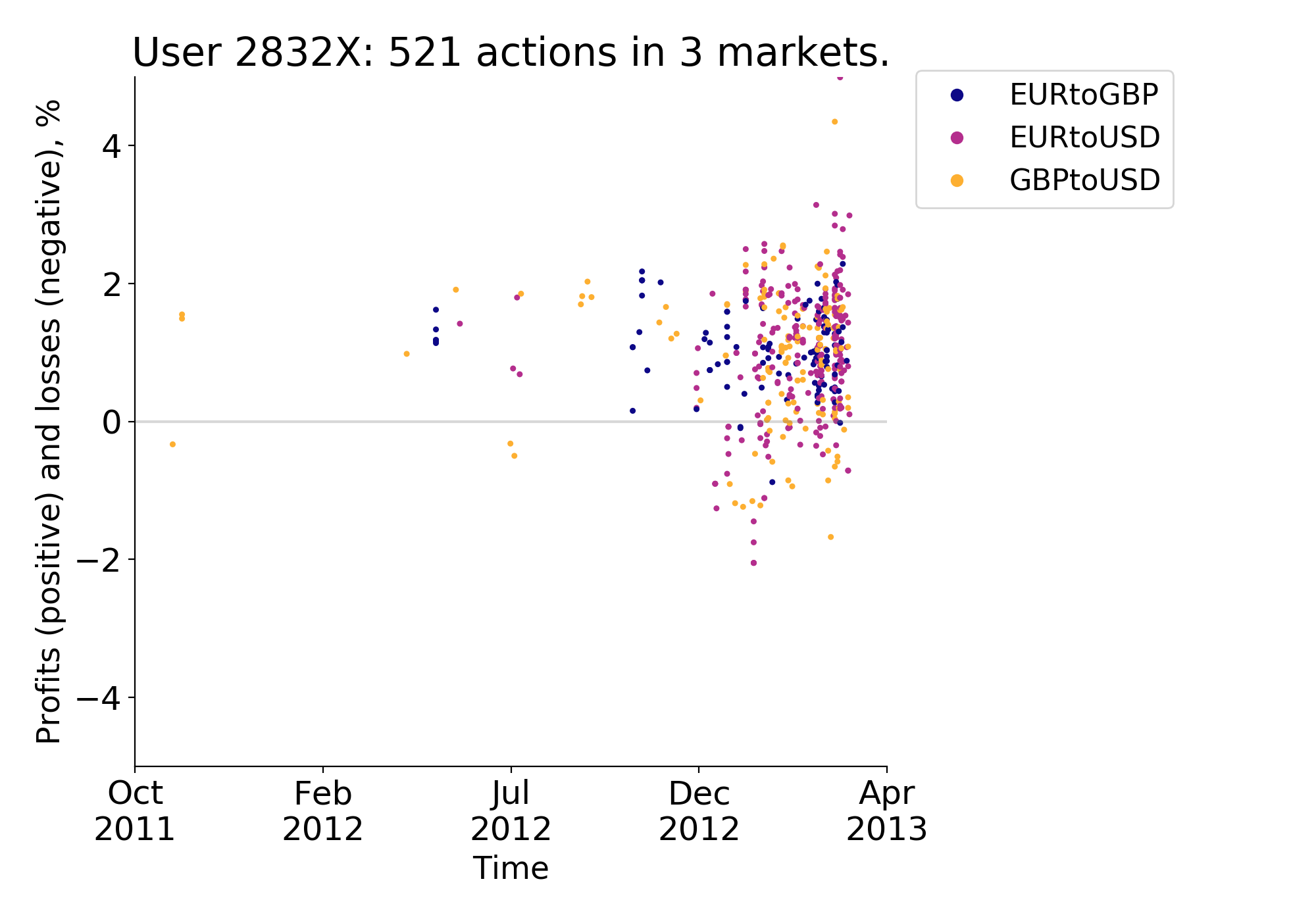}
		\caption{}
		\label{fig:2832X}
	\end{subfigure}
	\begin{subfigure}{0.475\textwidth}
		\includegraphics[width=\textwidth]{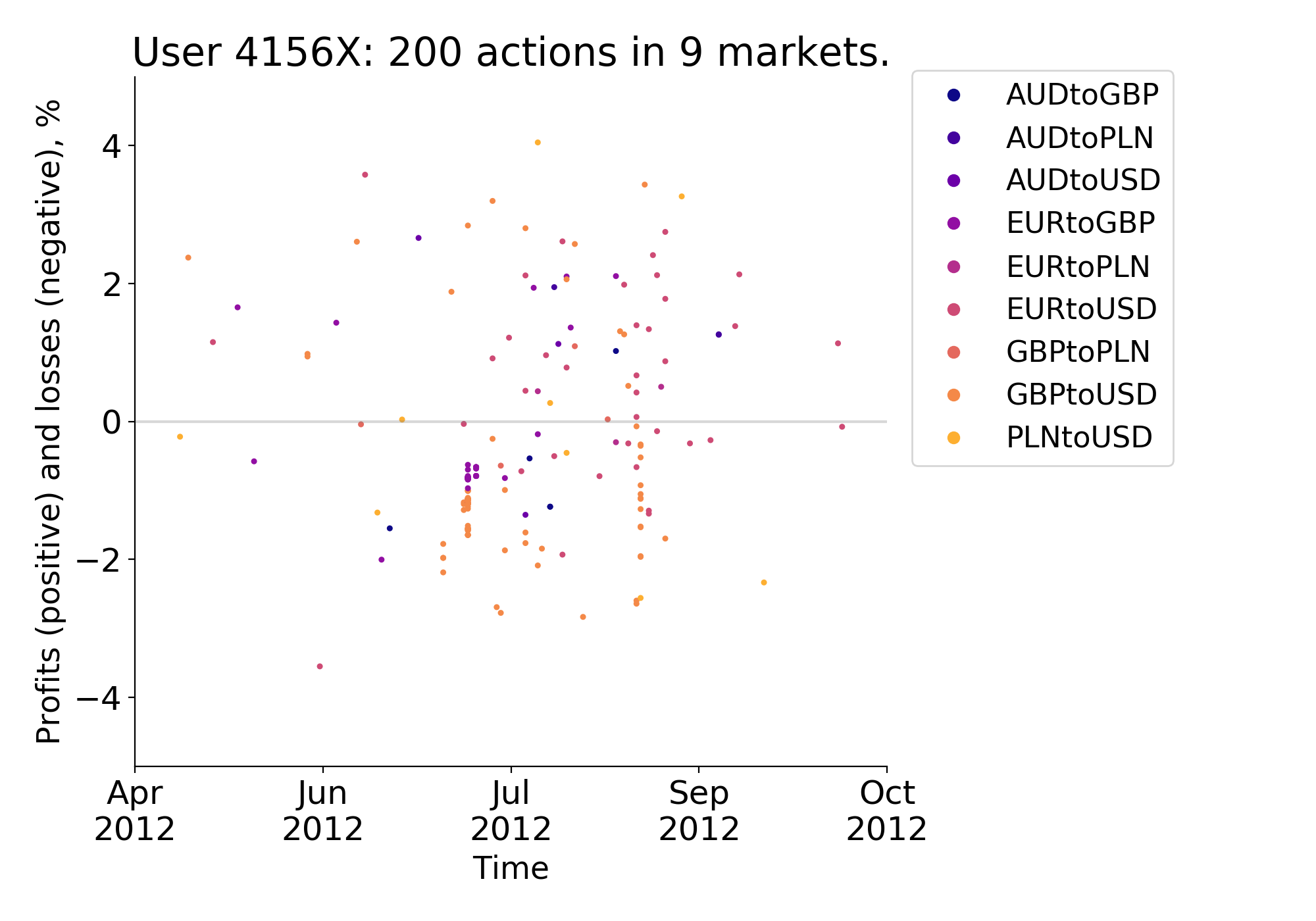}
		\caption{}
		\label{fig:4156X}
	\end{subfigure}
	\begin{subfigure}{0.475\textwidth}
		\includegraphics[width=\textwidth]{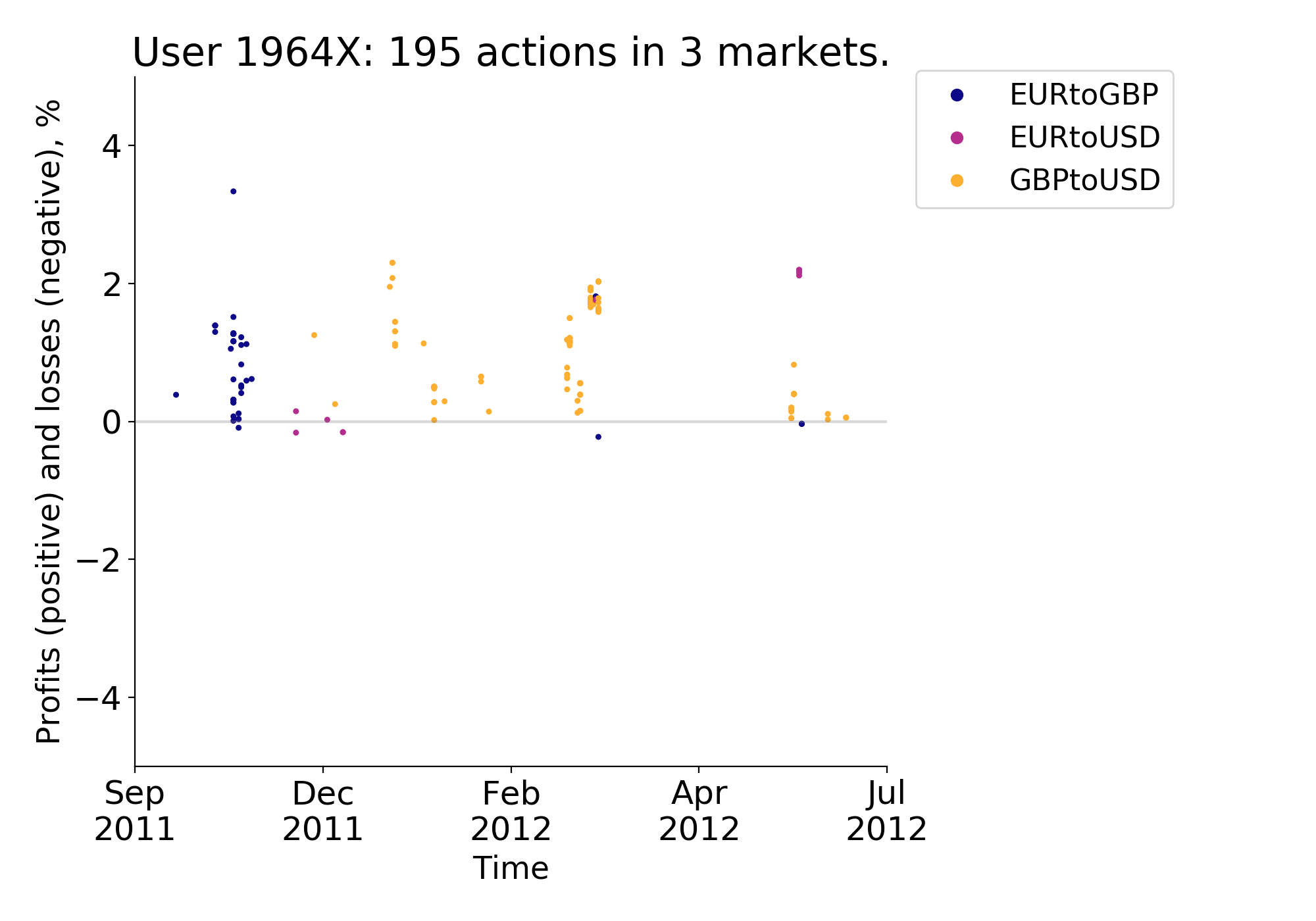}%
		\caption{}
		\label{fig:1964X}
	\end{subfigure}
	\begin{subfigure}{0.475\textwidth}
		\includegraphics[width=\textwidth]{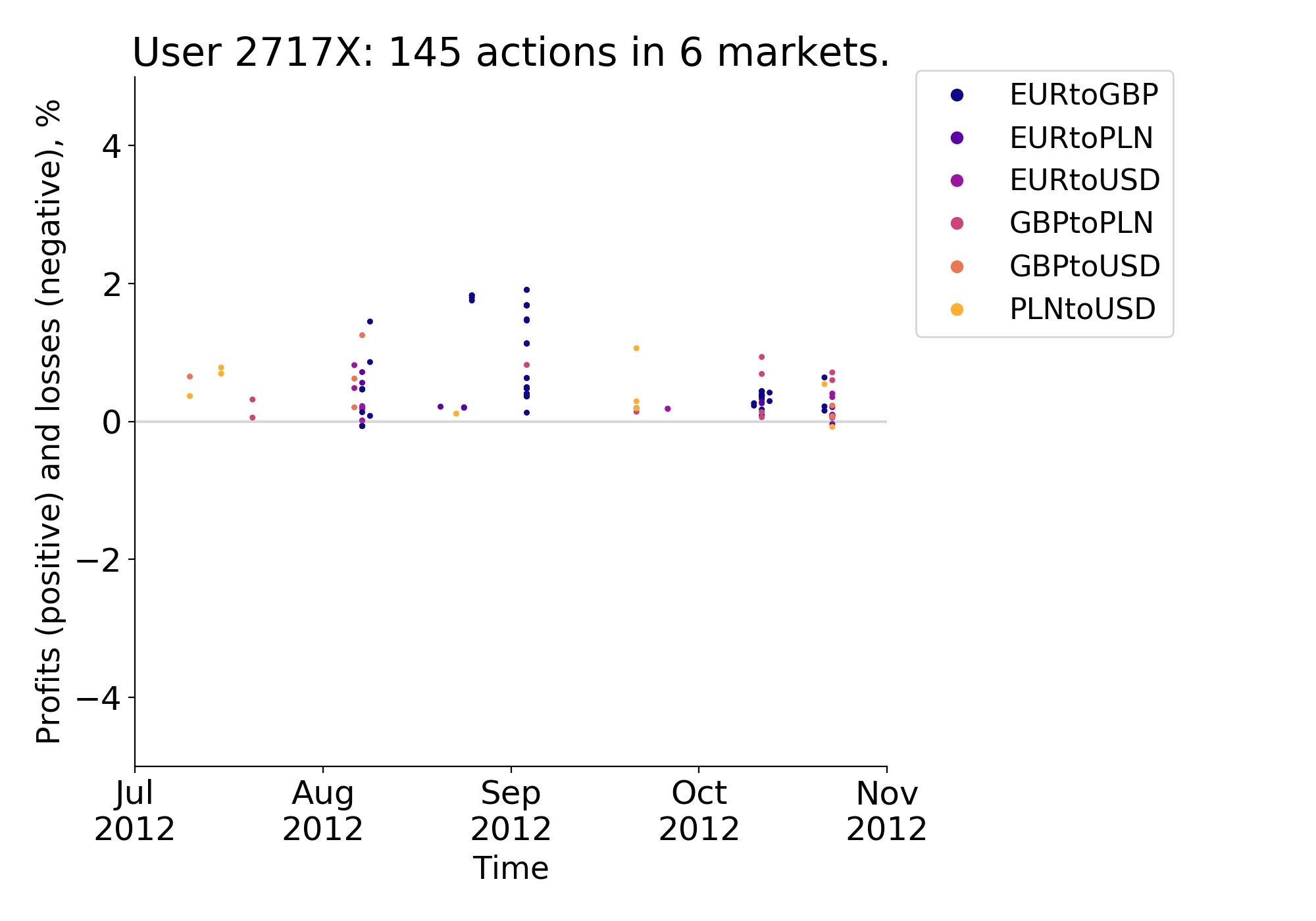}
		\caption{}
		\label{fig:2717X}
	\end{subfigure}
	\floatfoot{\emph{Notes:} the y-axis reports the profitability of the actions, depicted as dots, and the x-axis shows their evolution in time. The different colors correspond to actions conducted in different currency markets. We hide the last unit of each user identifier to preserve the anonymity.}
\end{figure}

\begin{figure}[htbp]
	\centering
	\includegraphics[width=\textwidth]{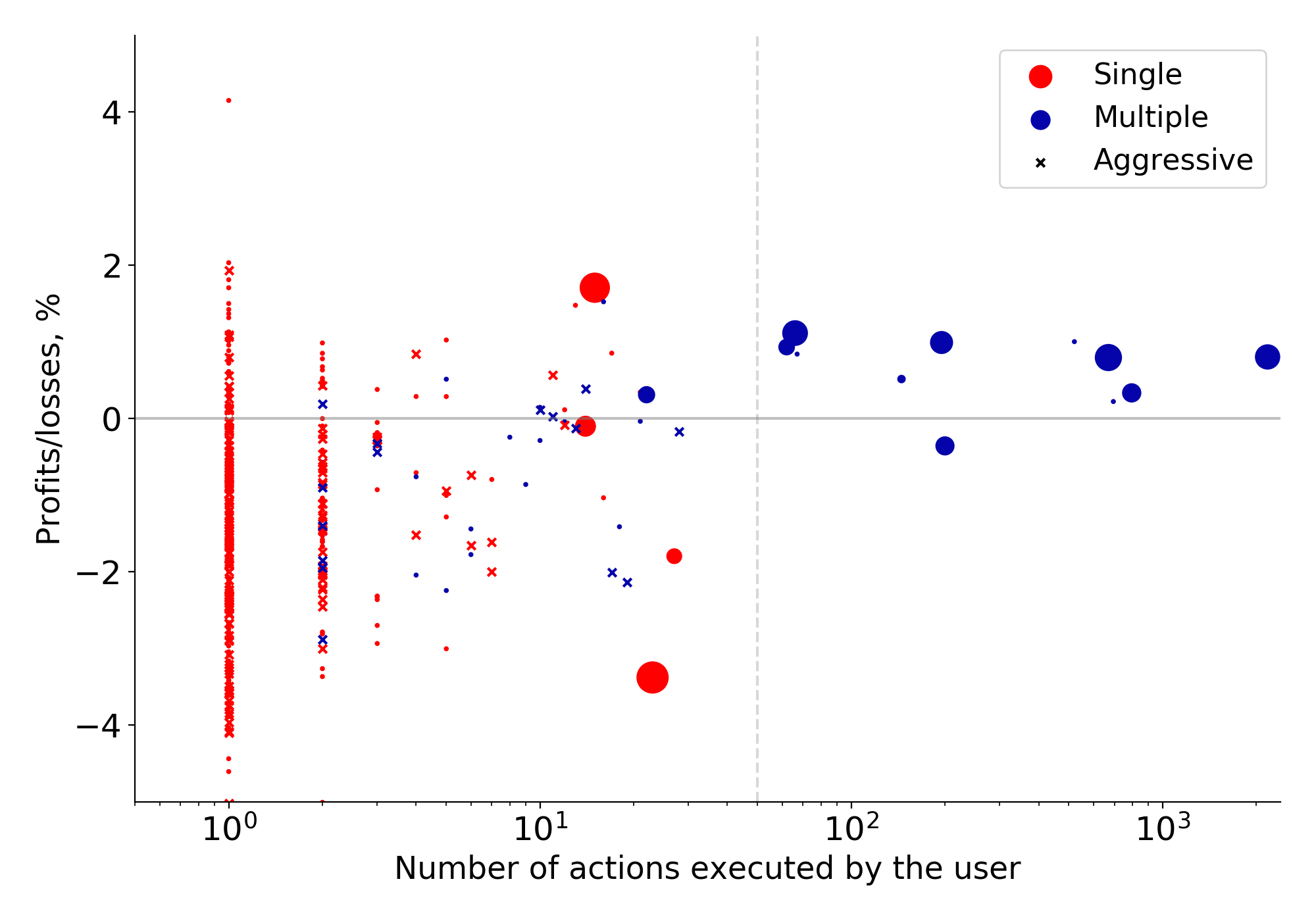}%
	\caption{Users actions: average profits as a function of the number of actions executed}
	\label{fig:main_indicators} 
	\floatfoot{\emph{Notes}: 
	each dot indicates on the x-axis the number of actions executed by a given user, and on the y-axis the average profits; the blue color is for the users that executed arbitrage actions on multiple markets, while those who exploited a single market are in red; dots whose size is increased correspond to users that executed metaorders, and the size is proportional to the percentage of activity executed through metaorders over the total number of arbitrage actions. The markers indicate whether the user executed or not aggressive actions (respectively cross and dot symbols). Most of the users who executed few actions implemented strategies on average non profitable. Striking differences arise when focusing on users that perform a high number of actions: all are active on multiple markets, most of them execute metaorders and all of them (except one) on average realize profits.
	}
\end{figure}

\begin{figure}[tp]
	\centering
	\begin{subfigure}{0.475\textwidth}
		\includegraphics[width=\textwidth]{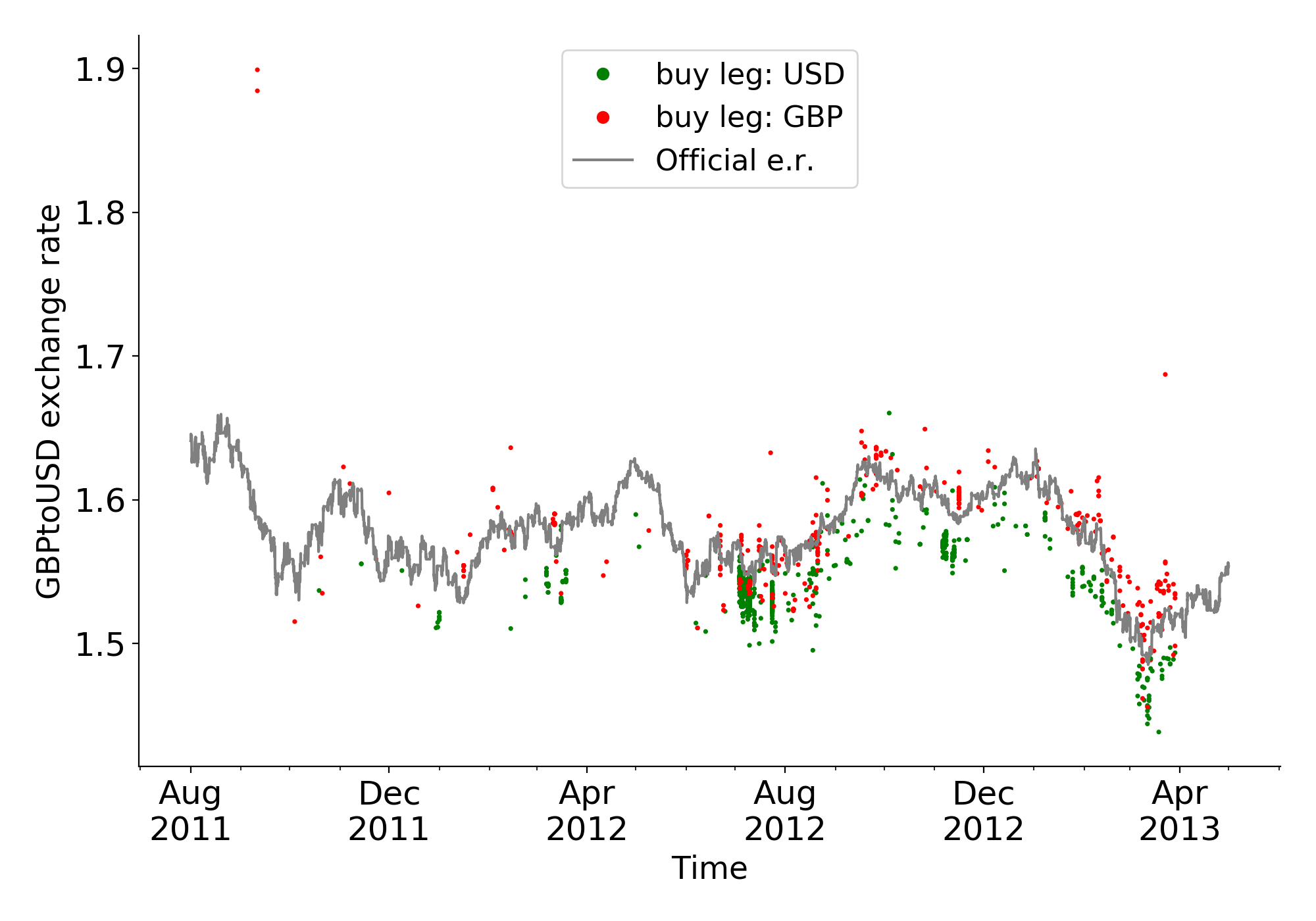}%
		\caption{{\small All arbitrage actions, GBPtoUSD market}}
		\label{fig:abs_GBPtoUSD}
	\end{subfigure}
	\begin{subfigure}{0.475\textwidth}
		\includegraphics[width=\textwidth]{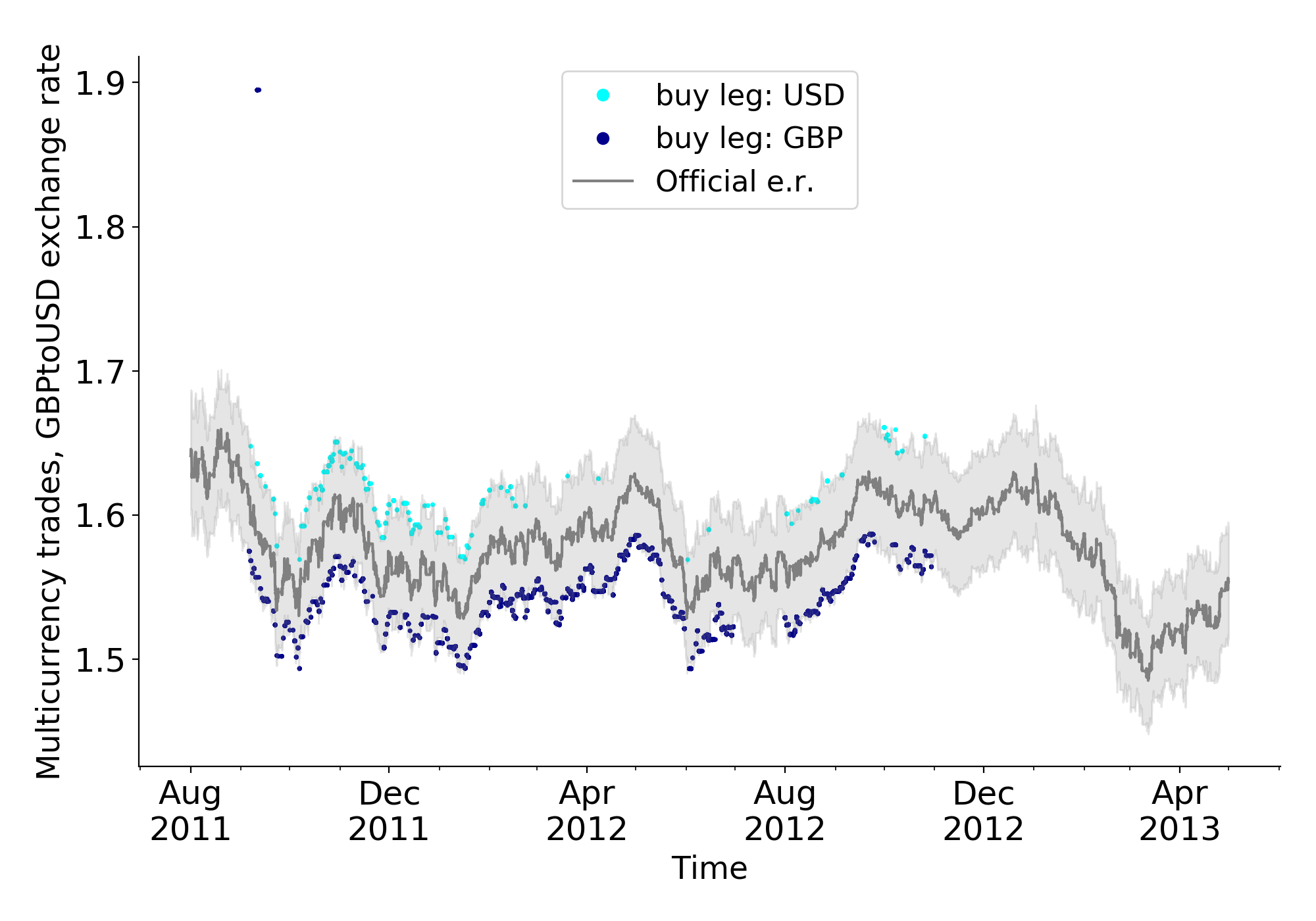}
		\caption{{\small Multi-currency trades, GBPtoUSD market}}
		\label{fig:tibanne_GBPtoUSD}
	\end{subfigure}
	\begin{subfigure}{0.475\textwidth}
		\includegraphics[width=\textwidth]{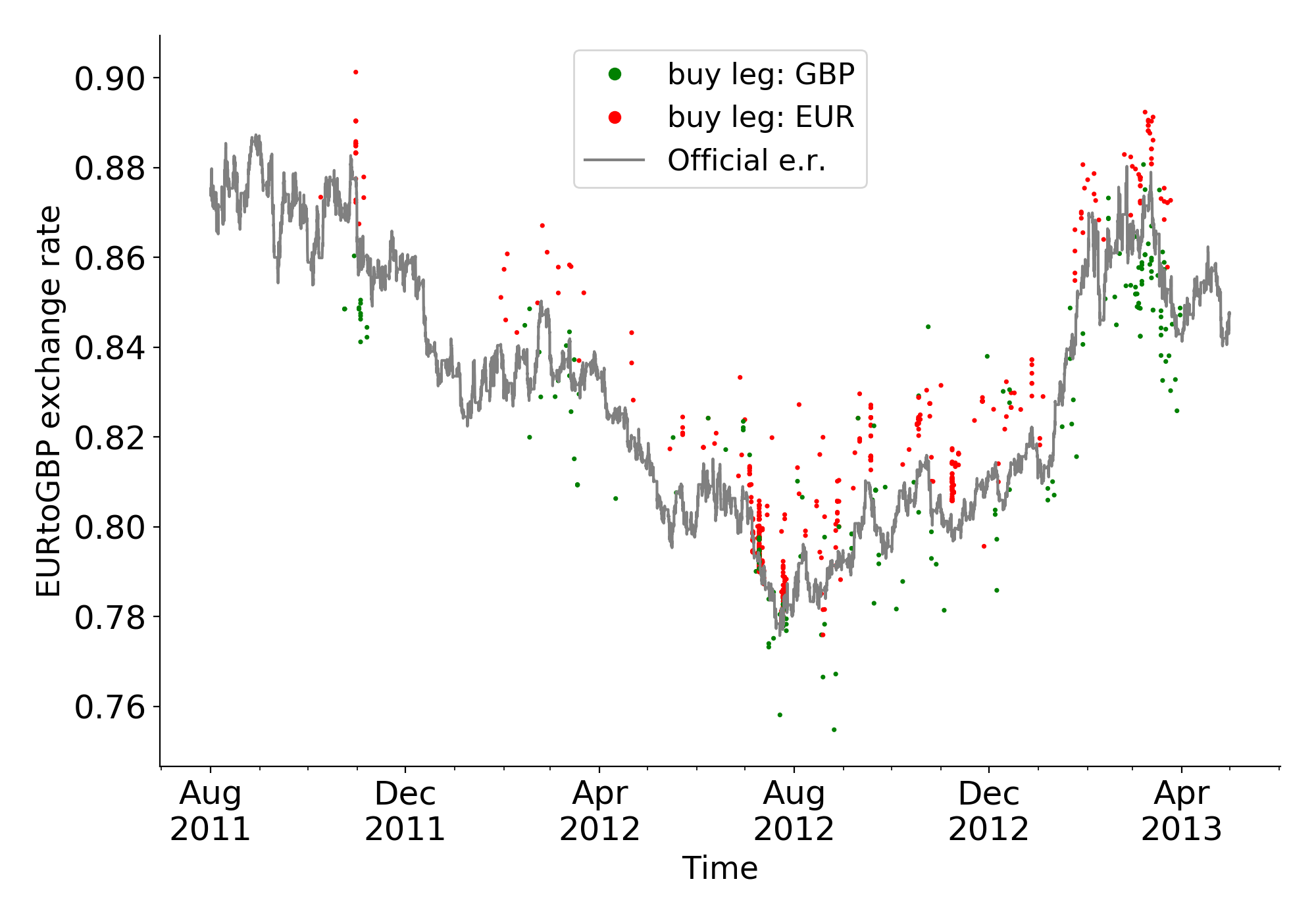}%
		\caption{{\small All arbitrage actions, EURtoGBP market}}
		\label{fig:abs_EURtoGBP}
	\end{subfigure}
	\begin{subfigure}{0.475\textwidth}
		\includegraphics[width=\textwidth]{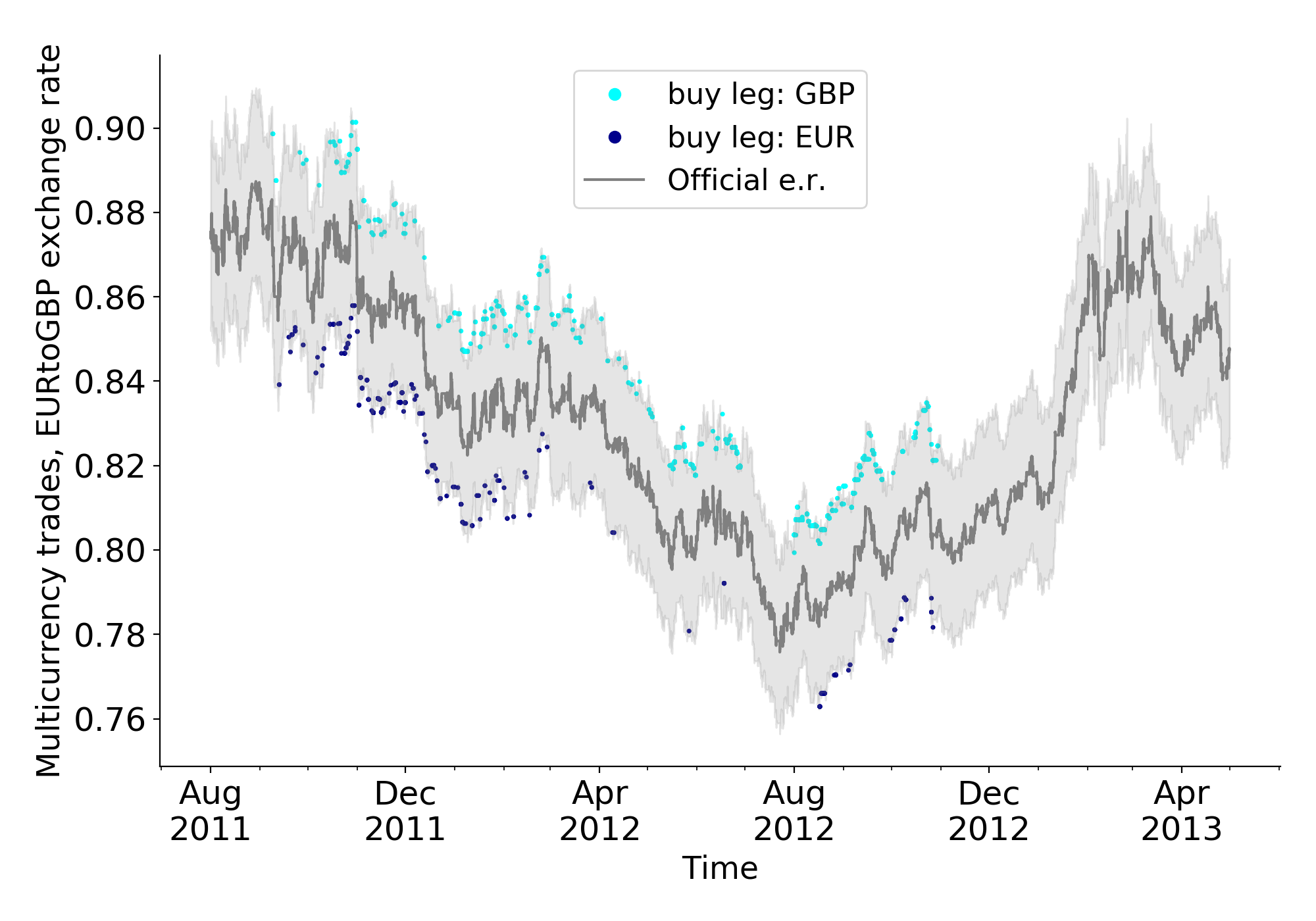}
		\caption{{\small Multi-currency trades, EURtoGBP market}}
		\label{fig:tibanne_EURtoGBP}
	\end{subfigure}
	\begin{subfigure}{0.475\textwidth}
		\includegraphics[width=\textwidth]{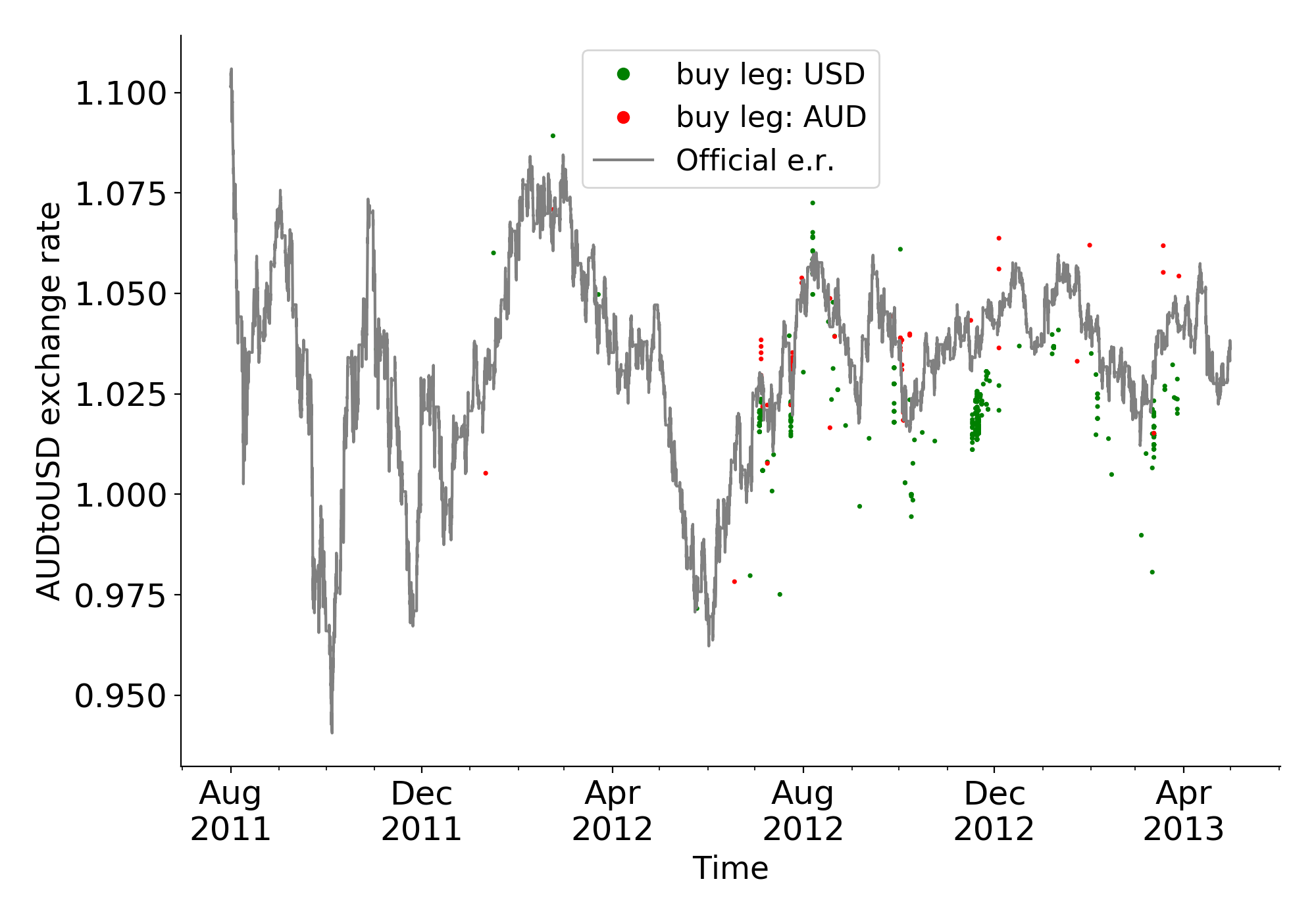}%
		\caption{{\small All arbitrage actions, AUDtoUSD market}}
		\label{fig:abs_AUDtoUSD}
	\end{subfigure}
	\begin{subfigure}{0.475\textwidth}
		\includegraphics[width=\textwidth]{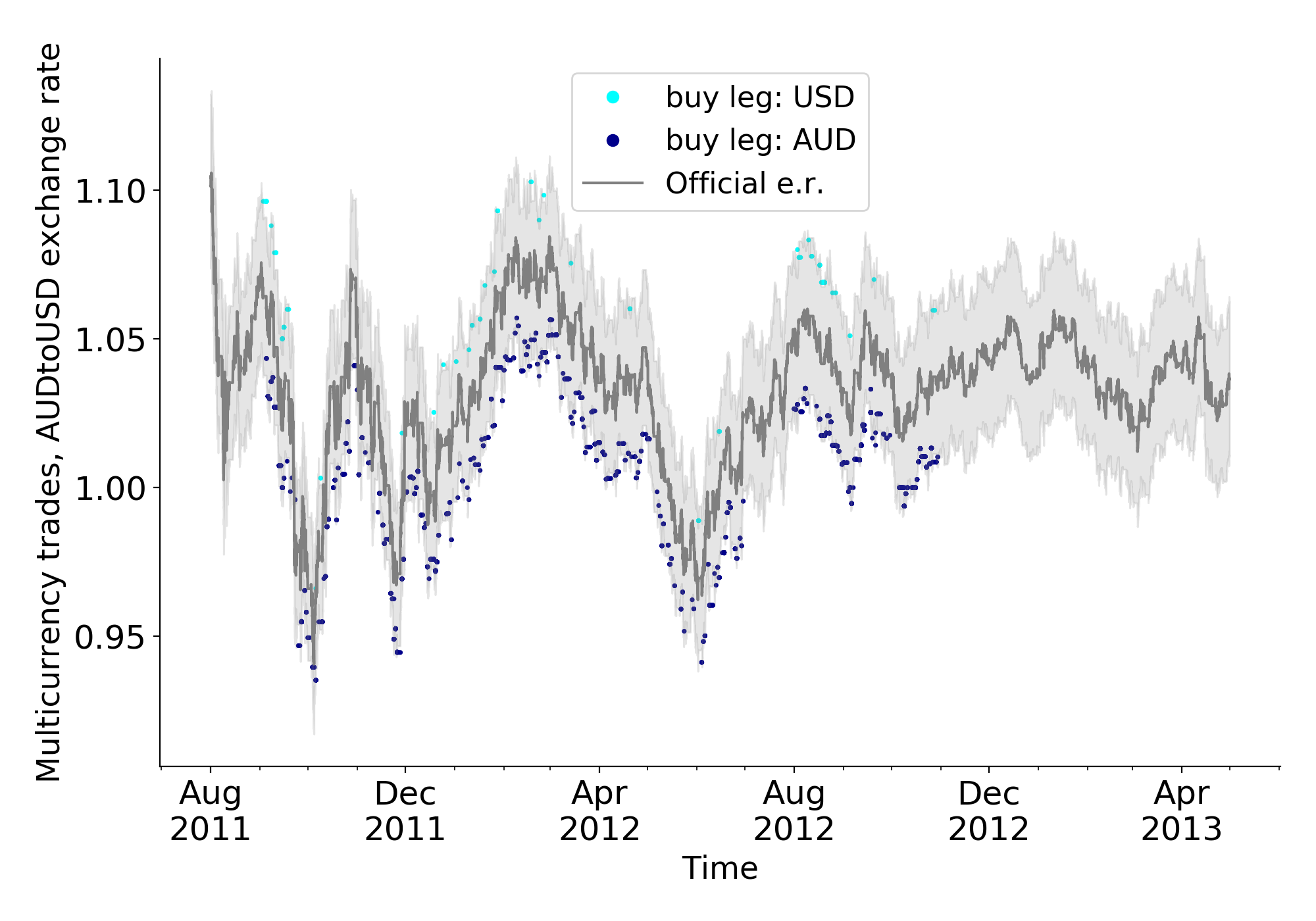}
		\caption{{\small Multi-currency trades, AUDtoUSD market}}
		\label{fig:tibanne_AUDtoUSD}
	\end{subfigure}
	\caption{Comparison between the implied and the official exchange rate for the detected arbitrage actions (left), and the multi-currency trades (right)}
	\label{fig:absolute_supplem}
	\floatfoot{\emph{Notes:} the Panels refer respectively to the GBPtoUSD, the EURtoGBP, and the AUDtoUSD markets. Left Panels report the arbitrage actions, right Panels the multi-currency trades. Values are reported in absolute terms, and the implied rates are computed without transaction costs; the gray line represents the official exchange rate.}
\end{figure}

\begin{figure}[tp]
	\centering
	\caption{Difference between the implied and the official exchange rate as a percentage of the exchange rate, actions executed only by users active in multiple markets}
	\label{fig:relative_GBPtoUSD}
	\begin{subfigure}{\textwidth}
		\includegraphics[width=0.5\textwidth]{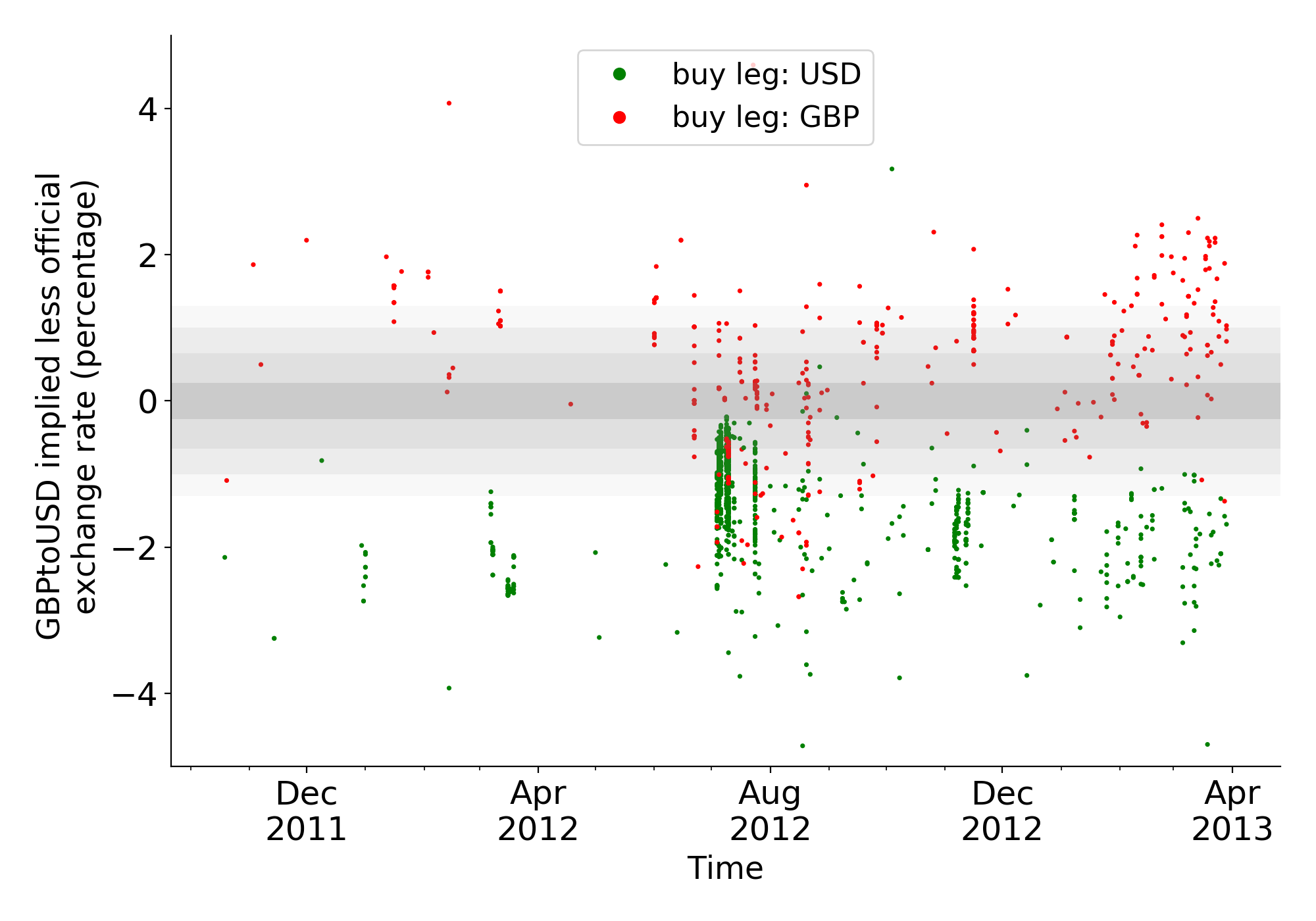}
		\includegraphics[width=0.5\textwidth]{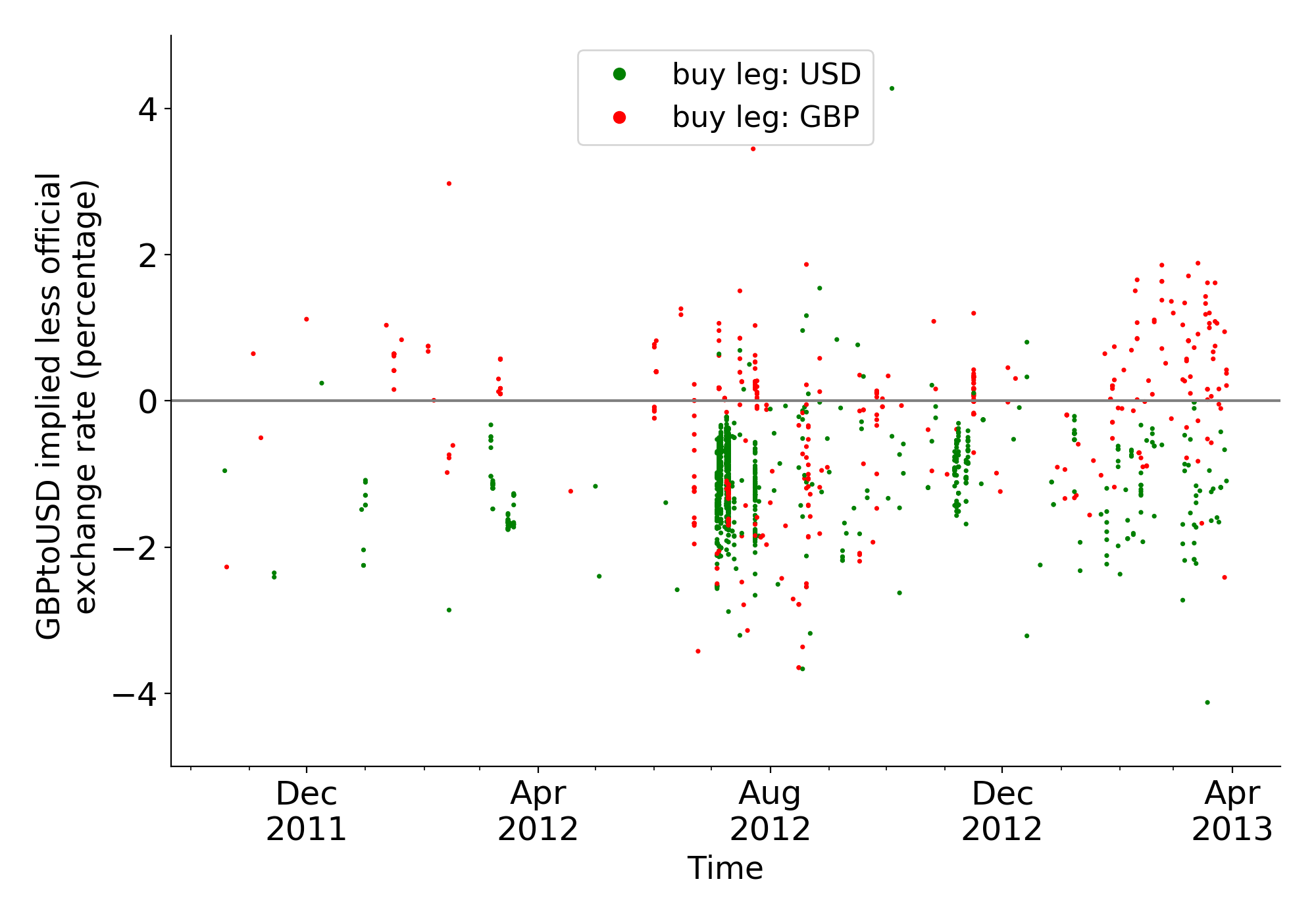}
		\caption{{\small GBPtoUSD. Without fees (left) \& including fees (right)}}
		\label{fig:manycurr_GBPtoUSD}
	\end{subfigure}
	\centering
	\begin{subfigure}{\textwidth}
		\includegraphics[width=0.5\textwidth]{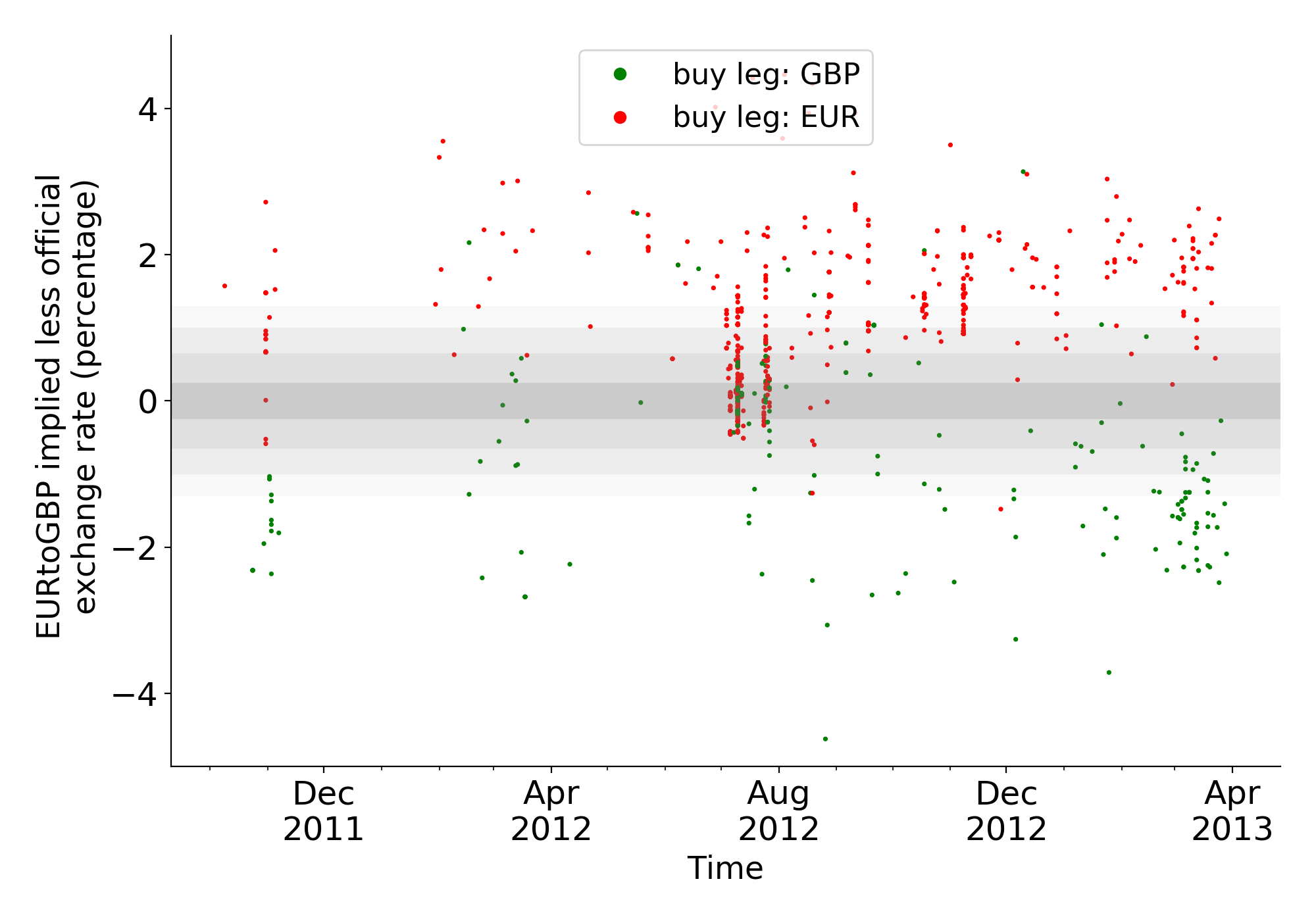}
		\includegraphics[width=0.5\textwidth]{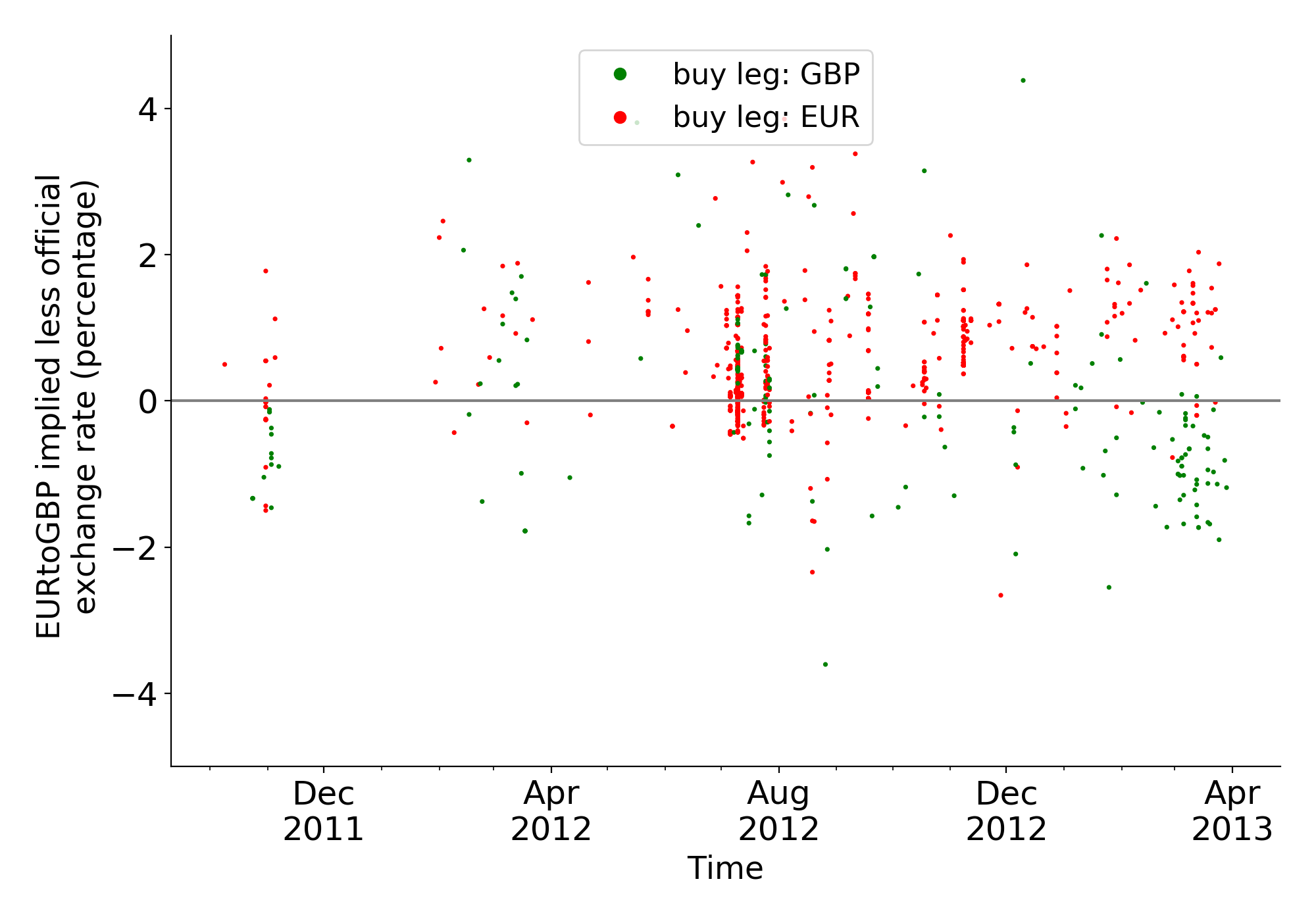}
		\caption{{\small EURtoGBP. Without fees (left) \& including fees (right)}}
		\label{fig:manycurr_EURtoGBP}
	\end{subfigure}
	\centering
	\begin{subfigure}{\textwidth}
		\includegraphics[width=0.5\textwidth]{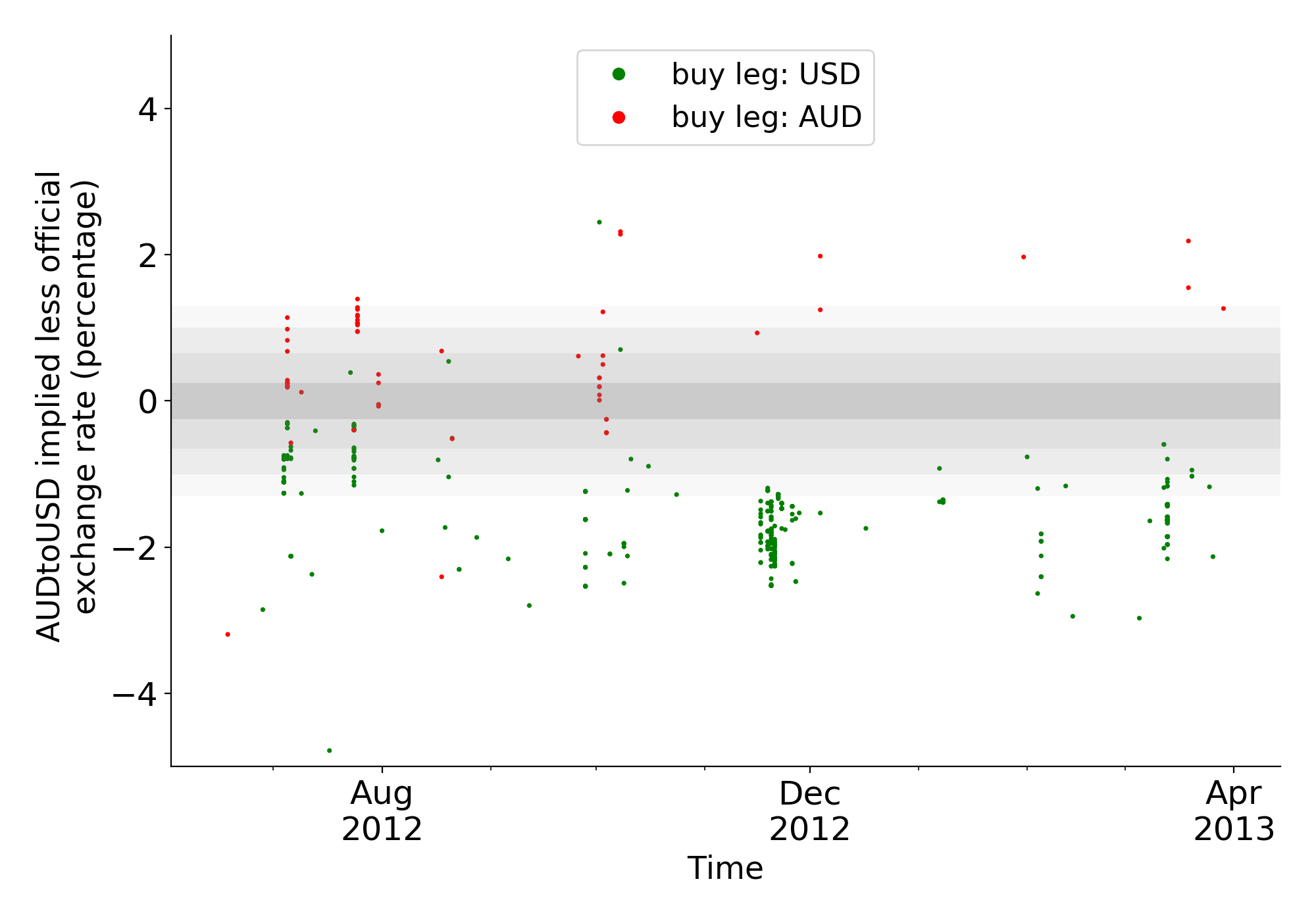}
		\includegraphics[width=0.5\textwidth]{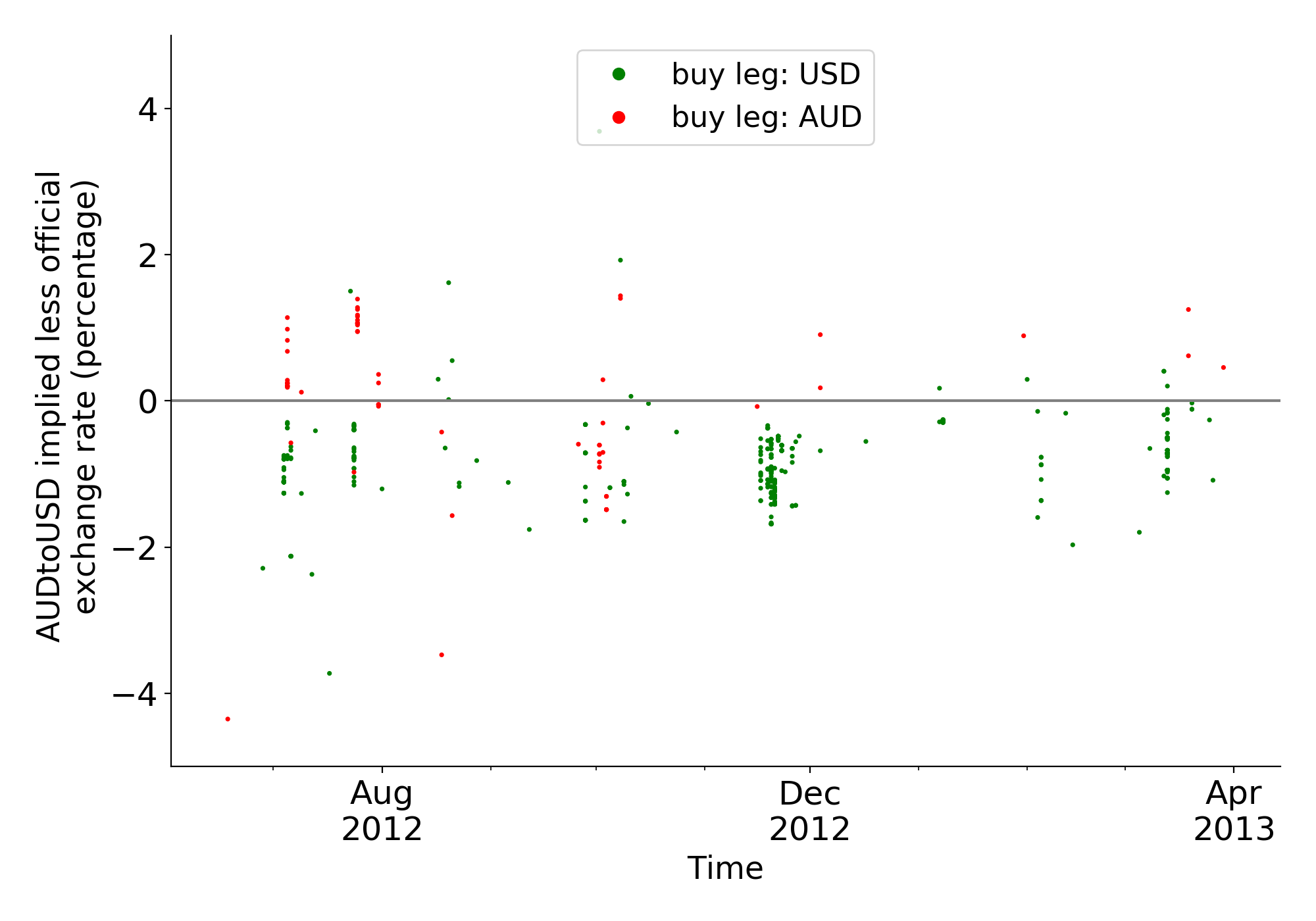}
		\caption{{\small AUDtoUSD. Without fees (left) \& including fees (right)}}
		\label{fig:manycurr_AUDtoUSD}
	\end{subfigure}
	\floatfoot{\emph{Notes:} Panels~(\subref{fig:manycurr_GBPtoUSD}) refer to the GBPtoUSD market, Panels~(\subref{fig:manycurr_EURtoGBP}) to the EURtoGBP, and Panels~(\subref{fig:manycurr_AUDtoUSD}) to the AUDtoUSD market. We report only the actions of users active in many markets as the other cases are negligible. On the right, fees are included; on the left, fees are excluded. The gray shaded bands represent the area in which transaction costs might exceed the potential profits.}
\end{figure}